\documentclass[3p]{elsarticle}

\usepackage{etoolbox}
\makeatletter
\patchcmd{\ps@pprintTitle}{\footnotesize\itshape
       Preprint submitted to \ifx\@journal\@empty Elsevier
       \else\@journal\fi\hfill\today}{\relax}{}{}
\makeatother 

\usepackage{lineno,hyperref}
\usepackage[usenames,dvipsnames,svgnames,table]{xcolor} 
\usepackage{bm} 
\usepackage{todonotes} 
\usepackage[cal=dutchcal]{mathalfa} 
\usepackage{amssymb} 
\usepackage{amsmath}   
\usepackage{chngcntr} 

\usepackage{mathtools}
\usepackage[latin1]{inputenc}
\usepackage{slashed}
\usepackage{graphicx}
\usepackage{tensor}
\usepackage{epstopdf}
\usepackage{setspace}

\usepackage[normalem]{ulem}

\usepackage[titletoc]{appendix}

 \newcommand{\vt}[1]{\mathbf{#1}} 
 \newcommand{\be}{\begin{equation}}
 \newcommand{\ee}{\end{equation}}
 \newcommand{\bea}{\begin{eqnarray}}
 \newcommand{\eea}{\end{eqnarray}}
 \newcommand{\dst}{\displaystyle}
 \newcommand{\fr}[2]{\frac{{\dst #1}}{{\dst #2}}}
 \newcommand*\chem[1]{\ensuremath{\mathrm{#1}}}
 \renewcommand{\Re}{\mathrm{Re }}
 \renewcommand{\Im}{\mathrm{Im }}

\newcommand\epigraph[3]{
\vspace{1em}\hfill{}\begin{minipage}{#1}{\begin{spacing}{0.9}
\small\noindent\textit{#2}\end{spacing}
\vspace{1em}
\hfill{}{#3}}\vspace{2em}
\end{minipage}}

 \def\lsim{\mathrel{\rlap{\lower4pt\hbox{\hskip1pt$\sim$}}
        \raise1pt\hbox{$<$}}}         
 \def\gsim{\mathrel{\rlap{\lower4pt\hbox{\hskip1pt$\sim$}}
        \raise1pt\hbox{$>$}}}         
        
\counterwithin{equation}{section}
        
\journal{Physics Reports}
 
\begin{document}

\begin{frontmatter}

\title{Theory and applications of free-electron vortex states}

\author[Riken,ANU]{K.~Y.~Bliokh}
\author[CFTP]{I.~P.~Ivanov}
\author[EMAT]{G.~Guzzinati}
\author[EMAT,Monash]{L.~Clark}
\author[EMAT]{R.~Van~Boxem}
\author[EMAT]{A.~B\'ech\'e}
\author[EMAT]{R.~Juchtmans}
\author[Rochester]{M.~A.~Alonso}
\author[Wien]{P.~Schattschneider}
\author[Riken,Michigan]{F.~Nori}
\author[EMAT]{J.~Verbeeck}

\address[Riken]{CEMS, RIKEN, 2-1 Hirosawa, Wako-shi, Saitama 351-0198, Japan}
\address[ANU]{Nonlinear Physics Centre, PSPE, The Australian National University, Canberra ACT 0200, Australia}
\address[CFTP]{CFTP, Instituto Superior T\'ecnico, Universidade de Lisboa, Lisbon, Portugal}
\address[EMAT]{EMAT, University of Antwerp, Groenenborgerlaan 171, 2020, Antwerp, Belgium}
\address[Monash]{School of Physics and Astronomy, Monash University, VIC, 3800, Australia}
\address[Rochester]{The Institute of Optics, University of Rochester, Rochester NY 14627, USA}
\address[Wien]{TU Wien, University Service Centre for Electron Microscopy, Wiedner Hauptstrasse 8-10, A-1040 Vienna, Austria}
\address[Michigan]{Physics Department, University of Michigan, Ann Arbor, MI 48109-1040, USA}

\begin{abstract}

Both classical and quantum waves can form {\it vortices}: with helical phase fronts and azimuthal current densities. These features determine the {\it intrinsic orbital angular momentum} carried by localized vortex states. In the past 25 years, optical vortex beams have become an inherent part of modern optics, with many remarkable achievements and applications. In the past decade, it has been realized and demonstrated that such vortex beams or wavepackets can also appear in {\it free electron waves}, in particular, in electron microscopy. Interest in free-electron vortex states quickly spread over different areas of physics: from basic aspects of quantum mechanics, via applications for fine probing of matter (including individual atoms), to high-energy particle collision and radiation processes. Here we provide a comprehensive review of theoretical and experimental studies in this emerging field of research. We describe the main properties of electron vortex states, experimental achievements and possible applications within transmission electron microscopy, as well as the possible role of vortex electrons in relativistic and high-energy processes. We aim to provide a balanced description including a pedagogical introduction, solid theoretical basis, and a wide range of practical details. Special attention is paid to translate theoretical insights into suggestions for future experiments, in electron microscopy and beyond, in any situation where free electrons occur.

\end{abstract}

\end{frontmatter}


\tableofcontents


\clearpage

\epigraph{4.5in}{ACIS. Even a vortex is a vortex in something. You can't have a whirlpool without water; and you can't have a vortex without gas, or molecules or atoms or ions or electrons or something, not nothing. \\
THE HE-ANCIENT. No: the vortex is not the water nor the gas nor the atoms: it is a power over these things.}{George Bernard Shaw ``Back to Methuselah''}



\section{Introduction}
\vspace{1mm}

\subsection{Wave-particle duality}
\vspace{2mm}

Electrons are elementary quantum particles which exhibit wave-particle duality inherent to all quantum objects \cite{Tonomura_book}. While their particle properties are known from classical electrodynamics (where electrons are considered as point charged particles experiencing Lorentz force and Coulomb interaction), the wave features of electrons are described in quantum mechanics by the Schr{\"o}dinger equation \cite{LandauLifshitz3,Cohen-Tannoudji}. 
Depending on the problem, either particle or wave picture could be more suitable. For example, confined electron states in atomic orbitals are clearly wave entities.

\begin{figure}[h]
\centering
\includegraphics[width=\linewidth]{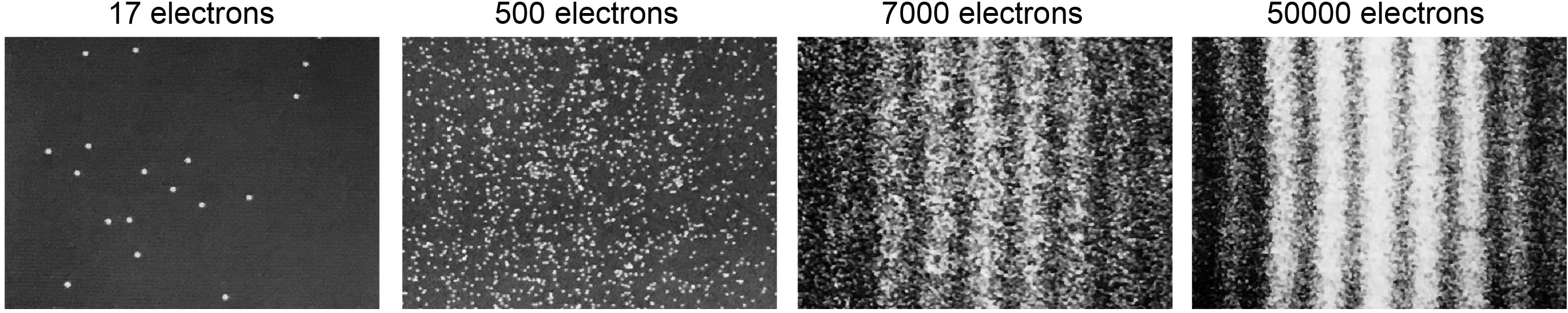}
\caption{Wave-particle duality of electrons. Single electrons, arriving one by one at the detector, build up the interference pattern in a two-slit interference experiment in a transmission electron microscope \cite{Tonomura_book}.} 
\label{fig:two-slit}
\end{figure}

The wave-particle duality of free electrons manifests itself naturally in electron microscopy \cite{Tonomura_book, DeGraefCTEM,tonomura_applications_1987}. Indeed, individual electrons are usually well separated from each other, and as a single electron hits the detector, it appears as a single bright point on the detector. At the same time, the signal accumulated from many electrons clearly exhibits interference patterns characteristic of waves: such as, e.g., two-slit interference \cite{Tonomura_book,Jonsson1961,Bach2013}, Fig.~\ref{fig:two-slit}. 
Therefore, the description of electron evolution in microscopes sometimes relies on classical equations of motion with the Lorentz force, and sometimes requires the use of the Schr{\"o}dinger wave equation.

In many cases it is sufficient to assume that the electron's wave nature reveals itself in the plane-wave-like phase acquired upon the electron propagation. To consider localized electrons, one usually implies semiclassical Gaussian-like wavepackets with spatial dimensions much larger than the de Broglie wavelength. The centroids of such wavepackets follow classical trajectories (according to the Ehrenfest theorem \cite{Cohen-Tannoudji,Tannor}), while their phase fronts can be locally approximated by a plane wave with the wave vector corresponding to the mean (expectation) value of the electron momentum.  

\subsection{Structured waves and vortices}
\vspace{2mm}

Plane waves are very basic wave entities, while generic wave fields can exhibit features drastically different from planar phase fronts propagating in the normal direction. Wave fields which are essentially different from plane waves (or smooth Gaussian wavepackets) are often called {\it structured} waves. 

Structured waves naturally appear in problems with external potentials, where plane waves are not solutions of the wave (Schr{\"o}dinger) equation. Examples include: atomic orbitals, modes of quantum dots or resonators, Landau states in a magnetic field, surface waves, to name a few. 
However, even {\it free-space} waves are generically structured. Of course, any free-space wave field is a superposition of multiple plane waves seen in the momentum (Fourier) representation. But interference of these plane-wave components can lead to rather non-trivial properties of the resulting wave field. This is because most of the important physical characteristics -- intensity, current, momentum, etc. -- are described by {\it quadratic} forms of the wave function, so that the superposition principle is applicable to wave fields, but not to their physical properties.

\begin{figure}[h]
\centering
\includegraphics[width=0.95\linewidth]{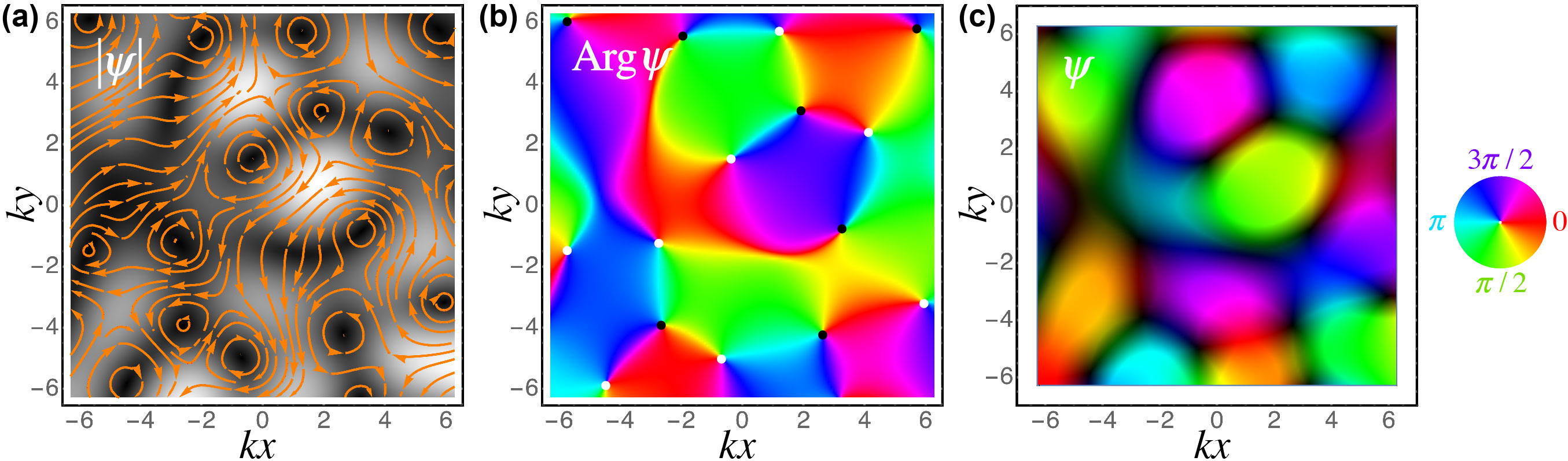}
\caption{Vortices (phase singularities) in a random 2D interference field \cite{Dennis2009}. The complex wave function $\psi({\bf r})$, ${\bf r} = (x,y)$ is obtained as a superposition of 100 plane waves with randomly-oriented wave vectors ${\bf k} = (k_x,k_y)$, fixed $k$, and random phases \cite{Dennis2009}. (a) The grayscale plot displays the absolute value of the wave function, $|\psi({\bf r})|$. Streamlines of the probability current density, ${\bf j} \propto \nabla {\rm Arg}\,\psi$, are shown in orange. (b) Color-coded phase of the wave function, ${\rm Arg}\,\psi({\bf r})$. (c) Combined representation of the complex $\psi({\bf r})$, where the brigtness is proportional to the amplitude, while the color indicates the phase \cite{Thaller_visual}. Black and white dots in (b) mark vortices with $\ell = 1$ and $\ell = -1$, respectively. The probability current in (a) forms whirlpools around these points. The typical distance between vortices in such interference patterns is of the order of the wavelength $2\pi/k$.} 
\label{fig:random}
\end{figure}

The interference of two plane waves can already be considered as a structured wave field. However, the most interesting and generic forms appear starting from {\it three}-wave intereference \cite{Masajada2001}. Namely, wave fields consisting of three or more interfering plane waves generically contain {\it phase singularities}, i.e., dislocations of phase fronts or {\it vortices} \cite{Nye1974a,Soskin2001,Dennis2009}. Such singularities appear in the points of destructive interference, ${\bf r} = {\bf r}_s$, where the amplitude of the wave function vanishes, $|\psi ({\bf r}_s)| = 0$, while its phase ${\rm Arg}\,\psi({\bf r}_s)$ is indeterminate. A vanishing amplitude of a complex field means two real conditions (vanishing of its real and imaginary parts), so that phase singularities generically appear as {\it points} in 2D plane or {\it lines} in 3D space. Most importantly, the phase of the wave function is well-defined around singular points/lines, and generically it has a nonzero increment for a countour enclosing the singularity: $\oint \nabla {\rm Arg}\,\psi({\bf r}) \cdot d{\bf r} = 2\pi\ell$. Here $\ell = 0, \pm 1, \pm 2, ...$ is an integer winding number (to provide continuity of the phase modulo $2\pi$), which is called the ``topological charge'' of the vortex. 
The typical behaviour of the wave function near the phase singularity is $\psi \propto |{\bf r} - {\bf r}_s|^{|\ell |} \exp (i \ell \varphi_s)$, where $\varphi_s$ is the azimuthal angle around the ${\bf r} = {\bf r}_s$ point.
Such wave forms are called vortices because the probability current density ${\bf j} \propto {\rm Im}\left(\psi^* \nabla \psi \right) = |\psi|^2 \nabla {\rm Arg}\,\psi$ swirls around phase singularities. For example, Figure~\ref{fig:random} shows multiple vortices in a 2D interference field obtained as a superposition of randomly-directed plane waves.
In the 3D case, vortex lines are {\it dislocation} lines for phase fronts (i.e., surfaces of constant phase) \cite{Nye1974a,Soskin2001,Dennis2009} (see Fig.~\ref{fig:vortex_beams} below). 

Since vortices are generic wave forms, they appeared in many early studies of various types of waves. In optical fields, an example of a vortex was described in 1950 for the total-internal-reflection of a light beam \cite{Wolter1950}, and a famous textbook \cite{Born1999} reproduces detailed figures from 1952 \cite{Braunbek1952} with multiple optical vortices in a plane wave diffracted by a half-plane.
For quantum matter waves, wave functions with vortices were known from the early days of quantum mechanics. Indeed, spherical harmonics, atomic orbitals with orbital angular momentum \cite{LandauLifshitz3,Cohen-Tannoudji}, and eigenmodes of the Schr{\"o}dinger equation in a magnetic field \cite{Fock1928} all contain the $\exp(i\ell\varphi)$ vortex factors. Furthermore, the seminal Dirac paper about magnetic monopoles \cite{dirac1931quantised} analyses the phase singularities in a wave function, and vortex eigenmodes appear in the related Aharonov--Bohm problem \cite{Aharonov_significance_1959}. 

Despite these multiple predecessors, the first systematic study of phase singularities was performed in 1974 by Nye and Berry \cite{Nye1974a} in the context of ultrasonic pulses. Almost simultaneously, Hirschfelder {\it et al.} \cite{Hirschfelder_I,Hirschfelder_II} analysed vortices in quantum wave functions. The seminal work by Nye and Berry gave birth to the field of {\it singular optics}, with thousands of studies in the past decades \cite{Berry1981,Soskin2001,Dennis2009}. Vortices were shown to be very important in the analysis of structured wave fields. They form a ``singular skeleton'', on which the phase and intensity structure hangs \cite{Freund1993,Dennis2009}. In particular, random wave fields, which are ubiquitous in nature, are pierced by numerous vortices \cite{Freund1993,Berry2001} (see Fig.~\ref{fig:random}) and even vortex knots (in the 3D case) \cite{Dennis_Knots,Taylor2016}. In this manner, vortices provide unique information about wave fields, both statistical and as ``fingerprints'' of individual realizations.

\subsection{Angular momentum and vortex beams}
\vspace{2mm}

The swirling current around phase singularities suggests that vortices should possess {\it angular-momentum} properties. Indeed, assuming cylindrical or spherical coordinates with the azimuthal angle $\varphi$, vortex wavefucntions $\psi \propto \exp(i\ell\varphi)$ are eigenmodes of the $z$-component of the quantum-mechanical {\it orbital angular momentum (OAM)} operator, $\hat{L}_z = -i\hbar \partial/\partial\varphi$, with the eigenvalues $\hbar\ell$ \cite{LandauLifshitz3,Cohen-Tannoudji}. In random wavefields, Fig.~\ref{fig:random}, the numbers of positive and negative vortices are approximately equal to each other, and the net OAM approximately vanishes. Moreover, only vortices with topological charges $\ell = \pm 1$ are generic in random wavefields: higher-order degeneracies split into several charge-1 degeneracies under small perturbations. Therefore, to have a field with noticable AM properties, one should produce an isolated vortex state, possibly with large $|\ell|$.

\begin{figure}[t]
\centering
\includegraphics[width=\linewidth]{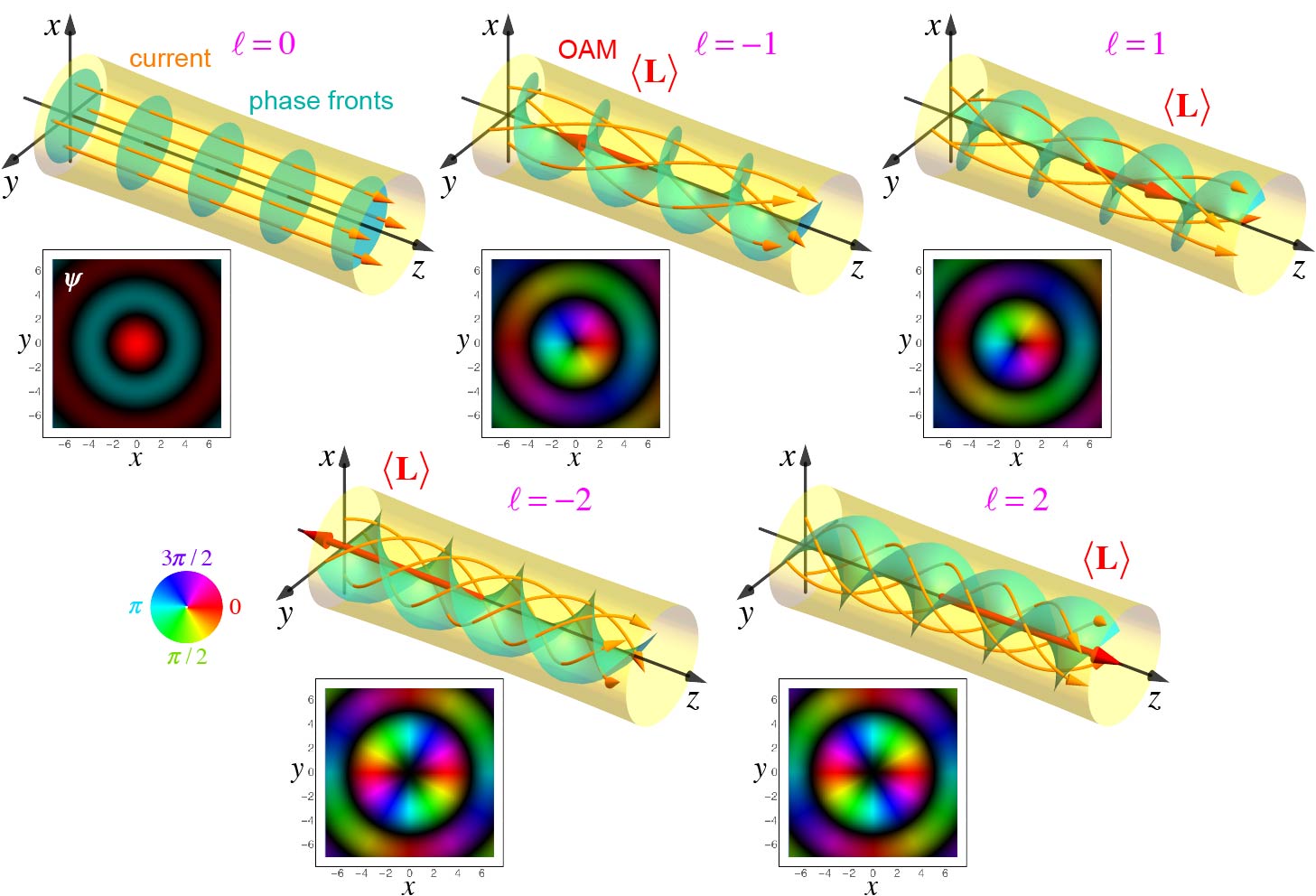}
\caption{Vortex beams are cylindrical solutions of the wave equation. They propagate along the $z$-axis and carry intrinsic longitudinal orbital angular momentum (OAM) $\langle {L_z} \rangle = \hbar\, \ell$ (assuming paraxial approximation) per particle. The 3D schematics show the phase fronts (cyan) and probability-current streamlines (orange) for beams with different vortex charges $\ell$. The 2D plots show the corresponding transverse wave function distributions $\psi(x,y)$ at $z=0$; the phase-amplitude representation is similar to Fig.~\ref{fig:random}(c). The radial profiles correspond to the Bessel modes analyzed in detail in Sections~\ref{sec:beams} and \ref{sec:OAM} below.} 
\label{fig:vortex_beams}
\end{figure}

Although the OAM eigenmodes with vortices have been known for many years in textbooks on quantum mechanics \cite{LandauLifshitz3,Cohen-Tannoudji}, only in 1992 Allen {\it et al.} \cite{Allen1992a} realized that such wave modes can be generated as free-space optical beams. Indeed, the {\it free-space} solutions of the wave equation in {\it cylindrical} coordinates $(r,\varphi,z)$, which propagate along the $z$-axis have a typical form $\psi({\bf r}) \propto f(r) \exp(ik_z z + i\ell\varphi)$, where $f(r)$ is the radial distribution (which can also slowly change with $z$ for diffracting beams), $k_z$ is the longitudinal wave number, and $\ell$ is the azimuthal quantum number. At $\ell=0$, such solutions describe usual Gaussian-like wave beams, while higher-order modes with $\ell \neq 0$ are the so-called {\it vortex beams}, shown in Fig.~\ref{fig:vortex_beams}. Such beams have isolated vortices of topological charge $\ell$ on their axes, helical phase fronts, and spiralling currents. Most importantly, being eigenmodes of $\hat{L}_z$, vortex beams carry a well-defined OAM $\hbar\ell$ per particle (photon in the case of optics) along their axes: $\langle L_z \rangle = \hbar\, \ell$.

The presence of a vortex and well-defined OAM dramatically modifies both geometrical and {\it dynamical} properties of the wave. Therefore, the description and generation of optical vortex beams in the beginning of the 1990s \cite{bazhenov1990laser,Allen1992a,heckenberg_laser_1992,beijersbergen_helical-wavefront_1994} caused enormous interest and initiated the rapidly-developing field of {\it optical angular momentum}. Since then, optical vortex beams have been intensively studied and have found numerous applications in diverse areas, including: optical manipulations of small particles or atoms \cite{Gahagan1996,Garces2003,grier2003revolution}, quantum information and communications \cite{Leach2002,Gibson2004,Wang2012,Bozinovic_2013,Krenn_2016}, quantum entanglement \cite{Mair_2001,Leach2010}, radio communications \cite{Thid__2007,Tamburini_2012}, astronomy and astrophysics \cite{harwit_photon_2003,Foo2005,Lee2006PRL,Berkhout2008,Tamburini2011}, optical solitons \cite{Kruglov1985,Swartzlander1992,Desyatnikov2005}, and Hall effects \cite{Bliokh2006,Bliokh_2009,Merano2010,Bliokh_2012}. In the past two decades, five books \cite{Allen_book,Bekshaev_book,Andrews_book2008,Torres_book,andrews2012angular} and many reviews \cite{Allen_review,Molina-Terriza2007,Franke-Arnold2008,Yao_OAM,Bekshaev2011,Bliokh2015PR,Bliokh2015NP} about optical vortex beams and OAM were published.

Several important physical points have to be made about the OAM of vortex wave states: 
\begin{itemize}
\item The $z$-directed OAM carried by vortex beams is {\it intrinsic} \cite{Berry1998,Bliokh2015PR}, i.e., independent of the choice of coordinates. This is in sharp contrast to the {\it extrinsic} mechanical OAM of classical point particles, ${\bf L} = {\bf r} \times {\bf p}$ (where {\bf r} and {\bf p} are the particle coordinates and momentum, respectively), which depends on the choice of the coordinate origin. 
\item Moreover, the mean (expectation) value of the OAM in vortex beams is {\it aligned} with the mean momentum: $\langle {\bf L} \rangle = \ell\, \langle {\bf p} \rangle / \langle p \rangle$. This is also in contrast to point-particle OAM, which is {\it orthogonal} to the momentum at every instant of time: ${\bf L} \perp {\bf p}$.
\item The intrinsic OAM and spiraling current density do not contradict the rectilinear propagation of either plane waves or classical particles in free space. Indeed, the centroid of a vortex state follows a {\it rectilinear} trajectory in free-space (e.g., lies on the axes of vortex beams). Also, vortex beams are superpositions of plane waves [see Figs.~\ref{fig:Bessel}(a) and \ref{fig:LG}(a) below], but the probability-current streamlines, i.e., {\it Bohmian trajectories} of the particles \cite{Bohm1987,Tannor,Bliokh2013}, can be {\it curvilinear} in free space \cite{Berry2008,Berry_OpticalCurrents}.
\item Vortex states carrying intrinsic OAM is not a collective effect, but a phenomenon that persists on the {\it single-particle} level \cite{Leach2002}. In other words, these are forms of the single-particle wave function. 
\end{itemize}

\subsection{From optics to electron waves}
\vspace{2mm}

Until recently, the majority of studies on phase singularities and free-space vortex beams dealt with optical fields and other {\it classical} waves. 
At the same time, the universal character of wave equations suggests that fundamental results of singular optics and optical angular momentum should be equally applicable to {\it quantum}, in particular electron, waves \cite{Dragoman_book}. Moreover, the concept of the OAM in vortex beams essentially relies on the quantum-mechanical operator $\hat{L}_z$. Nonetheless, until recently there were only a few studies of free-space quantum wave functions with vortices \cite{Hirschfelder_I,Hirschfelder_II,Bialynicki-Birula2000,Allen2001}.

In 2007 Bliokh {\it et al.} \cite{Bliokh2007} suggested that free electrons can be in vortex-beam (or vortex-wavepacket) states carrying intrinsic OAM. They also discussed basic interactions of such {\it vortex electrons} with external fields and possible ways they could be generated. In 2010, free-electron vortex beams were indeed produced in transmission electron microscopes (TEMs) by Uchida and Tonomura \cite{Uchida2010} and Verbeeck {\it et al.} \cite{Verbeeck2010}. One year later, McMorran {\it et al.} \cite{McMorran2011} demonstrated the generation of electron vortex beams with OAM up to $|\ell | = 100$. This is in enormous contrast with the {\it spin} angular momentum (SAM) of electrons, which is restricted to 1/2 (in units of $\hbar$). 
These studies initiated a new research area investigating free-electron vortex states, or, in a wider context, {\it structured quantum waves} \cite{Harris2015}. 
Electron vortex beams are currently attracting considerable interest, with potential applications in various fields, such as electron  microscopy, fundamentals of quantum mechanics, and high-energy physics. The present paper aims to provide the first comprehensive review of this emerging area of research. 

Thus, free-space vortex beams carrying OAM, based on the quantum-mechanical picture of angular momentum, were developed in classical optics, and recently returned to their quantum roots. While similarities between optical and electron waves are obvious, it is important to mention basic {\it distinctions} between optical and electron vortices. Apart from the huge difference in wavelengths, electrons, unlike photons, are {\it charged} particles, and therefore can directly {\it interact with each other as well as with external electric and magnetic fields}. Moreover, the presence of the OAM means the presence of a vortex-induced {\it magnetic moment} carried by vortex electrons.
Furthermore, electrons can interact with electromagnetic waves (photons), as well as radiate photons via the Vavilov--Cherenkov or other effects.
Vortex electrons can also participate in particle collisions in the context of high-energy physics. All these phenomena enrich the physics of structured electron waves, as compared to their optical counterparts. 
At the same time, some features, naturally present in optical waves, are practically absent in eletron optics. First of all, free-electron sources in electron microscopy generate {\it unpolarized} particles, which are described by the {\it scalar} wave function. This is in sharp contrast to optics, where the use of spin (polarization) degrees of freedom is ubiquitous both in regular and singular/OAM optics \cite{Azzam_book,Hasman2005,Bliokh2015NP}. 
In addition, electron beams in TEM are {\it highly-paraxial}, while modern nano-optics often deals with non-paraxial (tighly focused or scattered) fields with wavelength-scale inhomogenities \cite{Novotny_book}.

\subsection{Applications in electron microscopy}
\vspace{2mm}

The wave nature of electrons is naturally exploited in transmission electron microscopy and holography \cite{Tonomura_book,DeGraefCTEM,tonomura_applications_1987,WilliamsCarter,Reimer_book}. Electron microscopes can vastly outperform optical microscopes in terms of spatial resolution because of the extremely small wavelength (of the order of picometers) obtained in accelerated electron waves. This explains the tremendous success of TEMs in exploring the atomic structure of matter. 

On the one hand, in conventional TEM imaging and holography, a nearly-plane electron wave is produced to interact with a thin sample. The local interaction of the electron wave with microscopic electromagnetic potentials of the specimen leads to deformations of planar phase fronts and produces a {\it structured} transmitted wave. This wave contains information about the atomic structure and electromagnetic properties of the sample. Naturally, transmitted waves contain a multitude of vortices, and the well-developed methods of singular optics \cite{Soskin2001,Dennis2009} could provide a new insight and a toolbox for electron microscopy \cite{Allen2001,Lubk2013,lubk_electron_2015}.

On the other hand, recent progress in the deliberate creation of electron vortex beams \cite{Uchida2010,Verbeeck2010,McMorran2011} allows one to employ {\it incident} structured electrons and make use of the new {\it OAM degrees of freedom}. 

Actually, free-space vortex electron states by themselves offer unique opportunities of studying fundamental quantum-mechanical phenomena. In particular, interactions with external magnetic fields and structures bring about a number of fundamental effects involving vortices \cite{Bliokh2012e,Greenshields2012,Gallatin2012,Karimi2012}. Recent TEM experiments for the first time demonstrated free-electron {\it Landau states} (previously hidden in condensed-matter systems) and their fine internal dynamics \cite{Schattschneider2014NC}, as well as the interaction of electron waves with approximate {\it magnetic monopoles} \cite{beche_magnetic_2014} (previously only considered theoretically).

Most importantly, incident vortex electrons interacting with the specimen in a TEM can unveil new information about the sample or increase the resolution of the microscope \cite{Tamburini2006}.
In particular, recent experiments with electron vortex beams demonstrated their role in {\it chiral energy-loss spectroscopy} and {\it magnetic dichroism} \cite{Verbeeck2010,Lloyd2012_II,Schattschneider2013comment,Yuan:2013,Schattschneider2014,pohl_electron_2015,
Schachinger2016,Verbeeck2011}.
This is in sharp contrast to optics, where probing magnetic dichroism and chirality involve only polarization (spin) degrees of freedom of light and are mostly insensitive to vortices \cite{Babiker2002,Araoka2005,Loffler2011,Afanasev:2013}.
Moreover, focusing free-electron vortices makes them comparable in size and parameters with orbitals in atoms \cite{Verbeeck2011}. This opens a way for magnetic mapping with {\it atomic resolution} \cite{Schattschneider2012,Rusz_Achieving,rusz_magnetic_2016}.

\subsection{High-energy perspective: scattering and radiation}
\vspace{2mm}

Vortex electrons can also contribute to the study of fundamental interaction phenomena besides electron microscopy.
Two directions which can be pursued with current technology are: (i) the interaction
of vortex electrons with {\it intense laser fields} 
\cite{Karlovets:2012,Hayrapetyan-et-al:2014}
and (ii) {\it radiation processes} with vortex electrons (e.g., the Vavilov--Cherenkov and transition radiations), which were predicted to depart from their usual expressions 
\cite{Kaminer-2015,ISZ-2016,IvanovKarlovets:2013a,KonkovPotylitsyn:2013}.
For instance, vortex electrons (carrying large OAM and magnetic moment) can reveal the magnetic-moment contribution to the transition radiation, which has never been observed before.

Even more exciting is the possibility to bring vortex states in {\it quantum-particle collisions}
\cite{Jentschura:2011a,Ivanov:2011a}.
In all collider-like settings realized thus far, the colliding particles (electrons, positrons, or hadrons) behave as semiclassical Gaussian-like wavepackets, and their scattering processes can be safely calculated in terms of plane waves.
Collisions of vortex particles involve a completely new degree of freedom:
the intrinsic OAM. Therefore, in addition to the kinematical distributions and polarization dependences, one can study dependences of the scattering cross-sections on the OAM of the incoming particles.
This possibility is particularly tantalizing in hadronic physics, in the context of the {\it proton spin puzzle} \cite{Aidala:2012mv}.
Experimental data show that a significant part of the proton spin comes from
the orbital angular momentum of the quarks and gluons, but its exact contribution,
as well as the whole issue of the SAM--OAM separation, remains under hot debate 
\cite{Leader:2013jra,Wakamatsu:2014zza,Liu:2015xha}. 
Vortex electrons can serve as a new OAM-sensitive probe in this problem.

Recent theoretical investigations brought several examples
of quantities which have been inaccessible so far, but which could be revealed in vortex-particle collisions \cite{Ivanov:2012b,Ivanov:2016jzt,Ivanov:2016oue,Karlovets:2016dva,Karlovets:2016jrd}.
In particular, by scattering vortex electrons on a counterpropagating particle 
and observing the interference fringes in their joint angular distribution, 
one can directly probe its OAM state \cite{Ivanov:2016jzt}.
This phenomenon can be employed to probe the proton spin structure.
On the experimental side, major challenges still need to be overcome, 
such as the acceleration of vortex electrons to higher energies
and focusing them as tightly as possible onto the protons.
The generation and acceleration of vortex protons is another future milestone to be achieved experimentally.

\subsection{About this review}
\vspace{2mm}

The present paper aims to provide the first comprehensive review of the theory and applications of free-electron vortex states carrying OAM. The paper is organized as follows. This Introduction provides a pedagogical and historical overview of wave vortices and vortex OAM in classical and quantum waves. The mostly-theoretical Section~2 describes basic physical properties of free-electron vortex states: their wave functions, currents, AM properties, magnetic moment, evolution in external electric and magnetic fields, etc. Section~3 reviews the most important TEM experiments and proposals involving electron vortices: production methods, OAM measurements, elastic and inelastic interaction with matter, and transfer of mechanical angular momentum. Several applications and future prospects are also discussed. Readers interested in the mostly-experimental TEM part can skip Section~2 and read Section~3 right after the Introduction.
Section~4 describes the main problems involving vortex electrons in high-energy physics: relativistic effects and collisions of vortex particles. Section~5 explains peculiarities of the radiation processes (Vavilov--Cherenkov and transition radiation) with vortex electrons.
Finally, a brief outlook of future perspectives concludes the review. 

For the reader's convenience, in the Appendix A (Tables~\ref{table_I} and \ref{table_II}) we summarize the main abbreviations, conventions, and notations used in this paper.
Also the following conventions are used in figures throughout the review. Intensity (i.e., probability-density) distributions are mostly shown using 2D grayscale (or monochrome) plots, in arbitrary units, with brighter areas corresponding to higher intensities [as, e.g., in Figs.~\ref{fig:two-slit} and \ref{fig:random}(a)]. The phase distributions are shown using rainbow colors, as in Fig.~\ref{fig:random}(b). Similarly, rainbow-colored wave vectors [e.g., Fig.~\ref{fig:Bessel}(a)] indicate mutual phases of plane waves in the Fourier spectrum of the field. We also use combined intensity-phase (brightness-color) representations of complex wave functions \cite{Thaller_visual}, as in Figs.~\ref{fig:random}(c) and \ref{fig:vortex_beams}. Note also that in most theoretical figures the propagation $z$-axis is horizontal for the sake of convenience, while it is vertical in schemes related to electron-microscopy experiments, corresponding to the actual TEM setup.

\clearpage

\section{Basic properties of electron vortex states}
\label{sec:basic-properties}
\vspace{1mm}

\subsection{Plane waves, wave packets, beams}
\vspace{2mm}

As other quantum particles, electrons share both wave and particle properties. We start
with the simplest non-relativistic description of a scalar electron (i.e., without spin) in free space
(i.e., without external fields), which is based on the Schr{\"o}dinger wave equation \cite{LandauLifshitz3,Cohen-Tannoudji}:
\begin{equation}
\label{eq:schro}
i  \hbar \frac{\partial \psi}{\partial t} + \frac{\hbar^2}{2 m_e} \nabla^2 \psi = 0.
\end{equation}
Here $\psi (\mathbf{r},t)$  is the wave function, and $m_e$ is the electron mass. Most of the analysis below can
be applied to any quantum particle described by the Schr{\"o}dinger equation.

The wave properties of the electron reveal themselves in the \emph{plane wave} solution of the Schr{\"o}dinger equation:
\begin{equation}
\label{eq:planewave}
\psi = a\, \mathrm{exp}\! \left[ i \hbar^{-1}(\mathbf{p}\cdot \mathbf{r}-E\,t) \right],
\end{equation}
where $a$ is the constant amplitude, $\mathbf{p}$ is the wave momentum, and $E$ is the electron energy, which satisfy $E=p^2 / 2 m_e$. Plane waves (\ref{eq:planewave}) have well-defined momentum, but they are delocalized in space. Therefore, plane-wave solutions (\ref{eq:planewave})  cannot be normalized and cannot correspond to physical particles.

To describe \emph{localized} electron states, one has to consider superpositions of multiple plane waves (\ref{eq:planewave}) with different momenta $\mathbf{p}$, which produce an uncertainty in the momentum, $\delta \mathbf{p}$, and the corresponding finite uncertainty in the electron coordinate 
\cite{LandauLifshitz3,Cohen-Tannoudji}.
To model localized electrons, usually Gaussian distributions in both momentum and coordinate spaces are implied, which satisfy the Heisenberg uncertainty principle. Let the electron move along the $z$-axis with the mean momentum
$\langle \mathbf{p} \rangle = \mathbf{p}_0 = p_0 \bar{ \mathbf{z}}$
(hereafter, the overbar stands for the unit vector in the corresponding direction), and its mean coordinate at $t = 0$ is $\langle \mathbf{r}(0) \rangle = \mathbf{r}_0 = 0$.
Assuming the azimuthal symmetry of the electron's state about the $z$-axis, we can write the Gaussian amplitude envelope of the wave function, $a (\mathbf{r} ,t )$, as
\begin{equation}
\label{eq:distribution}
a(\mathbf{r},0) \propto \mathrm{exp}\! \left(- \frac{{\bf r}_{\perp}^{2}}{w^2} - \frac{z^2}{l^2}\right).
\end{equation}
Here $\mathbf{r}_\perp = (x,y)$ are the transverse coordinates, while $w$ and $l$ are the  width and length of the distribution.
If the electron is well-localized in momentum space, i.e., has small momentum uncertainy $| \delta \mathbf{p} | \ll p_0$, then the spatial dimensions of its probability density distribution are large as compared with the de Broglie wavelength: $w,l \gg \hbar / p_0$.
Under these conditions, the phase of the electron wave function approximately follows the plane-wave form (\ref{eq:planewave}) with $\mathbf{p}=\mathbf{p}_0$ and the distribution (\ref{eq:distribution}) propagates as a \emph{wavepacket} with velocity ${\bf v} = {\bf p}_0/m_e$, i.e., one can substitute $z \to (z - v t)$ in Eq.~(\ref{eq:distribution}) at $t \neq 0$.
 
The above consideration is valid only in the leading-order approximation in $| \delta \mathbf{p}| / p_0$ and neglects the diffraction and dispersion effects. These include a slow spread of both the transverse and longitudinal dimensions of the wavepacket, i.e., variations of $w$ and $l$ during the electron motion, as well as deformations of the phase front as compared to the plane wave 
\cite{Cohen-Tannoudji,Tannor}.

Note that a small uncertainty in the transverse momentum components, $\delta \mathbf{p}_{\perp}$, represent variations in the \emph{direction} of the momentum, while the longitudinal uncertainty $\delta \mathbf{p}_{\parallel} \simeq \delta p \, \bar{\mathbf{z}}$ represents variations in the \emph{absolute value} of the momentum, which is related to the energy uncertainty:
$\delta E \simeq p\, \delta p / m_e$.
Correspondingly, the $w$ and $l$ dimensions of the wavepackets are
linked to these uncertainties in the direction of propagation and energy:
$w \sim \hbar / | \delta {\bf p}_\perp |$ and $l \sim \hbar / \delta p$.
The latter relation can be written in terms of the temporal duration $\tau$ of the wavepacket, $ \tau = l / v$, and energy uncertainty: $\tau \sim \hbar / \delta E$.

In many problems, only the {\it transverse} localization of the electron is important. Then, one can consider states delocalized in the longitudinal dimension,
$l = \infty$, and hence \emph{monoenergetic}: $\delta p = \delta E = 0$.
Such states are called {\it wave beams} 
\cite{Siegman}. 
It should be noticed, however, that physical beams of electrons (e.g., in electron microscopes) consist of many wavepackets propagating one after another and having some finite energy uncertainty $\delta E$.
\footnote{This is a very accurate description for electrons in a TEM, except possibly for the source crossover region where electron-electron interaction can become non-negligble, especially at very high (pulsed) beam currents.}
Still, dealing with monoenergetic beam solutions significantly simplifies the
analysis of the problem and allows one to describe most of the phenomena related to the transverse structure of the electron wave function. Therefore, in most cases below, we will consider monoenergetic electron beams in the \emph{paraxial} approximation (i.e., $|\delta \mathbf{p}_\perp| \ll p_0$), analyzing the effects related to the energy uncertainty separately.

\begin{figure}[t]
  \centering
   \includegraphics[width=\linewidth]{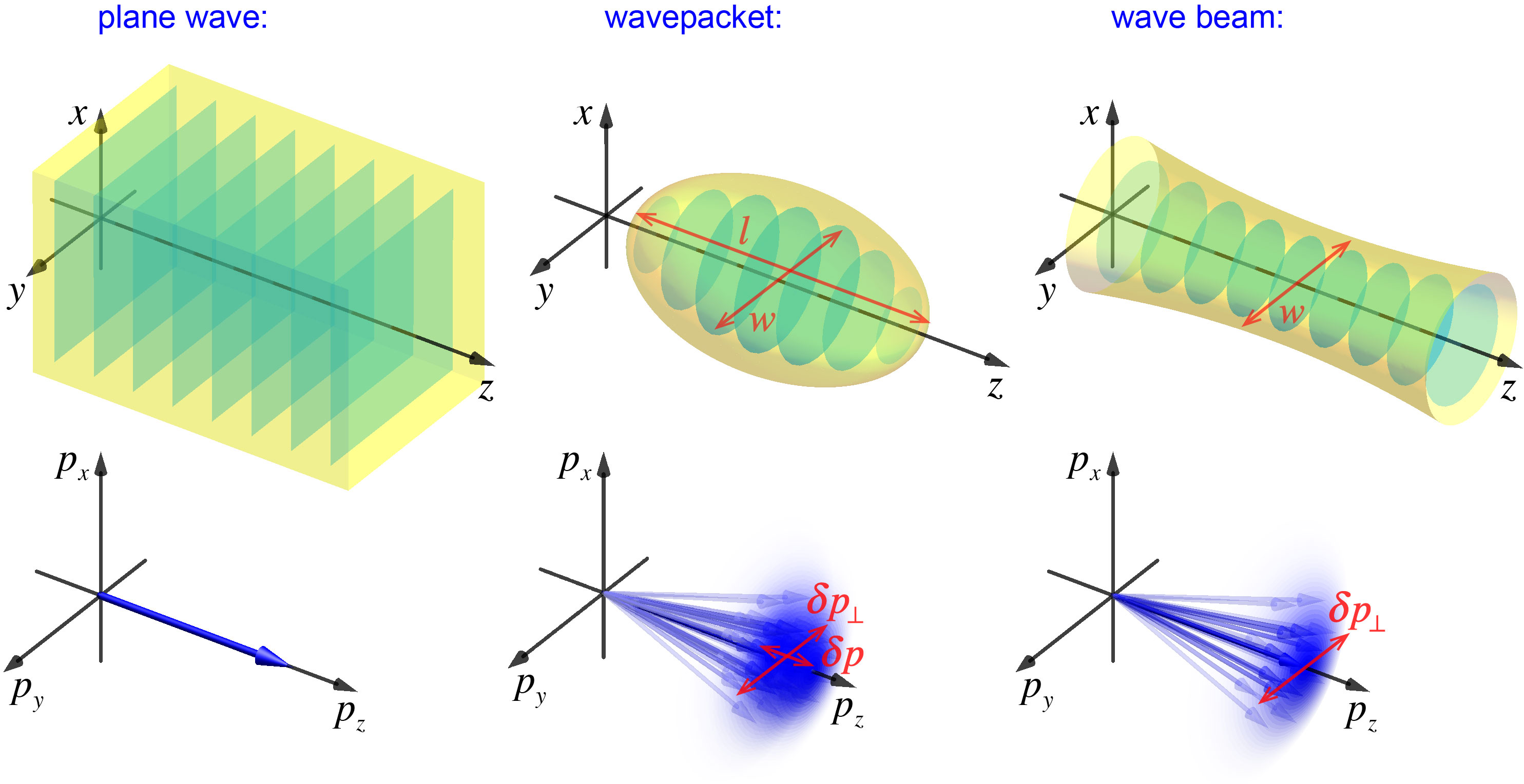}
  \caption{Schematic pictures of a plane wave, a wavepacket, and a wave beam in both real- and momentum-space representations. The real-space probability density distributions are schematically shown in yellow, while the phase fronts are shown in cyan, and azimuthal symmetry about the propagation $z$-axis is implied. 
Assuming the paraxial approximation $p_z \simeq p$, $p_{\perp} \ll p$, the characteristic dimensions of the real- and momentum-space distributions satisfy the uncertainty relations $l \sim \hbar / \delta p$ and $w \sim \hbar / \delta p_{\perp}$ (see explanations in the text).} 
\label{fig:wavebeamschematic}
\end{figure}

We also note that the localization or delocalization of electron states is directly related to the
continuous or discrete spectrum of the corresponding quantum parameters (in the case of a complete orthogonal set of modes). The plane waves (\ref{eq:planewave})
are delocalized in all three dimensions, and therefore are only described by three components of $\mathbf{p}$ with continuous spectra.
Wavepackets are localized in three dimensions and, correspondingly, can be characterized by three discrete quantum numbers. Gaussian wavepackets (\ref{eq:distribution}) correspond to the lowest-level state; higher-order states can be described, e.g., by Hermite-Gaussian modes \cite{Siegman}. In turn, wave beams (or spherical modes 
\cite{LandauLifshitz3,Cohen-Tannoudji}) 
are localized with respect to two dimensions and are described by two discrete quantum numbers related to the transverse modal structure of the beam \cite{Siegman,Andrews_book2008}.

\subsection{Vortex beams: Solutions of the Schr{\"o}dinger equation in cylindrical 
coordinates}
\label{sec:beams}
\vspace{2mm}

The solutions of the Schr{\"o}dinger equation (\ref{eq:schro}) can be decomposed via a complete set of orthogonal free-space modes. There are different sets of such modes, and the convenience of using one or another set is determined by symmetries and other conditions in each particular problem. Furthermore, solving the Schr{\"o}dinger equation in different representations and coordinates naturally leads to different modes. For example, planes waves (\ref{eq:planewave}) represent a complete set of orthogonal modes convenient in the momentum representation. These modes are delocalized and non-normalizable. 
In the coordinate representation, using Cartesian coordinates, one can obtain \emph{Hermite--Gaussian} modes with respect to the three dimensions.
However, these modes are not isotropic and are essentially attached to the directions of the Cartesian axes. Using spherical coordinates brings about \emph{spherical modes}, widely used in atomic physics 
\cite{Cohen-Tannoudji}.
These modes are suitable for localized electrons in atoms rather than for
electrons freely moving in the longitudinal $z$-direction in electron microscopes. 
Combining the $z$-direction of the electron motion with the isotropy of the free-space problem with respect to the transverse $(x,y)$-coordinates naturally results in the choice of \emph{cylindrical coordinates} $(r, \varphi, z)$ 
\cite{Bliokh2007}. 
The cylindrical solutions that we describe below allow a convenient analytical description and offer a good approximation to the electron states produced in electron microscopes.

\subsubsection{Bessel beams}

We seek monoenergetic beam eigenmodes of the Schr{\"o}dinger equation (\ref{eq:schro}), which correspond to the electron propagating along the $z$-axis. After substitution
$\psi (\mathbf{r}, t) \rightarrow \psi ( \mathbf{r}) \mathrm{exp} \left(i \hbar^{-1} E \, t \right)$, Eq.~(\ref{eq:schro}) in cylindrical coordinates takes the form:
\begin{equation}
\label{eq:cylinschro}
- \frac{\hbar^2}{2 m_e} \left[ \frac{1}{r}  \frac{\partial}{\partial r} \left( r \frac{\partial}{\partial r}\right) + \frac{1}{r^2} \frac{\partial^2}{\partial \varphi^2} + \frac{\partial^2}{\partial z^2}\right] \psi = E\, \psi.
\end{equation}
The axially-symmetric solutions of Eq.~(\ref{eq:cylinschro}) are
\cite{Bliokh2012e}:
\begin{equation}
\label{eq:Bessel}
\psi^B_\ell \propto J_{| \ell |} (\kappa r)\,  \exp\! \left[i\!\left(  \ell \varphi + k_z z \right)\right],
\end{equation}
where $J_{\ell}$ is the Bessel function of the first kind, $\ell = 0, \pm 1, \pm 2, ...$ is an integer number (azimuthal
quantum number), $k_z = p_z / \hbar$ is the longitudinal wave number, and $\kappa = p_\perp/ \hbar$ is the transverse (radial) wave number. Solutions (\ref{eq:Bessel}) satisfy Eq.~(\ref{eq:cylinschro}) provided that the following dispersion relation is fulfilled:
\begin{equation}
\label{eq:dispersion}
E= \frac{\hbar^2}{2 m_e} k^2 =  \frac{\hbar^2}{2 m_e}\left(  k^2_z  + \kappa^2 \right),
\end{equation}
where $k= p/ \hbar$.

\begin{figure}
\centering
\includegraphics[width=0.9\linewidth]{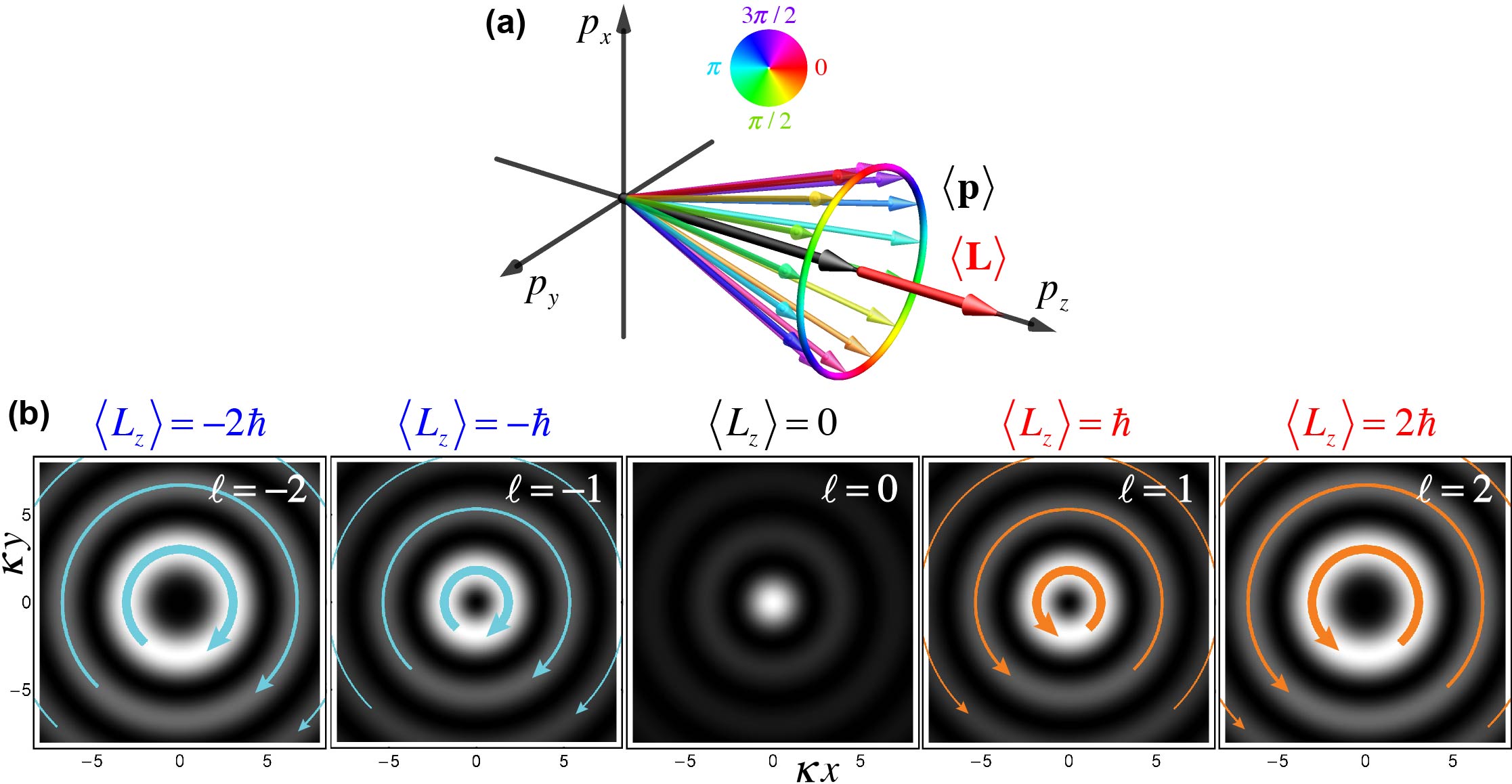}
\caption{Bessel-beam solutions (\ref{eq:Bessel}) and (\ref{eq:dispersion}) of the Schr{\"o}dinger equation. (a) The momentum spectrum consists of conically-distributed wave vectors with fixed $p_{\perp} = \hbar \kappa$ and $p_z = \hbar k_z$, i.e., forms a circle. The mean electron momentum is $\langle {\bf p} \rangle = \hbar\, k_z \bar{\bf z}$. The mutual phases (colour-coded) of the plane waves in the spectrum increase by $2\pi\ell$ around the circle. This forms a {\it vortex} of topological charge $\ell$ ($\ell=2$ here) and determines the intrinsic {\it orbital angular momentum (OAM)} of the electron: $\langle {\bf L} \rangle = \hbar\, \ell\, \bar{\bf z}$, Eq.~(\ref{eq:OAM}). (b) Transverse probability
density $\rho$ (grayscale plots) and current ${\bf j}$ (circular arrows) distributions in the Bessel beams with different values of $\ell$, Eq.~(\ref{eq:prob}). 
Here, the radii and thicknesses of the current circles correspond to the positions and values of the maxima (normalized in each panel independently) of the quantity $r j_{\varphi}$ that determines the contribution to the OAM, Eq.~(\ref{eq:expect2}).
}
\label{fig:Bessel}
\end{figure}

The cylindrically-symmetric modes (\ref{eq:Bessel}) and (\ref{eq:dispersion}) are called \emph{Bessel beams}  \cite{Durnin_NondiffractingBeams,Durnin_PRL,McGloin2005,Bliokh2012e,Grillo2014a}.
They have a cylindrical probability density distribution, independent of $z$, i.e., without diffraction. Most importantly, the azimuthal quantum number $\ell$ (also called the topological charge or vortex charge) determines the \emph{vortex} phase structure in Bessel beams and their \emph{orbital angular momentum (OAM)} properties \cite{Allen_book,Allen_review,Bliokh2007}, which are discussed in Section~\ref{sec:OAM} below.
The zero-order ($\ell=0$) beam has no vortex and maximal probability density on the axis, i.e., at $r = 0$.
The higher-order ($\ell \neq 0$) modes are characterized by the quantum vortex $\exp\!\left( i \ell \varphi \right)$, spiral phase structure, azimuthal probability current, and the probability density vanishing on the axis: $\left. \psi^{B}_{\ell} \right|_{r=0} = 0$.
Figure~\ref{fig:Bessel} shows the transverse probability density and current distributions in Bessel beams (\ref{eq:Bessel}) and (\ref{eq:dispersion}).

The Bessel beams represent the simplest theoretical example of vortex beams. Despite the probability density of Bessel modes decaying as $|\psi^B_\ell | \sim 1/r$ when $r \to \infty$, these solutions are not properly localized in the transverse dimensions. Indeed, the integral $\int^\infty_0  \left| \psi^B_\ell \right|^2 r\, dr$ diverges, and the function cannot be normalized with respect to the transverse dimensions.
\footnote{This means infinite number of particles or energy per unit $z$-length in the Bessel beams. Therefore, the exact solutions (\ref{eq:Bessel}) cannot be generated in practice, but a good approximation to this solution can be produced in experiments for finite radial apertures $r < r_{\rm max}$ and propagation distances $|z| < z_{\rm max}$ \cite{Durnin_PRL,McGloin2005,Grillo2014a}.}
The delocalized nature of Bessel beams is reflected in the absence of diffraction and a single transverse quantum number $\ell$ (instead of two transverse quantum indices in the properly-localized modes).

In terms of the plane-wave spectrum, the Bessel beam (\ref{eq:Bessel}) and (\ref{eq:dispersion}) represents a superposition of plane waves with \emph{conically}-distributed momenta: ${\bf p}_\parallel = {\bf p}_0= p_z \bar{\bf z}$ and $\left| {\bf p}_\perp \right| \equiv p_\perp = \hbar \kappa$, which can be characterized by the polar angle $\theta_0$, $\sin\theta_0 = \kappa/k$, Fig.~\ref{fig:Bessel}(a). This corresponds to the Fourier spectrum: 
\begin{equation}
\tilde{\psi}_{\ell}^{B}({\bf k}_\perp) \propto \delta(k_\perp - \kappa)\exp(i\ell\phi), \quad
\psi_{\ell}^{B}({\bf r}) = \int \tilde{\psi}_{\ell}^{B}({\bf k}_\perp)\, e^{i {\bf k}\cdot {\bf r}} d^2{\bf k}_\perp,
\label{eq:Fourier-Bessel}
\end{equation}
where ${\bf k} = {\bf p}/\hbar$ is a wave vector with transverse components ${\bf k}_\perp$ and azimuthal angle $\phi$ in ${\bf k}$-space, and $\delta$ is the Dirac delta-function.
The delocalization of the Bessel modes and absence of diffraction is a direct consequence of the fact that the wave vectors are distributed only azimuthally, while the radial transverse component $k_\perp$ is fixed.

\subsubsection{Laguerre--Gaussian beams}

To construct vortex beams properly localized (square-integrable) in the transverse
dimensions, one can use at least two alternative ways. First, considering superpositions of
multiple Bessel beams with the same fixed energy $E$ but different wave numbers $k_z$ and $\kappa$  (i.e., introducing some uncertainty $\delta \kappa$ 
in the radial momentum component), results in a general
integral form of such modes 
\cite{Schattschneider_2011}. 
However, to deal with analytical solutions, here we follow the second, simplified way. Namely, we make use of the paraxial approximation:
${p}_\perp \ll p$ and ${\bf p}_{\parallel}\simeq p \, \bar{\mathbf{z}}$ ($k_z \simeq  k$).
In the first-order approximation in $p_\perp /p = k_\perp /k \ll 1$,
the Schr{\"o}dinger equation (\ref{eq:schro}) or (\ref{eq:cylinschro}) can be simplified using the substitution $\partial^2 / \partial z^2 \simeq -k^2 +2 i k\, \partial / \partial z$.
In doing so, it takes the form of the so-called \emph{paraxial wave equation}, widely used in optics \cite{Siegman}:
\begin{equation}
\label{eq:paraxial}
2 i k \frac{\partial}{\partial z} + \left[ \frac{1}{r} \frac{\partial}{\partial r} \left(r \frac{\partial}{\partial r} \right) + \frac{1}{r^2} \frac{\partial^2}{\partial \varphi^2} \right] \psi=0.
\end{equation}
Interestingly, this equation has the form of a Schr{\"o}dinger-like equation with the time-like
coordinate $z$  and two space-like transverse coordinates $(r,\varphi)$. 

The solutions of equation (\ref{eq:paraxial}) in cylindrical coordinates are the \emph{Laguerre--Gaussian (LG) beams} 
\cite{Siegman,Allen1992a,Allen_review,Allen_book,Bliokh2012e}:
\begin{equation}
\label{eq:LG}
\psi^{LG}_{\ell, n} \propto \left(\frac{r}{w(z)} \right)^{| \ell |} L^{| \ell |}_n\! \left(  \frac{2r^2}{w^2(z)}\right) \mathrm{exp}\! \left(- \frac{r^2}{w^2 (z)}  + ik \frac{r^2}{2 R(z)} \right) e^{i(\ell \varphi +k z)} e^{-i\left(  2 n + | \ell | +1\right) \zeta (z)},
\end{equation}
where $L^{| \ell |}_n $ are the generalized Laguerre polynomials, $n=0,1,2,...$ is the radial quantum number, $w(z)=w_0 \sqrt{1+z^2 / z^2_R}$ is the beam width, which slowly varies with $z$ due to diffraction, $R(z)= z \left( 1+ z_R^2 /z^2  \right)$ is the radius of curvature of the wavefronts, and $\zeta (z) = \mathrm{arctan}(z / z_R)$.
Here, the characteristic transverse and longitudinal scales of the beam are the waist $w_0$ (the width in the focal plane $z = 0$) and the Rayleigh diffraction length $z_R$
\cite{Siegman}:
\begin{equation}
w_0 \gg 2 \pi / k, \qquad z_R = k w^2_0 /2 \gg w_0.
\end{equation}
The last exponential factor in Eq.~(\ref{eq:LG}) describes the \emph{Gouy phase} \cite{Gouy1890,Siegman,Boyd1980,Feng2001}; 
it yields an additional phase difference
\begin{equation}
\label{eq:Gouy}
\Phi_G = \left(2 n + | \ell | +1 \right) \pi
\end{equation}
upon the beam propagation through its focal point from $z/ z_R \ll -1$ to $z/ z_R \gg 1$.
The Gouy phase is closely related to the \emph{transverse confinement} of the modes 
\cite{Feng2001,Phillips1983}.
The dispersion relation for the LG beams is simply $E= \hbar^2 k^2 / 2m_e$ [cf. Eq~(\ref{eq:dispersion})], while the small transverse wave-vector components are taken into account in the $z$-dependent diffraction terms.

\begin{figure}[!t]
  \centering
   \includegraphics[width=0.9\linewidth]{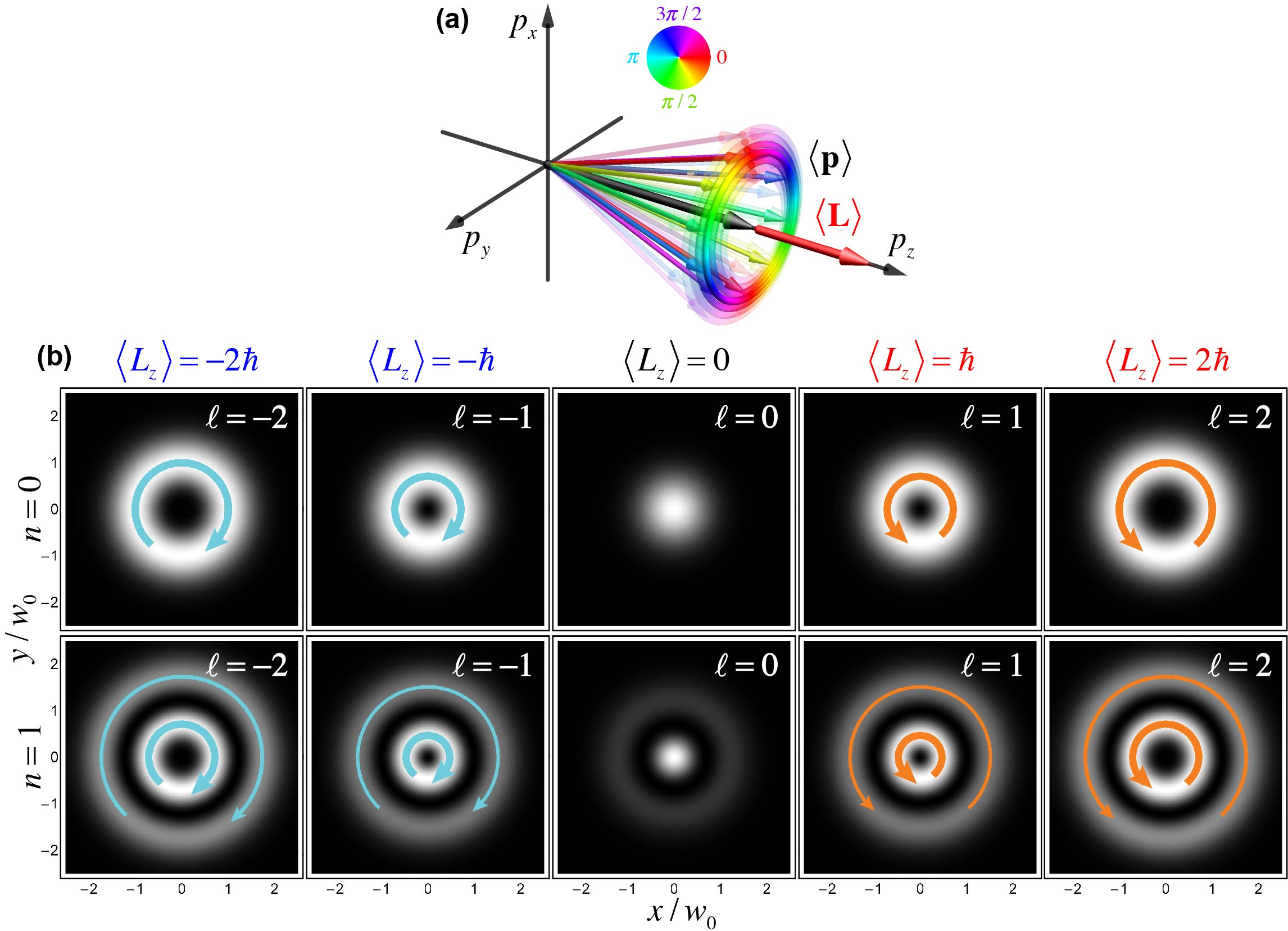}
  \caption{Same as in Fig.~\ref{fig:Bessel} but for the Laguerre--Gaussian (LG) beams (\ref{eq:LG})--(\ref{eq:Gouy}) and (\ref{eq:LGprobability}). 
The Bessel and LG beams have similar azimuthal and OAM properties. However, the LG beams are also characterized by the momentum spectrum with {\it radial} distribution of the wave vectors (a). This provides the radial confinement of the electron, which is characterized by the additional radial quantum number $n$. In this manner, $(n+1)$ is the number of radial maxima (rings) in the intensity distribution (b). 
\label{fig:LG}}
\end{figure}

The Laguerre--Gaussian beams (\ref{eq:LG})--(\ref{eq:Gouy}) are also vortex beams, characterized by the azimuthal quantum number $\ell$ and factor $\mathrm{exp}(i \ell \varphi)$.
However, in contrast to Bessel beams (\ref{eq:Bessel}) and (\ref{eq:dispersion}), they
are properly localized and normalizable in the two transverse dimensions. This is because the Fourier spectrum of LG beams is smoothly distributed over different radial wave-vector components $k_\perp$, Fig. \ref{fig:LG}(a) [cf. Eq.~(\ref{eq:Fourier-Bessel}) and Fig.~\ref{fig:Bessel}(a)].
This radial uncertainty of the momentum is related to the beam waist as
$\delta p_\perp \sim \hbar / w_0$.
The quantum number $n$ corresponds to the radial localization of the LG modes and determines the number of radial maxima in their probability density distributions (see Fig. \ref{fig:LG}).

Figure \ref{fig:LG}(b) shows the transverse spatial distributions of the probability densities and currents in the LG beams with different values of quantum numbers $(\ell, n)$.
The zero-order mode $\psi^{LG}_{0,0}$ is the standard \emph{Gaussian beam}, which can be regarded as an infinitely-long Gaussian wavepacket [cf. Eqs. (\ref{eq:planewave}) and (\ref{eq:distribution})] with $l \rightarrow \infty$ and $\delta E =  \delta p = 0$.
Gaussian beams or wavepackets are often implied in quantum models of free electrons, because they do not contain any intrinsic structures and degrees of freedom. In contrast to that, higher-order modes with $(\ell, n) \neq (0,0)$ exhibit a variety of structures related to the internal spatial degrees of freedom of localized electrons. In general, LG beams with
different $(\ell, n)$ or Bessel beams with different $\ell$, 
constitute a {\it complete set of orthogonal monoenergetic modes} for the free-space  Schr{\"o}dinger equation (the LG beams being restricted by the paraxial approximation).
Therefore, any free-electron state can be represented as a superposition of these modes. Vortex beams are the most convenient modes when one deals
with monoenergetic electrons with a well-defined propagation direction, and some sort of
azimuthal (cylindrical) symmetry in the problem. Importantly, the latter symmetry naturally involves the \emph{angular momentum} properties with respect to the propagation direction.

\subsection{Probability current and orbital angular momentum}
\label{sec:OAM}
\vspace{2mm}

We now describe the main observable characteristics of electron vortex beams. First, the
probability density and probability current density in quantum electron states are determined by 
\cite{LandauLifshitz3,Cohen-Tannoudji}:
\begin{equation}
\label{eq:properties}
\rho= \left|  \psi^2 \right|, \qquad \mathbf{j}= \frac{1}{m_e} \left(  \psi | \hat{\mathbf{p}} | \psi  \right) = \frac{\hbar}{m_e} \mathrm{Im} \left(   \psi^* \mathbf{\nabla} \psi \right).
\end{equation}
Here, $\hat{\bf p}= - i \hbar \nabla$ is the canonical momentum operator, and we use the notation $\left( \psi | ... | \psi  \right) \equiv \mathrm{Re\left(  \psi^\dagger ... \psi \right) }$ for the {\it local} expectation value of an operator.

Substituting the wave function (\ref{eq:Bessel}) into Eq.~(\ref{eq:properties}), we obtain the probability density and current in the Bessel beams (see Fig. \ref{fig:Bessel}):
\begin{equation}
\label{eq:prob}
\rho^B_{|\ell |}(r) \propto  \left| J_{| \ell |} (\kappa r) \right|^2, \qquad 
\mathbf{j}^B_\ell (r,\varphi) = \frac{\hbar}{m_e} \left( \frac{\ell}{r} \bar{{\bm \varphi}}  + k_z \bar{{\bf z}} \right) \rho^B_{| \ell |}(r),
\end{equation}
where $\bar{\bm \varphi}$ is the unit vector of the azimuthal coordinate. The $\ell$-dependent azimuthal component of the probability current (\ref{eq:prob}), together with its longitudinal component, result in a \emph{spiraling current}, Fig.~\ref{fig:vortex_beams}. 
This is a common feature of all vortex beams 
\cite{Allen1992a,Allen_review,Allen_book,Berry2008,Bliokh2007,Bliokh2012e}.

For LG beams (\ref{eq:LG}), the probability current density also has a radial component
related to the diffraction. The probability density and azimuthal $\ell$-dependent current component in the LG beams are (see Fig. \ref{fig:LG}):
\begin{equation}
\label{eq:LGprobability}
\rho^{LG}_{|\ell |, n}(r,z) \propto \left(  \frac{r^2}{w^2(z)}   \right)^{| \ell |}  \left| L^{| \ell |}_n  \left(  \frac{2r^2}{w^2(z)} \right) \right|^2 \! \mathrm{exp} \left( - \frac{2r^2}{w^2(z)} \right), \quad j^{LG}_{\ell, n \, \varphi}(r,z) = \ell \frac{\hbar}{m_e  r} \rho^{LG}_{| \ell  |, n}(r,z).
\end{equation}

The azimuthal probability current in vortex beams is directly related to the $z$-directed
\emph{orbital angular momentum} of such states.
The electron OAM can be defined either as the expectation value of the OAM operator or via the circulation of the probability current. Normalizing this per electron, we have:
\begin{equation}
\langle {\bf L} \rangle = \frac{\langle \psi | \hat{\bf L} | \psi \rangle}{\langle  \psi | \psi \rangle}
= \frac{m_e \int {\bf r} \times {\bf j} \, d^3 {\bf r}}{\int \rho \, d^3 {\bf r}},
\label{eq:expect}
\end{equation}
where $\hat{\bf L} = {\bf r} \times \hat{\bf p}$ is the canonical OAM operator
\cite{LandauLifshitz3,Cohen-Tannoudji}, 
and the inner product involves the volume integral $\int ... \, d^3 {\bf r}$. The definition (\ref{eq:expect}) is suitable for wavepackets localized in three dimensions. For 2D-localized beams one should use integrals over the two transverse dimensions: $\int ... \, d^3 {\bf r} \to \int ... \, d^2{\bf r}_{\perp}$. This means that for wave beams we deal with {\it linear densities per unit $z$-length} and normalize quantities per electron per unit $z$-length 
\cite{Allen_review,Allen_book}. 
In this manner, the longitudinal component of the OAM in a beam becomes in cylindrical coordinates:
\begin{equation}
\label{eq:expect2}
\langle L_z \rangle = \frac{\langle \psi | \hat{L}_z | \psi \rangle}{\langle  \psi | \psi \rangle}= \frac{m_e \int r j_\varphi d^2{\bf r}_{\perp}}{\int \rho \, d^2{\bf r}_{\perp}},
\end{equation}
where $\hat{L}_z = i \hbar \partial / \partial \phi$.

For any vortex beam with $\psi_{\ell} \propto \mathrm{exp}(i \ell \varphi)$ and $j_\varphi =  \ell \left(  \hbar / m_e r  \right) \rho$ (including the Bessel and LG beams), Eq~(\ref{eq:expect}) results in
\cite{Bliokh2007,Bliokh2012e}:
\begin{equation}
\label{eq:OAM}
\langle L_z \rangle = \hbar\, \ell.
\end{equation}
Thus, {\it an electron in a vortex-beam state carries a well-defined, longitudinal OAM, which is determined by the azimuthal quantum number $\ell$}. Furthermore, vortices $\psi_\ell$ are \emph{eigenmodes} of the OAM operator $\hat{L_z}$:  $\hat{L_z} \psi_\ell = \ell \, \psi_\ell$.
Notably, the OAM (\ref{eq:OAM}) is \emph{intrinsic}, 
i.e., independent of the choice of the coordinate origin 
\cite{Berry1998,Bliokh2015PR}. 
Although the radius vector $\mathbf{r}$ is present in the canonical OAM operator $\hat{\mathbf{L}}$ and in the local OAM density under the integral in Eq.~(\ref{eq:expect}), it disappears in the final expectation value $\langle L_z  \rangle$. 

Note that the \emph{extrinsic} OAM can be calculated as \cite{Bliokh2015PR}:
\begin{equation}
\label{extrinsic}
\langle \mathbf{L}^{\rm ext} \rangle = \langle \mathbf{r} \rangle \times \langle \mathbf{p} \rangle.
\end{equation}
Here 
\begin{equation}
\label{centroid}
\langle \mathbf{r} \rangle = \frac{\langle \psi | \mathbf{r} | \psi \rangle}{\langle  \psi | \psi \rangle} = 
\frac{\int \mathbf{r} \rho \, d^3 {\bf r}}{\int \rho \, d^3 {\bf r}}
\end{equation}
is the electron centroid, whereas 
\begin{equation}
\label{momentum}
\langle \mathbf{p} \rangle = \frac{\langle \psi | \hat{\bf p} | \psi \rangle}{\langle  \psi | \psi \rangle} = 
\frac{m_e \int \mathbf{j} \, d^3 {\bf r}}{\int \rho \, d^3 {\bf r}}
\end{equation}
is the expectation value of the electron momentum. For cylindrical vortex beams, $\langle \mathbf{p} \rangle \parallel \bar{\mathbf{z}}$ and $\langle \mathbf{r}_\perp \rangle =0$, so that the longitudinal component of the extrinsic OAM (\ref{extrinsic}) {\it vanishes}: $\langle L_{z}^{\rm ext} \rangle =0$.

\begin{figure} 
\centering
\includegraphics[width=\linewidth]{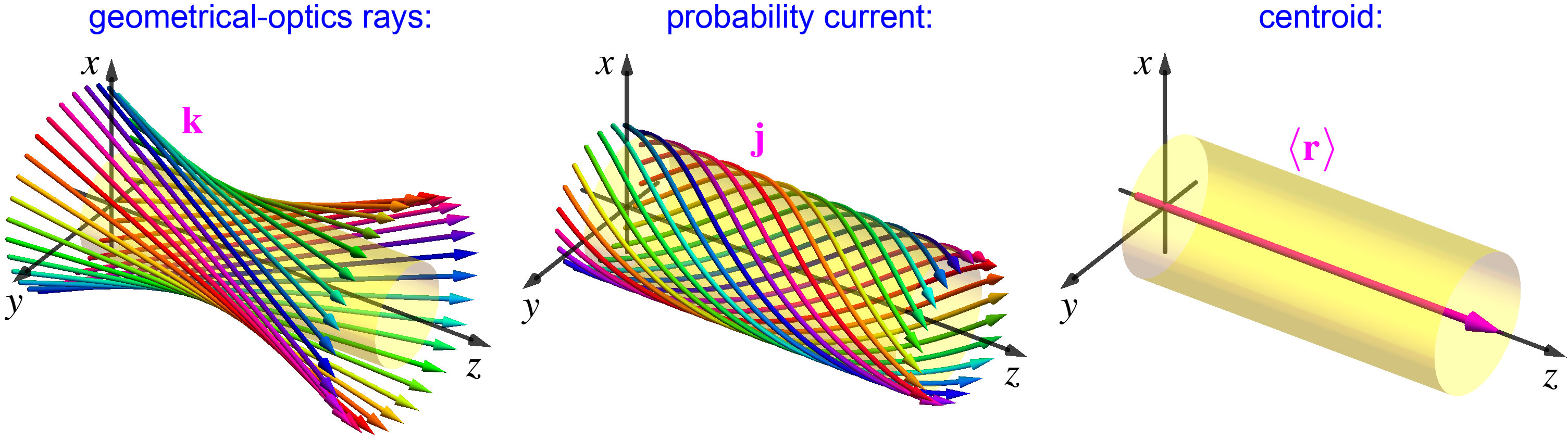}
\caption{Geometrical-optics rays, streamlines of the probability current, and centroid trajectory in a Bessel beam with $\ell=2$ (see Figs.~\ref{fig:vortex_beams} and \ref{fig:Bessel}). The rays form a two-parameter family of {\it straight} lines which are tangent to a cylindrical surface (here we show rays touching the cylinder at a given $z$ and different azimuthal angles $\varphi$) \cite{Berry2008}. The directions and color-coded phases of the rays correspond to the wave vectors ${\bf k}$ in the beam spectrum, Fig.~\ref{fig:Bessel}(a). In contrast, the streamlines of the probability-current ${\bf j}$, Eqs.~(\ref{eq:properties}) and (\ref{eq:prob}), (Bohmian trajectories) are {\it curvilinear} \cite{Bohm1987,Tannor}. For Bessel beams, these are spirals lying on the cylindrical surface \cite{Berry2008}. The azimuthal component of this current generates the OAM of the beam, Eq.~(\ref{eq:expect})--(\ref{eq:OAM}). Finally, the centroid $\langle {\bf r} \rangle$, Eq.~(\ref{centroid}), obviously coincides with the beam axis and corresponds to the rectilinear motion of the classical electron.
\label{fig:centroid}}
\end{figure}

The longitudinal intrinsic OAM of free electrons is a remarkable and somewhat
counterintuitive quantum property. Based on classical-mechanics intuition, one can expect that the angular momentum is produced by a rotational (i.e., curvilinear) motion. However, free-space electrons, in the absence of any external forces, must propagate along straight rectilinear trajectories. This apparent contradiction is removed if we carefully distinguish {\it local} and {\it integral} properties. For classical point particles, which cannot have any internal structure, there are no intrinsic properties. In contrast, quantum (wave) beams or wavepackets inevitably have finite sizes and inhomogeneous distributions of local densities. In particular, the local probability current density with the azimuthal component [see Eqs.~(\ref{eq:prob}) and (\ref{eq:LGprobability})] implies spiraling
streamlines, i.e., spiral \emph{Bohmian trajectories} $\mathbf{r}^{\rm Bohm} (t)$ of electrons \cite{Bohm1987,Tannor}, Fig.~\ref{fig:centroid}. 
At the same time, the quantum-classical correspondence (Ehrenfest theorem) requires only the trajectory of the electron \emph{averaged position (centroid)} to be rectilinear 
\cite{Cohen-Tannoudji,Tannor}. 
In agreement with this, $\langle \mathbf{r}_\perp \rangle = 0$, 
and the electron centroid coincides with the rectilinear beam axis. 
In the general case, the centroid (\ref{centroid}) always follows a rectilinear trajectory for any localized quantum state of free-space electrons.
Thus, internal spiraling streamlines of the probability current density generate the intrinsic OAM in free-electron vortex states, while electron's center of gravity always follows a rectilinear trajectory. Note also that vortex beams are superpositions of multiple \emph{plane waves}, and therefore are solutions of free-space wave (Schr{\"o}dinger) equation. In terms of geometrical optics, these plane waves determine a family of rectilinear \emph{rays} which are tangent to the rotationally-symmetric (e.g., cylindrical) surface, associated with the maximum probability density in the vortex beam 
\cite{Berry2008,McMorran2011}.
However, the superposition principle is valid for the wave functions but not for
the probability currents, which are quadratic forms. Therefore, streamlines of the probability current of a superposition of plane waves are curvilinear in the generic case \cite{Tannor,Berry2008}. 
Figure~\ref{fig:centroid} shows an example of rays, current streamlines, and centroid trajectory in a vortex Bessel beam.

Notably, non-relativistic scalar electrons in a vortex-beam state somewhat resemble  ``massless particles with spin $\ell$''.
Indeed, the spin angular momentum (SAM) of massless relativistic particles is aligned with their momentum, so that helicity is a well-defined quantum number. Vortex electrons carry similar OAM with well-defined longitudinal component, i.e., ``orbital helicity''. However, in contrast to the real SAM of the electron, which is limited by $\hbar/2$, the intrinsic OAM can take on arbitrarily large values of $\hbar\ell$. As such, the OAM of electron vortex states can have important consequences in the dynamics of electrons and their interactions with external fields, atoms, and other particles.

\subsection{Basic ways of generating electron vortex beams}
\label{sec:basic-ways}
\vspace{2mm}

After introducing the vortex-beam solutions of the Schr\"odinger equation, it is important to discuss the basic ways of how such states can be generated with accelerated electrons in electron microscopes.
Here we only briefly describe the main concepts, while a more detailed description of experimental techniques will be given in Section~\ref{sec:producing}.
Using analogies and differences of electron optics as compared to light optics, three ways of generating electron vortex beams were put forward in the original theoretical work \cite{Bliokh2007}.

\subsubsection{Spiral phase plate}

The first method is a straightforward analogy of \emph{spiral phase plates} used for photons in different frequency ranges \cite{beijersbergen_helical-wavefront_1994,Turnbull1996,peele_observation_2002}.
When free electrons move through a solid-state plate, they acquire an additional phase $\Delta\Phi$ as compared with free-space propagation 
\cite{tonomura_applications_1987,Tonomura_book}. 
This phase is proportional to the plate thickness $d$: $\Delta \Phi = \xi d$, and is analogous to the phase delay of an optical wave propagating through a dielectric plate. Therefore, a plate with spiral thickness varying with the azimuthal angle,
$d = \zeta \varphi$, will create the corresponding spiral phase in the transmitted wave: $\Delta \Phi=\xi\zeta\varphi$. Thus, if the incident wave was a plane wave, the transmitted wave will carry a vortex $\exp( i \xi\zeta\varphi)$ with topological charge $\ell = \xi\zeta$, see Fig. \ref{fig:5}(a). This idea was used in the first experiment \cite{Uchida2010} demonstrating the production of a free-electron vortex in a TEM, by employing a spiral-thickness-like region in a stack of graphite flakes. Since the phase change at the step was not an integer times $2\pi$  in that experiment, the output electron wave possessed a non-integer vortex 
\cite{Berry2004,Leach2004}, 
which can be regarded as a superposition of several vortex states with different quantum numbers $\ell$ 
\cite{Goette2008}.
Later, experiments with accurate spiral phase plates producing electron vortex beams with integer OAM were reported 
\cite{Shiloh2014,Beche2015a} 
(see Section~\ref{sec:phaseplate} below).

\subsubsection{Diffraction grating with a fork dislocation}

The second way of generating electron vortex states also represents a TEM adoption of the analogous optical method. The vortex structure in a wave field represents a \emph{screw dislocation} of the phase front 
\cite{Nye1974a,Soskin2001,Dennis2009}. 
Considering the diffraction of a basic Gaussian-like beam on a \emph{diffraction grating with an edge dislocation} (``fork''), the edge dislocation in the grating produces screw dislocations in the diffracted beams 
\cite{bazhenov1990laser,heckenberg_laser_1992}, see Fig.~\ref{fig:5}(b). 
If the dislocation in the grating is of order $\ell_0$, then the $N$th order of diffraction transforms the incident Gaussian-like beam ($\ell = 0$) into a vortex beam with ($\ell = N\ell_0$).
This method was first used for the efficient generation of high-quality electron vortex beams with integer $\ell$ in 
\cite{Verbeeck2010} (for the $\ell_0 = 1$ grating dislocation). 
Soon after, this method was extended up to $\ell_0 = 25$ in \cite{McMorran2011}.
Notably, this experiment demonstrated electron vortex beams with the topological charge up to $\ell = 100$ (in the $N = 4$ diffraction order). 
Thus, this technique allows the generation of quantum electron states with an intrinsic OAM of hundreds and even thousands of $\hbar$ \cite{Mafakheri2016}, which is impossible with spin angular momentum. 
For details and the state-of-the-art holographic techniques for the production of electron vortex beams see Section~\ref{gratings}.

\subsubsection{Magnetic monopole}

Finally, the third fundamental method of generating electron vortices has no
straightforward optical counterpart. Namely, it exploits the interaction of electrons with external magnetic fields and vector-potentials. Indeed, in contrast to photons, electrons are \emph{charged} particles, and this opens a route to interesting interactions of electron vortices with magnetic fields and structures (various examples of these are described below). 
Quantum phenomena of electron-field interactions appear in the electron phase and involve the vector-potential $\bf{A}(\bf{r})$. A famous example is the Aharonov--Bohm effect 
\cite{Aharonov_significance_1959,Tonomura_book}, intimately related to the so-called \emph{Dirac phase} \cite{dirac1931quantised}
\begin{equation}
\Phi_D=\frac{e}{\hbar c}\int_{C} {\bf A}({\bf r})\cdot d{\bf r}
\label{eq:17}
\end{equation}
for an electron moving along a contour ${C}$ in the presence of the vector-potential. 
Hereafter, $e = - |e|$ is the electron charge and $c$ is the speed of light.

\begin{figure} 
\centering
\includegraphics[width=0.95\linewidth]{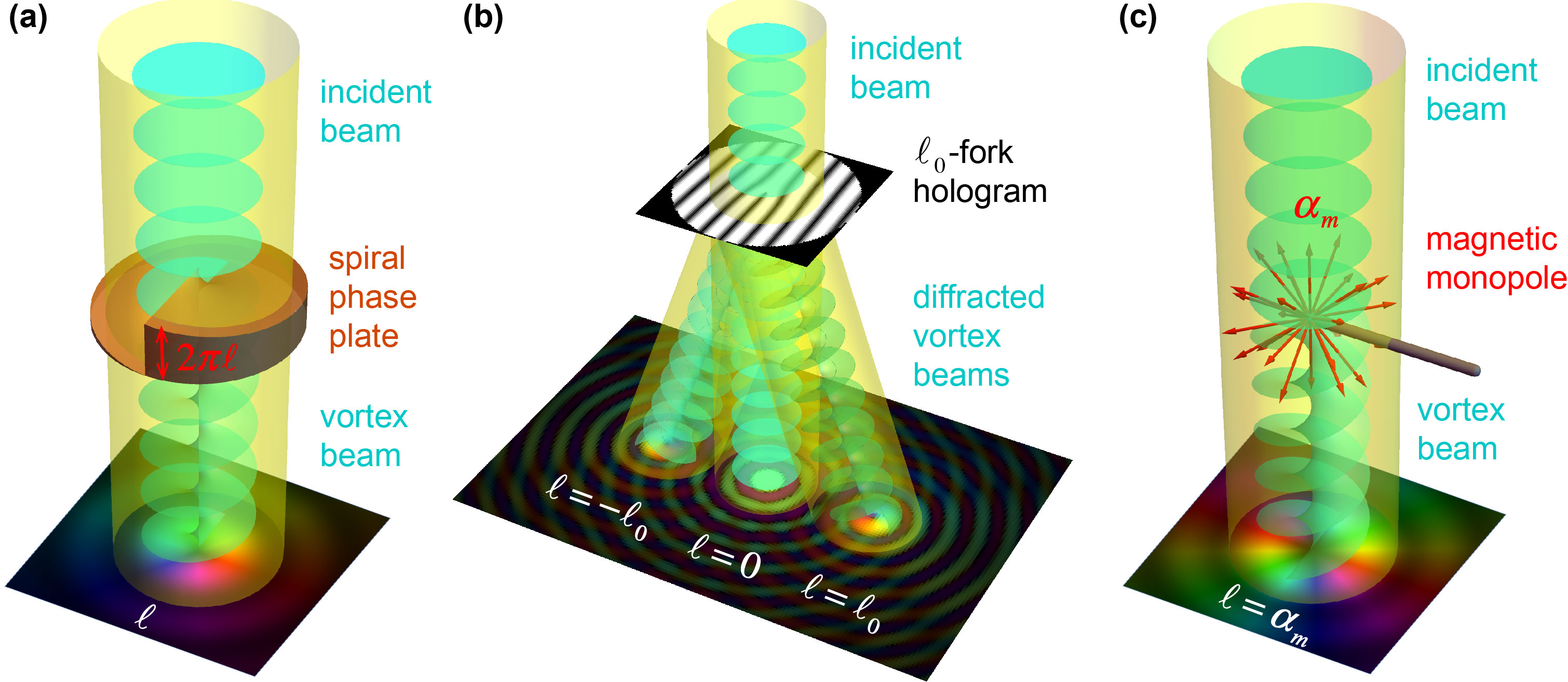}
\caption{Schematics of basic methods for the generation of electron vortex beams: (a) a spiral phase plate with the $2\pi\ell$ phase-shift increment around its center (here $\ell=1$); (b) a diffraction grating (hologram) with a fork-like edge dislocation of order $\ell_0$ (here $\ell_0=1$); and (c) a magnetic monopole of dimensionless charge $\alpha_m$ (here $\alpha_m=2$). See also explanations in the text.} 
\label{fig:5}
\end{figure}

Importantly, the vector-potential of a magnetic flux line (an infinitely thin solenoid) has the form of a \emph{vortex}: ${\bf A}({\bf r}) = (\hbar c\alpha_m / er )\bar{\bm \varphi}$, where $\alpha_m$ is the dimensionless magnetic-flux strength ($\alpha_m=1$ corresponding to two magnetic-flux quanta) \cite{Aharonov_significance_1959}.
This hints that the Dirac phase (\ref{eq:17}) from a magnetic flux line can produce a vortex phase 
$\exp(i \ell \varphi)$ with the quantum number $\ell=\alpha_m$.
To produce such vortex, one has to consider a transition of an electron without vortex ($\ell=0$) in the region without magnetic flux ($\alpha_m = 0$) to the region with the flux $\alpha_m \neq 0$. Notably, the end of the flux line represents nothing but a \emph{magnetic monopole} of strength $\alpha_m$ 
\cite{dirac1931quantised,Shnir_book}. 
Thus, \emph{scattering of an electron wave by a magnetic monopole generates an electron vortex of strength} $\ell = \alpha_m$ 
\cite{tonomura_applications_1987,Bliokh2007}, Fig.~\ref{fig:5}(c).
\footnote{Importantly, although in theoretical considerations it is convenient to consider the magnetic flux line (also called {\it Dirac string}) aligned with the propagation $z$-axis, observable quantities are independent of its orientation and involve only the monopole charge $\alpha_m$.}
Recently, this was demonstrated experimentally by using a thin magnetic needle with a sharp end, approximating a magnetic flux line with a monopole \cite{beche_magnetic_2014,Beche2016d}.

Generation of an electron vortex by a magnetic monopole can be
understood in terms of angular-momentum conservation. For simplicity, let us consider a {\it classical} point electron moving in a magnetic-monopole field. Although it might seem that the monopole is a spherically-symmetric object, the usual angular momentum of the electron, ${\bf L} = {\bf r} \times {\bf p}$, is not conserved. Indeed, the Lorentz force from the monopole is not central and it originates from the non-symmetric vector-potential. However, there exists another integral of motion, the {\it generalized} angular momentum 
\cite{Shnir_book}:
\begin{equation}
{\bf L}' = {\bf L} - \hbar \frac{\alpha_m}{2} \, \bar{\bf r}.
\label{eq:18}
\end{equation}
Here $\bar{\bf{r}}$ is the unit radius vector, and we assume that the monopole is located at the origin. 
The $z$-component of ${\bf L}'$ must be conserved in the electron scattering by the monopole. 
When the electron comes from $z\rightarrow -\infty$ and the scattered electron is observed at $z \rightarrow +\infty$, the radius-vector $\bar{\bf r}$ changes from $-\bar{\bf z}$ to $+\bar{\bf z}$, so that the electron OAM must change as $L_z^{\rm out}=L_z^{\rm in} + \hbar\,\alpha_m$. 

For classical electrons, this additional OAM can be explained by the Lorentz force from the  monopole. The monopole magnetic field can be written as ${\bf B} = \left( \hbar c \alpha_m / 2e \right) {\bf r} / |{\bf r}|^3$. In the eikonal approximation, a point-like electron approximately follows a straight-line trajectory passing at the radial distance $r_0$ from the monopole. Then, the Lorentz force from the monopole deflects the electron, so that it gains a transverse (azimuthal) momentum $p_{\varphi} = \hbar\, \alpha_m / r_0$. As a result, the electron acquires the OAM $L_z = p_{\varphi} r_0 = \hbar\, \alpha_m$.

This shift of the electron's angular momentum in the presence of a magnetic flux appears in both classical-particle and quantum-wave considerations 
\cite{Wilczek1982,Bliokh2012e}, provided we consider the {\it kinetic} rather than canonical OAM (see Section~\ref{sect:Landau} below). 

\subsection{Vortex electrons in electric and magnetic fields. Basic aspects. \label{sect:EVexternal}}
\vspace{2mm}

Electrons are charged particles which interact with electromagnetic fields. 
The classical equations of motion of a point electron in an external electric and magnetic fields are \cite{Jackson_book}:
\begin{equation}
\dot{\bf p} = e {\bf E} + \frac{e}{c} \dot{\bf r} \times {\bf B}, \qquad 
\dot{\bf r} = \frac{\bf p}{m_e}.
\label{eq:bliokh19}
\end{equation}
Here the overdot stands for the time derivative, ${\bf r}$ and ${\bf p}$ are the coordinates and momentum of the electron, while ${\bf E}$ and ${\bf B}$ are the electric and magnetic fields.

Quantum wavepacket or beam states of electrons have finite dimensions and therefore can possess internal properties, in addition to the electric charge $e$.
Finite-size electron states are characterized by the distributions of the \emph{charge density} $\rho_e = e\rho$ and \emph{electric current density} $\vt{j}_e = e\vt{j}$. 
Most importantly, the coiling current density in vortex electron states acts as a solenoid and
generates a \emph{magnetic moment}. The magnetic moment of a localized electron state can be defined as \cite{Jackson_book}:
\begin{equation}
{\bf M} = \frac{1}{2c} \frac{\int {\bf r} \times {\bf j}_e \, d^3 {\bf r}}{\int \rho \, d^3 {\bf r}} = \frac{e}{2m_e c} \langle {\bf L} \rangle.
\label{eq:bliokh20}
\end{equation}
In particular, the longitudinal $z$-directed magnetic moment of electron vortex beams (per unit $z$-length) in free space equals 
\cite{Bliokh2007,Bliokh_Relativistic}:
\begin{equation}
M_z = \frac{e\hbar}{2m_e c} \ell \equiv -\mu_B \ell,
\label{eq:bliokh21}
\end{equation}
where $\mu_B=|e|\hbar/(2 m_e c)$ is the Bohr magneton.

Thus, vortex electrons carry a longitudinal magnetic moment (\ref{eq:bliokh21}) proportional to the quantized OAM and anti-parallel to it.
Note that this magnetic moment corresponds to the gyromagnetic ratio with $g$-factor $g = 1$, while $g = 2$ for the magnetic moment generated by the spin \cite{Cohen-Tannoudji,QEDbook}. The presence of the magnetic moment should modify the equations of motion (\ref{eq:bliokh19}).

\begin{figure} 
	\centering
 \includegraphics[width=0.75\linewidth]{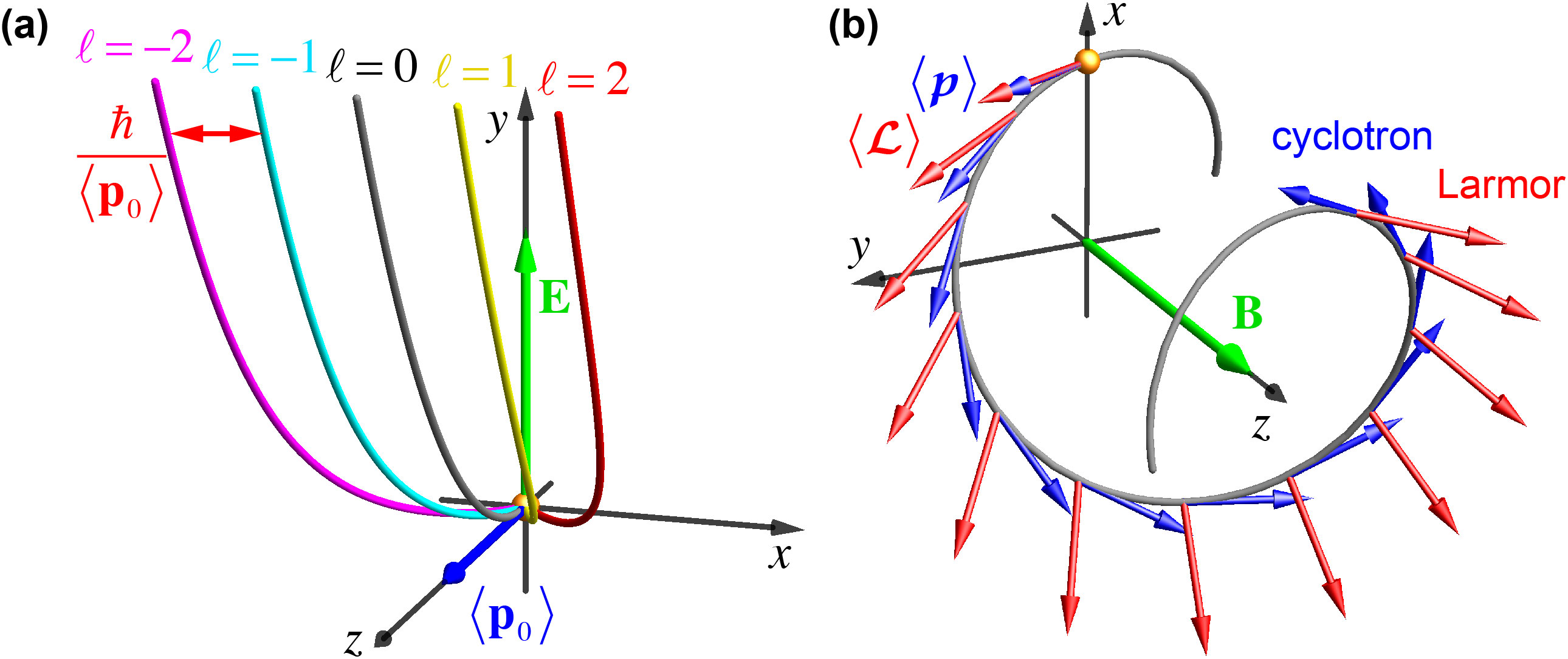}
	\caption{(a) The orbital Hall effect of vortex electron states in an electric field ${\bf E}$ \cite{Bliokh2007}, Eqs.~(\ref{eq:22a}). Here, $\langle {\bf p}_0 \rangle$ is the initial mean momentum of the electron. Bended by the electric field, the trajectories of the centroids of the vortex beams or wavepackets experience transverse $\ell$-dependent shifts of the order of the de Broglie wavelength. (b) Cyclotron motion of a vortex electron in a uniform magnetic field ${\bf B}$. The mean kinetic momentum $\langle {\mathbcal p} \rangle$ precesses about the magnetic field with the cyclotron angular frequency $\Omega_c$, Eq.~(\ref{eq:cyclotronmotion}). At the same time, the kinetic OAM $\langle {\mathbcal L} \rangle$ of a vortex electron state precesses with the Larmor frequency $\Omega_L = \Omega_c / 2$, Eq.~(\ref{eq:larmorprecession}). Therefore, the initially-alligned momentum and OAM change their mutual direction during this evolution \cite{Bliokh2007,Gallatin2012} (see also Fig.~\ref{fig:gallatin} below).}
\label{fig:cyclotron}
\end{figure}

We first consider the interaction of the magnetic moment or intrinsic OAM with an
external \emph{electric} field ${\bf E}$ (we set ${\bf B} = 0$ here); this can result in a spin-orbit-type interaction. In fact, since the intrinsic angular momentum has an orbital origin in our case, this should rather be called \emph{orbit-orbit interaction} between the intrinsic OAM (vortex) and extrinsic OAM (trajectory)
(see \cite{Bliokh2006,Bliokh_2009,Merano2010} for such effects in optical vortex beams). 
This interaction couples the intrinsic OAM $\langle {\bf L} \rangle$ and extrinsic trajectory parameters $\langle {\bf r} \rangle$ and $\langle {\bf p} \rangle$ in the equations of motion. 
To describe such semiclassical evolution, we consider localized (but sufficiently large) paraxial electron wavepacket and assume that the intrinsic OAM maintains its
form during the wavepacket propagation, $\langle {\bf L} \rangle = \hbar \ell \langle {\bf p} \rangle / p$ ($\left| \langle p \rangle \right| \simeq p$). In this case, the ``orbit-orbit interaction'' becomes equivalent to that of massless spinning particles with spin $\ell$ in an external scalar potential 
\cite{Berard2006}.
Using the Berry-connection formalism 
\cite{Xiao2010}, the semiclassical equations of motion take the form \cite{Bliokh2007}:
\begin{equation}
\langle \dot{\bf p} \rangle = e {\bf E}, \qquad 
\langle \dot{\bf r} \rangle = \frac{\langle {\bf p} \rangle}{m_e} + \hbar \ell \frac{\langle \dot{\bf p} \rangle \times \langle {\bf p} \rangle}{p^3},
\label{eq:22a}
\end{equation}
\begin{equation}
\langle \dot{\bf L} \rangle = - \frac{\left[ e {\bf E} \times \langle {\bf p} \rangle \right] \times \langle {\bf L} \rangle}{p^2}.
\label{eq:22b}
\end{equation}
The last term in Eq.~(\ref{eq:22a}) and equation (\ref{eq:22b}) describe the mutual influence of the intrinsic OAM and the trajectory. 
It can be readily shown that these equations are consistent with the assumed form $\langle {\bf L} \rangle = \hbar \ell \langle {\bf p} \rangle / p$, i.e., the ``orbital helicity'' $\langle {\bf L} \rangle \cdot \langle {\bf p} \rangle / p = {\rm const}$ is an integral of motion of Eqs.~(\ref{eq:22a}) and (\ref{eq:22b}). Equation (\ref{eq:22b}) is an analogue of the Bargman--Michel--Telegdi equation 
\cite{Bargmann1959,Bolte1999,Spohn2000} 
for the precession of the intrinsic angular momentum in an external electric field. 
In turn, the last term in Eq.~(\ref{eq:22b}) describes the OAM-dependent ($\ell$-dependent) transverse transport of the electron, Fig.~\ref{fig:cyclotron}(a). This is  an analogue of the intrinsic spin Hall effect, known in condensed-matter physics 
\cite{Murakami2003,Xiao2010}, high-energy physics 
\cite{Berard2006,Bliokh2005}, and optics (for photons) 
\cite{Bliokh2008,Bliokh2015NP}.
Here it should rather be called \emph{orbital Hall effect}. The typical value of the transverse $\ell$-dependent shift of the electron trajectory is $\hbar/ p$, i.e., the de Broglie wavelength of the electron \cite{Bliokh2007}.
Therefore, this effect is extremely small, and practically unobservable, for free electrons in electron microscopes. Note, however, that an analogous orbital Hall effect has been successfully measured for optical vortex beams interacting with dielectric inhomogeneities
\cite{Merano2010}, 
because subwavelength accuracy is quite achievable in modern optics. 
Furthermore, similar spin Hall effect of electrons in condensed-matter systems result in the observable accumulation of opposite spin polarizations on the opposite edges of the sample with an applied electric field \cite{Kato2004}. Thus, the orbital Hall effect for vortex electrons, Eq.~(\ref{eq:22a}), could still play a role, e.g., in condensed-matter phenomena \cite{Bernevig2005,Tanaka2008}.

Let us now consider the interaction of the intrinsic OAM with an external magnetic field
${\bf B}$ (we set ${\bf E} = 0$ for simplicity). One could expect that the interaction between the magnetic moment of the electron (\ref{eq:bliokh20}) and the external magnetic field is described by a Zeeman-like energy $\Delta E = - {\bf M} \cdot {\bf B}$ 
\cite{Bliokh2007}. 
However, expression (\ref{eq:OAM}) for the intrinsic OAM and the corresponding Eq.~(\ref{eq:bliokh21}) for the magnetic moment are derived using the \emph{free-space} vortex-beam solutions, i.e., without any external fields. 
In the presence of external fields one has to find a solution of the corresponding Schr\"odinger equation, and it will contain a self-consistent distribution of charges,
currents, and fields, including their interactions 
\cite{Bliokh2012e}. 
As we show in the next Section~\ref{sect:Landau}, this drastically modifies the values of the electron OAM and its magnetic moment in the presence of a magnetic field.

Moreover, in the presence of a magnetic field, the dynamical evolution of the electron is described by the {\it kinetic} momentum $\langle \mathbcal{p} \rangle$ and OAM $\langle \mathbcal{L} \rangle$. These quantities differ from their {\it canonical} counterparts, $\langle {\bf p} \rangle$ and OAM $\langle {\bf L} \rangle$, by the vector-potential contribution \cite{Bliokh2012e} (such that kinetic and canonical quantities coincide in the absence of the vector-potential). 
Below we formally introduce kinetic characteristics, and here only make one important point. Namely, generically, an electron in a magnetic field {\it cannot} keep its intrinsic OAM
parallel to its momentum, i.e., $\langle \mathbcal{L} \rangle \parallel \langle \mathbcal{p} \rangle$, which was assumed in all solutions considered above \cite{Bliokh2007}.
Indeed, as follows from Eqs.~(\ref{eq:bliokh19}), semiclassical electrons approximately follow cyclotron trajectories, and the mean momentum evolves (at least, in the classical limit $\hbar \rightarrow 0$) as
\begin{equation}
\langle \dot{\mathbcal p} \rangle = \frac{e}{m_e c} \langle {\mathbcal p} \rangle \times {\bf B}.\label{eq:cyclotronmotion}
\end{equation}
Thus, the momentum precesses around the magnetic-field direction with the \emph{cyclotron} frequency $\Omega_c = e B / (m_e c)$. At the same time, the evolution of the intrinsic OAM $\langle {\mathbcal L} \rangle$ in the magnetic field is described by the Zeeman energy term and the corresponding Larmor precession equation (Bargman--Michel--Telegdi equation with $g = 1$ factor in a magnetic field)
\cite{Cohen-Tannoudji}:
\begin{equation}
\langle \dot{\mathbcal L} \rangle = \frac{e}{2m_e c} \langle {\mathbcal L} \rangle \times {\bf B}.\label{eq:larmorprecession}
\end{equation}
This means that the electron OAM precesses about the magnetic-field direction with the \emph{Larmor} frequency $\Omega_L = eB / (2m_e c)$. Since the Larmor and cyclotron frequencies differ by a factor of two, $\Omega_c = 2\,\Omega_L$, the momentum $\langle {\mathbcal p} \rangle$ and OAM $\langle {\mathbcal L} \rangle$ cannot be parallel in such evolution, with the exception of the case $\langle {\mathbcal L} \rangle \parallel \langle {\mathbcal p} \rangle \parallel {\bf B}$, Fig.~\ref{fig:cyclotron}(b). This means that the initial free-space form of the electron vortex states, $\langle {\mathbcal L} \rangle = \hbar\ell \, \langle {\mathbcal p} \rangle / p$, cannot survive in a magnetic field with a non-zero transverse component \cite{Bliokh2007,Gallatin2012}.

The two-frequency evolution of Eqs.~(\ref{eq:cyclotronmotion}) and (\ref{eq:larmorprecession}) results in very interesting dynamics of vortex electrons in a magnetic field which is considered in detail below. Note that the evolution of the \emph{SAM} of the electron does not face such a problem. Because of the $g = 2$ factor for spin, the frequency of its precession becomes twice the Larmor frequency, i.e., the cyclotron one \cite{Bargmann1959,Bolte1999,Spohn2000,Cohen-Tannoudji}. 
Therefore, in contrast to the OAM, the SAM precession is synchronized with the momentum evolution, and the helicity (projection of the spin onto the momentum direction) is conserved.

\subsection{Longitudinal magnetic field. Landau states.}
\label{sect:Landau}
\vspace{2mm}

We now provide a self-consistent quantum treatment of electron vortex modes in a magnetic field ${\bf B}$. The free-space electron Hamiltonian underlying the Schr{\"o}dinger equation (\ref{eq:schro}) is modified in a magnetic field as:
\begin{equation}
\hat{H} = \frac{\hat{\bf p}^2}{2m_e} \rightarrow \frac{\hat{\bm {\mathbcal{p}}}^2}{2m_e}=\frac{1}{2m_e}\!\left(\hat{\bf p} - \frac{e}{c} {\bf A}\right)^2,
\label{eq:bliokh25}
\end{equation}
where $\hat{\bf p} = - i \hbar \nabla$ is the \emph{canonical} momentum operator, $\hat{\bm{\mathbcal{p}}} = \hat{\bf p} - \frac{e}{c} {\bf A}$ is the \emph{kinetic} (or covariant) momentum shifted by the vector-potential ${\bf A}({\bf r})$ generating magnetic field ${\bf B} = {\nabla} \times {\bf A}$. 

The presence of the coordinate-dependent solenoidal vector-potential considerably complicates the Schr\"o\-dinger equation, and it allows a simple analytical solution only in some cases, such as the following case of a uniform and constant magnetic field ${\bf B}$. Choosing the $z$-axis to be directed along the field, ${\bf B} = B\, \bar{\bf z}$, 
\footnote{Note that here we choose $B=B_z$ rather than $B=|{\bf B}|$, so that quantities $B$, $\Omega_L$, and $\Omega_c$ can be either positive or negative depending on the direction of the magnetic field, $\sigma = {\rm sgn}(B)$.}
the problem acquires the cylindrical symmetry natural for vortex-beam solutions. Moreover, in this geometry, the vector-potential can be chosen to have only an azimuthal component, i.e., to form a \emph{vector-potential vortex}:
\begin{equation}
{\bf A} = \frac{Br}{2} \bar{\bm \varphi}\,.
\label{eq:26}
\end{equation}
The corresponding stationary Schr\"odinger equation (\ref{eq:cylinschro}) with a uniform magnetic field in cylindrical coordinates becomes:
\begin{equation}
-\frac{\hbar^2}{2m_e}\left[\frac{1}{r}\frac{\partial}{\partial r}\left(r\frac{\partial}{\partial r}\right)+\frac{1}{r^2}\left(\frac{\partial}{\partial\varphi}+i\sigma\frac{2r^2}{w^2_m}\right)^2+\frac{\partial^2}{\partial z^2}\right] \psi = E\, \psi.
\label{eq:27}
\end{equation}
Here $w_m=2\sqrt{\hbar c/\left|eB\right|}=\sqrt{2\hbar/m_e |\Omega_L|}$ is the \emph{magnetic length} parameter, and $\sigma= {\rm sgn}(B) = \pm 1$ indicates the direction of the magnetic field. Note that the Larmor frequency $\Omega_L$ (rather than the cyclotron frequency $\Omega_c=2\,\Omega_L$) is the fundamental frequency in the quantum-mechanical problem \cite{Brillouin1945,Bliokh2012e}. 
This is related to Larmor's theorem, the conservation of angular momentum, and this will be clearly seen below from the quantum picture of the electron evolution. 

\begin{figure}[!t] 
	\centering
   \includegraphics[width=0.9\linewidth]{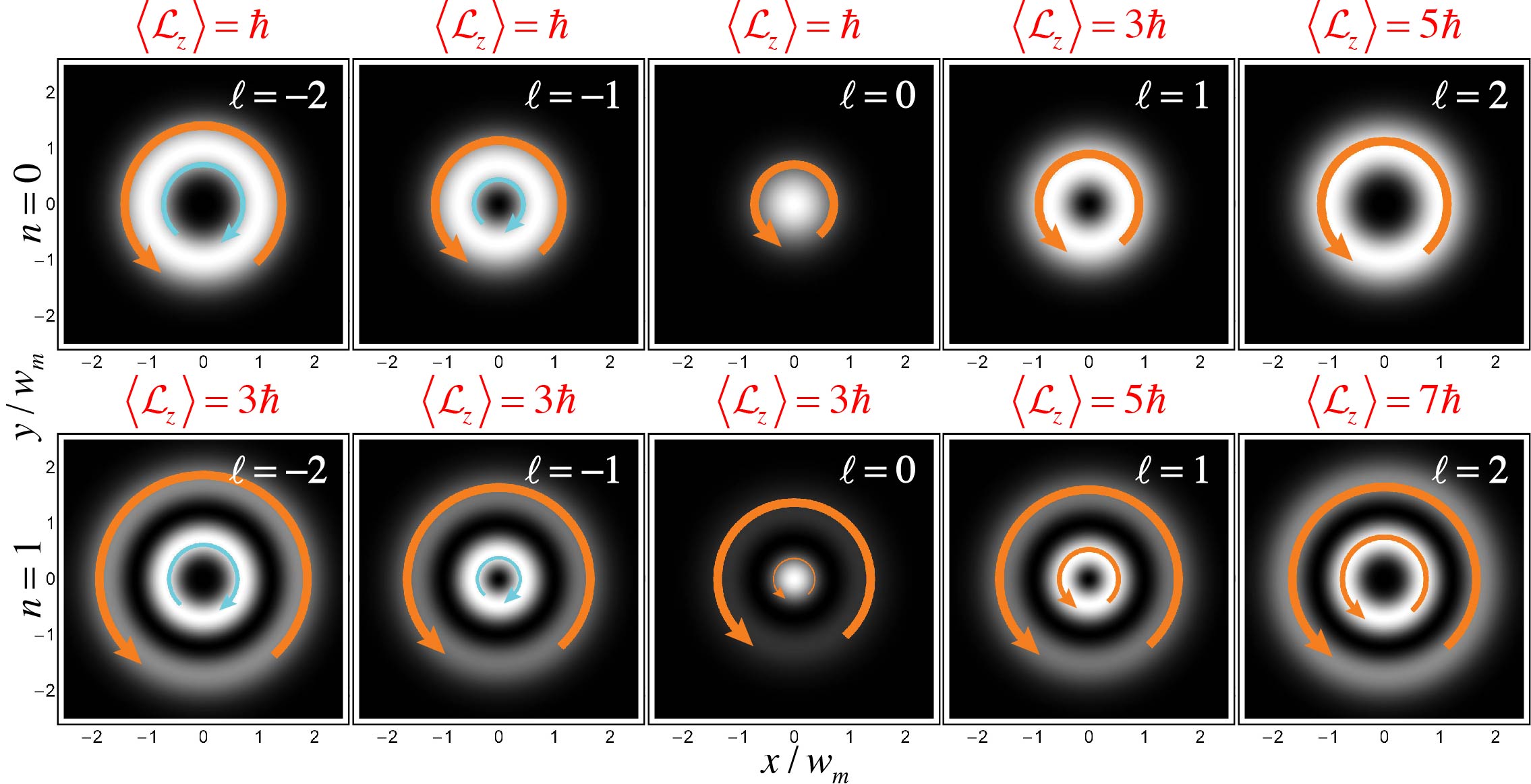}
\caption{Same as in Figs.~\ref{fig:Bessel}(b) and \ref{fig:LG}(b) but for the Landau states (\ref{eq:28})--(\ref{eq:37}) with different values of $(\ell,n)$ in a uniform magnetic field ${\bf B} = B\, \bar{\bf z}$ [here $\sigma = {\rm sgn}(B) = +1$]. Although the probability density distributions of the Landau modes is entirely similar to those of the LG beams, Fig.~\ref{fig:LG}(b), their current densities and angular-momentum properties differ significantly. In particular, the {\it kinetic OAM} of Landau states, $\langle {\mathcal L}_z \rangle$, cannot be zero or negative even for $\ell < 0$, and its minimum value is $\hbar$. This is because of the always-positive (for $\sigma = +1$) vector-potential contribution to the azimuthal probability current, Eqs.~(\ref{eq:bliokh33}) and (\ref{eq:bliokh34}).} 
\label{fig:7}
\end{figure}

The solutions of Eq.~(\ref{eq:27}) are known as \emph{Landau states} 
\cite{LandauLifshitz3,Cohen-Tannoudji,Fock1928,Landau1930}, 
and they have the form of {\it non-diffracting LG beams} (see Fig.~\ref{fig:7}) \cite{Bliokh2012e,Greenshields2012}:
\begin{equation}
\psi^L_{\ell,n} \propto \left(\frac{r}{w_m}\right)^{|\ell|}\! L_n^{|\ell|}\!\left(\frac{2r^2}{w^2_m}\right) \exp\!\left(-\frac{r^2}{w^2_m}\right) \exp\left[i\!\left(\ell\varphi + k_z z \right)\right],
\label{eq:28}
\end{equation}
where the wave number $k_z$ must obey the dispersion relation considered below, Eq.~(\ref{eq:30}). The Landau states (\ref{eq:28}) are identical to the LG beams (\ref{eq:LG}) with the beam waist $w_0 = w_m$ at $z = 0$. 

We also introduce a longitudinal scale $z_m=v/|\Omega_L|$ determined by the Larmor frequency and the electron velocity $v = \sqrt{2E/m_e}$. The transverse magnetic length $w_m$ and longitudinal {\it Larmor length} $z_m$ represent counterparts of the beam waist and Rayleigh length of the free-space LG beams (\ref{eq:LG}) but here they are uniquely determined by the electron energy and magnetic field strength:
\begin{equation}
w_m=\frac{2\sqrt{\hbar c}}{\sqrt{\left|eB\right|}}, \qquad 
z_m=\frac{2c\sqrt{2Em_e}}{\left|eB\right|}, \quad 
\textrm{i.e.,} \quad z_m=\sqrt{\frac{E}{\hbar\left|\Omega_L\right|}}\,w_m.
\label{eq:29}
\end{equation}
The fact that eigenmodes (\ref{eq:28}) in the magnetic field are non-diffracting and transversely confined (i.e., possess a discrete radial quantum number $n$) reflects the boundedness of classical electron orbits in a magnetic field.
\footnote{In optics, non-diffracting LG modes entirely analogous to Eq.~(\ref{eq:28}) appear in parabolic-index optical fibers 
\cite{Dooghin1992}. This is related to the fact that the Schr\"odinger equation in a uniform magnetic field can be mapped onto a two-dimensional quantum-oscillator problem \cite{Cohen-Tannoudji}}

While the diffracting LG beams (\ref{eq:LG}) represent \emph{approximate} paraxial solutions of the Schr\"odinger equation, Landau LG modes (\ref{eq:28}) yield {\it exact} solutions of the problem with magnetic field. In doing so, the wave numbers satisfy the following dispersion relation \cite{Bliokh2012e}:
\begin{equation}
E = \frac{\hbar^2k_z^2}{2m_e} - \hbar\Omega_L\ell + \hbar\left|\Omega_L\right| (2n + \left|\ell\right| + 1) \equiv E_\parallel + \underbrace{E_Z+E_G}_{E_\perp}.
\label{eq:30}
\end{equation}
Here $E_\parallel = \hbar^2k_z^2/2m_e$ is the energy of the free longitudinal motion, while the quantized transverse-motion energy in Eq.~(\ref{eq:30}) can be written as
\begin{equation}
E_\perp = \hbar\left|\Omega_L\right|(2N_L +1),\qquad 
N_L = n + \frac{1}{2}\left|\ell\right|\left[1+\textrm{sgn}(\sigma\ell)\right] = 0,1,2,...~.
\label{eq:31}
\end{equation}
Thus, Eq.~(\ref{eq:31}) describes the structure of quantized Landau energy levels 
\cite{LandauLifshitz3,Cohen-Tannoudji,Fock1928,Landau1930,Bliokh2012e,Greenshields2012}. 
Equation~(\ref{eq:30}) shows that Landau energies consist of two terms \cite{Bliokh2012e}: $E_\perp = E_Z + E_G$. The first one, $E_Z = - \hbar\,\Omega_L \ell = - M_z B$, represents the \emph{Zeeman energy} of the free-space magnetic moment (\ref{eq:bliokh21}) in a magnetic field $B$. The second term $E_G = \hbar\,|\Omega_L |\left( 2n + \left|\ell\right| + 1\right)$ can be associated with the \emph{Gouy phase} (\ref{eq:Gouy}) of the diffractive LG modes. (Recall that the Gouy-phase term is related to the transverse kinetic energy of spatially-confined modes \cite{Boyd1980,Feng2001}, which shifts the propagation constants and eigenfrequencies of the waveguide and resonator modes \cite{Siegman}.) As we show below, the Zeeman and Gouy-phase contributions are separately observable and lead to a remarkable behavior of the electron probability density distributions in a magnetic field.

Obviously, the transverse probability-density distributions of Landau modes (\ref{eq:28}) are entirely analogous to those of the LG modes (\ref{eq:LG}) [Fig. \ref{fig:7}, cf. Eq.~(\ref{eq:LGprobability}) and Fig.~\ref{fig:LG}]:
\begin{equation}
\rho^L_{|\ell|,n} (r) \propto \left(\frac{r^2}{w^2_m}\right)^{|\ell|} \left|L_n^{|\ell|}\left(\frac{2r^2}{w^2_m}\right)\right|^2 \exp\!\left(-\frac{2r^2}{w^2_m}\right).
\label{eq:32}
\end{equation}
However, their probability-current and AM properties differ significantly from the free-space solutions. This is because the definitions of the gauge-invariant probability current density and of the kinetic momentum/AM are essentially modified by the presence of the vector-potential. Namely, the probability current density is now determined as the local expectation value of the \emph{kinetic} (covariant) momentum operator (\ref{eq:bliokh25}) 
\cite{LandauLifshitz3,Cohen-Tannoudji}:
\begin{equation}
{\bf j} = \frac{1}{m_e}\left(\psi\left|\hat{{\mathbcal{p}}}\right|\psi\right) = \frac{\hbar}{m_e}\Im\left(\psi^*{\nabla}\psi\right) - \frac{e}{m_e c}{\bf A}\,\rho.
\label{eq:bliokh33}
\end{equation}
This means that the vector potential ${\bf A}$ produces an additional probability current in quantum electron states. For Landau states (\ref{eq:28}) the current (\ref{eq:bliokh33}) yields:
\begin{equation}
{\bf j}^L_{\ell,n} (r,\varphi)= \frac{\hbar}{m_e} \left[\frac{1}{r}\left(\ell+\sigma\frac{2r^2}{w^2_m}\right)\! \bar{\bm{\varphi}} + k_z \bar{\bf z} \right] \rho^L_{|\ell|,n}(r).\label{eq:bliokh34}
\end{equation}
Here, the $\sigma$-dependent term is the vector-potential contribution. It is worth noticing that for the counter-circulating vortex $\exp(i \ell \varphi )$ and vector-potential $A_\varphi$, $\ell\sigma < 0$, the azimuthal current in (\ref{eq:bliokh34}) changes its sign at $r = r_{|\ell|} \equiv |\ell| w_m/\sqrt{2}$, i.e., around the first radial maximum of the LG mode. For $r < r_{|\ell|}$ the current from the vortex $\exp(i \ell \varphi)$ prevails, whereas for $r > r_{|\ell|}$ the contribution from the vector-potential $A_\varphi$ becomes dominant (see Fig. \ref{fig:7}).

Taking into account the vortex-like form of the vector-potential (\ref{eq:26}) and its appearance in the azimuthal probability current (\ref{eq:bliokh34}), it should also contribute to the OAM of the electron. In fact, one can define {\it two} OAM quantities in the presence of a magnetic field. The first one is the \emph{canonical} OAM, which is determined by the canonical OAM operator $\hat{\bf L} = {\bf r} \times \hat{\bf p}$. Its longitudinal component $\hat{L}_z = - i \hbar \partial / \partial\varphi$ acts only on the vortex phase factor $\exp(i\ell\varphi)$ in Landau modes (\ref{eq:28}). Hence, similar to free-space vortex beams, Landau states are eigenmodes of the canonical OAM operator and have the same expectation value of the canonical OAM as in Eq.~(\ref{eq:OAM}) 
\cite{Bliokh2012e}:
\begin{equation}
\hat{L}_z \psi^L_{\ell,n} = \ell\, \psi^L_{\ell,n}, \qquad 
\left<L_z\right> = \frac{\left<\psi\left| \hat{L}_z \right| \psi \right>}{\left< \psi \left| \psi \right.\right>} = \hbar\,\ell.
\label{eq:35}
\end{equation}
The second OAM of the electron in a magnetic field is the \emph{kinetic} OAM determined by the kinetic momentum operator: $\hat{\mathbcal{L}}= {\bf r} \times \hat{\mathbcal{p}}$ or the probability current density:
\begin{equation}
\left<\mathbcal{L}\right> = \frac{\left< \psi \left| \hat{\mathbcal{L}} \right| \psi \right>}{\left< \psi \left| \psi \right. \right>} = \frac{m_e \int {\bf r} \times {\bf j}\, d^3 {\bf r}}{\int \rho\, d^3 {\bf r}}.\label{eq:36}
\end{equation}
It is kinetic OAM (\ref{eq:36}) that describes the mechanical action of the electron OAM and observable rotational dynamics in electron states.

Substituting characteristics of Landau modes, Eqs.~(\ref{eq:28}), (\ref{eq:32}), and (\ref{eq:bliokh34}), into Eq.~(\ref{eq:36}) (with the beam substitution $d^3 {\bf r} \to d^2 {\bf r}_{\perp}$) and using Eq.~(\ref{eq:31}), we arrive at 
\cite{Bliokh2012e}
\begin{equation}
\left<\mathcal{L}_z\right>=\hbar\left[\ell+\sigma\left(2n+|\ell|+1\right)\right]=\hbar\,\sigma\left(2N_L +1\right).
\label{eq:37}
\end{equation}
Equation (\ref{eq:37}) reveals nontrivial properties of the electron OAM in a magnetic field, Fig.~\ref{fig:7}. First, it shows that the sign of the kinetic OAM is solely determined by the direction of the magnetic field, $\sigma$, and is independent of the vortex charge $\ell$. This is because after the integration (\ref{eq:36}) the vector-potential contribution to the azimuthal current always exceeds the vortex one. Note also that for parallel OAM and magnetic field, $\sigma\ell> 0$, the canonical OAM $\hbar\ell$ is enhanced (in absolute value) by the magnetic-field contribution: $\left<\mathcal{L}_z^{\uparrow\uparrow}\right>=\hbar\left[2\ell+\sigma(2n+1)\right]$. At the same time, in the opposite case of anti-parallel OAM and magnetic field, $\sigma\ell < 0$, the kinetic OAM takes the form $\left<\mathcal{L}_z^{\uparrow\downarrow}\right>=\hbar\sigma(2n+1)$, i.e., becomes {\it independent} of the vortex charge $\ell$. This is caused by the partial cancellation of the counter-circulating azimuthal currents produced by the vortex $\exp(i \ell \varphi)$ and by the magnetic vector-potential $A_\varphi$.
\footnote{The vector-potential contribution to the kinetic OAM is sometimes called ``diamagnetic angular momentum'' \cite{Greenshields2015}.} 
Second, the value of $\left<\mathcal{L}_z\right>$ is independent of the magnitude of the magnetic field, $|B|$. This is because the radius of the beam changes as $w_m\propto 1/\sqrt{|B|}$, Eq.~(\ref{eq:28}), the angular velocity $\Omega_L\propto |B|$, whereas the mechanical OAM behaves as $\mathcal{L}_z \propto\Omega_L w_m^2$. Third, in contrast to the classical electron motion in a magnetic field, which can have zero OAM, Eq.~(\ref{eq:37}) shows that there is a \emph{minimal kinetic OAM} of quantum Landau states: $\left|\left<\mathcal{L}_z\right>\right|_{\rm min} = \hbar$.

Importantly, modified definitions of the probability current density (\ref{eq:bliokh33}) and kinetic OAM (\ref{eq:36}) also affect the value of the electron {\it magnetic moment} in the presence of a magnetic field. Indeed, using the definition (\ref{eq:bliokh20}) with the modified current density (\ref{eq:bliokh33}), we obtain \cite{Bliokh2012e}:
\begin{equation}
{\bf M} = \frac{e}{2c} \frac{\int {\bf r} \times {\bf j}\, d^3 {\bf r}}{\int \rho\, d^3 {\bf r}} 
= \frac{e}{2m_e c} \left<\mathbcal{L}\right>.
\label{eq:bliokh38}
\end{equation}
Thus, the magnetic moment of the electron in a magnetic field is determined by the \emph{kinetic} OAM. Since $\left<\mathbcal{L}\right>$ is always aligned with ${\bf B}$ and $e<0$, the magnetic moment (\ref{eq:bliokh38}) is {\it anti-parallel} to the magnetic field. This determines the {\it diamagnetic} response of free scalar electrons in a magnetic field \cite{Landau1930,Darwin1931}. 

The magnetic moment of the Landau states, $M_z$, shares all the unusual properties of the kinetic OAM (\ref{eq:37}) and differs strongly from the magnetic moment of free-space vortex electrons, Eq.~(\ref{eq:bliokh21}). Using the magnetic moment (\ref{eq:bliokh38}), the transverse energy of the electron, Eqs.~(\ref{eq:30}) and (\ref{eq:31}), can be written as a single Zeeman term:
\begin{equation}
E_\perp = - M_z B.
\end{equation}
It now includes both the ``pure'' Zeeman term from the coupling of the free-space magnetic moment (\ref{eq:bliokh21}) with the field as well as the Gouy-phase term. Notably, the latter term can be considered as a \emph{nonlinear} (with respect to the field) effect of the interaction of the vector-potential current $- (e/m_e c) {\bf A} \rho$ (``diamagnetic angular momentum'') with the magnetic field ${\bf B}$ \cite{Bliokh2012e}.

Most peculiarities of the electron Landau states in a magnetic field are contained in their dispersion relation (\ref{eq:30}) and (\ref{eq:31}), depending on quantum numbers $\ell$ and $\sigma$, and the corresponding OAM values (\ref{eq:37}). These quantities bring about rather nontrivial rotational dynamics when various superpositions of Landau modes propagate in a magnetic field \cite{Bliokh2012e,Greenshields2012,Gallatin2012,Guzzinati2013,Schattschneider2014NC,
Greenshields2014,Greenshields2015,Schachinger2015}. Below we show some examples of peculiar rotations of electron vortex states in a magnetic field.

\subsection{Unusual rotational dynamics in a magnetic field}
\label{sec:magnetic-rotations}
\vspace{2mm}

In terms of the cylindrically-symmetric Landau modes, the rotational dynamics of asymmetric electron states in a magnetic field appear from the interference of different modes (\ref{eq:28})--(\ref{eq:30}) acquiring different phases during propagation. Assuming that all the modes have the same fixed energy $E$ and that they are paraxial, i.e.,  $E_\perp \ll E$, $w_m \ll z_m$, one can write the longitudinal wave number as $k_z \simeq k + \Delta k _z$, where $\hbar k = \sqrt{2Em_e}$ and
\begin{equation}
\Delta k_z = - \left[\sigma \ell + \left( 2n+|\ell|+1\right) \right] z_m^{-1} .
\label{eq:bliokh40}
\end{equation}
Thus, the Larmor length $z_m$, Eq.~(\ref{eq:29}), determines the characteristic longitudinal scale of the beam evolution. During the propagation along the $z$-axis, the correction (\ref{eq:bliokh40}) to the wave number $k$ yields an additional phase
\begin{equation}
\Phi_{LZG} = \Delta k_z z .
\label{eq:bliokh41}
\end{equation}
This phase depends on both vortex and magnetic-field properties. We call it Landau--Zeeman--Gouy phase \cite{Bliokh2012e} because of its intimate relation to the Landau levels, Zeeman coupling [the $\sigma\ell$-term in (\ref{eq:bliokh40})], and Gouy phase [the $\left( 2n+|\ell|+1\right)$-term in (\ref{eq:bliokh40})]. The interplay between the Zeeman and Gouy terms results in rich dynamics of various Landau-mode superpositions.

\begin{figure}[tb]
\centering
\includegraphics[width=\linewidth]{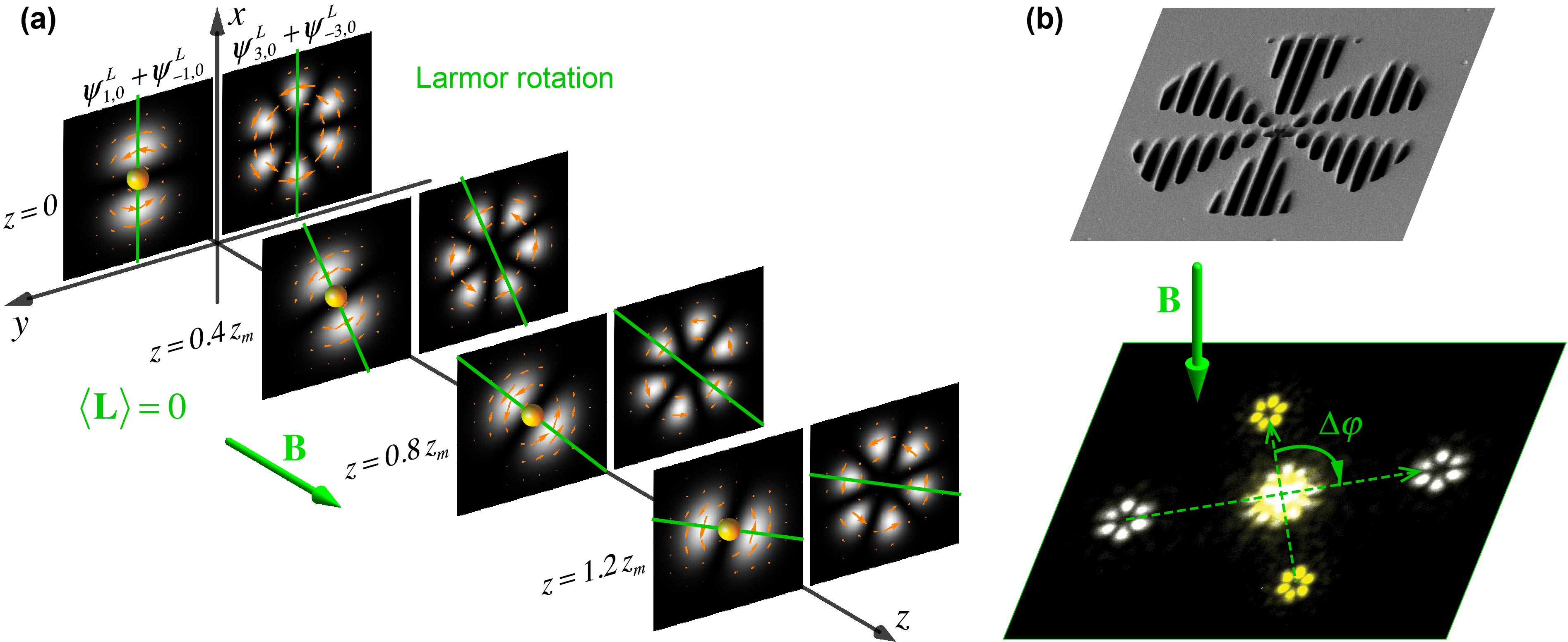}
\caption{(a) Superpositions of the Landau states (\ref{eq:28})--(\ref{eq:30}) $\psi^L_{\ell,n} + \psi^L_{-\ell,n}$ (with $n=0$ and $\ell = 1,3$ here) undergo the {\it Larmor rotation} (\ref{eq:bliokh42}) during the propagation in the magnetic field ${\bf B}$ \cite{Bliokh2012e,Greenshields2012}. The transverse probability-density (grayscale) and probability-current (orange arrows) distributions are shown for different propagation distances $z$. The net canonical OAM of such superpositions vanishes, $\langle {\bf L} \rangle = 0$, and their centroids (shown by yellow spheres for the $\ell=1$ superposition) obey rectilinear trajectories parallel to the magnetic field. (b) Experimental demonstration of this Larmor rotation in a TEM \cite{Guzzinati2013}. The holographic aperture (top) produces superpositions of the LG modes with $(\ell,n)=(\pm 3,0)$ in the first diffraction orders (bottom). Changing magnification in the imaging magnetic lens (i.e., the effective magnetic field ${\bf B}$) from $41\cdot 10^3 \times$ (yellow) to $55 \cdot 10^3 \times$ (white) produces a rotation of the image by the angle $\Delta \varphi = 106^{\circ}$ corresponding to the Larmor rotation (\ref{eq:bliokh42}). This is an example of image rotation which is well known in electron microscopy \cite{DeGraefCTEM}.}
\label{fig:bliokh8}
\end{figure}    

We first consider the simplest superposition of two Landau modes (\ref{eq:28}) with equal amplitudes, the same radial index $n$, and opposite vortex charges $\pm \ell$: 
$\psi = \psi^L_{-\ell,n} + \psi^L_{\ell,n}$. Such superposition carries no net canonical OAM, $\left\langle L_z \right\rangle = 0$, and its transverse probability density distribution represents a flower-like pattern with $2|\ell |$ petals: $|\psi|^2 \propto \cos^2(\ell \varphi)$ (see Fig.~\ref{fig:bliokh8}). The phases (\ref{eq:bliokh41}) of the two interfering modes $\psi^L_{\pm\ell,n}$ differ only in their {\it Zeeman terms}: $\Delta \Phi_{LZG}=\mp \ell \sigma z/z_m$. Combining these terms with the azimuthal vortex dependencies as $\exp\left( \pm i \ell \varphi \right) \to \exp\left[ \pm i \ell (\varphi - \sigma z / z_m) \right]$, one can see that this results in the rotation of the interference pattern by the angle 
\cite{Bliokh2012e,Greenshields2012} (Fig.~\ref{fig:bliokh8})
\begin{equation}
    \Delta \varphi^{(0)} = \sigma \frac{z}{z_m}.
    \label{eq:bliokh42}
\end{equation}
Since $z/z_m = |\Omega_L| z / v $, the rotation (\ref{eq:bliokh42}) is characterized by the  \emph{Larmor frequency} $\Omega_L$. Such rotation of a superposition of two opposite-$\ell$ vortex modes in a magnetic field was recently observed in 
\cite{Guzzinati2013}. 
In fact, the Larmor rotation of images in a magnetic field is well known in transmission electron microscopy 
\cite{DeGraefCTEM}. 
The above theory provides a convenient quantum-mechanical description of this effect. Indeed, any superposition (image) carrying no net angular momentum and consisting of pairs of opposite-$\ell$ modes will undergo the same Larmor rotation (\ref{eq:bliokh42}).

\begin{figure}[t]
\centering
\includegraphics[width=\linewidth]{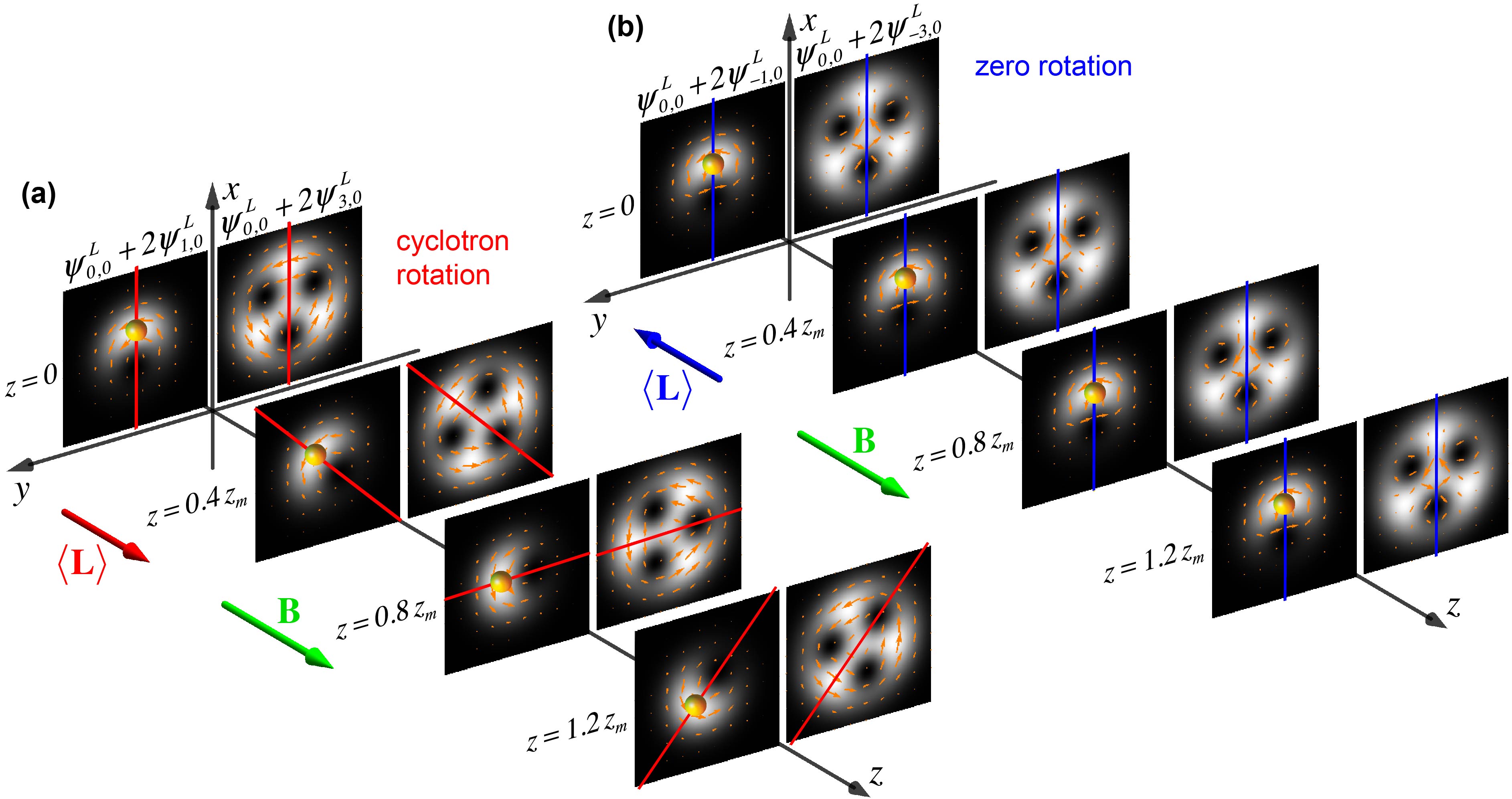}
\caption{(a) Same as in Fig.~\ref{fig:bliokh8}(a) but for superpositions $\psi^L_{0,n} + a \psi^L_{\ell,n}$ (with $a=2$, $n=0$, and $\ell = 1,3$ here), which carry nonzero net canonical OAM $\langle {\bf L} \rangle$ {\it parallel} to the magnetic field ${\bf B}$. These superpositions undergo {\it cyclotron (double-Larmor) rotation} (\ref{eq:bliokh43}) during the propagation \cite{Bliokh2012e}. The superposition with $\ell = 1$ has an off-axis centroid, which follows a classical cyclotron trajectory, in agreement with Eqs.~(\ref{eq:bliokh45}). 
(b) Analogous superpositions with $\ell = -1,-3$ and the net canonical OAM {\it anti-parallel} to the magnetic field experience {\it zero rotation}, Eq.~(\ref{eq:bliokh44}) \cite{Bliokh2012e}. This is due to the cancellation of the vortex and vector-potential contributions to the azimuthal probability current (\ref{eq:bliokh33}).}
\label{fig:bliokh9}
\end{figure}

As another example, we now consider a superposition of two Landau modes (\ref{eq:28}) with the same radial index $n$, and vortex charges $0$ and $\ell$: $\psi = \psi^L_{0,n} + a\, \psi^L_{\ell,n}$, where $a$ is some constant amplitude \cite{Bliokh2012e}. Such a superposition has a nonzero net canonical OAM $\left\langle L_z \right\rangle \propto \ell$, and is characterized by a pattern of $|\ell|$ off-axis vortices (Fig.~\ref{fig:bliokh9}). 
Landau modes with different $|\ell|$ involve the {\it Gouy term} in the difference of phases  (\ref{eq:bliokh40}) and (\ref{eq:bliokh41}). Namely, the $\psi^L_{\ell,n}$ mode acquires an additional phase $\Delta \Phi_{LZG} = -\left( \ell \sigma + | \ell | \right) z/z_m$ as compared with the $\psi^L_{0,n}$ mode. From here, it follows that the superposition $\psi$ with parallel OAM and magnetic field, $\ell \sigma > 0$, exhibits a rotation of the interference pattern by the angle
\begin{equation}
\Delta \varphi^{\uparrow \uparrow} = 2 \sigma \frac{z}{z_m}.
\label{eq:bliokh43}
\end{equation}
In contrast to this, the superposition with anti-parallel OAM and magnetic field, $\ell\sigma<0$, shows no rotation at all:
\begin{equation}
\Delta \varphi^{\uparrow \downarrow} = 0.
\label{eq:bliokh44}
\end{equation}
Equation (\ref{eq:bliokh43}) describes the rotation of the image with the \emph{double-Larmor (i.e., cyclotron) frequency} $\Omega_c = 2 \Omega_L$. Note also that non-rotating superpositions, Eq.~(\ref{eq:bliokh44}), consist of modes with the kinetic OAM values independent of the vortex charge: $\left\langle {\mathcal L}_{z}^{\uparrow \downarrow} \right\rangle = \hbar \sigma \left(2 n + 1 \right)$. For $n=0$, these correspond to the lowest Landau energy level. Figure~\ref{fig:bliokh9} shows examples of the cyclotron and zero rotations described by Eqs.~(\ref{eq:bliokh43}) and (\ref{eq:bliokh44}).

Equations (\ref{eq:bliokh42})--(\ref{eq:bliokh44}) represent an intriguing result. Namely, the rotational dynamics of quantum electron states with OAM in a magnetic fields is characterized by {\it three} frequencies: (i) Larmor, (ii) cyclotron (double-Larmor), and (iii) zero frequency 
\cite{Bliokh2012e}. 
This is in sharp contrast to the classical electron evolution (\ref{eq:cyclotronmotion}), which is described by a single cyclotron rotation \cite{Jackson_book}. 
In spite of such difference, the quantum evolution (\ref{eq:bliokh42})--(\ref{eq:bliokh44}) is fully consistent with the classical evolution (\ref{eq:cyclotronmotion}). Indeed, according to the Ehrenfest theorem, the expectation values of the electron coordinates and momentum must obey classical equations of motion 
\cite{Cohen-Tannoudji,Tannor}. 
Importantly, one should take the expectation values of the {\it kinetic} quantum quantities, which correspond to classical trajectories. Using the kinetic momentum, Eq.~(\ref{eq:bliokh25}), the equations of motion (\ref{eq:bliokh19}) and (\ref{eq:cyclotronmotion}) become:
\begin{equation}
\left\langle \mathbcal{\dot{p}} \right\rangle  = 
\frac{e}{m_e c} \left\langle \mathbcal{p} \right\rangle \times\mathbf{B}, \quad
\left\langle \mathbf{\dot{r}} \right\rangle  = 
\frac{\left\langle \mathbcal{p} \right\rangle}{m_e} .
\label{eq:bliokh45}
\end{equation}
Explicit calculations for the superpositions of Landau states considered above show that for Larmor-rotating and non-rotating states, Eqs.~(\ref{eq:bliokh42}) and (\ref{eq:bliokh44}), the mean kinetic momentum is always {\it aligned} with the magnetic field: $\langle \mathbcal{p}\rangle \parallel {\bf B} \parallel \bar{\bf z}$, i.e., $\langle\mathbcal{p}_{\perp}\rangle =0$.
Therefore, the centroid $\langle {\bf r}_{\perp}\rangle$ of such states lies on a {\it rectilinear} trajectory parallel to the magnetic field.
In contrast to this, states rotating with the cyclotron angular velocity, Eq.~(\ref{eq:bliokh43}), can have a non-zero transverse mean momentum, $\langle\mathbcal{p}_\perp\rangle \neq 0 $, and their centroids $\langle \mathbf{r}_\perp \rangle$ trace classical cyclotron orbits along the beam propagation, see Figs.~\ref{fig:bliokh8} and \ref{fig:bliokh9}.

The nontrivial rotational dynamics of quantum electron states is closely related to the {\it summation of the vortex and vector-potential contributions to the probability current} (\ref{eq:bliokh33}) and (\ref{eq:bliokh34}).
For parallel OAM and magnetic field, the two contributions produce azimuthal currents of the same sign, which result in the double-Larmor (cyclotron) rotation.
For anti-parallel OAM and magnetic field, the two azimuthal contributions cancel each other, which produces a non-rotating state \cite{Bliokh2012e}.

\begin{figure}[!t]
\centering
\includegraphics[width=\linewidth]{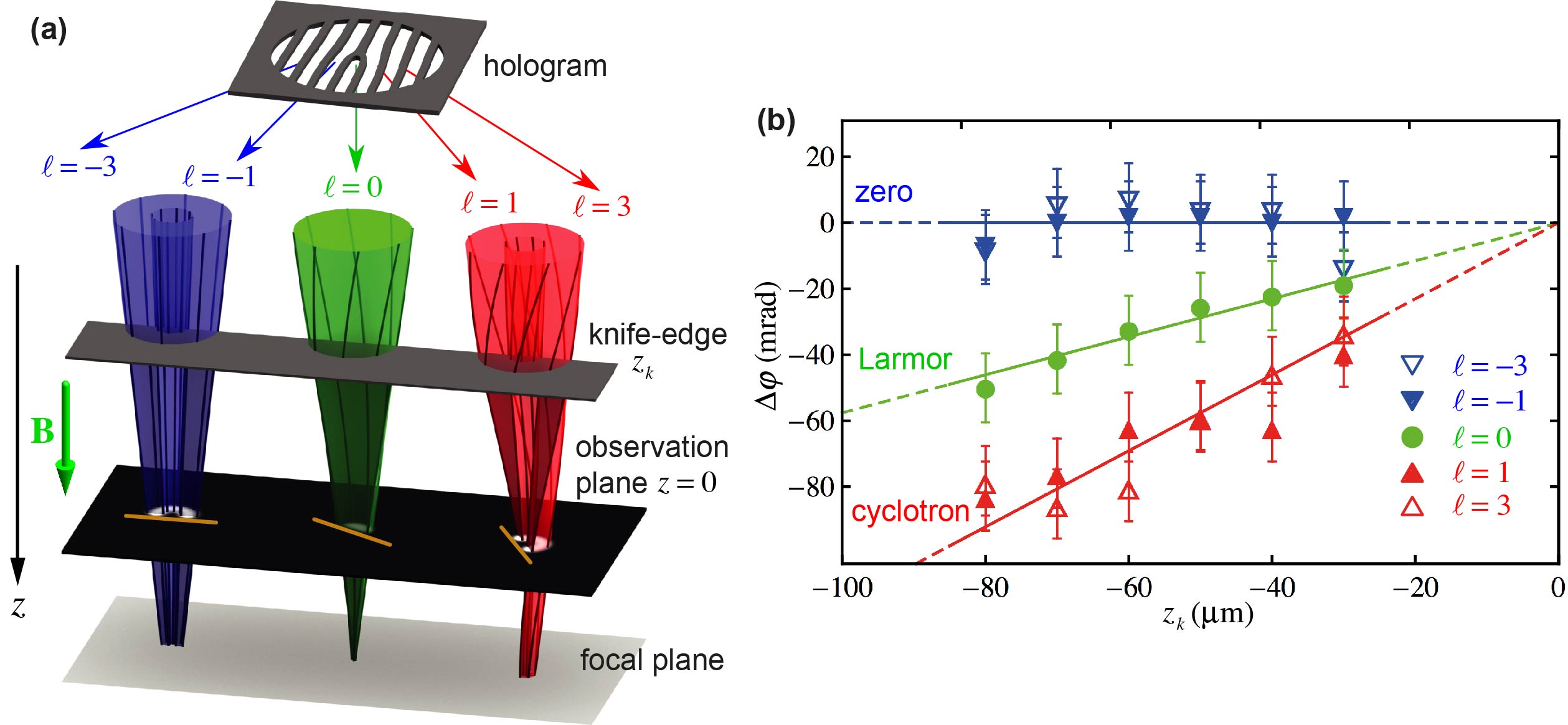}
\caption{Experimental observation \cite{Schattschneider2014NC} of the three-frequency rotational dynamics of electron vortex states in a magnetic field, Eqs.~(\ref{eq:bliokh42})--(\ref{eq:bliokh44}) and (\ref{eq:bliokh46}).
(a) Schematics of the TEM experimental setup. A holographic fork mask generates a row of vortex beams with different azimuthal indices $\ell = ..., -3, -1, 0, 1, 3, ...$. These beams are focused by a magnetic lens and are studied in the region of maximal quasi-uniform magnetic field. The focal plane is shifted few Rayleigh ranges below the observation plane $z=0$ to reduce the Gouy-phase rotation. A knife-edge stop is placed at $z_k < 0$, where it blocks half of each of the beams (to break the azimuthal symmetry of the probability density distributions). The spatial rotational dynamics of the cut beams propagating to the observation plane was observed by varying the position $z_k$ of the knife edge (see details in \cite{Schattschneider2014NC}). (b) A quantitative analysis of the $\ell$-dependent beam rotations. The azimuthal orientations of the cut modes $\Delta \varphi$ with respect to the extrapolated reference azimuth $\varphi_0 = \left. \varphi \right|_{z_k=0}$ are plotted versus $z_k$. Three lines correspond to the zero, Larmor and cyclotron rotations,~Eq.~(\ref{eq:bliokh46}).}
\label{fig:bliokh11}
\end{figure}

Furthermore, the above ``three-frequency dynamics'' immediately follows from the expression (\ref{eq:bliokh34}) for the probability current in Landau states. Indeed, one can define the local value of the electron angular frequency 
as $\Omega(r) = v_\varphi(r) / r = j_\phi(r) / \left( r\rho(r) \right)$. Calculating the expectation value of this quantity, we obtain \cite{Schattschneider2014NC}:
\begin{align}
\left\langle \Omega \right\rangle = \frac{\int_{0}^{\infty} \Omega(r) \rho(r)\, r\,dr}{\int_{0}^{\infty} \rho(r)\, r\,dr} =
\begin{cases}
        \;0 & {\rm for}\;\; \ell \sigma <0 \\
        \;\Omega_L & {\rm for}\;\; \ell =0 \\
        \;2 \Omega_L & {\rm for}\;\; \ell \sigma > 0. 
\end{cases}
\label{eq:bliokh46}
\end{align}
These expressions can be regarded as internal angular velocities of electrons in pure Landau states. Notably, they correspond exactly to expressions (\ref{eq:bliokh42})--(\ref{eq:bliokh44}) for the rotations of Landau-mode superpositions with similar OAM properties (i.e., zero, parallel, and anti-parallel OAM with respect to the magnetic field). Recenty, all three kinds of rotations (\ref{eq:bliokh46}) in Landau modes were observed experimentally 
\cite{Schattschneider2014NC}. 
In that experiment, the cylindrically-symmetric probability density distribution of Landau modes was broken by a sharp aperture stop, which cut half of the beam, and then the rotational evolution of such half-beams was traced, Fig.~\ref{fig:bliokh11}. Truncated modes can be considered as superpositions of multiple pure Landau modes, and this explains the exact correspondence between the internal dynamics (\ref{eq:bliokh46}) and superposition rotations (\ref{eq:bliokh42})--(\ref{eq:bliokh44}).

Other remarkable aspects of the electron vortex beams dynamics in a magnetic field were considered in recent works 
\cite{Gallatin2012,Greenshields2014,Greenshields2015,Schachinger2015}. 
In particular, the radial dynamics of vortex mode superpositions and the angular momentum conservation was analysed in \cite{Greenshields2014}. 
Afterwards, Ref. \cite{Greenshields2015} investigated the evolution of vortex beams shifted and tilted with respect to the $z$-axis. Of course, such shifted/tilted beams can also be considered as superpositions of multiple pure Landau modes. However, in this case it is instructive to separate the internal vortex properties and external dynamics of the vortex/beam centroid. Notably, the shifted/tilted Landau vortex mode preserves its shape with respect to the centroid, while the centroid follows a classical cyclotron trajectory, Eqs.~(\ref{eq:bliokh45}). This allows separating not only canonical (vortex) and vector-potential contributions to the OAM, but also its \emph{intrinsic} and \emph{extrinsic} parts. 

Using the electron centroid (\ref{centroid}), the intrinsic and extrinsic parts of the kinetic OAM are [cf. Eq.~(\ref{extrinsic})] \cite{Bliokh2015PR}:
\begin{equation}
    \left\langle\mathbcal{L}^{\rm ext}\right\rangle =
    \left\langle\mathbf{r}\right\rangle \times
    \left\langle\mathbcal{p}\right\rangle, \qquad
    \left\langle\mathbcal{L}^{\rm int}\right\rangle =
    \left\langle\mathbcal{L}\right\rangle -
    \left\langle\mathbcal{L}^{\rm ext}\right\rangle.
    \label{eq:bliokh48}
\end{equation}
Recall that $ \left\langle\mathbcal{L}\right\rangle = \left\langle {\bf r} \times \mathbcal{p} \right\rangle $, while $\left\langle\mathbcal{p}\right\rangle$ is the expectation value of the kinetic momentum, defined similarly to Eq.~(\ref{momentum}) but with the operator $\hat{\mathbcal{p}}$. It follows from Eqs.~(\ref{eq:bliokh48}) that the intrinsic OAM does not change its value under transverse spatial translations, while the extrinsic OAM is transformed according to the ``parallel-axis theorem'' of classical mechanics:
\begin{equation}
    \mathbf{r} \to \mathbf{r} + \mathbf{r}_{0\perp}: \quad
    \left\langle\mathbcal{L}^{\rm int}\right\rangle \to
    \left\langle\mathbcal{L}^{\rm int}\right\rangle, \quad
    \left\langle\mathbcal{L}^{\rm ext}\right\rangle \to
    \left\langle\mathbcal{L}^{\rm ext}\right\rangle +
    \mathbf{r}_{0\perp} \times \left\langle\mathbcal{p}\right\rangle.
    \label{eq:bliokh49}
\end{equation}
For example, the superposition $\psi_{0,0}^{L} + a\, \psi_{1,0}^{L}$ of the Landau modes, shown in Fig.~\ref{fig:bliokh9}(a), has non-zero transverse momentum $\langle {\mathbcal p}_{\perp} \rangle$, shifted off-axis centroid $\langle {\bf r}_{\perp} \rangle$, and, hence, non-zero longitudinal component of the extrinsic OAM (\ref{eq:bliokh48}): $\langle {\mathcal L}^{\rm ext}_{z} \rangle = \left( \langle {\bf r}_{\perp} \rangle \times \langle {\mathbcal p}_{\perp} \rangle \right)_z$.
We note that the intrinsic-extrinsic separation (\ref{eq:bliokh48}) and properties (\ref{eq:bliokh49}) are generic and independent of the presence of a magnetic field.

\begin{figure}[t]
\centering
\includegraphics[width=0.6\linewidth]{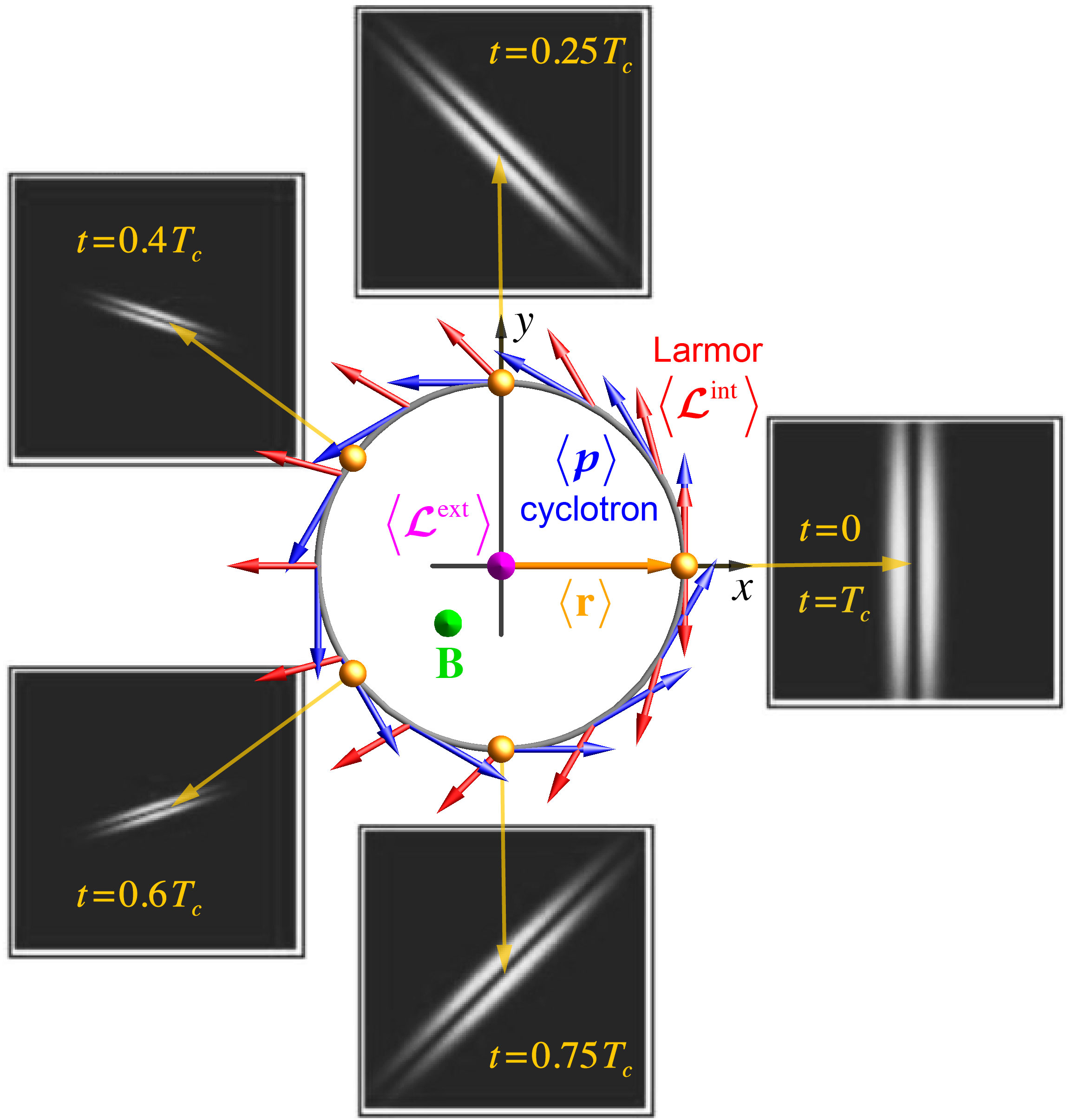}
\caption{Temporal evolution of an electron wavepacket with a vortex ($\ell =1$ here) in an {\it orthogonal} magnetic field ${\bf B}$ \cite{Gallatin2012}. The centroid of the wavepacket follows the classical cyclotron orbit in agreement with Eqs.~(\ref{eq:bliokh45}). Accordingly, the mean momentum $\langle {\mathbcal p} \rangle$ also undergoes the {\it cyclotron precession} (\ref{eq:cyclotronmotion}) with period $T_c = 2\pi / |\Omega_c|$. At the same time, the intrinsic OAM $\langle {\mathbcal L}^{\rm int} \rangle$ due to the vortex experiences the {\it Larmor (half-cyclotron) precession} (\ref{eq:larmorprecession}), cf. Fig.~\ref{fig:cyclotron}(b). Because of this, the vortex orientation rotates by an angle $\pi$ during the $2\pi$ cyclotron rotation of the electron, and the initially {\it parallel} $\langle {\mathbcal p} \rangle$ and $\langle {\mathbcal L}^{\rm int} \rangle$ become {\it anti-parallel} after one period $T_c$. The cyclotron motion of the electron centroid also produces an {\it extrinsic} OAM $\langle {\mathbcal L}^{\rm ext} \rangle$, Eq.~(\ref{eq:bliokh48}) parallel to the magnetic field (see explanations in the text).} 
\label{fig:gallatin}
\end{figure}

So far we considered only vortex beams propagating along the magnetic field or slightly tilted with respect to it. As the opposite limiting case, one can consider an electron vortex in the \emph{orthogonal} magnetic field. Such problem was analyzed in detail 
\cite{Gallatin2012} 
for paraxial electron \emph{wavepackets} with vortices. This yielded a remarkable example of the intrinsic (vortex) and extrinsic (centroid) evolution associated with the Larmor and cyclotron rotations, see Fig.~\ref{fig:gallatin}. Let the uniform magnetic field be still aligned with the  $z$-axis, ${\bf B} = B\, \bar{\bf z}$, whereas the vortex evolution occurs in the transverse  $\left(x,y\right)$-plane. For instance, let the vortex wavepacket be oriented along the  $x$--axis, with some initial momentum along this axis,
$\left\langle \mathbcal{p}(t = 0) \right\rangle \equiv
 \mathbcal{p}_0 \parallel \bar{\bf x}$.
Then, the wavepacket undergoes rotational evolution (in time) in the $(x,y)$-plane. Namely, in agreement with the Ehrenfest theorem, the centroid of the wavepacket follows the {\it cyclotron} orbit, Eq.~(\ref{eq:bliokh45}). At the same time, the orientation of the wavepacket, together with the vortex core and associated intrinsic OAM $\langle {\mathbcal L}^{\rm int} \rangle$, experiences the {\it Larmor} precession (\ref{eq:larmorprecession}) with half the cyclotron frequency. In doing so, the $\pi$-angle rotation of the wavepacket orientation (during the $2\pi$ rotation of its centroid) brings its probability density distribution back to the original distribution, but now with the intrinsic OAM pointing in the opposite direction, Fig.~\ref{fig:gallatin}. 

This example also provides a nice illustration of different types of the electron OAM: canonical, kinetic, intrinsic, and extrinsic. Let the wavepacket centroid lie in the $z=0$ plane: $\langle z \rangle = 0$. First, since the vector-potential does not have a $z$-component, $A_z = 0$, the in-plane OAM has a purely {\it canonical} origin (the ``diamagnetic angular momentum'' has only the $z$-component): $\langle {\mathbcal L}_{\perp} \rangle = \langle {\bf L}_{\perp} \rangle$ (here we keep the $\perp$ subscript to denote the $(x,y)$-plane). Second, the cyclotron motion of the centroid implies that $\langle {\mathcal p}_z \rangle = \langle p_z \rangle = 0$. It follows from here that the in-plane OAM also has purely {\it intrinsic} origin: $\langle {\mathbcal L}_{\perp} \rangle = \langle {\mathbcal L}^{\rm int} \rangle$ and $\langle {\mathbcal L}^{\rm ext}_{\perp} \rangle = 0$, Eqs.~(\ref{eq:bliokh48}). At the same time, the cyclotron motion of the electron centroid produces the $z$-directed {\it extrinsic} OAM: 
$\langle {\mathbcal L}^{\rm ext} \rangle = \langle {\bf r}_{\perp} \rangle \times \langle {\mathbcal p}_{\perp} \rangle = \sigma r_0 {\mathcal p}_0\, \bar{\bf z}$, where $r_0 = |\langle {\bf r}_{\perp} \rangle |$ is the radius of the cyclotron orbit, ${\mathcal p}_0 = |\langle {\mathbcal p}_{\perp} \rangle |$ is the absolute value of the kinetic momentum of the electron, and $\sigma = {\rm sgn}(B)$ is the direction of the magnetic field, Fig.~\ref{fig:gallatin}. This extrinsic OAM has both canonical and ``diamagnetic'' (vector-potential)  contributions because $\langle {\mathbcal p}_{\perp} \rangle \neq \langle {\bf p}_{\perp} \rangle$.

\subsection{Spin-orbit interaction phenomena}
\label{spinorbit}
\vspace{2mm}

Until now we considered electrons described by a scalar wave function $\psi$. However, real electrons are fermions, i.e., \emph{vector} particles with intrinsic \emph{spin} degrees of freedom. Since spin produces intrinsic angular momentum, it is interesting to consider its interplay with the OAM due to vortices. Here we only briefly describe spin properties of electrons, because most of electron-microscopy systems use unpolarized electron beams, i.e., essentially the scalar electrons considered above.

Spin is a fundamental relativistic property, and its self-consistent description requires the use of the Dirac equation rather than the Schr{\"o}dinger equation (\ref{eq:schro}) 
\cite{QEDbook,Thaller_book}. 
Proper consideration of vortex solutions of the Dirac equation will be given in Section~\ref{sect:highenergy}, and here we only list the main results following from the proper relativistic description of spin degrees of freedom of the electron 
\cite{Bliokh_Relativistic}. 
First, the Dirac electron wave function $\Psi$ has four components (it is a bi-spinor), and the spin operator is a $4\times4$ matrix, which acts on the components of this wave function \cite{QEDbook,Thaller_book}:
\begin{equation}
    \mathbf{\hat{S}} = \frac{\hbar}{2}
        \left(\begin{array}{cc}
        \bm{\hat{\sigma}} & 0 \\
        0 & \bm{\hat{\sigma}}
        \end{array} \right).
   \label{eq:bliokh50}
\end{equation}
Here $\bm{\hat{\sigma}}$ is the vector of $2 \times 2$ Pauli matrices. In the non-relativistic limit,
\footnote{Unike previous sections using nonrelativistic kinetic energy, in this section, we imply relativistic energy $E$, including the rest-mass contribution.}
$pc \ll E \simeq m_e c^2$, only the upper two components of the wave function, $\Psi^{+}$, play a role (the other two components describe positron states). In this case, the spin is described by Pauli matrices:
\begin{equation}
    \mathbf{\hat{s}} = \frac{\hbar}{2} \bm{\hat{\sigma}}.
    \label{eq:bliokh51}
\end{equation}

The canonical spin operators (\ref{eq:bliokh50}) and (\ref{eq:bliokh51}) seem to be completely independent of the spatial (orbital) degrees of freedom. However, the vector and spatial degrees of freedom are essentially coupled in the Dirac equation (where differential operators are multiplied by matrix operators), and, hence, in its spinor solutions ${\Psi}(\mathbf{r},t)$. Using a plane-wave solution of the Dirac equation, ${\Psi}_{\bf p}(\mathbf{r},t)$, with a well-defined momentum $\bf{p}$ and energy $E=\sqrt{p^2 c^2 + m_{e}^{2} c^4}$, the expectation value of the spin operator (\ref{eq:bliokh50}) becomes 
\cite{QEDbook,Thaller_book,Bliokh_Relativistic,Bliokh_2015}:
\begin{equation}
    \mathbf{S} = \frac{{\Psi}_{\bf p}^\dagger \mathbf{\hat{S}} {\Psi}_{\bf p}}{{\Psi}_{\bf p}^\dagger {\Psi}_{\bf p}}
    = \frac{m_e c^2}{E} \mathbf{s}+\left(1-\frac{m_e c^2}{E}\right) \frac{{\bf p}\,({\bf p} \cdot {\bf s})}{p^2}.
    \label{eq:bliokh52}
\end{equation}
Here $\bf{s}$ is the expectation value of the non-relativistic spin (\ref{eq:bliokh51}); it can be regarded as spin in the electron rest frame. Equation (\ref{eq:bliokh52}) clearly indicates a coupling between spin and momentum properties of the Dirac electron, i.e., the  \emph{spin-obit interaction} (SOI). While the non-relativistic spin $\bf{s}$ can have arbitrary direction, independently of the electron momentum, the relativistic spin has a $\bf{p}$-dependent correction. In the ultra-relativistic (or massless) limit $m_e c^2 / E \rightarrow 0$, the spin becomes ``enslaved'' by the momentum direction: $\mathbf{S \parallel p}$.

The relativistic SOI manifests itself even in the {\it free-space} solutions of the Dirac equation. For example, one can construct vortex Bessel-beam solutions of the Dirac equation \cite{Bliokh_Relativistic,Karlovets:2012}, 
i.e., vector analogues of the scalar Eq.~(\ref{eq:Bessel}) and Fig.~\ref{fig:Bessel}. We choose the non-relativistic spin to be parallel or anti-parallel to the propagation $z$-axis, $\mathbf{s} = s\, \mathbf{\bar{z}}$, $s=\pm 1/2$, whereas the spatial vortex properties are characterized by the vortex charge $\ell$ as well as the radial and longitudinal momentum components $p_\perp = \hbar\kappa$ and $p_z = \hbar k_z$. Calculating the expectation values of the spin and orbital AM (defined with the suitable spatial integration for wavepackets or beams), we obtain \cite{Bliokh_Relativistic}:
\begin{equation}
    \left\langle \mathbf{L} \right\rangle =  \hbar (\ell + \Lambda s) \mathbf{\bar{z}}, \qquad 
    \left\langle \mathbf{S} \right\rangle =  \hbar ( s   - \Lambda s) \mathbf{\bar{z}},
    \label{eq:bliokh53}
\end{equation}
where 
\begin{equation}
    \Lambda = \left(1-\frac{m_e c^2}{E}\right) \left(\frac{\kappa}{k}\right)^2
    \label{eq:bliokh54}
\end{equation}
is the dimensionless parameter, which determines the strength of the SOI effects. 

\begin{figure}[t]
\centering
\includegraphics[width=\linewidth]{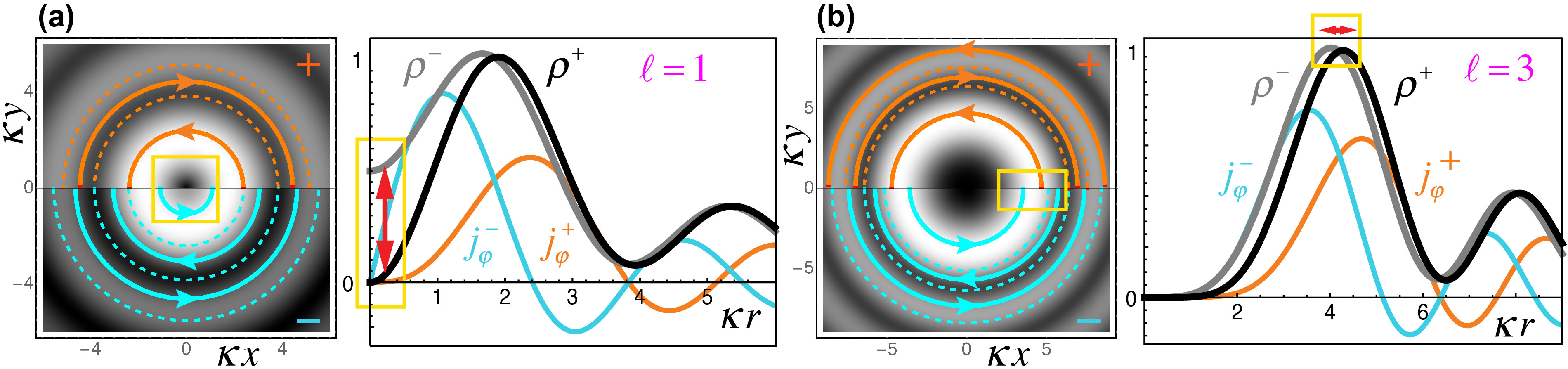}
\caption{Transverse spin-dependent probability-density and probability-current distributions in Bessel-beam states of relativistic Dirac electrons \cite{Bliokh_Relativistic} [cf., scalar non-relativistic Bessel beams in Fig.~\ref{fig:Bessel}(b)]. These distributions are shown for $\ell = 1$ (a) and $\ell = 3$ (b), for the opposite spin states: $s =\pm 1/2$ (marked by the ``$+$'' and ``$-$'' signs), see Eqs.~(\ref{eq:bliokh53})--(\ref{eq:bliokh55}). The parameters are: $p = \hbar k = 2.4\, m_e c$ and $\kappa = 0.7\, k$, i.e., $\Lambda \simeq 0.3$. One can see that the probability-density distributions (\ref{eq:bliokh55}) differ for the beams with parallel and anti-parallel SAM and OAM, i.e., $\ell\sigma > 0$ and $\ell\sigma < 0$. This signals the {\it spin-orbit interaction (SOI)} in free-space relativistic vortex electrons. In particular, states with $|\ell| =1$ and $\ell\sigma < 0$ have nonzero probability density in the vortex center $r=0$.}
\label{fig:bliokh13}
\end{figure}

Equations (\ref{eq:bliokh53}) demonstrate that the SAM and OAM of relativistic vortex solutions are inevitably coupled with each other, and part of the non-relativistic SAM is converted to the OAM. This phenomenon is called  \emph{spin-to-orbital AM conversion}, and it is well known in non-paraxial (e.g., focused or scattered) optical fields 
\cite{Dogariu2006,Zhao2007,Nieminen2008,Bliokh2010,Bliokh2011,Bliokh2015NP}.\footnote{Since photons are massless particles, relativistic SOI phenomena are inherent in optics \cite{Bliokh2015NP}. For optical couterparts of Eqs.~(\ref{eq:bliokh53})--(\ref{eq:bliokh55}) see \cite{Bliokh2010}.} 
This effect manifests itself in the spin-dependent spatial distributions of the probability density and current in the beams. In particular, the probability density distribution in the Dirac Bessel beams becomes (cf. Eq.~(\ref{eq:prob})) 
\cite{Bliokh_Relativistic}:
\begin{equation}
    \rho^B_{\ell,s}(r)\propto
    \left( 1-\frac{\Lambda}{2} \right) \left| J_{|\ell |} (\kappa r) \right|^2 +
    \frac{\Lambda}{2} \left| J_{\left|\ell+2s\right|} (\kappa r) \right|^2.
    \label{eq:bliokh55}
\end{equation}
Thus, the probability density distributions differ for the beams with the same $\ell$ and opposite $s$, as shown in Fig.~\ref{fig:bliokh13}. Namely, the beams with $\ell \sigma >0$ (parallel SAM and OAM) have larger radii as compared to the analogous beams with $\ell \sigma < 0$ (anti-parallel SAM and OAM) 
\cite{Bliokh2010,Bliokh_Relativistic}. 
The most interesting consequence of this is that for the $|\ell|=1$ vortex and anti-parallel spin $s=-\ell/2$ the probability density becomes finite even on the beam axis $r=0$, Fig.~\ref{fig:bliokh13}(a) (this effect is known and observed in optics 
\cite{Bokor2005,Gorodetski2008,Bliokh2011}).
The SOI properties of relativistic vortex electrons, described in \cite{Bliokh_Relativistic}, were recently confirmed in \cite{Bialynicki-Birula2017,Barnett2017}.
    
Equation (\ref{eq:bliokh54}) shows that this SOI effect has two independent sources of smallness: it becomes small (i) in the {\it non-relativistic} case $(E-m_e c^2) \ll E$ and (ii) in the {\it paraxial} case $\kappa \ll k$. In modern TEMs, electrons are often accelerated to relativistic energies, so that the first factor in (\ref{eq:bliokh54}) is not necessarily small. However, electron beams are always highly paraxial, and the strongest focusing currently possible in TEMs is lower than $\kappa/k \sim 10^{-1}$, i.e., $\Lambda \sim 10^{-2}$.
Therefore, the relativistic SOI effects are very difficult to observe in TEM experiments. 
Even the most noticeable effect of non-zero probability density in the center of a vortex mode with $s = -\ell/2$, Fig.~\ref{fig:bliokh13}(a), in practice becomes non-observable \cite{Boxem2013} due to decoherence effects related to the source-size broadening, which blurs the vortex core even in the non-relativistic scalar case \cite{Schattschneider2012b} (see details in Section~\ref{practical_issues}).

Dealing with the Dirac equation also allows rigorous calculations of the {\it magnetic moment} of relativistic vortex elecrons. For the Dirac--Bessel electron beams carrying SAM and OAM (\ref{eq:bliokh53}), this yields [cf. Eqs.~(\ref{eq:bliokh20}) and (\ref{eq:bliokh21})] 
\cite{Bliokh_Relativistic}:
\begin{equation}
     \mathbf{M} = 
    \frac{e\, c}{2 E}\left[ \left\langle \mathbf{L} \right\rangle + 2 \left\langle \mathbf{S} \right\rangle \right].
    \label{eq:bliokh56}
\end{equation}
This reflects the well known fact that the orbital and spin AM contribute to the magnetic moment with the $g=1$ and $g=2$ factors, respectively 
\cite{Cohen-Tannoudji}. 
Note that the magnetic moment (\ref{eq:bliokh56}) corresponds to {\it free-space} Dirac--Bessel solutions. In the presence of a magnetic field, one has to solve the relativistic Landau problem and find relativistic counterparts of Eqs.~(\ref{eq:37}) and (\ref{eq:bliokh38}) with the kinetic OAM \cite{Thaller_book,Kruining2017}.

Importantly, intrinsic relativistic spin-orbit interactions can be strongly enhanced in artificial structures. In optics, various anisotropic structures enable efficient manipulations of spin (polarization) degrees of freedom and couple these to the orbital properties of light 
\cite{Hasman2005,Marrucci2011,Brasselet2009,Bliokh2015NP}. Recently, there was a very interesting proposal to use similar inhomogeneous {\it magnetic} structures (in particular, Wien filters) for the spin-to-orbital angular momentum conversion and spin filtering of electron beams 
\cite{Karimi2012}. 
This can create a new platform for exploring electron SOI phenomena (which are so far mostly restricted to solid-state and condensed-matter electrons).  

\subsection{Electron-electron interactions}
\vspace{2mm}

One of the crucial differences between electrons and photons is that electrons are charged particles, and, therefore, can interact with each other even in free space. Since such interaction is of electromagnetic nature, the self-consistent description of the electron dynamics should involve the Maxwell equations for electromagnetic fields. One can describe the collective behaviour of electrons in a beam using a plasma model, which typically couples the classical equations of motion of electrons and Maxwell equations. However, in our case of coherent electron waves, we need to use the Schr{\"o}dinger wave equation (\ref{eq:schro}) (we again assume non-relativistic scalar electrons) instead of the classical equations of motion.

The most straightforward way to describe the ``electrons + fields'' system is to consider the electric charge and current distributions associated with the quantum probability distributions (\ref{eq:properties}):
\begin{equation}
    \rho_e=e|\psi|^2, \qquad
    \mathbf{j}_e=\frac{e \hbar}{m_e}\Im\left( \psi^{*} \nabla \psi \right).
    \label{eq:bliokh57}
\end{equation}
These electric charge and current densities are sources of electromagnetic fields according to Maxwell equations 
\cite{Jackson_book}:
\begin{align}
    & \nabla\cdot \mathbf{E} = 4 \pi \rho_e, \qquad
    \nabla \times \mathbf{E} = -\frac{1}{c}\frac{\partial \mathbf{B}}{\partial t},
    \nonumber \\
    & \nabla\cdot \mathbf{B} = 0, \qquad
    \nabla \times \mathbf{B} = -\frac{1}{c}\frac{\partial \mathbf{E}}{\partial t} + \frac{4 \pi}{c} \mathbf{j}_e.
    \label{eq:bliokh58}
\end{align}
In turn, electric and magnetic fields can be expressed via the electromagnetic potentials: ${\bf B} = \nabla\times\mathbf{A}$ and ${\bf E} = -\nabla V - c^{-1}\partial \mathbf{A}/ \partial t$, which enter the Schr\"odinger equation:
\begin{equation}
    i\hbar \frac{\partial \psi}{\partial t} - \left[ \frac{1}{2 m_e} \left(-i\hbar \nabla - \frac{e}{c} \mathbf{A}\right)^2 + eV \right] \psi = 0.
    \label{eq:bliokh59}
\end{equation}
Equations (\ref{eq:bliokh57})--(\ref{eq:bliokh59}) represent a complete self-consistent set for the electron wave function $\psi$ and electromagnetic potentials $\bf{A}$ and $V$. In particular, considering stationary solutions and neglecting magnetic interactions (which have additional relativistic smallness), the interaction is described via the scalar potential $V$, which fulfils the Poisson equation following from Eqs.~(\ref{eq:bliokh57}) and (\ref{eq:bliokh58}): $\Delta V = - 4 \pi e |\psi |^2$. Coupled to the Schr\"odinger equation (\ref{eq:bliokh59}) (with $\mathbf{A} = 0$), it describes the Coulomb interaction between electrons. As a result, the Schr\"odinger equation becomes effectively {\it nonlinear}. This may potentially lead to electron vortex solitons and other interesting nonlinear phenomena 
\cite{Desyatnikov2005}. 

Some examples of the electron-electron interactions in vortex beams were considered in \cite{Lloyd2012}. 
Moreover, the simplest consequences of Coulomb repulsion (``the space-charge effect''), such as additional defocussing of electron beams, are known in electron microscopy 
\cite{King2005,Zhu2015}. 
However, it should be noticed that the model of coupled Schr\"odinger--Maxwell equations (\ref{eq:bliokh57})--(\ref{eq:bliokh59}) has a significant drawback. Namely, it describes a non-zero interaction even for a {\it single} electron in a wavepacket or beam state described by the wave function $\psi$. But an electron cannot interact with itself, and only the interaction with other electrons makes sense \cite{Schattschneider2014comment}. 
Thus, a proper description of the electron-electron interactions in a beam should involve some pair characteristics and quantum {\it multi-body} methods. In particular, the interaction must depend on the average distance between individual electrons in the beam, transverse and longitudinal sizes of individual electron wavepackets, etc. Such accurate description of interacting electrons and fields remains a challenge. Recently, this problem was analysed using a Hartree--Fock approach in \cite{Mutzafi2015}, 
where non-diffracting vortex-beam solutions with balanced electron-electron interaction were found.

\clearpage

\section{Vortex beams in electron microscopy}
\vspace{1mm}

\subsection{Introduction}
\vspace{2mm}

Electron microscopes are popular instruments, used to characterize materials on the micro, nano and atomic scales. A typical electron microscope is designed to impinge a beam of accelerated electrons onto a sample to produce an image resulting from the interaction of the electrons with the material. Many types of interactions can be exploited, and the electrons can either be made to interact with the surface of materials in scanning electron microscopy (SEM) or to interact with the internal structure of a layer of the material, thin enough to allow electrons to pass through, in {\it transmission electron microscopes} (TEMs) \cite{DeGraefCTEM,WilliamsCarter,Reimer_book}. 

\begin{figure}[!h]
\centering
\includegraphics[width=0.64\columnwidth]{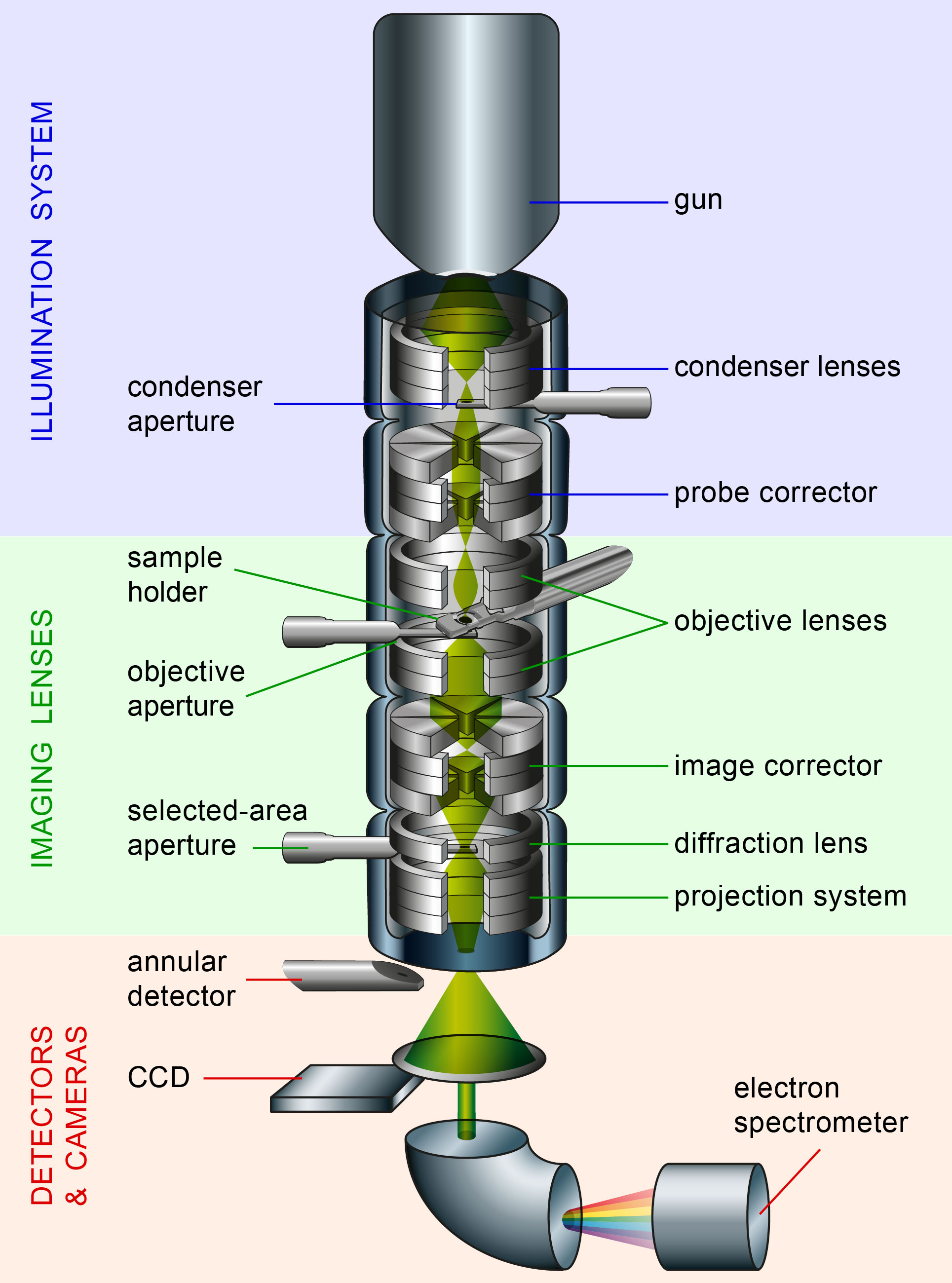}
\caption{Simplified schematic of a transmission electron microscope (TEM), indicating the essential planes and aperture positions that will be referred to when describing various experiments. The microscope can broadly be divided into an {\it illumination stage} and an {\it imaging stage}, each comprising several lenses and apertures, with a subsequent section containing the detectors and cameras used for the collection of  data.}
\label{microscopesketch}
\end{figure}

At present, the majority, if not all, of the research on electron vortex beams uses the TEM setup. In hindsight, this is partially a coincidence of suitable instruments being available, but if one realizes that a modern TEM is constructed to offer highly-coherent electron beams to obtain information about materials, one can see the close connection with laser optics --- a standard platform to explore optical vortex beams. Indeed, a schematic of a TEM (Fig.~\ref{microscopesketch}) shows that it can provide a fixed {\it optical bench for electron optics}, with a large number of adjustable magnetic lenses, the ability to insert custom-designed apertures, and choice of what is placed in the sample stage. 

Transmission electron microscopes are typically used in one of two main operating modes. The {\it conventional} TEM technique illuminates the sample with a broad, {\it planar} wavefront and either the image or far-field diffraction pattern is recorded, where all parts of the image are recorded simultaneously. Alternatively, the electron beam can be {\it focused} onto a small spot in the sample plane. Varying the position of such an electron ``{\it probe}'' generates the image of the sample formed in a raster scanning fashion. This technique is known as {\it scanning}-TEM (STEM). It enables one to gather additional information about the sample on a point-wise basis, e.g., by spectroscopic techniques
\cite{WilliamsCarter}.

In TEM and STEM, typical electron acceleration (kinetic) energies are $E \sim$ 80--300~keV, leading to de Broglie wavelengths $\lambda = 2\pi/k \sim$ 4--2~pm. However, unlike modern optical microscopes, electron microscopes are not wavelength-limited, but rather {\it aberration-limited} due to the severe spherical aberrations that are intrinsic to cylindrically-symmetric magnetic lenses \cite{ErniBook,Scherzer1936}. 
These spherical aberrations can be corrected with non-cylindrically-symmetric lenses, but at the expense of a vast increase in complexity, cost, and stability requirements. These aberration correctors can be placed in the illumination and/or imaging system of the microscope.
Nevertheless, severe higher-order aberrations remain, and the best resolution obtained in an electron microscope to date is around 40--50~pm, still much larger than the wavelength \cite{ErniPRL,batson2002,sawada_stem_2009}. 
Despite these limitations, electron microscopy offers one of the highest resolution imaging and spectroscopy techniques, with great flexibility to study various properties of a wide range of materials. 

Below we will discuss some of the ways in which electron optics of commercially available TEMs can be used to produce, measure and study electron vortex beams and their interaction with matter. We will review the generation of the electron vortex beams, the measurements of the OAM carried by such beams, and peculiarities of their interaction with materials. We will also focus on emerging applications and examine the potential of vortex electrons to provide novel characterization methods in TEM and beyond.

\subsection{Generation of electron vortex beams}
\label{sec:producing}
\vspace{2mm}

Since the first experimental demonstration of electron vortex beams in 2010
\cite{Uchida2010,Verbeeck2010}, a veritable zoo of methods for the production of electron beams with OAM has been developed. Some of these methods have achieved greater popularity, while others remain more exotic. Here we comprehensively review these methods, compare their efficiencies, OAM-mode purity, as well as experimental advantages and drawbacks. We will also discuss which methods are most appropriate for certain categories of experiment or prospective applications.

Several important points should be made about electron beams in TEMs. 
    First, electron beams are always highly {\it paraxial}. Even strongly-focused STEM probes are characterized by convergence angles $k_\perp /k \simeq \theta \sim 10^{-3}$--$10^{-2}$. Therefore, the paraxial approximation $k_z \simeq k$ is always justified in TEM.
    
Second, TEM electrons can achieve kinetic energies $E$ comparable with their rest-mass energy $m_e c^2$. Therefore, weak {\it relativistic} effects can become noticable. At the same time, the TEM electrons are {\it unpolarized}, and no observable spin effects occur. In practice, the analysis of Section~\ref{sec:basic-properties}, based on the non-relativistic Schr\"odinger equation and classical equations of motion, remains valid at TEM energies. The only noticable effects are corrections to the electron wavelength (momentum) from the relativistic dispersion $E=\sqrt{m^2_e c^4 + p^2c^4} - m_e c^2$ and the relativistic modification of the electron mass $m_e \to \gamma m_e$, where $\gamma = 1/\sqrt{1-v^2/c^2}$ is the Lorentz factor. 

    Third, due to instrumental factors, the wave beams produced in TEMs are {\it not} exactly Bessel or Gaussian beams described in Section~\ref{sec:beams}. Instead, their transverse Fourier spectrum is characterised by a uniform intensity for all wave vectors below a certain cutoff frequency $\kappa_{\rm max}$ (corresponding to a circular aperture): $\tilde{\psi}({\bf k}_\perp) \propto \Theta(\kappa_{\rm max} - k_\perp)$, where $\Theta$ is the Heaviside step function.
The real-space distribution $\psi({\bf r}_\perp)$ for the spectrum (\ref{eq:step-function}) is the well-known Airy disc \cite{Born1999}. 
Electron vortex beams are produced from the incoming beams (\ref{eq:step-function}), and, therefore, are characterized by a similarly abrupt Fourier spectrum with an additional azimuthal phase: 
\footnote{The actual spectrum of electron vortex beams can behave differently and vanish in a small area around the center $k_\perp =0$, but it is well approximated by a uniform amplitude in most of the circle $k_\perp < \kappa_{\rm max}$.}
\begin{equation}
\tilde{\psi}({\bf k}_\perp) \propto \Theta(\kappa_{\rm max} - k_\perp) \exp({i\ell\phi}).
\label{eq:step-function}
\end{equation}
Such circular-aperture beams look similar to the Laguerre--Gaussian beams (Fig.~\ref{fig:LG}) but they also exhibit additional rings (radial maxima) similar to those seen in Bessel beams (Fig.~\ref{fig:Bessel}), although with smaller quickly-decaying amplitudes (see the far-field patterns in Figs.~\ref{fig:5} and \ref{fig:holoreconstruct}).
One can design special holograms to generate beams approximating any desired radial distrubution, e.g., Bessel beams \cite{Grillo2014a}, but the natural and most efficient TEM beams have a radial spectrum corresponding to the uniformly-illuminated circular aperture.

We finally note that, similarly to optical systems, the Fourier transform corresponds to the transition from near-field to the far-field zone (e.g., in the diffraction from an aperture). Therefore, both the real-space [$\psi({\bf r})$] and Fourier [$\tilde{\psi}({\bf k})$] distributions mentioned above can appear as real-space distributions $\psi({\bf r})$ at different planes of the TEMs.

\subsubsection{Phase plates}
\label{sec:phaseplate}

When electrons travel through a thin material in the TEM, they experience a {\it phase shift} depending on the average electrostatic potential inside the material, which changes the electron momentum as compared to travelling in vacuum \cite{Reimer_book}. This global phase shift is only a zero-order effect, describing the interaction of the electron with the {\it space-averaged} potential $\bar{V}$ produced by the atoms making up the material. The actual microscopic electrostatic potential $V({\bf r})$ causes additional scattering effects, leading, e.g., to atomic-resolution images and a variety of dynamical Bragg scattering effects which are commonly exploited in the TEM to gain information about the sample. In the following, we neglect the microscopic-field effects, assuming that the TEM has been setup such that either the high-frequency atomic-field information is discarded (e.g., by using a restrictive objective aperture which acts as a low-pass filter) or averaged out by undersampling and/or averaging in the detection plane. In this case, the phase shift caused by a homogeneous material can be written as \cite{Reimer_book}:
\begin{equation}
\Delta \Phi({\bf r}_\perp)= C_E \int_{0}^{d({\bf r}_\perp)}\! V({\bf r})\, dz \equiv C_E\, \bar{V}({\bf r}_\perp)\, d({\bf r}_\perp) ,
\label{eq:phase-shift}
\end{equation} 
where $d$ is the thickness of the material (which can vary with the transverse coordinates ${\bf r}_\perp$), $\bar{V}$ is the mean inner potential (MIP) (which is assumed to be ${\bf r}_\perp$-independent for homogeneous materials), and $C_E = k\, e (E+E_0)/E(E+2 E_0)$ is the interaction constant. Here  $E_0 = m_e c^2$ is the electron rest-mass energy, and $E$ is the acceleration energy of the microscope. For example, $C_E=0.007288~{\rm V^{-1}nm^{-1}}$ for $E=200$~keV. For many materials, the MIP $\bar{V}$ is in the range of 5--30~V \cite{gajdardziska-josifovska_accurate_1993}. 
Then, for $E= 200$~keV, an amorphous \chem{SiO_2} layer of $d \simeq 85$~nm causes a phase shift (\ref{eq:phase-shift}) $\Delta \Phi = 2 \pi$. 

Similar phase shifts are produced in optics using transparent materials with refractive index $n \neq 1$. This allows one to shape the phase front of the transmitted wave with plates of space-varying thickness $d({\bf r}_\perp)$. In particular, photonic vortex beams were generated in optics, millimeter waves, and X-rays using {\it spiral phase plates} with azimuthally-varying thickness $d \propto \varphi$ corresponding to the phase shift $\Delta \Phi = 2 \pi \ell \varphi$ \cite{beijersbergen_helical-wavefront_1994,Turnbull1996,peele_observation_2002}. 
The same method can be employed for electrons, Fig.~\ref{fig:5}(a).

Historically, the first deliberately-made electron vortex beam was demonstrated by exploiting the fortuitous helical stacking of three graphite flakes \cite{Uchida2010}. Even though this setup is only a rudimentary approximation of a spiralling thickness profile, it was sufficient to demonstrate the presence of typical vortex characteristics. Progressing from this first demonstration, several researchers made attempts to perfect the setup in order to obtain high-purity electron vortex states \cite{Shiloh2014,Beche2015a}. In order to avoid directional scattering due to the Bragg diffraction, amorphous materials are preferred. Also preferrable are low-density materials with weak MIP, because the higher thickness needed to obtain a $2\pi$ phase shift requires less precise thickness control, and the discrete atomic nature or surface roughness of the material are less significant over the higher volume.

\begin{figure}[!t]
\begin{center}
\includegraphics[width=\columnwidth]{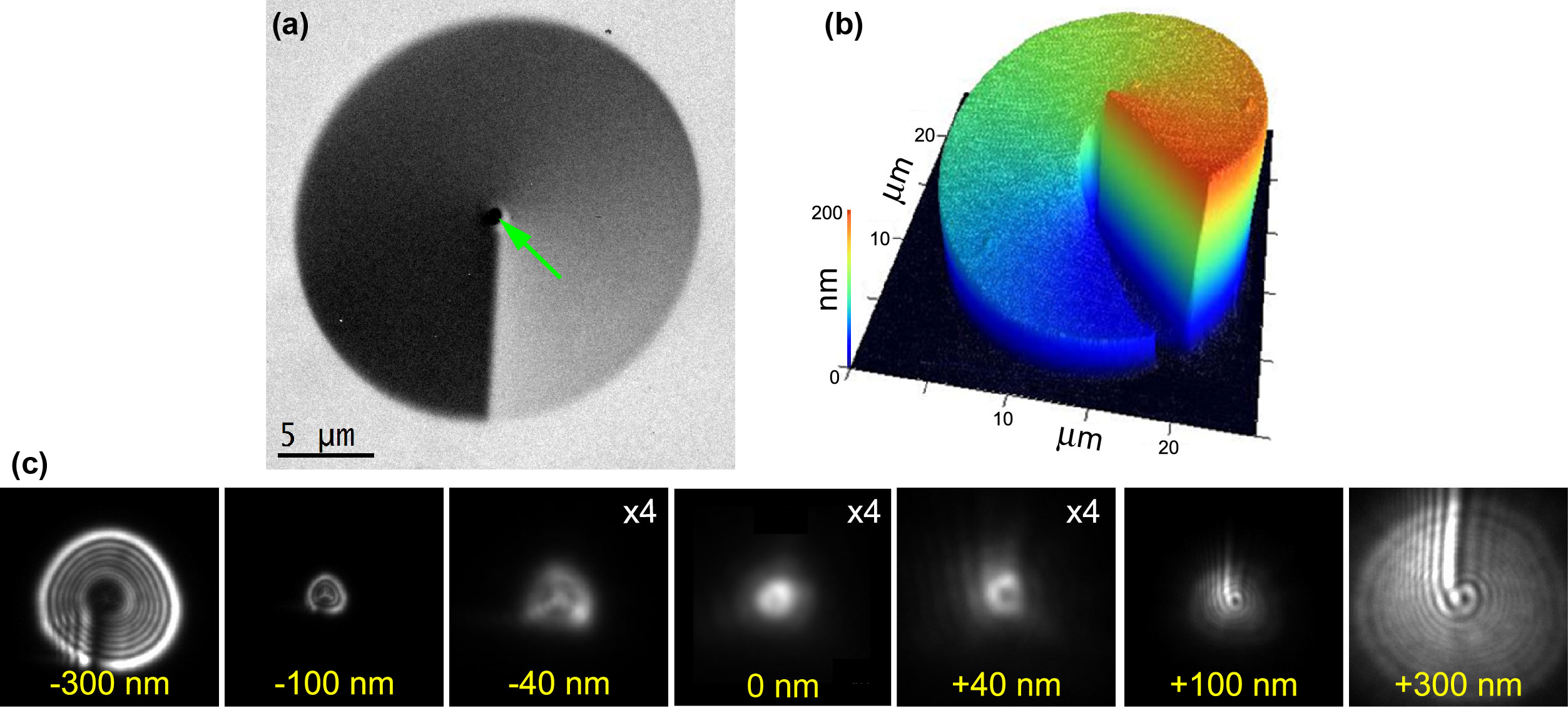}
\caption{Spiral phase plates for electron optics [see also Fig.~\ref{fig:5}(a)]. These devices exploit the internal potential of a material, which alters the momentum of the electron travelling inside the material, thereby imprinting a phase shift (\ref{eq:phase-shift}). 
(a) Experimental TEM image of a spiral phase plate, produced by the focused-electron-beam-induced deposition (FEBID) of \chem{SiO2} \cite{Beche2015a}. The black central spot is made of platinum and acts as a beam stopper (green arrow). 
(b) Atomic force microscopy height profile of the plate revealing its spiral shape. Since the material is uniform, the phase shift is proportional to the thickness. 
(c) A through-focus series of the transverse intensity distributions for the electron vortex beam produced by the spiral phase plate (i.e., the far-field pattern of the plate) \cite{Beche2015a}. A dark core caused by the destructive interference in the beam center does not vanish upon focusing or defocusing: a typical signature of a vortex beam.}
\label{Fig_Spiral}
\end{center}
\end{figure}

Focused ion beam (FIB) milling was used in \cite{Shiloh2014} to create an accurate spiral phase plate by milling away material from a commercially available \chem{SiN} grid. In addition to making high-purity vortex beams, the technique allows the fabrication of many different types of phase plates. Instead of milling material away, one can also use additive manufacturing offered by ion- or electron-beam-induced deposition of amorphous materials. In Ref. \cite{Beche2015a} the focused-electron-beam-induced deposition (FEBID) of \chem{SiO_2} on an ultrathin substrate of \chem{SiN} was used to create another version of the spiral phase plate, shown in Fig.~\ref{Fig_Spiral}(a,b). In order to remove a part of the beam that could pass through the hole in the center of the spiral, a small platinum beam stopper was deposited via FEBID. Since \chem{SiO_2} is an insulator, the phase plate had to be coated with a thin layer of carbon to prevent unwanted charging effects when used in the microscope.
The phase plate was then introduced in the sample plane of a TEM (operating in Lorentz mode).
\footnote{This technique requires a very high lateral coherence length, see Section~\ref{practical_issues}.}
Figure~\ref{Fig_Spiral}(c) shows a through-focus series of the focused electron beam (probe) in the far-field plane of the spiral phase plate. The characteristic destructive-interference area at the vortex-beam core is clearly visible and does not vanish upon focusing.

Even though the above spiral phase plates are highly versatile (the thickness profile can in principle encode for any desired phase profile), the technique has several significant drawbacks. Namely, charging and contamination need to be carefully avoided and the phase plate has to be designed for a {\it specific} acceleration voltage ($C_E$ depends on the acceleration energy $E$ and electron wave number $k$). The thickness of the phase plate also has to be carefully calibrated, since the MIP is not exactly known due to the less-than-perfect quality of the deposited or milled material. In practice, this means that the fabrication procedure has to go through several iterations before a reliable and predictable deposition procedure is obtained. Furthermore, the unavoidable electron scattering by microscopic fields inside the phase plate leads to imperfections, especially when aiming for atomic-size vortex beams.
\footnote{This is, however, not a physical limit but a technological one, as the effect could be counteracted with a set of limiting apertures to erase the high-frequency information, and demagnifying lenses to reduce the size of this purified vortex.} 
A potentially significant advantage of the phase-plate method is the fact that no magnetic materials are used, and the plate can work well even in the very high magnetic fields inside electron-optical lenses.

\subsubsection{Holographic gratings: binary and phase}
\label{gratings} 

The complexity of precise height control to create a spiral phase plate, as well
as the limitation of using only one acceleration voltage, can be lifted by employing {\it holographic gratings}, as in Fig.~\ref{fig:5}(b).
Such gratings with dislocations, generating different vortex beams in different diffraction orders, were first introduced in optics \cite{bazhenov1990laser,heckenberg_laser_1992}.

\begin{figure}[!t]
\centering
\includegraphics[width=0.87\textwidth]{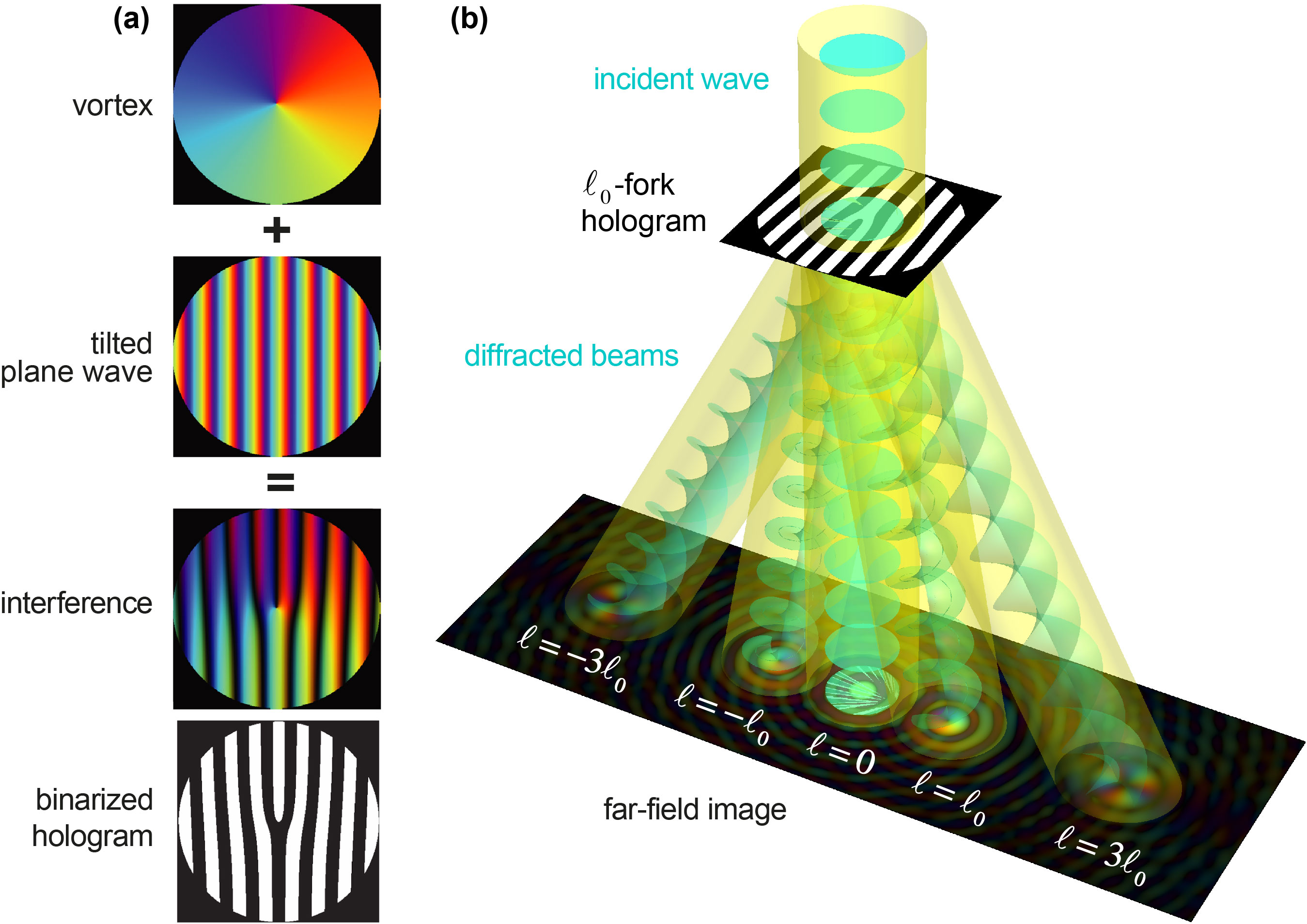}
\caption{Schematics of the holographic reconstruction concept, Eqs.~(\ref{eq:target})--(\ref{eq:binarized}), and generation of electron vortex beams using binarized fork holograms. (a) The interference of the target wave (vortex beam with $\ell_0=1$ here) with a reference wave (tilted plane wave) yields a characteristic interference pattern with an edge dislocation (fork) of order $\ell_0$. For practical reasons, this pattern is usually binarized, Eq.~(\ref{eq:binarized}). (b) The resulting pattern is then milled in a thin film that is placed as a holographic aperture in a TEM. When illuminated with a coherent electron beam, this aperture produces an array of vortex beams of order $\ell = N \ell_0$ in different diffraction orders $N=0,\pm 1,\pm 3,...$. For non-binarized holograms only the first orders $N=0,\pm 1$ are present, Fig.~\ref{fig:5}(b).}
\label{fig:holoreconstruct}
\end{figure}

To describe their holographic reconstructive action, let us start from the desired {\it target wave} (a vortex beam with topological charge $\ell_0$ in our case), which can be written as:
\begin{equation}
\psi_\mathrm{target}(r,\varphi,z) = a(r,z)\exp (i \ell_0 \varphi + i k z),
\label{eq:target}
\end{equation}
where $a(r,z)$ is the amplitude (assumed to be real-valued, for simplicity), and $k$ is the wave number of the paraxial beam. Consider the interference of this target wave with a {\it reference wave}, which can be chosen as, e.g., a plane-like wave slightly tilted in the $x$-direction:
\begin{equation}
\psi_\mathrm{ref}(r,\varphi,z) = a(r,z) \exp (i k_x x + i k z),
\label{eq:reference}
\end{equation}
where $k_x \ll k$, and for the sake of simplicity we assume the same amplitude $a(r,z)$ as in the target wave. The interference of the two waves (\ref{eq:target}) and (\ref{eq:reference}) leads to the following intensity pattern in the reference plane $z=0$:
\begin{align}
\rho_{\rm int}(r,\varphi) &= |\psi_\mathrm{target}(r,\varphi,0)+\psi_\mathrm{ref}(r,\varphi,0)|^2 = a^2(r,0) \left[1+\cos(\ell_0 \varphi-k_x r\cos\varphi) \right]. 
\label{eq:interference}
\end{align}
This equation describes a {\it fork-like} interference pattern with an {\it edge dislocation of the order} $\ell_0$ and grating period $x_g= {2 \pi}/{k_x}$, as shown in Fig.~\ref{fig:holoreconstruct}(a).

Now consider an aperture with a transmission function $T (r,\varphi)$ mimicking the interference pattern (\ref{eq:interference}) (in amplitude, phase, or both). Illuminating it with the untilted reference plane wave $\psi \propto \exp{(i k z)}$ allows the reconstruction of the tilted target vortex wave. Indeed, assuming an amplitude-modulating aperture with $T(r,\varphi) \propto \rho_{\rm int}(r,\varphi)$, the transmitted wave becomes:
\begin{align}
& \psi_\mathrm{reconstruct} (r,\varphi,z) = T(r,\varphi) \exp{(i k z)} \nonumber \\
& \propto a^2(r,0) \exp{(i k z)} \left[ \exp (i \ell_0 \varphi - i k_x x) + 2 + \exp(-i \ell_0 \varphi + i k_x x) \right].
\end{align}
Thus, the electron wave after interaction with such an aperture is a superposition
of: (i) the target wave, but now with the tilt, (ii) the untilted reference wave, and (iii) the complex conjugate of the target wave, tilted in the opposite direction, see Fig.~\ref{fig:5}(b). Moving away from the $z=0$ plane containing the aperture separates the three terms due to their tilts. The beams are typically observed in the far field ($z=\infty$) of the aperture plane (essentially, the Fourier transform of the aperture-plane waves), where the three components are well separated as long as their Fourier components contain spatial frequencies below $k_x/2$:
\begin{equation}
\tilde{\psi}_\mathrm{reconstruct}(\vt{k_\perp}) \, \propto\, \tilde{a}(\vt{k_\perp},0)* [ \tilde{\psi}_\mathrm{target}(\vt{k_\perp}+k_x\vt{\bar{x}})+
2\tilde{a}(\vt{k_\perp},0)+\tilde{\psi}_\mathrm{target}^*(\vt{k_\perp}-k_x\vt{\bar{x}})].
\end{equation}
where $\tilde{\psi}({\bf k}_\perp)$ and $\tilde{a}({\bf k}_\perp,0)$ are the Fourier transform of ${\psi}({\bf r}_\perp,0)$ and $a({\bf r}_\perp,0)$, while ``$*$'' denotes the convolution of functions.

A problem with the above reconstruction method is that currently it is near-impossible to make a smooth sinusoidally-modulating hologram for electrons. An often-used approximation is to {\it binarize} the sinusoidal pattern (\ref{eq:interference}) to either block ($T=0$) or transmit ($T=1$) the electrons. Mathematically, such binarized transmission function can be written as:
\begin{eqnarray}
T(r,\varphi) = \frac{1}{2} \Theta(r_{\rm max} - r) \left\{1+\mathrm{sgn}\left[\cos(\ell_0\varphi - k_x r \cos \varphi)\right]\right\},
\label{eq:binarized}
\end{eqnarray}
where we assumed that $a(r,0)=\Theta(r_{\rm max}-r)$, corresponding to the circular aperture and Eq.~(\ref{eq:step-function}), and the resulting binarized hologram is shown in Fig.~\ref{fig:holoreconstruct}(a).
This binarization leads to higher-order diffraction harmonics with increasing vortex charge $\ell = N\ell_0$ and tilt $N k_x$, where $N=\pm1, \pm3, \pm 5, ...$ \cite{Janicijevic2008, Topuzoski2011}, see Fig.~\ref{fig:holoreconstruct}(b). The intensity of the $N$th-order diffraction spot scales as $1/N^2$.
In some experiments using this technique, also faint even-order harmonics are present due to imperfections in the binary aperture \cite{McMorran2011,Yuan:2013}.
Note also that the above considerations on the intensity of the diffracted spots are true as long as the bars and slits in the grating have the same widths. The general case is similar to the well known problem of multiple-slit diffraction, and the relative intensities of different diffraction orders are determined by the ratio between the width of bars and slits \cite{Born1999}.

To achieve high values of the OAM ($\hbar\ell$ per one electron) carried by electron vortex beams, one should (i) use the grating with high-order dislocation $|\ell_0| \gg 1$ and (ii) consider higher diffraction orders with $|N| >1$. This provided the first demonstration of high-OAM electron vortex beams up to $\ell=100$ \cite{McMorran2011}, while currently similar methods reached $\ell=1000$ \cite{mafakheri_holograms_2015,Mafakheri2016}.
Such high-order vortex electrons carry huge magnetic moments, Eqs.~(\ref{eq:bliokh20}) and (\ref{eq:bliokh21}), exceeding the usual spin magnetic moment (Bohr magneton) by several orders of magnitude, which can be important in experiments involving magnetic interactions.

\begin{figure}[!t]
\centering
\includegraphics[width=0.75\textwidth]{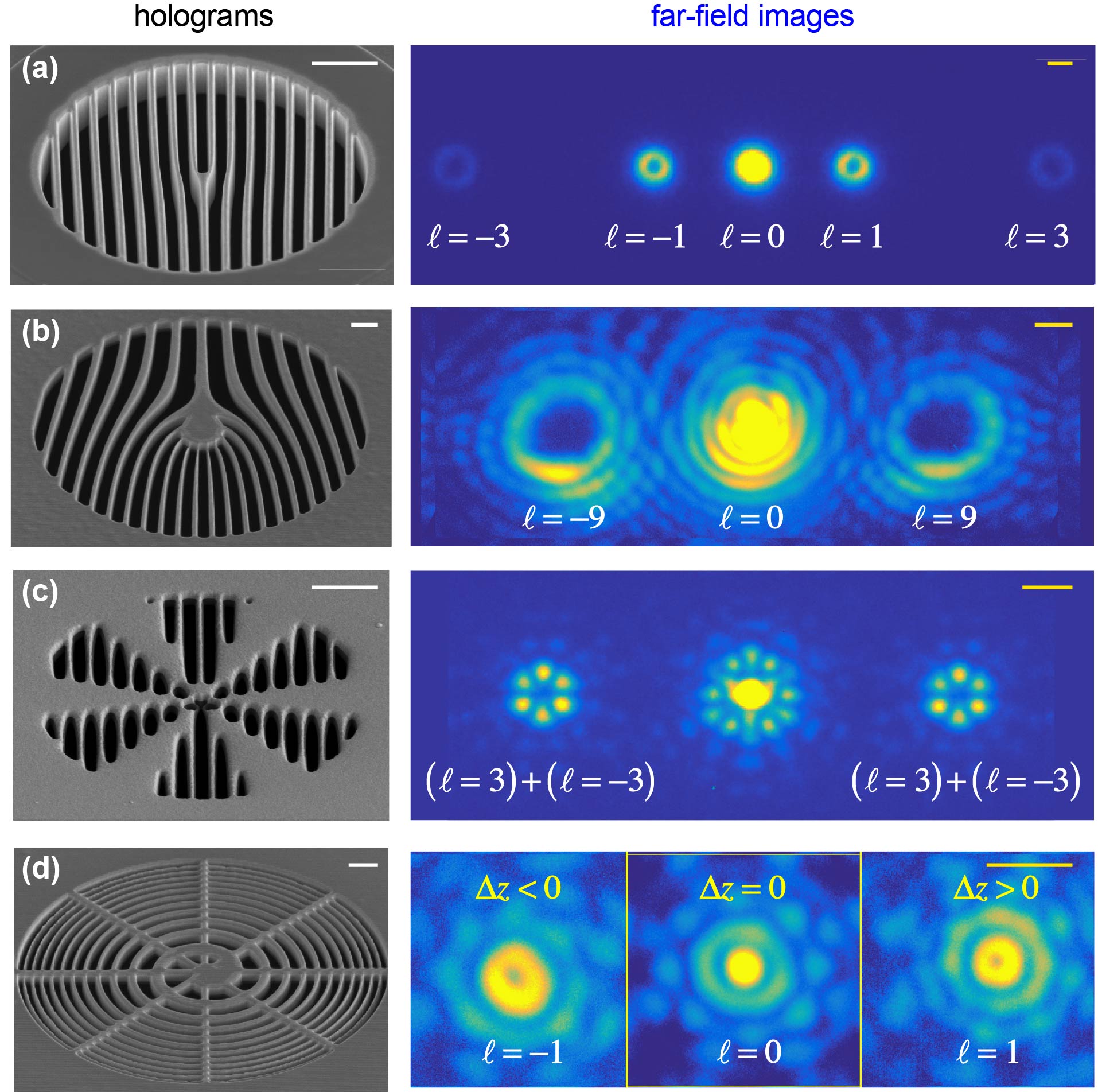}
\caption{Examples of vortex beams and other structured modes generated using holographic reconstructions. The SEM micrographs of the holograms are shown alongside the beams they produce in TEM. The $2\,\mathrm{\mu m}$ white scalebars and $10$~nm yellow scalebars are shown in the hologram and far-field beam images, respectively.
(a) A fork hologram with $\ell_0=1$. The $\ell=0$ beam is visible in the center, with the first sidebands with $\ell=\pm 1$, and the higher-order beams with $\ell=\pm 3$ at the borders of the image \cite{Verbeeck2014a}.
(b) A similar fork hologram with $\ell_0=9$, which generates $\ell=\pm 9$ vortex modes in the first diffraction orders \cite{Verbeeck2014a}.
(c) Hologram creating flower-shaped superpositions of $\ell=3$ and $\ell=-3$ vortex modes in the first diffractions orders \cite{Guzzinati2013}.
(d) Spiral hologram corresponding to the interference of the target vortex beam (with $\ell_0=1$) with a {\it spherical} reference wave \cite{Verbeeck2011a}. The generated probes with $\ell=1$ and $\ell=-1$ are separated {\it along the propagation $z$-axis}, i.e., appear at certain planes with defocusing $\Delta z > 0$ and $\Delta z < 0$.}
\label{fig:holography}
\end{figure}

A typical binary, amplitude modulating, holographic grating is shown in Fig.~\ref{fig:holography}(a).
The aperture has a typical diameter ranging between $10 \, \mathrm{\mu m}$
(for optical-bench-like experiments) to $30\, \mathrm{\mu m}$ (used to generate STEM probes), and is made out of a 0.5--1~$\mathrm{\mu m}$ thick sputtered gold film deposited on a commercial 200~nm thick \chem{Si_3 N_4} support film. The period of the bars, limited by the resolution reachable by the focused ion beam milling (typically of the order of the gold-film thickness), is about $x_g=500$~nm . Note that this period and the wavelength $\lambda = 2$~pm correspond to a very small diffraction
angle $\theta_d = k_x/k = 0.1$~mrad. However, the lenses between the condenser plane (where the apertures are typically placed 
\footnote{The aperture can also be placed in other aperture planes
like, e.g., the sample plane or the selected-area plane, as long as the lateral coherence length of the incoming electron beam extends over the whole diameter of the aperture in that plane.}) and the sample allow one to vary the magnification and separation of the generated beams
in the sample plane. Typically, in the STEM mode, the convergence angle of the probe is calibrated for a given radius of the aperture $r_{\rm max}$  as $\theta_{\rm max} = {r_{\rm max}}/{L}$, where $L$ is the equivalent camera length of the lens system in a given configuration.
Using a forked grating aperture of the same diameter produces probes which are
separated by the distance $L \theta_d$. Then, taking typical parameters $r_{\rm max}=15$~$\mu$m and $\theta_{\rm max}=20$~mrad, yields $L=0.75$~mm and probes  separated by $L\theta_d=75$~nm. 
This is a factor of $r_{\rm max}/x_g$ larger than the diffraction limit of the probes. A larger
separation can be achieved by altering the condenser lenses to obtain a higher effective $L$ but this at the same time sacrifices the size of the diffraction-limited probe by lowering the
convergence angle. In general, holographic apertures allow one to produce electron vortex beams, focused on the sample, ranging in size from atomic scale to hundreds of nanometers.

An example of a resulting far-field pattern for the hologram with $\ell_0=1$ is presented in Fig.~\ref{fig:holography}(a). The central beam is surrounded by two vortex beams with $\ell=\pm1$ and the 3rd-order harmonics with $\ell = \pm 3$ being considerably fainter.
Figure~\ref{fig:holography}(b) shows a similar hologram but with an $\ell_0=9$ dislocation \cite{Verbeeck2014a}. It generates $\ell = \pm 9$ vortex beams in the first diffraction orders.
Importantly, in addition to pure vortex beams, the holographic-reconstruction method allows the generation of {\it any structured modes}. For example, Fig.~\ref{fig:holography}(b) shows a holographic aperture and the corresponding far-field intensity distributions, which correspond to the superposition of $\ell=3$ and $\ell=-3$ vortex modes. Such superpositions were used in studies of the rotational dynamics of electron modes in a magnetic field \cite{Guzzinati2013,Bliokh2012e,Greenshields2012}, Fig.~\ref{fig:bliokh8}.

Alongside with these impressive demonstrations, the holographic-reconstruction method
has some shortcomings. First, the binary mask blocks a significant fraction of
the intensity (in our case 50\%), with the remaining part being subdivided between different diffraction orders. As a result, the first-order beam contains at most 10\% of the impinging intensity.
Second, multiple diffraction orders produce undesired beams which are focused on the sample at the same time. This leads to a complicated mixing of the OAM-dependent signals which cannot be easily untangled. However, this can be avoided by blocking the undesired beams with additional apertures \cite{Krivanek2014} or employing alternative mechanisms which are discussed below.

Instead of the plane wave (\ref{eq:reference}), one can use a {\it spherical} wave as a reference wave \cite{Heckenberg1992,Verbeeck2011a}. In this case, the corresponding hologram takes the form of a (spiral) zone plate, and the different diffracted orders are separated {\it along the $z$-axis}, i.e., can be brought into focus one at a time by varying the defocus $\Delta z$, Fig.~\ref{fig:holography}(d). This approach has been applied for the
production of vortex beams for STEM imaging, where only one vortex beam is in focus on the sample while the other beams generate an out-of-focus background~\cite{Verbeeck2011a}. An obvious drawback of this technique
is that all out-of-focus diffraction orders still remain present at all times. Therefore, an OAM-dependent interaction with the target can only be studied if its spatial scale is small enough to be neglected for the out-of-focus waves. Such focus-dependent vortex beams can be used for studying crystals, where the structure of the lattice can be probed with a single diffraction order, if the probe can be made small enough to resolve atomic distances.

So far, we considered only holograms, which modulate the {\it intensity} of the incident wave. Holograms fabricated with electron-transparent materials, such as carbon or silicon nitride, modulate the {\it phase} of the electron beam, similar to phase plates in Section~\ref{sec:phaseplate}. This approach has an advantage that the whole incoming beam is transmitted through the aperture, doubling the intensity with respect to the binary
amplitude grating approach. Such holograms also allow one to minimize or even cancel the intensity in the undesired reference part of the wave, thereby obtaining a higher fraction of the diffracted intensity in the desired first-order beams~\cite{Harvey2014}.
In addition, since this method uses thinner films and relies on structures with a depth of only several (or tens of) nanometers, the lateral resolution of the milling can be much higher, allowing finer and more sophisticated masks. Lastly, using a phase hologram with a {\it blazed} profile, one can enhance the fraction of the intensity directed in a single first-order diffracted beam (e.g., $N=1$) while suppressing the opposite ($N=-1$) beam~\cite{Grillo2014a,Harvey2014}. The experimentally-demonstrated efficiency of such blazed phase holograms is up to $40\%$ of the incoming beam ending up in the desired vortex state.

Phase holograms also have practical limitations. Carbon has a tendency to migrate under the beam, compromising the quality of the hologram and altering the phase shift, while the \chem{Si_3 N_4} films tend to charge due to the emission of secondary electrons. A gold or chromium coating with a thickness of few nanometers can be used to reduce the charging, at the cost of a lower diffraction efficiency. Furthermore, since the phase shift depends on the electron's energy, the performance of a phase-modulated mask depends on the acceleration voltage used, which is not the case for binary amplitude
holograms (with the exception that the positions of different diffraction orders depend on the electron wavelength).

As an alternative approach to holographic reconstruction -- holograms producing {\it on-axis} structured electron beams -- have also been demonstrated \cite{Shiloh2014}. Such  holograms are calculated from the target wave through an iterative Fourier-transform algorithm and are imprinted on a transparent film to obtain a pure phase mask. While this approach allows one to sculpt the intensity distribution of the beam with a high flexibility, its efficiency is low. Due to the impossibility of producing subwavelength structures for fast electrons at TEM energies, most of the wave intensity still forms an on-axis central spot, which is about 400 times more intense than the generated target wave. Speckling and multiple diffraction orders also occur in the pattern.

\subsubsection{Mode conversion using astigmatism} 
\label{EVastig}

An alternative method to produce electron vortex beams relies on earlier experiments in optics using the linear relation between the so-called Hermite--Gaussian (HG, which in the low-order HG$_{01}$ mode resembles p-orbitals) and Laguerre--Gaussian (LG) beams \cite{Allen1992a,beijersbergen_astigmatic_1993}. Indeed a proper linear combination of two HG beams can create an LG beam and vice versa. Starting from a single-mode laser that produces a HG beam, it is possible to create a vortex LG mode using a mode converter based on two {\it cylindrical lenses} \cite{beijersbergen_astigmatic_1993}. 

A similar approach can be applied to electrons \cite{Schattschneider2012a}. It starts with a phase plate which changes the phase of the electron wave by $\pi$ in a half-plane [e.g., $\Delta\Phi = \pi \Theta(x)$], generating a HG--like mode in the electron microscope. 
Then, exploiting the tuneability of {\it astigmatism} through quadrupolar lenses available in any TEM, one can introduce cylindrical distortions to electron optics to make it analogous to the optical setup with cylindrical lenses. The conversion of electron vortex beams into HG--like beams and vice versa was demonstrated in \cite{Schattschneider2012a}.

The advantage of this method is that almost all the electrons of the beam end up in the desired mode, and the sign of the OAM can be easily reversed using only tuneable lenses, which are part of any standard electron microscope. The drawback of the technique is that it
still uses a half-plane phase plate which suffers from charging and contamination and needs to be carefully tuned in thickness for each acceleration voltage. 

Recently an alternative setup producing HG-like beams was demonstrated using the Aharanov--Bohm effect around a {\it magnetized needle} (an analogue of the magnetic-flux line) that divides an aperture into two half-planes \cite{Guzzinati2016}. When carefully tuning the magnetic flux in the needle to an odd number of flux quanta, a phase difference between the two half planes becomes exactly $\pi$, independently of the acceleration voltage. This setup has an additional benefit: charging and contamination are strongly reduced, while only a small part of the beam is absorbed when hitting the needle. Combining this superior setup to generate HG beams in combination with two astigmatic lenses could be an attractive way to produce vortex electrons of high purity and intensity.

\subsubsection{Detuning the aberration corrector}

In visible-light optics, spatial light modulators allow flexible manipulations of both the phase and amplitude of the wave. Unsurprisingly, this tool has become indispensable for many tasks involving structured waves, such as optical vortex beams
\cite{andrews2012angular,curtis2002,grier2003revolution,kumar2013stability}. 
An electron analogue of spatial light modulators would be an ideal tool to deal with structured electron waves, but, unfortunately, it does not exist yet \cite{Guzzinati2015}.

In recent years, hardware {\it aberration correctors} have been developed \cite{rose2006aberration,haider2009current,hetherington2004aberration}, to counteract spherical aberrations which are intrinsic to cylindrically-symmetric magnetic lenses \cite{Scherzer1936}.
Aberration correctors consist of a sequence of adjustable multipolar lenses connected by transfer doublets. By modifying the relative strength of these magnetic multipoles, the electron wavepacket can be deformed, adjusting the lower-order Seidel aberrations of the lens system.
The aberration corrector is typically adjusted such that the electron beam impinging on the
sample approximates either a planar wavefront (for the TEM work), or a spherical wave focusing to a small probe (for the STEM imaging). Negative spherical aberration can also be used, to enhance contrast when imaging some challenging specimens \cite{JiaNegativeCs}.

Recently, it was shown that a more radical adjustment of the aberrations can be used to directly create electron vortex beams \cite{Clark2013}.
The vortex electron phase should exhibit the linear azimuthal depedence $\Phi( r,\varphi ) \simeq \ell \varphi$. 
However, the control software of the aberration corrector is designed to measure the wavefront of the beam in terms of the Seidel aberrations \cite{haider2008prerequisites}, none of which depends linearly on $\varphi$. 
Nonetheless, the $\Phi \simeq \ell \varphi$ dependence can be neatly expanded into a sawtooth Fourier series, with a period of $2 \pi$. Such an expansion reveals that the aberrations of the image shift, along with the $n$-fold orders of astigmatism, can be manipulated into a spiral phase structure, but only for a limited radial range.

The radial range of the beam can be easily limited by inserting an annular aperture into the
column. In doing so, the angular size of the aperture should be tuned to the desired mode purity (a broader annulus provides a less ideal vortex phase structure, but higher intensity) and to the obtainable aberrations in the microscope (the required astigmatism values vary with the aperture size).
In the FEI Titan$^3$ microscope used in \cite{Clark2013}, an annular aperture
provided an angular selection from $5.7$ to $8.3$~mrad, such that the required aberrations, up to three-fold astigmatism, stayed within obtainable limits. This resulted in a vortex beam with over $60 \%$ mode weighting in the $\ell=1$ state, and almost $50 \%$ of the initial beam intensity transmitted. 

The advantage of this method is its flexibility: the opposite-order vortex beam can be obtained by switching between aberration settings, as long as hysteresis and drift in the corrector are not too significant. The drawbacks are: the annular aperture severely limiting the total beam current, the non-atomic size of the vortex probe, and the less-than-perfect mode purity.

\subsubsection{Monopole-like field at the tip of a magnetic needle}

Close relations between {\it magnetic monopoles} and phase singularities in electron waves were first emphasized in the pioneering Dirac paper \cite{dirac1931quantised}. Later, this led to the suggestion that an electron plane wave interacting with a magnetic monopole can be converted into a vortex state \cite{fukuhara_electron_1983,tonomura_applications_1987,Bliokh2007}.
At first glance, this might seem to be a fantasy, given that no magnetic monopole has ever been observed in nature. 
However, recently it was shown that an {\it approximate} monopole,
manufactured ad-hoc, works very well in practice \cite{beche_magnetic_2014,Beche2016d}. 

\begin{figure}[!t]
\begin{center}
\includegraphics[width=0.95\columnwidth]{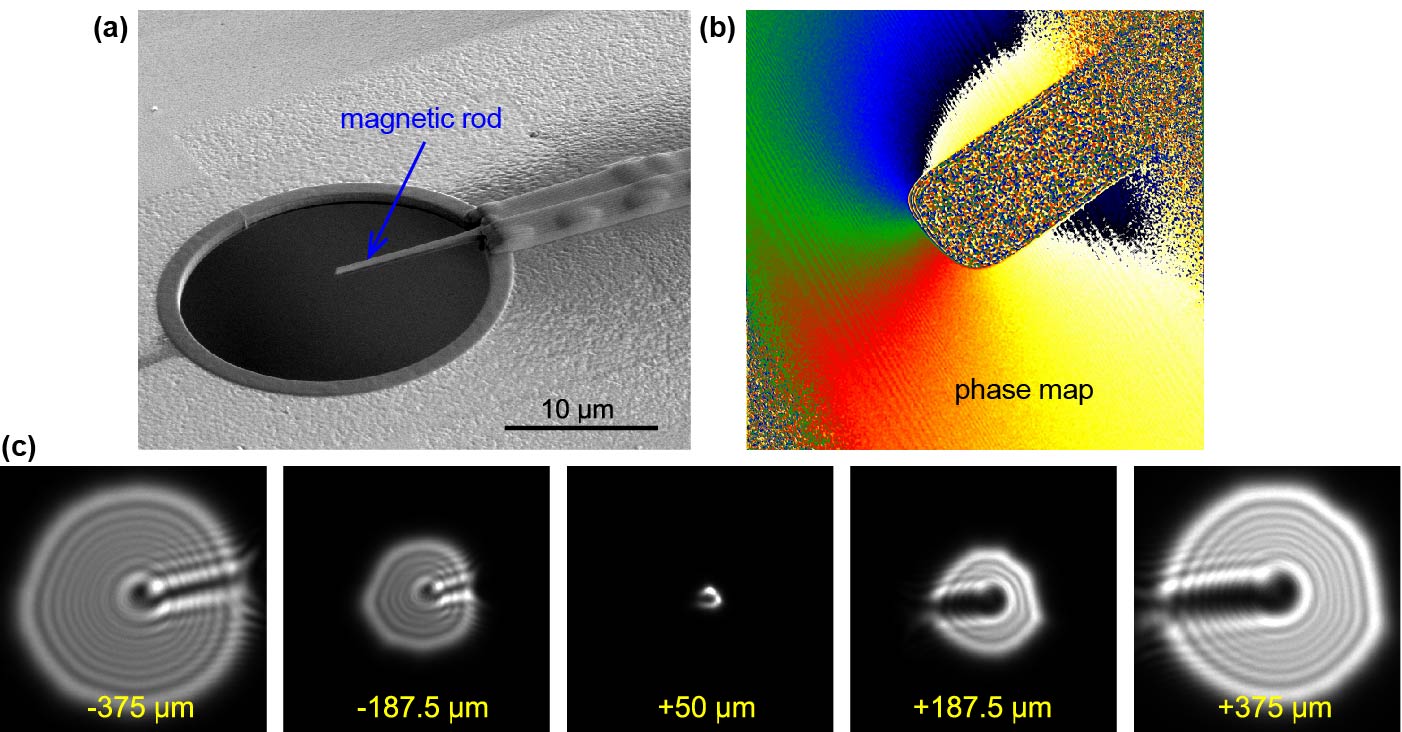}
\caption{Generation of electron vortex beams using the magnetic monopole-like field of a magnetized-needle tip \cite{Beche2016d}, see also Fig~\ref{fig:5}(c). (a) SEM view of a thin magnetic rod placed over a round aperture. (b) Color-coded phase shift around the tip of the needle measured and reconstructed through electron holography. (c) A through-focus series of the far-field images of the aperture illuminated with a plane electron wave. The dark spot  in the center indicates the vortex character of the transmitted beam, and the radial ``shadow'' from the magnetic needle is also seen.}
\label{Fig_Magnetic_needle}
\end{center}
\end{figure}

The experiments \cite{beche_magnetic_2014,Beche2016d} employed a {\it ferromagnetic needle}, magnetized along the needle's length in a single-domain magnetic state, and with a thickness chosen so that the magnetic flux through the section equals an integer number $\alpha_m$ of {\it double} magnetic-flux quanta. Then, the magnetic field around a tip of the needle approximates the monopole field, while the magnetic flux passing through the needle plays the role of the so-called Dirac string (ideally, an infinitely-thin magnetic-flux line) \cite{dirac1931quantised}. Placing such an approximate monopole in the center of the electron beam produces a transmitted vortex beam with topological charge $\ell = \alpha_m$, see Section~\ref{sec:basic-ways} and Fig.~\ref{fig:5}(c). Making use of an extra round aperture ensures that incoming electrons interact with the region close to the chosen tip of the needle, leaving the other tip of the needle (the opposite $-\alpha_m$ monopole) outside the electron beam, Fig.~\ref{Fig_Magnetic_needle}(a).

This provides a highly efficient method for the production of electron vortex beams, with several important advantages.
First, the phase profile is independent of the acceleration voltage. 
Second, the transmittance is near 100\% as the only part that absorbs electrons is the needle, whose ``shadow'' is small compared to the area of the aperture.
Third, the purity of the generated vortices can be very high, as long as the length of the needle is much larger than the aperture diameter. This assures that the magnetic field around the tip approximates well the monopole one.
Finally, the magnetization of the needle can in principle be reversed, allowing the dynamical control of the sign of the topological charge $\ell$.

The results of the most recent magnetic-monopole experiment \cite{Beche2016d} are presented in Fig.~\ref{Fig_Magnetic_needle}. A 60~nm thick pure nickel (\chem{Ni}) film was deposited on a silicon-nitride (\chem{Si_3N_4}) film and covered by a 1~$\mu$m sputtered gold layer. A focused ion beam was used to cut a very elongated bar out of this ferromagnetic film. Then, this needle was extracted from its original film and deposited on another gold-plated \chem{Si_3N_4} grid, with one of its ends positioned over the centre of a precut 20~$\mu$m round aperture, Fig.~\ref{Fig_Magnetic_needle}(a).
The needle was fabricated with a very high aspect ratio (300~nm wide by 50~$\mu$m long) in order to increase the shape anisotropy, which forces the needle to be in a single-domain magnetic state, with the magnetization along the needle axis. The cross-section of the needle, in combination with the saturation field of the material, was chosen to create a magnetic flux equalling two flux quanta, which corresponds to $\ell=\alpha_m=1$ and a $2\pi$ phase shift between electrons passing on the two opposite sides of the needle.

In practice, finding the exact width needed for the needle cross-section requires iteratively measuring the phase in a TEM with electron holography
[see Fig.~\ref{Fig_Magnetic_needle}(b)] and going back to the FIB.
The reason for this difficulty is likely caused by the FIB, where a fraction of the \chem{Ga} ions used for the milling ends up implanted in the \chem{Ni} film altering its magnetic properties and the thickness of the films used, which limits the milling resolution.
The lithographic production of highly controlled \chem{Ni} needles seems very attractive for future exploration.

Once this artificial ``magnetic monopole'' was fabricated, it was placed in the sample plane of the TEM (operating in Lorentz mode to avoid the presence of the strong magnetic field in the lenses), and a series of through-focus images was recorded in the far-field plane (i.e., in the diffraction mode). The results are shown in Fig.~\ref{Fig_Magnetic_needle}(c). One can easily recognize the typical doughnut-like profile of the transmitted vortex beam with a dark central region of the destructive interference. The electron hologram acquired at the tip of the magnetic needle, Fig.~\ref{Fig_Magnetic_needle}(b), also provides the measured phase shift. In this particular case, an interpolation of the phase over the whole aperture showed the expectation value of the OAM per electron to be $\langle L_z \rangle \simeq1.2\hbar$.
Placing this aperture in the condenser aperture plane allows the production of high-quality atomic-size focused vortex beams for STEM experiments \cite{Beche2016d}.

\subsubsection{Practical considerations and comparison of different methods} 
\label{practical_issues}

The application of the methods described above imposes several practical constraints to be kept in mind. Electron guns emit electrons from a small but {\it finite} area of a sharp tip. Electrons emanating from different positions are typically mutually-incoherent and the emitting tip behaves as an {\it extended incoherent source} \cite{Born1999,Dwyer2010}. 
Suppose that each individual electron can be considered as a perfectly coherent wavepacket. 
If the goal is to create the smallest possible electron vortex beam, one uses the setup employed to generate the STEM electron probe, where the sample plane is conjugated with the source plane.
This means that the ideal vortex probe formed by a single emitted electron will be superimposed incoherently with a large number of such vortices, each slightly displaced with respect to the other. This causes a {\it blurring} of the vortex intensity profile, leading to the appearance of a finite intensity in the dark $r=0$ core of the vortex \cite{Schattschneider2012b}.

This effect can be reduced by demagnifying the source with the condenser lenses either by changing the experimental ``spot size'' (lens settings) and/or by varying the strength of the ``gun lens'', depending on the exact technical characteristics of the microscope.
In any case, this demagnification will always require a trade off in intensity unless a brighter electron source can be chosen. For this reason, the field emission guns are currently preferred, but work is underway to develop superior options \cite{houdellier_development_2015}. Current probe-corrected microscopes tend
to have a source-size broadening of the order of the smallest probe achievable on that instrument. Indeed, as far as the probe size is concerned, the source-size broadening does
not considerably deteriorate the spatial resolution (for STEM imaging), as long as it remains smaller than the diffraction limit imposed by the aberrations. However, for vortex beams, the same effect destroys the vortex by blurring the dark core, and the source-size broadening becomes a limiting factor.
For larger-diameter electron vortices, the source-size broadening becomes insignificant and
near-perfect coherence can be achieved.

If the aperture is placed in either the sample plane or the selected-area plane, the situation is different and the most important factor becomes how {\it paraxial} the incoming beam can be made. Indeed the spatial coherence in the aperture plane needs to extend over the size of the aperture, which is inversely proportional to the angular spread of the beam. In the sample plane, this requires extremely small angular spread or very small aperture sizes. Typically, the low-magnification modes in TEMs can reach spatial coherence covering a 50~$\mu$m aperture. This allows one to test the apertures in the sample plane, which provides higher flexibility and speed in terms of transferring the aperture in and out of the vacuum.

Next, the {\it material} of the apertures is important for the electron energy loss spectroscopy (EELS). Indeed, fast electrons excite transitions in the infrared to ultraviolet range in the material, experiencing the corresponding energy loss.
These losses can be significant for trajectories of electrons close to the edge of the aperture (interacting with the metallic film) or travelling through the light phase-plate material. This generates a background spectrum exclusively due to the aperture (not the sample), with spectral features depending on the material chosen and mixing with the actual EELS spectrum of the sample. This effect can be particularly noticable for the MIP-based phase plates and fork binary holograms, while the magnetic-needle setup is preferred in this respect as only a small amount of material is present in the aperture.

{\it Charging} is another factor which complicates the design of phase apertures. Secondary-electron emission typically causes the aperture to charge positively, if there is no way for an electric current to flow from the aperture to ground, replenishing the lost electrons.
Metallic materials are preferred but are often crystalline, which leads to unwanted Bragg diffraction, and are generally heavier, which leads to higher scattering.
Amorphous insulating materials can be used, but coating with a thin layer of metal can solve or reduce this issue. Different work functions can also make a difference in the secondary-electron emission and, hence, should also be considered.

To compare different methods of the generation of electron vortex beams, we summarize their main features in Table~\ref{tab:creationmethods}. Undoubtedly, other methods for electron vortex production can be envisioned, and this list is likely to expand in the future.

\begin{table}[h]
\centering
\begin{tabular}{l l l l l}

\hline
\parbox[c]{1.8cm}{{\bf Method} } &  \parbox[c]{2.5cm}{\vspace{5mm} {\bf Efficiency \\ theory/exp.} \vspace{1mm}} & \parbox[c]{4.0cm}{\bf Advantages} & 
\parbox[c]{4.2cm}{\bf Disadvantages} & \parbox[c]{1.8cm}{\bf Papers} \\
\hline
\noalign{\vskip 2mm}

\parbox[c]{1.8cm}{Holograms (Binary)} & $\sim$10\% / $\sim$10\% & \parbox[c]{4.0cm}{Versatile, straightforward,\\ high quality EVBs} &
\parbox[c]{4.2cm}{\vspace{3mm} Inefficient, multiple \\ diffracted beams \vspace{3mm}} &
\parbox[c]{2.0cm}{\centering \cite{Verbeeck2010, McMorran2011, Harvey2014}, \\ \cite{Grillo2014,VolochBloch2013}, \\ \cite{Verbeeck2011a,Verbeeck2011, Krivanek2014} }\\
        
\parbox[c]{1.8cm}{Holograms (Phase)} & 100\% / $\sim$40\% & \parbox[c]{4.0cm}{Efficient, versatile, \\ high quality EVBs }&
        \parbox[c]{4.2cm}{\vspace{3mm} Charging, multiple \\ diffracted beams, apertures are specific to a given value of kinetic energy \vspace{3mm}} &
        \parbox[c]{1.8cm}{\centering \cite{Grillo2014,Harvey2014} \cite{Shiloh2014,Grillo2014a} }\\

\parbox[c]{1.8cm}{MIP phase plate} & 100\% / $\sim$55\% & \parbox[c]{4.0cm}{Efficient, single output \\ beam} & \parbox[c]{4.2cm}{\vspace{3mm} Charging, difficult fabrication, aperture is specific to a given kinetic energy \vspace{3mm}} &
\parbox[c]{1.8cm}{\centering \cite{Shiloh2014,Beche2015a}} \\

\parbox[c]{1.8cm}{Magnetic phase plate }& $>$95\% / $\sim$92\%  & \parbox[c]{4.0cm}{Efficient, single output \\ beam, independent of \\ kinetic energy } &
\parbox[c]{4.2cm}{\vspace{3mm} Difficult fabrication, cannot be used inside a strong magnetic field such as the objective back focal plane\vspace{3mm}} &
\cite{beche_magnetic_2014,Blackburn2014,Beche2016d} \\
        
\parbox[c]{1.8cm}{Aberration tuning }& $\sim$50\% / $\sim$32\% & \parbox[c]{4.0cm}{Single output beam, \\ more efficient than \\ binary holograms} & \parbox[c]{4.2cm}{\vspace{3mm} Limited by aberration \\ correction technology \vspace{3mm}}&
\parbox[c]{1.8cm}{\centering \cite{Clark2013,Guzzinati2015}} \\
        
\parbox[c]{1.8cm}{Astigmatic mode \\ conversion}& 100\% / N/A & \parbox[c]{4.0cm}{Efficient, single output \\ beam, tuneable in real \\ time} & \parbox[c]{4.2cm}{\vspace{3mm} Charging, apertures are \\ specific to a given value of \\ kinetic energy \vspace{3mm}}& \parbox[c]{1.8cm}{\centering \cite{Schattschneider2012a,Guzzinati2016}} \\

\noalign{\vskip 2mm}
\hline

\end{tabular}
\caption{Summary of different methods for the generation of electron vortex beams. 
Presented are: theoretically expected and experimentally achieved efficiencies (defined as the fraction of the impinging intensity to end up in the desired vortex state), the main advantages and drawbacks, and relevant bibliographic references.}
\label{tab:creationmethods}
\end{table}

\subsection{Measurements of the orbital angular momentum of electron beams}
\label{sec:oammeasure}
\vspace{2mm}

\subsubsection{General problem}

In many cases, in addition to the production of electron vortex beams, one also needs to measure the OAM $\langle L_z \rangle$ carried by the beam. It is important to remember that the result of such measurements depend, in the generic case, on the position of the $z$-axis with respect to the electron beam.
So far, we mostly assumed cylindrical symmetry of the electron vortex beams,
but real-life conditions often deviate from this symmetry, e.g., due to the occurence of optical aberrations or scattering by non-cylindrically-symmetric objects. In this case, the
direct proportionality between the topological charge $\ell$ and OAM breaks down \cite{ONeil2002}, meaning that measuring the topological charge is not generically equivalent to measuring the OAM \cite{amaral_characterization_2014}.

At the quantum-mechanical level, the $z$-component of the OAM of a paraxial wave can be calculated by integrating the OAM density, proportional to $r j_{\varphi}$, Eqs.~(\ref{eq:expect}) and (\ref{eq:expect2}).
Therefore, an ideal method should be able to measure this quantity independently of the radial component of the current, $j_r$, and independently of the shape of the wave. As we will see, all the methods developed so far fall short of this requirement. 
Note that wave propagation, whether free or operated by lenses, is intimately linked to the Fourier transform and, as such, allows the efficient mapping of the transverse momentum components ${\bf p}_{\perp}$ into coordinates ${\bf r}_{\perp}$. This simple relation lies as the heart of diffractive techniques, that reveal the linear momentum spectrum of the outgoing wave in the diffraction pattern, as used, e.g., in TEM studies of reciprocal crystal lattices. 
If a similar process could be devised for the longitudinal OAM, it would facilitate studies of a variety of phenomena.

An efficient far-field sorting of different OAM states was suggested in optics \cite{Berkhout2010}. By employing a transverse-coordinate transformation ${\bf r}_{\perp} \to {\bf r}^{\prime}_{\perp}$ that converts every $r={\rm const}$ circle into a straight line (e.g., $y^\prime = {\rm const}$), it is in principle possible to transform the $2\pi\ell$ azimuthal phase into a linear phase ramp, thereby causing a shift in the far-field intensity proportional to $\ell$. Mathematically this can be done using a conformal map mimicking the transition from Cartesian to log-polar coordinates:
\begin{equation}
x^{\prime} =  - a \arctan\left(\frac{y}{x}\right), \qquad y^{\prime} = -a \ln \left(\frac{\sqrt{x^2+y^2}}{b} \right),
\end{equation}
where $a$ and $b$ are constant parameters.
Such conformal mapping can be realized via ad-hoc phase plates, and efficient OAM sorting was demonstrated for optical beams \cite{Berkhout2010,Mirhosseini2013}. However, the limitations of the phase-plate technology for electrons hampered the adoption of this method in the TEM.
Only recently the first demonstration of this method has been realised using holographic gratings (with all the ensuing limitations in terms of efficiency) \cite{grillo2016measuring}. There were also proposals to employ more efficient electrostatic optical components, but these have yet to be realised experimentally \cite{mcmorran2016efficient}.

Several other approaches to OAM measurement have been developed in optics, with sensitivity up to the single-photon level \cite{Leach2002,Lavery2011b}. However, these methods cannot be directly adopted by electron microscopy due to the limited flexibility of electron-optical elements (e.g., because of the absence of beam splitters for electrons). Therefore, the techniques that have so far been demonstrated with electrons offer a much lower level of generality and/or detection efficiency than those available for photons \cite{Guzzinati2014,grillo2016measuring}.

\begin{figure}[!t]
\centering
\includegraphics[width=0.85\textwidth]{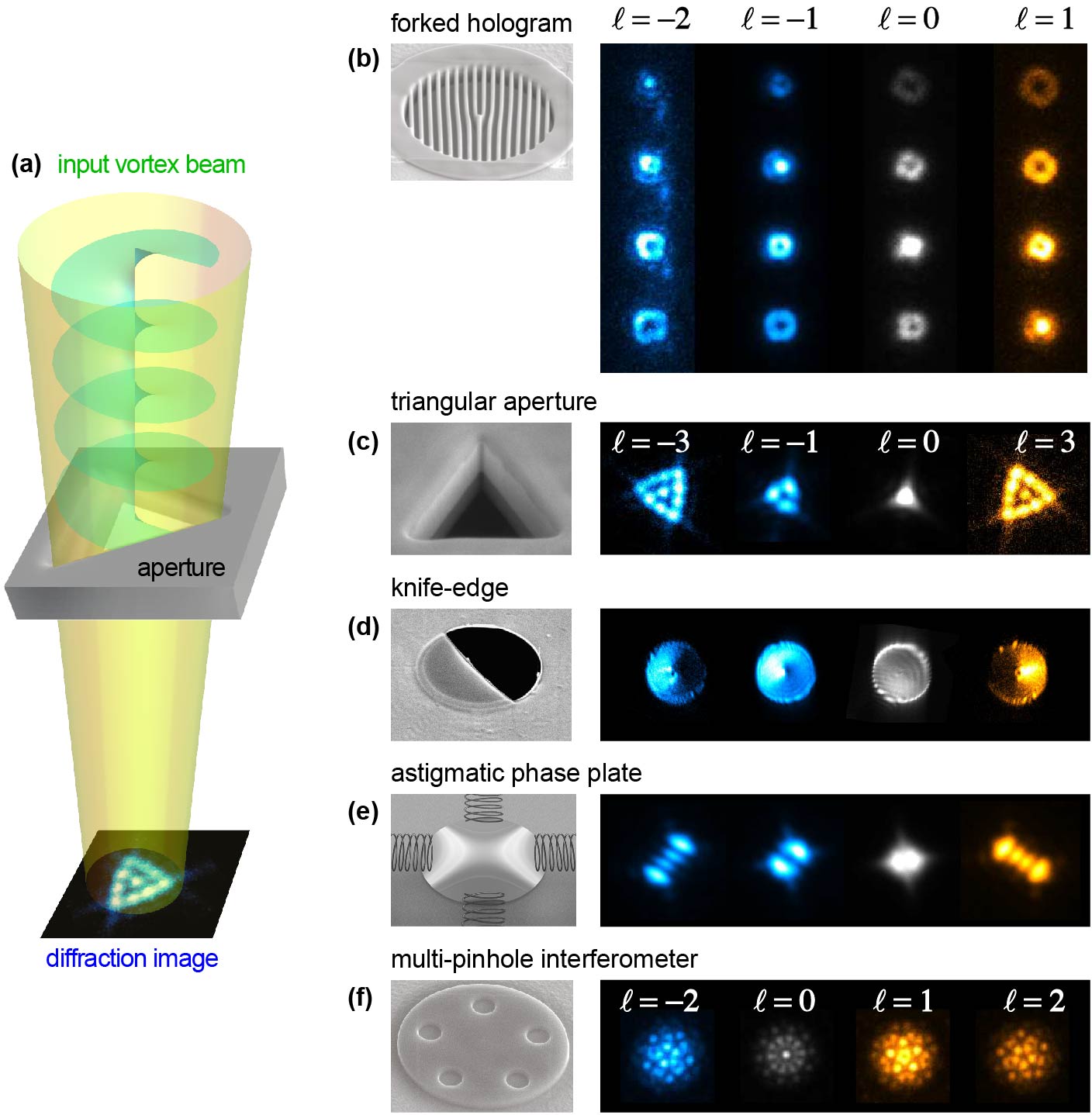}
\caption{Diffraction-based methods for measurements of the topological charges of electron vortex beams \cite{Guzzinati2014,Clark2014}. (a) Schematic diagram of the main idea. The vortex beam impinges on a specially-shaped aperture (or a phase plate) producing the far-field diffraction pattern which reveals the value of $\ell$. In (b--f) the patterns produced by different elements are shown, with the yellow and blue far-field images corresponding to the $\ell>0$ and $\ell<0$ beams. 
The fork hologram (b) reveals the topological charge of the beam by observing which diffraction order does not have a doughnut profile. The triangular aperture (c) yields a triangular lattice of spots where the number of spots along one side is $\ell+1$. The knife-edge aperture (d) causes a transverse deformation and shift of the beam intensity that depends on the sign and value of the OAM. The astigmatic phase plate (e) also causes a multi-lobed pattern where the number of lobes is proportional to $\ell+1$. The multi-pihole interferometer (f) produces patterns which are related to the topological charge of the original wave function. Furthermore, the autocorrelation function of these patterns allows one to obtain an approximate OAM spectrum of the incoming wave \cite{Clark2014}.}
\label{fig:OAM_measurement}
\end{figure}

\subsubsection{Diffraction from holograms and apertures}

The first measurement of the topological charge has been demonstrated by projecting a vortex beam on a {\it fork-grating hologram} \cite{Saitoh2013,Guzzinati2014}, Fig.~\ref{fig:OAM_measurement}(a,b). The far-field pattern generated by the hologram in this condition is similar to the one obtained for an impinging plane wave (Fig.~\ref{fig:holoreconstruct}) and is formed by a one-dimensional array of vortex beams. The relative intensities of the diffracted beams are, to a good approximation, unaltered \cite{Topuzoski2011,Saitoh2013}, while their topological charges $\ell^\prime$ are now given by $\ell^\prime = N \ell_0 +\ell$ where $\ell_0$ is the dislocation order of the fork and $\ell$ is the topological charge of the impinging beam.
In particular, a diffracted beam that satisfies $N \ell_0 = -\ell$ will be a non-vortex $\ell^\prime=0$ mode with the maximum intensity in the center. This allows the determination of the topological charge of the impinging beam via the dislocation charge of the fork hologram, $\ell_0$, and the diffraction order $N$ of the beam without the typical doughnut-like intensity distribution Fig.~\ref{fig:OAM_measurement}(b). A sufficiently small pinhole centered on a diffracted spot can readily discriminate between the vortex and non-vortex modes and detect the $\ell^\prime=0$ component with the maximal intensity in the center. This method has been suggested as possible basis for automated measurements of the topological charge of the impinging beam using a set of appropriately-placed pinholes. Such measurements, however, would be highly inefficient as most of the beam intensity would be discarded due to the limited transparency of the mask, division of the intensity into several diffraction orders, and the subsequent limited transmittance of the pinhole \cite{Guzzinati2014}.

The topological charge of a vortex beam can also be revealed by diffraction from
{\it geometric apertures}. The common feature of these
methods is that the diffraction pattern given by the geometrical aperture is altered by the unique phase profile of the vortex beam in a way that reveals the topological charge. For example, a triangular aperture generates a triangular lattice of spots, with the side of the triangle having $| \ell | + 1$ spots and the orientation of the triangle depending on ${\rm sgn}(\ell)$ via the Gouy phase \cite{Guzzinati2014}, see Fig.~\ref{fig:OAM_measurement}(c). While such geometric apertures have not been widely used to measure the OAM of
electron vortex beams, the diffraction patterns reveal a variety of remarkable physical phenomena \cite{Clark2016}.

A particular case of geometric aperture is a {\it knife-edge aperture} that blocks half (or a finite segment) of the vortex beam at its waist, Fig.~\ref{fig:OAM_measurement}(d). This produces a C-shaped beam, which rotates, upon propagation to the far field, as an effect of the Gouy phase, in the direction determined by $\mathrm{sgn}(\ell)$ \cite{Arlt2003,Hamazaki2006,Guzzinati2013}. Importantly, this effect can be regarded as a direct manifestation of the azimuthal probability current $j_\varphi$ in vortex beams. Indeed, the spiraling current density in a cylindrically-symmetric vortex beam produces zero transverse momentum: $\langle {\bf p}_{\perp} \rangle = 0$, Eq.~(\ref{momentum}). Blocking part of the beam breaks this symmetry, and the resulting C-shaped beam acquires non-zero transverse momentum $\langle {\bf p}_{\perp} \rangle \neq 0$, which leads, upon propagation, to the transverse shift of the diffracted-beam centroid $\langle {\bf r}_{\perp} \rangle$ \cite{Bliokh2010}, see Fig.~\ref{fig:OAM_measurement}(d). 
Variations of this method have been employed to confirm the vortex character
of doughnut-shaped beams \cite{beche_magnetic_2014} and to investigate rotational vortex-beam dynamics in a magnetic field \cite{Schattschneider2014NC} (Fig.~\ref{fig:bliokh11}).

A more complete method for the analysis of the OAM spectrum is offered by a {\it multi-pinhole interferometer} (MPI)~\cite{Clark2014}. This is an interferometer comprising a set of small holes; for the OAM measurement the most interesting configuration is $n$ pinholes evenly distributed around a circle, Fig.~\ref{fig:OAM_measurement}(f). When a wave is projected on such an interferometer, the far-field intensity distribution is determined by  relative phases between the pinholes, which allows one to retrieve the value of $\ell$.
The MPI, however, is more than just another type of geometrical aperture, as it can be used to obtain a {\it quantitative} OAM decomposition of the beam, even for mixed OAM
states. Namely, recording the diffraction pattern allows one to determine the autocorrelation
function of the interferometer by employing the Wiener--Khinchin theorem:
\begin{equation}
\mathcal{A} \left( \psi \right) = \mathcal{F}^{-1}\! \left(|\tilde{\psi}|^2\right),
\end{equation}
where $\mathcal{A}$ denotes the autocorrelation, and $\mathcal{F}^{-1}$ indicates the inverse Fourier transform. This means that the autocorrelation function of the MPI-transmitted wave $\psi$ can be obtained by the inverse Fourier transformation of the far-field diffraction patterns $|\tilde{\psi}|^2$ shown in Fig.~\ref{fig:OAM_measurement}(f).
Since $\mathcal{A}(\psi)$ exhibits peaks at positions corresponding to the distances between similar objects, it will show peaks at displacements from the centre equivalent to the
displacement vector between different pinholes (in addition to the central peak which is always present). Furthermore, each peak will be characterized by the phase equal to the phase difference between the two contributing pinholes. Once the phase differences are obtained from the autocorrelation function, it is possible to perform the OAM-harmonic decomposition to obtain the OAM spectrum of the original beam \cite{Clark2014}. 
While such quantitative analysis is a major step forward, the MPI method suffers from a few limitations. First, the pinholes need to be small enough to consider the phase as approximately constant inside each pinhole. Second, they should be distant enough so that the peaks in the autocorrelation function do not overlap. Third, since we are sampling the wave at only $n$ positions, the MPI cannot distinguish between the vortex
mode of the orders $\ell$ and $\ell+n$, as they yield identical phase differences in
the $n$ pinhole positions.

\subsubsection{Astigmatic phase plate} 
\label{ssec:modeconversion_measure}

The topological charge can also be measured by directly manipulating the wave phase, as in the conformal-map method \cite{Berkhout2010}. For electrons, a much simpler approach was demonstrated using mode conversion with an astigmatic plate \cite{Schattschneider2012a,Guzzinati2014}.
When a quadratic phase plate is applied to a vortex mode by tuning the quadrupolar electron stigmators, the doughnut-like intensity profile of the beam is split into a number of linearly-arranged intensity lobes as shown in Fig.~\ref{fig:OAM_measurement}(e), where the number of lobes is equal to $|\ell|+1$. Furthermore, the orientation of the pattern with respect to the phase plate (at the $\pm \pi/4$ angle), reveals ${\rm sgn}(\ell)$ \cite{Vaity2012,Vaity2013}.

Due to the ubiquitous presence of electron stigmators, this method is particularly convenient to employ within the TEM: it requires the manual adjustment of only one freely tunable parameter. Therefore, this technique can be an ideal way to confirm the vortex-beam order during the preparation of a more complex experimental setup (which can be then readjusted to the astigmatism-free condition).

\subsubsection{Nondestructive measurements using induced currents}

Recently, a method has been proposed \cite{Larocque2016} to measure the OAM of a vortex electron, by exploiting its magnetic moment, Eqs.~(\ref{eq:bliokh20}) and (\ref{eq:bliokh21}). Namely, when a vortex electron carrying magnetic moment passes through a hollow metallic cylinder, it induces eddy currents in the cylinder without changing the electron vortex state. Simulations performed for a 20~$\mu$m long and 1~$\mu$m wide  platinum cylinder, and an electron with a kinetic energy $E=100$~keV, showed the eddy currents to be of the order of several picoamperes: an amount which would be measurable with current technology. The same simulations have estimated the energy loss of the electron to be extremely small, $\sim 10^{-20}$~eV, supporting the nondestructive nature of this approach \cite{Larocque2016}. However, since the current pulse in the cylinder contains the same amount of energy, this extremely low value also means that it would be extremely difficult, if not impossible, to measure it in practical experimental conditions.

\subsection{Spiral phase plate imaging} 
\label{sec:spp_imaging}
\vspace{2mm}

Beside producing and detecting electron vortices, one might wonder if we can also use spiral phase plates as an {\it imaging filter} by changing the topological charge of the transfer function of the microscope. This can, in principle, be obtained by mounting a spiral phase plate in the back focal plane of the objective lens of the microscope. In practice, this could be a true spiral phase plate or an alternative means to add a vortex phase factor to this plane, changing the conventional transfer function of the microscope as:
\begin{equation}
\exp [\,i \Delta\Phi(\vt{k_\perp})\,] \to \exp\!\left[\, i \Delta \Phi(\vt{k_\perp}) + i \ell \phi\, \right], 
\end{equation}
where $\Delta\Phi ({\bf k}_\perp)$ is the aberration function describing the phase behaviour of the objective lens, $\ell$ 
is the topological charge of the spiral phase plate, and $\phi$ is the azimuthal angle in the ${\bf k}_\perp$-plane. Doing so results in a redistribution of the image intensities (i.e., probability densities):
\begin{equation}
\rho_0 ({\bf r}_\perp) \to \rho_\ell({\bf r}_\perp) .
\end{equation}
In real experimental conditions, spiral phase plates do not conserve the total current. Indeed, the center of the plate, where the phase (and the thickness) is indeterminate, is usually blocked by putting a small amount of an opaque material, which effectively removes the lowest frequency components from the filtered image, see Fig.~\ref{Fig_Spiral}(a).

Applying such a spiral phase filter results in an image where the intensity in each point is directly proportional to the weight of the $\ell$th OAM component measured {\it with respect to the same point} \cite{juchtmans_spiral_2016}. Recording a series of such images $\rho_\ell$ with $\ell$ covering a range of values provides a point-by-point OAM spectrum of the conventional image wave $\psi_0$. This can, in principle, allow the reconstruction of the whole complex wave field $\psi_0$ (otherwise unknown, as we typically record only $\rho_0 = | \psi_0 |^2 $).
However, in practice this would require some form of a programmable phase plate which could easily change its topological charge to allow a rapid recording of the image series. Such plates currently do not exist in TEM \cite{Guzzinati2015}.

Lacking such an $\ell$-varying phase plate one, could resort to only measuring the image intensities with two opposite topological charges: e.g., $\ell=\pm 1$. In this case, the difference and sum of the image intensities have been demonstrated to yield
\cite{juchtmans_spiral_2016}:
\begin{align}
{\rho_1+\rho_{-1}} \propto |\nabla_\perp \psi_0 |^2, \qquad {\rho_1 - \rho_{-1}} 
\propto (\nabla_\perp \times {\bf j}_{0\perp})\cdot \bar{\bf z},
\label{eq:probability_curl}
\end{align}
where ${\bf j}_{0\perp}$ is the probability current density for the original wave function $\psi_0$ in the plane of interest (typically the plane immediately after the sample, containing the so-called exit wave). 
Remarkably, the second Eq.~(\ref{eq:probability_curl}) enables one to reconstruct the curl of the transverse current density, which is typically omitted in conventional {\it transport of intensity} reconstructions \cite{Lubk2013}.
Furthermore, the first Eq.~(\ref{eq:probability_curl}) includes the gradient of the phase of the wave function $\psi_0$, and, hence, can improving the contrast in highly transparent \emph{weak phase objects} (where intensity is practically uniform).

A practical realization of the vortex filtering setup remains difficult due to the
unwanted scattering and contamination issues with phase plates in the objective back focal plane.
The most attractive option might seem to be placing a magnetic needle (with the monopole-like field acting as spiral phase plate) in the back focal plane, but this is hampered by the presence of a strong magnetic field in that plane which would likely magnetize the needle in the unwanted up/down direction rather than along the needle. 
A workable alternative is the use of the reciprocity theorem, turning the electron trajectories upside down (time reversal) and creating a bright-field STEM setup with an incoming vortex probe. This setup works reasonably well, relying on different vortex-probe production schemes, but is highly dose-inefficient, as it requires a very limited acceptance angle, discarding the majority of the electrons.

\subsection{Elastic interaction of vortex electrons with matter}
\vspace{2mm}

When electron vortex beams interact with matter, the cylindrical symmetry is typically broken, and the beam evolves in a highly nontrivial way through the material. In this process, the electron OAM changes significantly for all but the thinnest sample and can even change its sign for certain thickness~\cite{Loffler2012,mendis_dynamic_2015,mendis_electron_2015}. 
This process can be effectively simulated using standard multislice numerical calculations  solving the non-relativistic Schr\"odinger equation for a fast electron wavepacket travelling through the potential produced by the atoms. Extending these simulations to include the incoming vortex states is trivial as long as the source code of the software is open
\cite{rosenauer_stemsimnew_2008,lobato_multem:_2015,lobato_progress_2016}.

This thickness dependence seems to undermine all hopes for using vortex electron beams to obtain unique information through scattering by materials except for very thin samples. However, even though the electron OAM varies upon propagation through a crystal, the phase singularity in the vortex centre is far more stable. Indeed, the net topological charge is conserved during the wave evolution, and the phase singularity has the tendency to follow a crystallographic channel of atoms \cite{Lubk2013a}. This leads to the peculiar effect: a focused vortex beam entering the sample along a column of atoms remains captured in this potential channel much longer than a beam with the same opening angle but without the vortex phase structure \cite{Lubk2013a}.
Thus, electron vortex beams could potentially be used to suppress the unwanted
dechanneling effects that occur as conventional focused beams travel through the sample. Such dechanneling leads to the leaking of the beam's intensity to atomic columns neighbouring the initial one, which is detrimental for the localization of, e.g., spectroscopic signals, such as EELS or energy dispersive X-ray (EDX) spectroscopy.
So far, this attractive proposition has not yet been demonstrated experimentally, likely due to the effect of source-size broadening (see Section \ref{practical_issues}).

\begin{figure}[!t]
\centering
\includegraphics[width=0.8\textwidth]{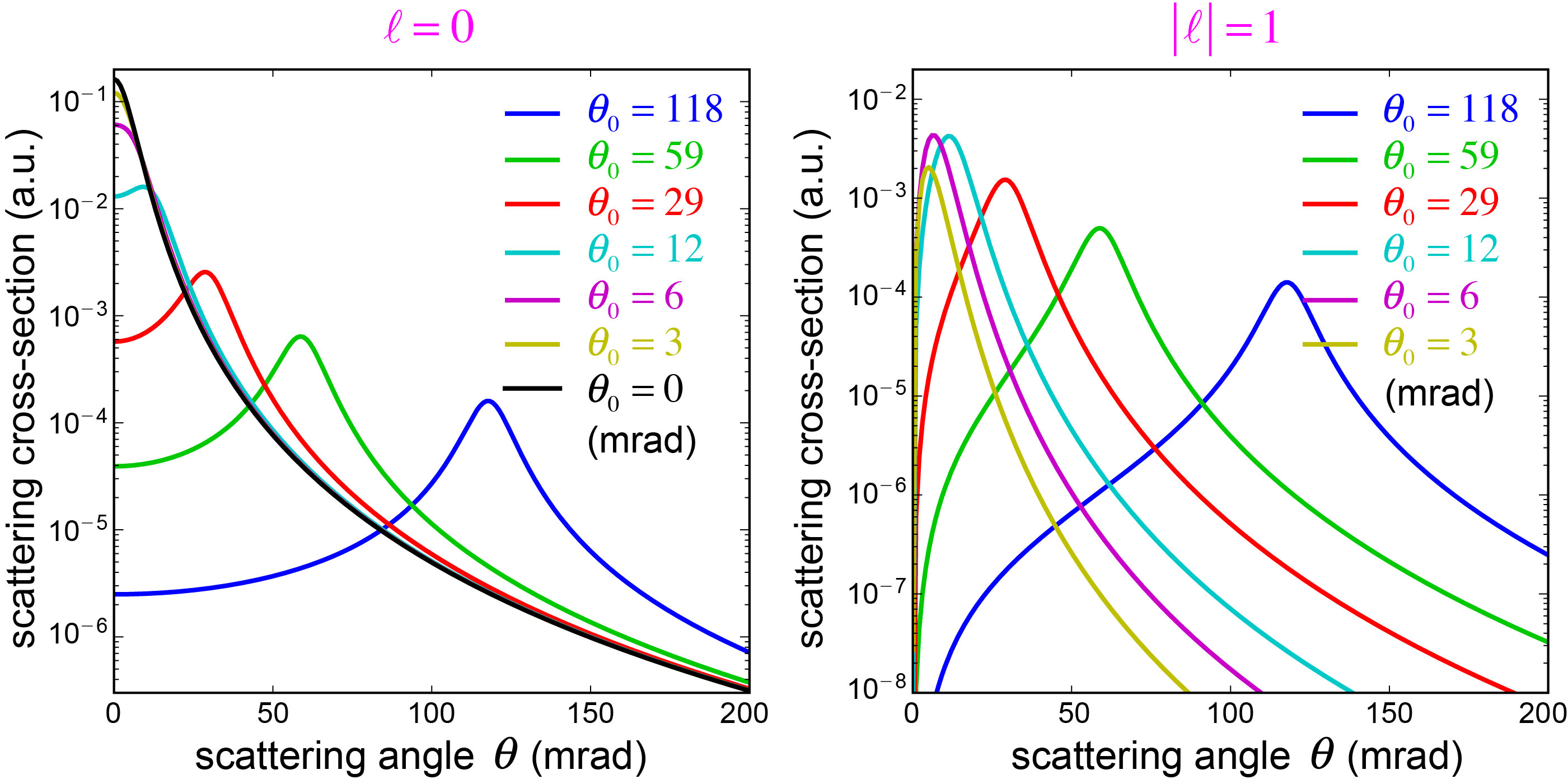}
\caption{The scattering cross-section for electron Bessel beams with an energy of 300~keV, the vortex charges $\ell =0$ and $\ell =1$, and different values of the transverse momentum $\kappa$ (expressed via the opening angle $\theta_0$: $\sin \theta_0 = \kappa/k$), on a screened Coulomb potential corresponding to an iron atom~\cite{VanBoxem_Rutherford,Jablonski}. The scattering amplitude is cylindrically symmetric (up to the vortex phase factor), and the plots show its dependence on the polar scattering angle $\theta$. Positions of the maxima approximately correspond to the opening angles $\theta_0$ of the incident Bessel beams.} 
\label{fig:rutherford_vortex}
\end{figure}

In order to gain further insights in the elastic scattering of a vortex probe on materials, it is
useful to have an analytical description of the elastic-scattering amplitude.
In Refs. \cite{VanBoxem_Rutherford, Berkes}, the {\it Rutherford scattering} amplitude of Bessel beams by a screened Coulomb potential has been analytically derived. 
This showed that the transverse momentum structure of an electron beam (with or without OAM) can have a significant impact on elastic scattering, even by a simple Coulomb potential. Figure~\ref{fig:rutherford_vortex} shows the scattering amplitudes for beams with $\ell=0$ and $\ell=1$, and different values of the transverse momentum $\kappa$. 
The scattering amplitude of a Bessel beam can be seen as the convolution of a ring of tilted plane waves [see Fig.~\ref{fig:Bessel}(a)] with the plane-wave scattering amplitude by the spherically-symmetric potential. This coherent superposition of the plane-wave Rutherford scattering peaks leads to a ring-like scattering intensity structure (depending on the polar scattering angle $\theta$) even for non-vortex beams ($\ell=0$). For $\ell=1$, the intensity vanishes in the forward direction $\theta =0$, and this dip becomes more visible as the transverse momentum $\kappa$ increases.

\begin{figure}[!t]
\centering
\includegraphics[width=\textwidth]{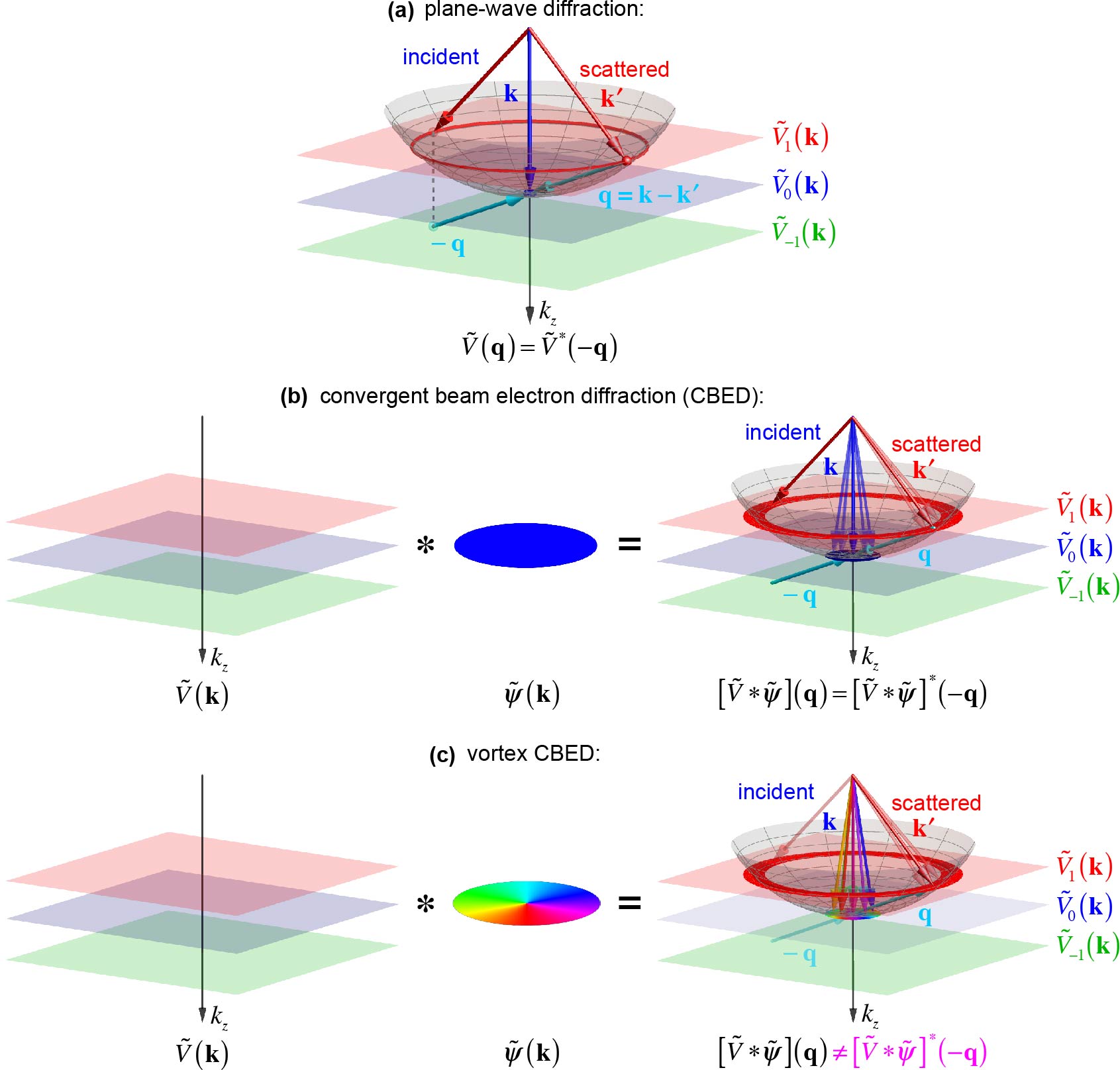}
\caption{Visual representation of Eq.~(\ref{Convolution}) for elastic electron scattering ${\bf k} \to {\bf k}'$, Eqs.~(\ref{eq:elastic}) and (\ref{Convolution}), in a periodic crystal potential $V({\bf r})$. 
(a) In the {\it plane-wave} regime, the scattering is determined by the intersection of the Fourier transform (FT) of the potential, $\tilde{V}({\bf k})$ (here shown as discrete planes for a $z$-periodic potential), with the Ewald sphere of radius $k$ (shown in gray) \cite{WilliamsCarter,Reimer_book}. The corresponding zero- and first-order Laue zones are indicated by the small blue circle near the pole of the sphere and by the red ring, respectively. According to Friedel's law, $\tilde{V}({\bf q}) = \tilde{V}^*(-{\bf q})$, and scattering events with opposite ${\bf q} = {\bf k} - {\bf k}'$ vectors are equally probable and the diffraction pattern is centrosymmetric. (b,c) For non-plane-wave probes, the diffraction condition is determined by the convolution of $\tilde{V}({\bf k})$ with the FT of the incoming wave, $\tilde{\psi}({\bf k} )$. (b) In convergent beam electron diffraction (CBED) $\tilde{\psi} ({\bf k})$ is a disc with a constant phase and Friedel's law remains valid after the convolution. (c) For a {\it vortex} probe, $\tilde{\psi} ({\bf k})$ is a disc with an azimuthally dependent phase. Convolving it with $\tilde{V}({\bf k})$ breaks Friedel's law and generically results in different probabilities for the ${\bf q}$ and $-{\bf q}$ event. In particular, the corresponding non-centrosymmetric diffraction patterns allow one to probe {\it chirality} of crystals \cite{Juchtman2015-II,Juchtmans2015a}.} 
\label{FigEwald}
\end{figure}

To calculate the elastic scattering in the case of an arbitrary potential, produced by a configuration of atoms in a material, one can adopt the so-called {\it kinematic approximation}. Assuming the scattered part of the wave to be much smaller than the incoming part, it considers the potential as a small
perturbation and takes into account only single-scattering events. In this approximation,
the scattering amplitude for an incoming wave $\psi ({\bf r})$ scattered on a potential
$V({\bf r})$ to a plane wave $\psi'({\bf r}) = \exp(i\, {\bf k}' \cdot {\bf r})$ with the wave vector ${\bf k}'$ can be written as \cite{DeGraefCTEM}:
\begin{equation}
f(\vt{k}') = \left< {\bf k}' \left| V \right| {\psi} \right>
\propto \int d^3{\bf r}\, e^{-i\,{\bf k}' \cdot {\bf r}}\, V({\bf r})\, \psi({\bf r})\, .
\label{eq:elastic}
\end{equation}
This scattering amplitude can be presented as a convolution of the Fourier transform (FT) of the potential and that of the incoming wave:
\begin{equation}
f({\bf k}') = \mathcal{F}[V\cdot\psi](-{\bf k}')= [\tilde{V}*\tilde{\psi}](-{\bf k}')\,.
\label{Convolution}
\end{equation}
For an incoming {\it plane wave}, $\psi({\bf r}) = \exp(i\, {\bf k}\cdot {\bf r})$, the
FT $\tilde{\psi} ({\bf k})$ is a delta-function and the diffraction pattern will be determined solely by the FT of the potential: $\tilde{V}({\bf k})$. 
For elastic scattering, the energy of the scattered electron (and, hence, the
length of the wave vector) is conserved, $k = k'$, and the diffraction pattern is determined in ${\bf k}$-space by the intersection of $\tilde{V}({\bf k})$ and a sphere with radius $k$, called the {\it Ewald sphere} \cite{WilliamsCarter,Reimer_book}. This means that the momentum transferred in the scattering process, ${\bf q} = {\bf k} - {\bf k}'$, must satisfy $\tilde{V}(\mathbf{q}) \neq 0$, which is the {\it momentum matching} condition.
A schematic example for a potential ${V}({\bf r})$ periodic along the propagation $z$-axis is given in Fig.~\ref{FigEwald}(a).
The FT of the potential, $\tilde{V}({\bf k})$, corresponds to discrete planes perpendicular to the $z$-axis, so that the resulting diffraction pattern consists of rings which coincide with the zeroth and higher-order Laue zones in conventional electron diffraction \cite{WilliamsCarter,Reimer_book}.

The diffraction pattern can be significantly altered when modifying the probe (incoming wave) $\psi ({\bf r})$. For instance, in {\it convergent-beam electron diffraction} (CBED) the scattering amplitude is convolved with the FT of the probe, $\tilde{\psi} ({\bf k}) \propto \Theta(\kappa_{\rm max} - k_\perp)$, Eq.~(\ref{eq:step-function}), i.e., a disc with uniform phase and amplitude \cite{WilliamsCarter,Reimer_book}, Fig.~\ref{FigEwald}(b). In turn, the FT of a converging {\it vortex} electron beam is a disc with the azimuth-dependent phase: $\tilde{\psi}({\bf k}_{\perp}) \propto \exp(i \ell\phi)\, \Theta(\kappa_{\rm max} - k_\perp)$ \cite{Juchtmans2015a}, see Fig.~\ref{FigEwald}(c).

To show the effect of the vortex-probe diffraction, now consider a 3D crystal model. The FT of the potential, $\tilde{V}({\bf k})$, can now be modelled by a lattice of delta-functions (not explicitly shown in Fig.~\ref{FigEwald}) giving rise to {\it Bragg spots} in the {\it plane-wave} diffraction pattern. 
Since the symmetry of the crystal potential is constrainted by Friedel's law, requiring  $\tilde{V}({\bf q}) = \tilde{V}^* (-{\bf q})$, the zero-order Laue zone must be {\it centrosymmetric} \cite{DeGraefCTEM}. In other words, scatterings with ${\bf q}$ and ${- \bf q}$ momentum transfers always have equal probabilities $|f({\bf k}')|^2$, Fig.~\ref{FigEwald}(a), which yields centrosymmetric diffraction patterns (even for non-centrosymmetric, e.g., chiral, cystals).
In the case of CBED, the Bragg spots become discs because of the convolution with the FT of the probe, $\tilde{\psi} ({\bf k})$. However, Friedel's law is still satisfied, $[\tilde{V}*\tilde{\psi}]({\bf q}) = [\tilde{V}*\tilde{\psi}]^*(-{\bf q})$, and opposite-${\bf q}$ scatterings are still equiprobable, Fig.~\ref{FigEwald}(b), and the resulting diffraction pattern is always centrosymmetric.
In contrast to this, the {\it vortex} probe circumvents Friedel's law, and non-centrosymmetric crystals produce {\it non-centrosymmetric} diffraction patterns 
\cite{Juchtman2015-II,juchtmans_extension_2016}. This is because for the vortex wave function $\tilde{\psi}({\bf k})$, generically $[\tilde{V}*\tilde{\psi}]({\bf q}) \neq [\tilde{V}*\tilde{\psi}]^*(-{\bf q})$, and opposite-${\bf q}$ scatterings can have different probabilities, Fig.~\ref{FigEwald}(c).
This opens up the vortex-beam probing of complex crystal symmetries, which are hidden for standard techniques.

\subsubsection{Chirality in crystals}
\label{chiral}

An important application of the elastic scattering of vortex beams on crystals is the probing of the {\it chirality} of crystalline materials. Chirality is the property of any 3D object that cannot be superimposed with its mirror image \cite{Barron_book}. A chiral object and its mirror image are said to be the two \emph{enantiomers} of the same structure. Chirality is ubiquitous in nature, and it underpins many physical phenomena, such as, e.g., optical activity. The two enantiomers of the same compound often have different physical, chemical or biological properties. A chiral crystal is characterized by a space group possessing only proper symmetry elements [rotations and screw-axes] and no improper
symmetry elements [mirror(-glide) planes and (roto-)inversion centers]. Mirroring the crystal
with respect to any plane results in a fundamentally different crystal.

Determining the chirality of a sample in a TEM is challenging because high-resolution
imaging techniques record only a 2D projection of the atomic arrangement. If the sample is mirrored with respect to the $(x,y)$ plane, the crystal is replaced with its enantiomer, while the projection remains identical. Moreover, as we discussed above, the exact space group (specifically, chirality) of the crystal cannot be determined using the conventional (plane-wave or CBED) electron diffraction. In recent works \cite{Juchtman2015-II,Juchtmans2015a} electron-vortex probing of crystal chirality was suggested and demonstrated experimentally.  A kinematical treatment of an electron vortex beam of order $\ell$ focused on a $Q$-fold screw axis of the chiral crystal, shows that the $N$th-order Laue zone is centrosymmetric if and only if \cite{Juchtman2015-II}
\begin{equation}
\ell - \chi N = n Q.
\end{equation}
Here, $\chi = \pm 1$ indicate the opposite chiralities of the crystal, while $n$ is an integer number. In Ref.~\cite{Juchtman2015-II} this method was demonstrated experimentally by focusing electron vortex beam on the three-fold screw axis of a ${\rm Mn_2Sb_2O_7}$ crystal.

\subsubsection{Magnetic contrast}

Elastic electron scattering on {\it magnetic fields} forms another interesting proposal for experiments. We have already seen that a magnetic monopole converts a plane wave into a vortex beam, and actual magnetic fields around (nano) materials can be interpreted as emanating from ``local magnetic monopoles'', inside and on the surface of the material with a density $\rho_{m} \propto \nabla \cdot {\bf m}$, where ${\bf m}$ is the local magnetization of the material.
Even though the actual magnetic field is strictly divergence-free, the scattered electron wave nevertheless exhibits a current density which curls near the ``magnetic poles'', in a way similar to how an approximate magnetic monopole can be used to generate electron vortex beams. 

This effect can be exploited to directly visualize the ``monopole density'' in a magnetic material in the TEM by obtaining the curl of the in-plane probability current, $\nabla_{\perp}\times {\bf j}_{\perp}$, as detailed in Section~\ref{sec:spp_imaging} (i.e., using the difference of two TEM images obtained with opposite-$\ell$ spiral phase plates). As mentioned above, this setup is currently difficult to obtain, but the experiment \cite{verbeeckunpublished} suggests that the reciprocal setup, with an incoming vortex beam directed on the sample and a bright-field detector accepting electrons scattered along the optical axis, behaves as predicted. This effect is highly attractive as it localizes the measured signal at the virtual monopoles. This leads to a potentially superior signal-to-noise ratio in the measurement as the fields can be derived at a later stage by convolution with the field kernel of a single monopole. Still, further experiments are needed to explore this effect, but this prediction highlights once more why an image filter based on a tunable spiral phase plate would be highly desirable.

\subsection{Inelastic interaction of vortex electrons with matter}
\vspace{2mm}

The fast electrons in a TEM can also interact with the sample {\it inelastically}, leaving both
the sample and the fast electron with energies different from the initial ones. This
phenomenon is actively exploited in the technique of electron energy loss spectroscopy (EELS), where the kinetic energy distribution of an ensemble of fast electrons is recorded after passing through the sample \cite{Egerton}. Such a spectrum contains information about all excitation events in the sample and their respective probability, revealing information such as: which elements are in the sample, what is their concentration, and how are they bound to each other. All this information can be combined with the atomic resolution provided by a focused STEM probe, which explains the widespread use of this technique.

\subsubsection{Inelastic scattering by atoms in the Born approximation}

A simple model, that can illuminate a number of important features  of inelastic vortex-electron scattering, is the Bessel-beam scattering on a single-electron atom in the Born approximation. For plane waves, this scenario provides a building block of electron-matter interaction, which is used in EELS.
The results presented below can be deduced analytically, as shown in \cite{van_boxem_inelastic_2015}.

We consider the same geometry as for the elastic scattering: a centered $z$-propagating Bessel beam scatters on a single-electron atom.
Using the same mathematical formalism as developed for elastic Bessel-beam scattering \cite{VanBoxem_Rutherford}, one can determine analytical expressions for {\it all} the partial scattering amplitudes for this system.
The Bessel beam represents a coherent superposition of plane waves with conically distributed wave vectors ${\bf k}$, see Fig.~\ref{fig:Bessel}(a). Therefore, its scattering into a plane wave with the fixed wave vector ${\bf k}^\prime$ is determined as an integral of the plane-wave scattering amplitude $f({\bf k},{\bf k}^\prime)$ over incoming ${\bf k}$ vectors or momentum-transfer vectors ${\bf q} = {\bf k} - {\bf k}^\prime$, Fig.~\ref{fig:inelastic_bessel_scattering_fourier} (see Section~\ref{sec:fixed-target} below).

\begin{figure}[t]
\centering
\includegraphics[width=0.45\textwidth]{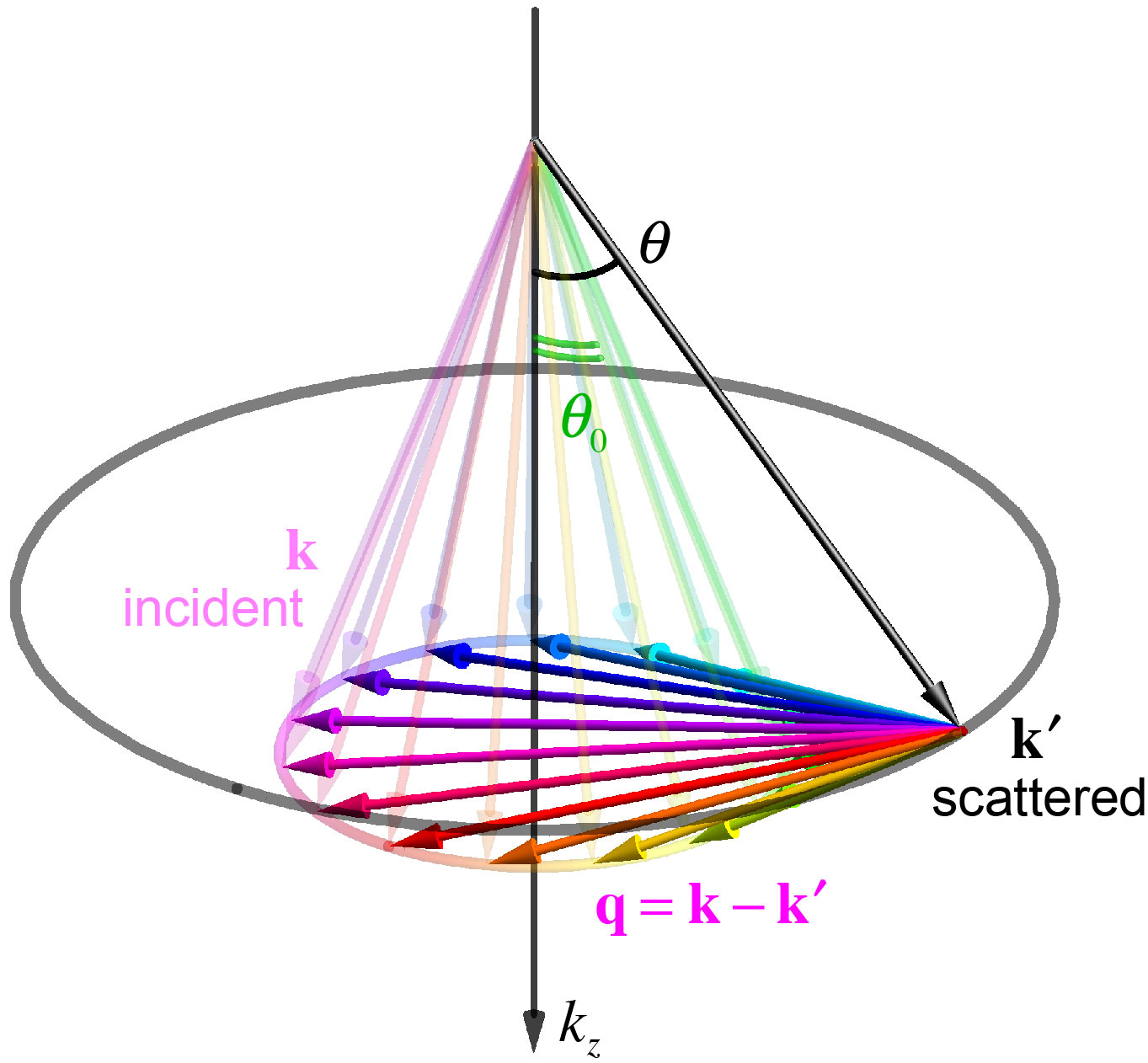}
\caption{Schematic of the electron Bessel-beam scattering by an atom into a plane wave. The incident Bessel beam consists of multiple plane waves with the ${\bf k}$-vectors conically-distributed with the polar (convergence) angle $\theta_0$ (shown semitransparent here). These are scattered into a single plane wave with the wave vector ${\bf k}^\prime$ (polar scattering angle $\theta$). The corresponding momentum-transfer vectors are ${\bf q} = {\bf k} - {\bf k}^\prime$.} \label{fig:inelastic_bessel_scattering_fourier}
\end{figure}

\begin{figure}
\centering
\includegraphics[width=0.60\textwidth]{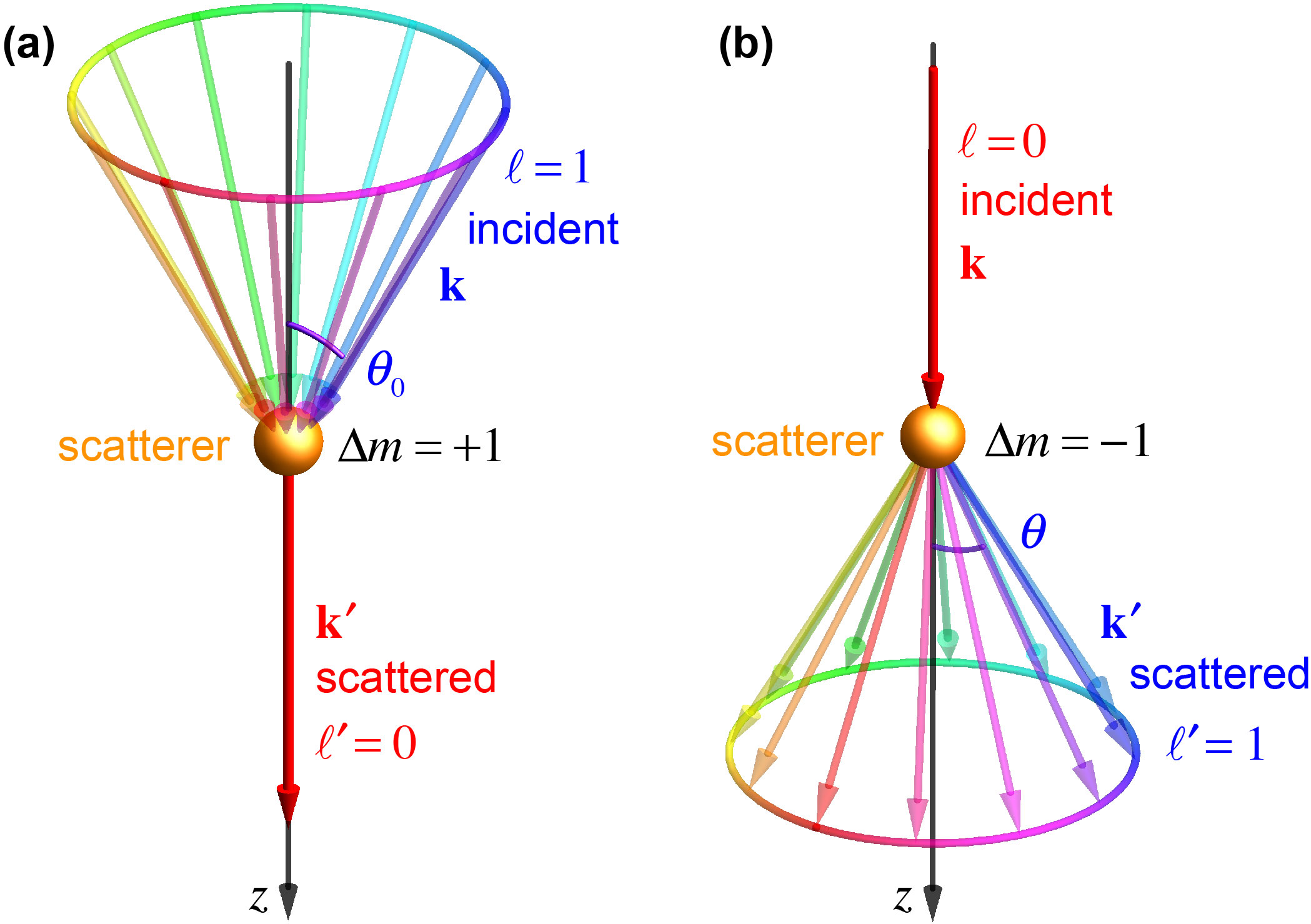}
\caption{Examples of the inelastic electron scattering accompanied by the angular-momentum exchange with the atom \cite{van_boxem_inelastic_2015,VanBoxem_PhDThesis}. The OAM difference between the incident and scattered electron waves, $\ell - \ell^\prime$, corresponds to the change in the magnetic quantum number of the atom, $\Delta m$. The examples shown here correspond to the reciprocal cases of ``vortex $\to$ plane wave'' and ``plane wave $\to$ vortex'' scatterings with $\theta = \theta_0$ and equal probability amplitudes.} 
\label{fig:inelastic_symmetry}
\end{figure}

The resulting scattering amplitudes for various atomic transitions, e.g. $1{\rm s} \to 2{\rm p}^{\pm}$, can be written down in an analytic (athough rather cumbersome) form \cite{van_boxem_inelastic_2015,VanBoxem_PhDThesis}.
In particular, these vortex-scattering amplitudes describe properties of the angular-momentum exchange with the atom, Fig.~\ref{fig:inelastic_symmetry}. 
Namely, when the scattering of a vortex electron with topological charge $\ell$ is accompanied by the change of the magnetic quantum number of the atom (with respect to the $z$-axis), $\Delta m$, the scattered wave will correspond to the vortex state with $\ell^\prime = \ell - \Delta m$. The cases $\ell = \Delta m$ and $\ell^\prime = - \Delta m$ correspond to the ``vortex $\to$ plane wave'' and ``plane wave $\to$ vortex'' scatterings, as shown in Fig.~\ref{fig:inelastic_symmetry}. Importantly, the scattering amplitudes for these reciprocal cases are exactly equal to each other, assuming the same parameters in the incoming and outgoing vortex states. Such ``inelastic OAM reciprocity'' enables one to chose between the two equivalent probing schemes involving vortex states of either incident or scattered electrons.

\subsubsection{EELS cross-sections and selection rules. Magnetic dichroism.}

In the dipole approximation, the plane-wave scattering of a fast electron by an isotropic 
point scatterer can be characterized by the following scattering amplitude \cite{Reimer_book,Egerton}:
\begin{equation}
f({\bf k},{\bf k}') \propto \langle i |\, {\bf q}\cdot {\bf r}\,| f \rangle,
\label{dipole-approx}
\end{equation}
where ${\bf q} = {\bf k} - {\bf k}'$ is the momentum-transfer vector, whereas $\left| i \right>$  and $\left| f \right>$ are the initial and final quantum states of the scatterer. In the $q \ll k$ approximation, the selection rules for the angular-momentum ($l$) and magnetic ($m$) quantum numbers of the atom read \cite{Egerton}:
\begin{align}
\Delta l = \pm 1, \qquad
\Delta m = -1,0,+1.
\end{align}
For the ${\rm 1s} \to {\rm 2p}$ atomic transition (corresponding to the K absorption edge), three separate transitions to ${\rm p}_x$, ${\rm p}_y$, and ${\rm p}_z$ are possible, involving a momentum transfer ${\bf q}$ in the corresponding $x$, $y$, and $z$ directions. 
Using projective measurements, one can selectively measure excitations to these states attached to Cartesian coordinates. This method involves the so-called {\it linear dichroism}, (also known as ``anisotropic scattering cross section'') which gives access to the population of the different ${\rm 2p}$ states in anisotropic materials. A typical example can be found in the studies of graphite and graphene, where the in-plane $\sigma^*$ bonds can be selectively probed off-axis, while the excitation to the $\pi^*$ states leaves fast electrons close to the $z$-axis of the incoming electrons \cite{jouffrey_magic_2004,hebert_elnes_2006}.

The above linear dichroism works well for anisotropic materials, but fails to identify the {\it angular-momentum} properties of the ${\rm 2p}$ orbitals. The linear combinations ${\rm p^\pm} = {\rm p}_x \pm i\, {\rm p}_y$  describe the orbital states with opposite OAM. Conventional EELS, with an incoming plane wave, cannot distinguish between excitations to the ${\rm p}^+$ and ${\rm p}^-$ states, because both scatterings lead to the same ring-like probability distribution in the far field.
\footnote{One might think that the two states could be distinguished
in energy. But even if the energy splitting is high enough to be resolved in the EELS spectrum, still one cannot know which energy corresponds to which transition.}
Having access to probabilities of these angular-momentum excitations would be highly desirable as it could provide an important information about {\it magnetic} properties of the sample.

One way to overcome this limitation is to allow \emph{two} coherent plane waves to
interact with the sample simultaneously, so that constructive or destructive interference in the diffraction plane reveals the difference between exciting to the ${\rm p}^+$ or ${\rm p}^-$ states \cite{hebert_proposal_2003,schattschneider_detection_2006,Schattschneider_Book}. This technique is called {\it energy loss magnetic chiral dichroism} (EMCD). It allows one to obtain magnetic information similar to that given by the X-ray magnetic chiral dichroism \cite{schutz_absorption_1987}, but using EELS and now at a nanometer-scale resolution \cite{schattschneider_detection_2008,rusz_magnetic_2016}.
The drawback of this technique is the difficulty in setting up the two plane waves,
which need to be operated by a precisely controlled diffraction condition of the crystal lattice which also contains atoms. This restricts the use of the EMCD to crystalline materials with precisely known thickness and where the correct orientation can be predicted with precision.
On the other hand, by using the specimen crystal itself as a beam splitter for the two (or more) coherent plane waves, we have locally and temporally extremely stable site selectivity by the intrinsic phase lock in the periodic lattice \cite{Ennen2012}.

The atomic transitions ${\rm 1s} \to 2{\rm p}^\pm$ are characterized by changes
in the angular momentum of the atom by $\Delta m = \pm 1$. In this case, the plane-wave scattering amplitude (\ref{dipole-approx}) takes the form of \cite{van_boxem_inelastic_2015}:
\begin{equation}
f^{\pm}({\bf k},{\bf k}') \propto f(q_\perp) \exp\!\left(- i\, \Delta m\, \phi_{\bf q}\right),
\label{3.19}
\end{equation}
where $q_\perp = | {\bf q}_\perp |$, and $\phi_{\bf q}$ is the azimuthal angle of the ${\bf q}$ vector. Equation (\ref{3.19}) is equivalent to the transmission function of a spiral phase plate, which means that the {\it outgoing electron is in the vortex state with $\ell^\prime = -\Delta m$}, Fig.~\ref{fig:inelastic_symmetry}(b). The total scattering probability $|f|^2$ is independent of the $m$-dependent phase term and therefore does not allow distinguishing between the excitation of the ${\rm p}^+$ and ${\rm p}^-$ states. But detecting the OAM phase properties of the outgoing electrons will provide the desired magnetic information.
In particular, if the densities of unoccupied states in the ${\rm p}^+$ or ${\rm p}^-$ orbitals are different, there should be a preferential scattering channel to the corresponding $\ell^\prime =\mp 1$ vortex states of the scattered electron.
\footnote{In practice, the $2{\rm p} \to 3{\rm d}$ transitions are often investigated for the \chem{L_{2,3}} edges of transition metals. This complicates the above description with the spin-orbit splitting, but essentially the same imbalance remains (albeit less strong than in the case considered above). For a detailed treatment, see  \cite{schattschneider_real-space_2010}.}

Thus, retrieving the desired magnetic information about the sample requires an efficient far-field OAM sorter for transmitted electrons. Such an ideal sorter is not currently available (as discussed in Section~\ref{sec:oammeasure}), although promising setups have been proposed \cite{mcmorran2016efficient}. The first hints of this effect have been demonstrated on a ferromagnetic iron sample, using a forked holographic grating as an OAM filter before the electron spectrometer \cite{Verbeeck2010}. 
However, further simulations of this setup have revealed that there is a very complex interplay between chromatic aberrations, the position of the scattering atom in the sample, elastic scattering, and precise focus, complicating the interpretation of the experiment \cite{rusz_boundaries_2013}.

Nevertheless, the experiment \cite{Verbeeck2010} ignited significant interest in the reciprocal scheme where a {\it focused electron vortex probe with topological charge $\ell = \pm 1$} impinges on the atom, breaking the symmetry of the excitation to the ${\rm p}^\pm$ states, Fig.~\ref{fig:inelastic_symmetry}(a). In this case, considering electron scattering in the {\it forward} direction (i.e., into a $z$-directed plane wave), the approximate EELS selection rules become \cite{van_boxem_inelastic_2015}:
\begin{align}
\Delta l =\pm 1, \qquad 
\Delta m =\ell.
\label{eq:selection}
\end{align}
These selection rules signify {\it vortex-dependent magnetic dichroism} for electrons. 
This is in sharp contrast to optics, where probing magnetic dichroism and chirality involve only polarization (spin) degrees of freedom of light and are mostly insensitive to vortices \cite{Babiker2002,Araoka2005,Loffler2011,Afanasev:2013}.
Note that the limited validity of the dipole approximation for electron scattering leads to a limited validity of the above selection rules~\cite{Saldin,Loffler_Dipole,Afanasev2016}.

The selection rules (\ref{eq:selection}) should in principle give access to the magnetic state of an atom with a resolution defined by the probe size, even down to the {\it single atomic column}.
The setup was shown to be viable on theoretical grounds, with an increasing level of details included \cite{schattschneider_real-space_2010,rusz_boundaries_2013,Rusz_Towards}. 
At the same time, numerical simulations revealed that elastic electron scattering on the crystal lattice (Bragg diffraction, channeling effects) is rather detrimental to the vortex
character of the incoming probe (it breaks the cylindrical symmetry, making $L_z$ a bad
quantum number). This sets an upper limit for the thickness of the crystal used. Moreover, the magnitude of the magnetic signal in the EELS spectrum is expected to be at best $\sim 10\%$ of the EELS signal in the ${\rm L}_{2,3}$ edge of the transition metal. This makes the experiments extremely challenging in view of the low signal-to-noise ratios often encountered in atomic-resolution EELS. Note also that this method has the desired sensitivity to the magnetic properties only when the probe is positioned precisely on the atom columns and is of atomic size.
Deviations from these parameters will further reduce the magnetic sensitivity and can even switch the sign of the effect when the probe is displaced from the atom column or is too large \cite{Rusz_Towards}.

Despite the tremendous experimental efforts that were put into realizing the OAM-magnetic probing experiment by several groups \cite{pohl_electron_2015,Rusz_Towards,Beche_private,idrobo_detecting_2017,ercius_atomic-resolution_2014}, the results are still inconclusive. 
Nonetheless, the inelastic scattering of shaped beams has been succesfully applied to the study of surface plasmons \cite{Guzzinati2016}, and this success in a similar case supports the validity of the description outlined above. 
However, as compared to the case of surface plasmons, experiments on inelastic scattering by atoms are heavily limited by two factors. First, the source-size broadening is much more important due to the much smaller scale of the scattering objects (atomic scale rather than 10s of nm). Second, the scattering cross-sections are several orders of magnitude smaller, resulting in a weaker signal.
Limiting the incoherent source-size broadening without losing too much intensity is currently the main challenge (see Section~\ref{practical_issues}). 
Making significant progress on this issue requires the development of brighter electron sources \cite{houdellier_development_2015}.

Some groups have proposed and demonstrated other symmetry-breaking mechanisms in the electron probe to reveal the atomic-scale magnetic signal. In particular, Refs.~\cite{idrobo_mapping_2016,idrobo_detecting_2017} deliberately added four-fold astigmatism through the probe aberration corrector, while the works \cite{Schattschneider2012,rusz_magnetic_2016} dealt with off-axis inelastic scattering as a function of the probe displacement with respect to the atomic column. These experiments showed an atomically-resolved magnetic signal albeit at very low signal-to-noise ratios.

For the actual application of the above techniques, it is crucial to make a significant
improvement of the signal-to-noise ratio. The most likely approach to this challenge is to deal with the ``plane wave $\to$ vortex'' scattering with the subsequent effective OAM sorting in the far field, prior to the EELS spectrometer. 

\subsubsection{Interaction with plasmons. Probing 	of chirality.}
\label{sec:plasmons}

Electron vortex beams have also been considered as a possible tool in the spectroscopic investigation of localised {\it surface plasmon resonances} (SPRs).
SPRs are collective excitation of electrons (in metals) and electromagnetic fields, which appear due to the confinement of the conduction electron gas in metals and small nanoparticles \cite{Maier_book}. 
Due to their peculiar properties, such as strong localized electromagnetic fields and high sensitivity to nanometer-scale environmental changes, they offer a unique platform for sub-wavelength optics, nano-photonics, and optoelectronic devices \cite{Maier_book,Zayats2005,Ozbay2006}.
EELS plays an important role in this research because it allows the intensity of the optical electric field produced by SPRs to be mapped \cite{Nelayah2007}:
\begin{equation}
\Gamma ({\bf r}_\perp,\omega) \propto \int \left| E_z\! \left({\bf r}_\perp, z, \omega \right) \right|^2 dz ,
\end{equation}
where $\Gamma({\bf r}_\perp,\omega)$ is the probability of the electron beam focused on the position ${\bf r}_\perp$ to lose an energy $\hbar \omega$, whereas $E_z$ is the $z$-component of the electric SPR field, and $\omega$ is the SPR frequency.
While this is a powerful tool to investigate SPRs, the loss of information caused by recording
only the electric-field intensity does not allow one to obtain the charge distribution or the direction of the electric field (see, e.g., \cite{Schmidt2012}). Furthermore, important phenomena, such as circular dichroism of chiral plasmonic structures cannot be investigated at the nanoscale using electron plane waves.
Electron vortex beams have been proposed as a potential tool to address these limitations.

\begin{figure}[!t]
\centering
\includegraphics[width=0.77\textwidth]{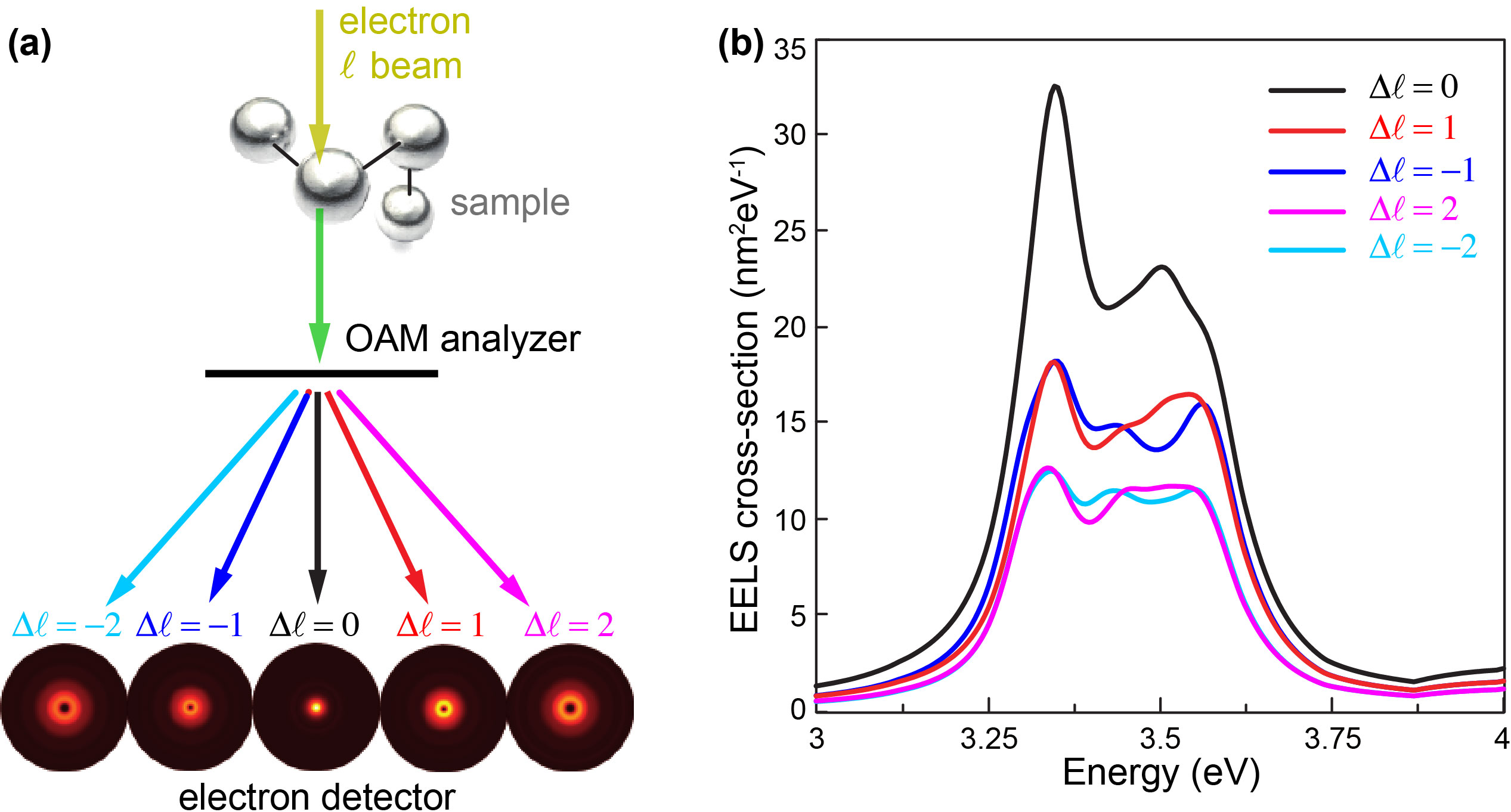}
\caption{OAM-resolved inelastic scattering cross-section for an electron beam impinging on a chiral cluster of plasmonic nanoparticles. (a) Schematic representation of the proposed experiment. An electron beam impinges on a nanoparticle cluster, and then the inelastically-scattered electron wave is filtered with an OAM analyzer. (b) The energy loss spectra for each OAM component, with differences clearly visible for the $\Delta\ell = \pm1$ or $\Delta\ell = \pm 2$ components. Adopted with permission of the authors from \cite{Asenjo-Garcia2014}, copyrighted by the American Physical Society.}
\label{fig:plasmondichroism}
\end{figure}

In Ref.~\cite{Asenjo-Garcia2014} the OAM exchange in the inelastic interaction
between fast electron beams and {\it chiral} assemblies of metal nanoparticles was studied.
Calculations of the OAM- and energy-resolved EELS cross-section
showed that there is a significant OAM transfer between the electron beam and the SPRs of the nanoassembly, Fig.~\ref{fig:plasmondichroism}
\cite{Asenjo-Garcia2014}.
Performing the OAM-resolved EELS on such structures would reveal strong (up to $\sim 10\%$) dichroism between transitions with opposite handedness, i.e., with opposite amount of the transferred OAM: $\Delta \ell = \pm 1$, $\Delta \ell = \pm 2$, etc. This effect strongly depends on the axis with respect to which the OAM is measured. This can be used for the local probing of the chirality of the plasmon resonances on the nanometer scale, including biomolecules \cite{Asenjo-Garcia2014}. Preliminary experiments have provided encouraging results, although the interpretation requires further clarification \cite{Harvey2015}.

We again note that this is in contrast to optical probing of chiral nanoparticles and molecules \cite{Barron_book}. Circular dichroism enables optical probing of chiral objects using only polarization (spin or helicity) degrees of freedom of light \cite{Tang2010,Bliokh2011_II,Tang2011,Bliokh2014}. At the same time, optical vortices, are  insensitive to chiral properties of matter and do not produce dichroism \cite{Babiker2002,Loffler2011,Araoka2005}.

The interplay between the $\ell$-fold azimuthal symmetry of electron vortex beams and the discrete rotational symmetry of SPR modes was theoretically investigated in \cite{Ugarte2016}. The interaction between these two symmetries could allow selective probing of desired SPR modes. For instance, the $\ell=1$ vortex beam preferentially excites the {\it dipole} resonance, while the $\ell=2$ vortex couples more strongly to the {\it quadrupole} one.
More generally, it was recently shown that, in the inelastic interaction between an electron beam and a plasmonic nanostructure, a signature of the scalar electromagnetic potential potential of SPR, $V_{\rm SPR}$, is imprinted in the {\it phase} of the scattered electron wave \cite{Guzzinati2016}.
By accepting only electrons scattered along the propagation $z$-axis (i.e., the non-vortex mode with $\ell^\prime =0$), it becomes possible to detect only resonances whose potential possesses the same symmetry as the impinging electron beam. This was also
tested experimentally using a two-lobed beam, reminiscent of an HG$_{01}$ Hermite--Gaussian mode. In agreement with the theory and numerical simulations, the experiments \cite{Guzzinati2016} showed that such a beam can selectively detect the dipole excitation mode while suppressing the higher-order excitations.

\begin{figure}[!t]
\centering
\includegraphics[width=0.99\textwidth]{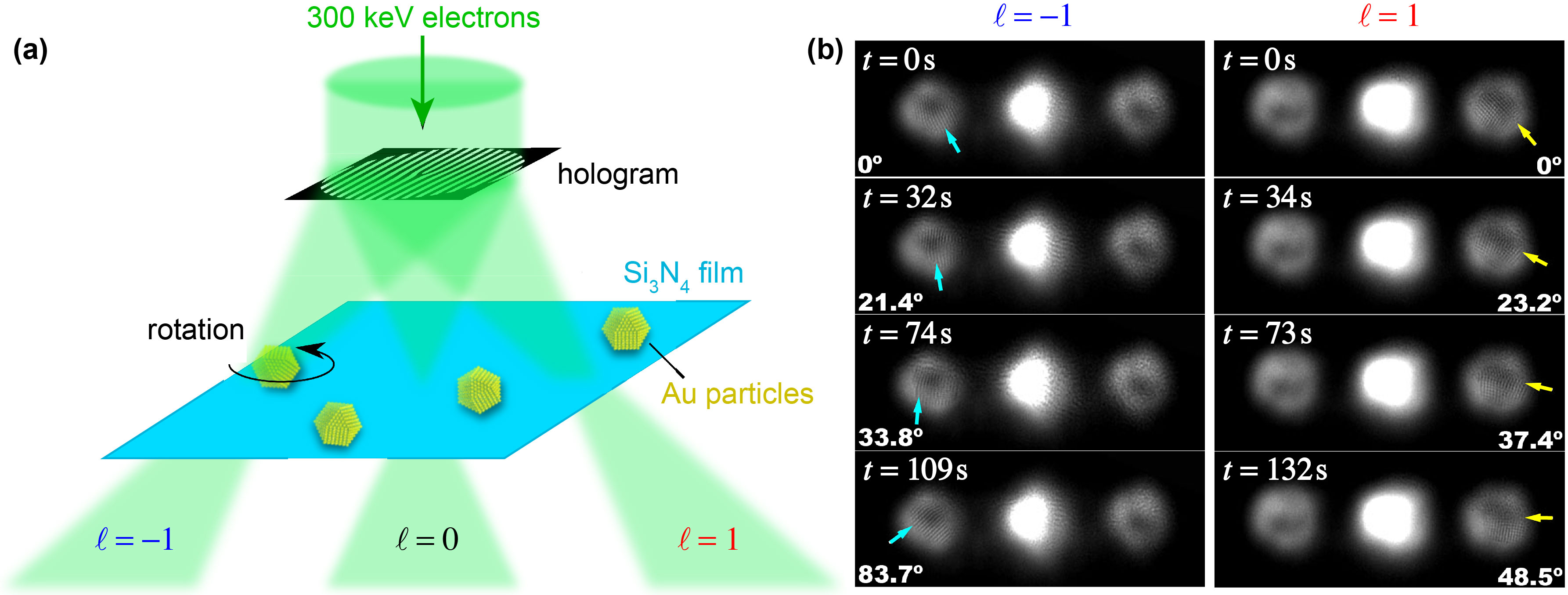}
\caption{Angular-monentum transfer from electron vortex beams to nanoparticles \cite{Verbeeck2013}. (a) Schematic of the experiment. Vortex beams with different topological charges $\ell$, generated with a holographic mask, are focused on the sample, so that only one of the beams impinges on the selected gold nanoparticle. (b) Experimental images displaying clockwise and counter-clockwise rotations of a nanoparticle (indicated by arrows) in the $\ell=-1$ and $\ell=1$ vortex beams.}
\label{fig:OAMtransfer}
\end{figure}

\subsubsection{Angular momentum transfer to nanoparticles}

The OAM exchange between electron waves and the sample in inelastic scattering can also be revealed by observing the sample state. Indeed, in the extreme case when the electron is absorbed by the sample, the total angular momentum of the sample is increased by the OAM carried by the electron. This was used in optics for the detection of mechanical properties of optical OAM beams, which produced the rotation of absorptive particles \cite{He1995,Garces2003,ONeil2002,Curtis_2003,Bliokh2015PR}. 
Similar angular-momentum transfer from electron vortex beams to small particles can become noticable when the current of the electron beam is high, while the moment of inertia of the particle is low. In the absence of friction, the particle would start to rotate and the rotational velocity would keep increasing as more and more electrons are absorbed. In reality, any particle is in contact with a substrate (typically a support film), and significant friction exists between the support film and the particle, causing at best a low rotational speed that is dominated by a balance between the friction and the flux of the incoming OAM quanta. Recent experiments have demonstrated this effect \cite{Verbeeck2013,Gnanavel2012}, as shown in Fig.~\ref{fig:OAMtransfer}.
A quantitative interpretation is complicated due to many factors: unknown friction between particle and support; strong dependence of the effect on the orientation of crystalline particles; simultaneous effects of phonon, plasmon, core-loss and elastic scatterings; and rather unpredictable effect of carbon contamination and beam damage effects occurring at the high electron currents \cite{Verbeeck2013,lloyd_mechanical_2013}. Therefore, it is very desirable to preform such experiments on particles in a liquid medium in an environmental sample holder or using diamagnetically-levitated particles.
\footnote{Charging due to secondary-electron generation will likely make this difficult even
though we have demonstrated diamagnetic sample levitation inside an SEM chamber.} 
This might offer direct mechanical measurements of the angular momentum carried by electron vortex states.

\clearpage

\section{High-energy processes with vortex electrons}
\label{sect:highenergy}
\vspace{1mm}

\subsection{Vortex solutions of the Dirac equation: Dirac--Bessel beams}
\vspace{2mm}

High-energy processes, such as collisions of electrons, require that the electrons be treated in a fully relativistic manner. We start this section with a reminder of this formalism and with a review of the exact vortex solutions of the free Dirac equation,
which were constructed in \cite{Bliokh_Relativistic,Karlovets:2012,Serbo:2015}.
These expressions have different forms but are equivalent to each other.
Depending on the specific problem, one of these forms may be more appropriate than the others. Below we will describe these solutions in detail, with the aim to provide
a convenient reference for future calculations of vortex electron scattering processes.
Throughout this section, we will use the relativistic units $\hbar = c = 1$.

Before going into details, we mention that exact solutions of the Dirac
equation in cylindrical coordinates, either free or in the presence of external fields or potentials, 
have been known since decades. In particular, the first edition of the monograph \cite{Bagrov:1990} published back in 1982 \cite{Bagrov:1982} already contained
free solutions of the Dirac equation exhibiting vortex-like azimuthal dependences.
Various formal aspects of such solutions, with different boundary conditions, were studied later by several authors, see, e.g., \cite{Villalba:1990,OUYANG1999297,Bagrov:2005nf,Mihaila:2006hi,Leary2008}.
However it was only in \cite{Bliokh_Relativistic} and later publications that the exact vortex solutions of the Dirac equation were written in a way convenient both for theoretical exploration of its angular-momentum ``anatomy'' and for usage in scattering processes.
In this section we expose these recent developments.

We start with the {\it scalar} Bessel-beam solutions (\ref{eq:Bessel}) in the Fourier-integral form and with the restored $\exp(-i E t)$ factor:
\begin{equation} 
 \psi_{k_z \kappa \ell}({\bf r}, t) = e^{-i Et}
 \int \frac{\mathop{d^2 {\bf k}_\perp}}{(2\pi)^2}\, 
\tilde \psi_{\kappa \ell}({\bf k}_\perp)\, e^{i {\bf k} \cdot {\bf r}},
\label{scalarlike-again}
\end{equation}
where the Fourier amplitude corresponds to the conical distribution of wave vectors, Eq.~(\ref{eq:Fourier-Bessel}) and Fig.~\ref{fig:Bessel}(a): 
\begin{equation} 
\tilde\psi_{\kappa \ell}({\bf k}_\perp) =  (-i)^\ell\, \frac{\delta(k_\perp - \kappa)}{\kappa}\, e^{i \ell \phi}.
\label{a-kappa-ell}
\end{equation}
Here $(k_{\perp},\phi,k_z)$ are the cylidnrical coordinates in ${\bf k}$-space, and the subscripts explicitly indicate all continuous- and dicrete-spectrum parameters of the solutions. The wave functions (\ref{scalarlike-again}) and (\ref{a-kappa-ell}) correspond to a complete orthogonal set of solutions of the scalar wave equation with a definite energy $E = \sqrt{k^2+m_e^2}$, longitudinal momentum $p_z = k_z$, 
and $z$-component of the OAM $L_z = \ell$. 
These solutions are normalized as
\be
\int d^3 {\bf r}\, \psi_{k_z \kappa \ell}^*({\bf r}, t)\, \psi_{k'_z \kappa' \ell'}({\bf r}, t) = 
{1 \over \kappa}\, \delta(\kappa-\kappa')\, \delta(k_z-k'_z)\,\delta_{\ell \ell'}\,,
\ee
where $\delta_{ab}$ is the Kronecker delta.

A relativistic electron with spin is described by the {\it multi-component} (bi-spinor) wave function $\Psi({\bf r}, t)$. In this case, the Bessel states can be introduced as a straigthforward generalization of scalar Bessel modes:
\begin{equation} 
\Psi_{k_z \kappa \ell s}({\bf r}, t) = e^{-i Et}
\int \frac{\mathop{\mathrm{d}^2 {\bf k}_\perp}}{(2\pi)^2}\, 
\tilde\psi_{\kappa \ell}({\bf k}_\perp)\, u_{{\bf k} s}\, e^{i {\bf k}\cdot {\bf r}}.
\label{Dirac-twisted}
\end{equation}
This equation differs from Eq.~(\ref{scalarlike-again}) only by the presence of the bispinor $u_{{\bf k} s}$ which corresponds to the plane-wave solution with momentum ${\bf k}$ and in the spin state which we generically denote by $s$.
Note that Eq.~(\ref{Dirac-twisted}) includes the assumption
that the Fourier amplitude $\tilde\psi_{\kappa \ell}$ is the same as in (\ref{a-kappa-ell}) 
and, in particular, that it does not depend on the spin state of the electron.
\footnote{This assumption is not mandatory. One can build vortex electron solutions as superpositions of plane-wave electrons with continuously varying ${\bf k}_\perp$ accompanied with the varying spin state, i.e., with inhomogeneous polarization \cite{Karimi2012}. We will not discuss such solutions, apart from the mere possibility of introducing a spin- and azimuthal-angle-dependent phase factor, see the discussion after Eq.~(\ref{bispinor-PW-helicity}).}

There are two crucial aspects in which Eq.~(\ref{Dirac-twisted}) differs from the scalar case.
First, the quantity $\ell$ can no longer be interpreted as the $z$-component of the electron OAM.
In fact, the solution (\ref{Dirac-twisted}) is not an eigenstate of the OAM operator $\hat{L}_z$ because even the individual plane-wave components in the superposition do not possess a well-defined OAM [see Eq.~(\ref{eq:bliokh53})].
This is a manifestation of the fact that the OAM and spin operators, $\hat{L}_z$ and $\hat{S}_z$, do not commute with the Dirac Hamiltonian \cite{QEDbook,Thaller_book}, which leads to the intrinsic spin-orbit interaction \cite{Bliokh_Relativistic} (see Section~\ref{spinorbit}). 
The solution (\ref{Dirac-twisted}) has a well-defined $z$-component of the {\it total} angular momentum, though, $\hat{J}_z = \hat{L}_z + \hat{S}_z$, and we will describe below its relation with the parameter $\ell$. 
Second, we have the freedom to choose the basis for describing the polarization state $s$.
One possibility exploited in \cite{Bliokh_Relativistic} is to define polarization states with respect to the same {\it $z$-axis}. Another possibility is to use the {\it helicity basis} \cite{Karlovets:2012,Serbo:2015},
which is especially convenient for scattering processes involving high-energy electrons.

Let us first take the former option and consider the plane-wave bispinor $u_{{\bf k}s}$,
whose polarization state is defined with respect to the fixed $z$-axis.
The bispinor has the following form \cite{QEDbook}:
\be
u_{{\bf k} s}=\left(\begin{array}{c}
\sqrt{E_+}\;W_{s} \\
\sqrt{E_-} (\hat{\bm \sigma} \cdot \bar {\bf k})\;W_{s} \end{array}\right),
\label{bispinor-PW-fixed}
\ee
where $\hat{\bm \sigma}$ is the vector of $2\times 2$ Pauli matrices, with $\hat{\bf s} = \hat{\bm \sigma}/2$ being the spin-1/2 operator in the electron rest frame, Eq.~(\ref{eq:bliokh51}), $E_\pm=E\pm m_e$, $\bar {\bf k} = {\bf k}/k$ is the unit vector in the ${\bf k}$-direction. The basis spinors $W_s$ are eigenvectors of the non-relativistic spin $\hat{s}_z$ with the eigenvalues $s = \pm 1/2$:
\be
W_{+1/2} =  \left(\!\begin{array}{c} 1 \\ 0 \end{array}\!\right)\,,\quad
W_{-1/2} =  \left(\!\begin{array}{c} 0 \\ 1 \end{array}\!\right)\,.\label{spinor-basis}
\ee
It is instructive to express $\hat{\bm \sigma} \cdot \bar {\bf k}$ using spherical coordinates $(k,\phi,\theta)$ in ${\bf k}$-space:
\be
\hat{\bm \sigma} \cdot \bar {\bf k} =  \hat{\sigma}_+ \sin\theta e^{-i\phi} + \hat{\sigma}_- \sin\theta e^{i\phi} + \hat{\sigma}_z \cos\theta,
\label{helicity-operator}
\ee
where $\sigma_{\pm} = (\sigma_x \pm i\sigma_y)/2$.

Substituting Eqs.~(\ref{bispinor-PW-fixed})--(\ref{helicity-operator}) into Eq.~(\ref{Dirac-twisted}), we obtain \cite{Bliokh_Relativistic}:
\bea
\Psi_{k_z \kappa \ell, +{1\over2}}({\bf r}, t) = {e^{i (k_z z- Et)} \over 2\pi}
\left[\left(\begin{array}{c}
\sqrt{E_+}\\
0 \\
 \sqrt{E_-} \cos\theta_0 \\
 0 \end{array}\right) e^{i\ell \varphi} J_{\ell}(\kappa r)
+ i  
\left(\begin{array}{c}
0\\
0 \\
0 \\
\sin\theta_0 \sqrt{E_-} \end{array}\right) e^{i(\ell+1) \varphi} J_{\ell+1}(\kappa r_\perp)\right] ,\nonumber\\
\Psi_{k_z \kappa \ell, -{1\over2}}({\bf r}, t) = {e^{i (k_z z- Et)} \over 2\pi}
\left[\left(\begin{array}{c}
0 \\
\sqrt{E_+}\\
0 \\
- \sqrt{E_-} \cos\theta_0 \end{array}\right) e^{i\ell \varphi} J_{\ell}(\kappa r)
- i  
\left(\begin{array}{c}
0\\
0 \\
\sin\theta_0 \sqrt{E_-} \\
0 \end{array}\right) e^{i(\ell-1) \varphi} J_{\ell-1}(\kappa r)\right] ,
\label{spinors-bliokh}
\eea
where $r=|{\bf r}_\perp|$, as before, and $\sin\theta_0 = \kappa/k$ fixes the polar angle of the conical Bessel plane-wave spectrum, Fig.~\ref{fig:Bessel}(a).
The solutions (\ref{spinors-bliokh}) represent Bessel beams for Dirac electrons, i.e., the electron counterparts of optical {\it vector} Bessel beams \cite{Jauregui2005,Bliokh2010}. 
The probability density and current distributions for these {\it Dirac--Bessel beams} with different $\ell$ and $s$ are shown in Fig.~\ref{fig:bliokh13}.
The presence of two terms with distinct $\ell$ for any spin state $s = \pm 1/2$
in Eqs.~(\ref{spinors-bliokh}) makes it evident that such solutions are not eigenmodes of the OAM operator $\hat{L}_z = -i \partial/\partial\varphi$. These are not eigenmodes of the $z$-component of the full relativistic ($4\times 4$) SAM operator (\ref{eq:bliokh50}) $\hat{\bf S} = {\rm diag}(\hat{\bm \sigma},\hat{\bm \sigma})$ either. Instead, vector Bessel beams (\ref{spinors-bliokh}) are eigenmodes of the total angular momentum operator $\hat{J}_z = \hat{L}_z + \hat{S}_z$ with the eigenvalues $J_z = \ell + s$. 
As described in Section~\ref{spinorbit}, one observes
the intrinsic {\it spin-orbit interaction} (SOI) in the free Bessel electron beam, whose
strength is determined by the dimensionless parameter $\sqrt{\Lambda} = \sin\theta_0 \sqrt{E_-/E} $, Eq.~(\ref{eq:bliokh54}).
This interaction becomes weak, $\Lambda \ll 1$, both in the {\it non-relativistic} limit $E_- \ll E$ 
and in the {\it paraxial} approximation $\theta \ll 1$; the vortex quantum number $\ell$
corresponds to the approximate OAM $L_z \simeq \ell$ in either of these limits.
For non-paraxial relativistic electron vortex beams, the SOI is significant and leads to spin-dependent probability densities (see Fig.~\ref{fig:bliokh13}) \cite{Bliokh_Relativistic}. 

We now explore the second option, choosing the {\it helicity} basis for the electron spin states, which is more convenient for high-energy electrons. Consider again the plane-wave spinor (\ref{bispinor-PW-fixed}) but now use, instead of $W_s$, the eigenstates of the helicity operator $\hat{\chi} = \hat{\bf s} \cdot \bar {\bf k}=  \hat{\bm \sigma} \cdot \bar {\bf k}/2$: 
%
\be
\hat{\chi}\, W^{(\chi)} = \chi\, W^{(\chi)},\quad \chi = \pm 1/2 .
\ee
%
These spinors can be explicitly written as 
\be
W^{(+1/2)}({\bf k}) = \left(\begin{array}{c}
\cos \frac{\theta}{2} \\
 \sin \frac{\theta}{2}e^{i\phi} 
\end{array}\right) ,
\quad 
W^{(-1/2)}({\bf k}) = \left(\begin{array}{c}
-\sin \frac{\theta}{2}e^{-i\phi} \\
 \cos \frac{\theta}{2} 
\end{array}\right).
\label{spinors-helicity}
\ee
Then, we can replace $\hat{\bm \sigma} \cdot \bar {\bf k} \to 2\chi$ inside the bispinor (\ref{bispinor-PW-fixed}) to arrive at
 \be
u_{{\bf k} \chi}=\left(\begin{array}{c}
\sqrt{E_+}\;W^{(\chi)} \\
2\chi \sqrt{E_-} \;W^{(\chi)} \end{array}\right).
\label{bispinor-PW-helicity}
\ee
The $\phi$-dependence is still present inside the spinors $W^{(\chi)}$, 
and it is different for its upper and lower components, displaying once again
that the Dirac electron is not in a fixed-OAM state.
Next, we obtain electron Bessel beams with a given helicity
by substituting the bi-spinors (\ref{bispinor-PW-helicity}) into Eq.~(\ref{Dirac-twisted}) with the scalar Fourier components $\tilde\psi_{\kappa \ell}$, Eq.~(\ref{a-kappa-ell}) \cite{Karlovets:2012}.
Akin to the Bessel beams with a well-defined non-relativistic spin component $s_z = s$, Eqs.~(\ref{spinors-bliokh}), these solutions are eigenmodes of the total angular momentum operator $\hat{J}_z$ with the eigenvalues $J_z=\ell + \chi$, involving the helicity $\chi$ instead of $s$.
Thus, alternatively, one can use the helicity-dependent Fourier components $\tilde\psi_{\kappa (J_z-\chi)}$, where $J_z$ denotes the total angular momentum of the Bessel electron irrespective of its helicity $\chi$ \cite{Serbo:2015}. 
Note that in the paraxial approximation $\theta \ll 1$, the SOI effects become negligible and the $s$-based and $\chi$-based solutions approximately coincide with each other.

Finally, the link between the fixed-spin basis and helicity basis can be established with the aid of Wigner's $D$-function \cite{LandauLifshitz3}:
\be
W^{(\chi)}({\bf k}) = \sum_{s=\pm 1/2} D^{1/2}_{s\chi}(\phi,\theta,-\phi) W_{s} ,
\ee
where $W^{(\chi)}$ and $W_s$ are given by (\ref{spinors-helicity}) and (\ref{spinor-basis}), respectively.
The explicit form of the Wigner $D$-function for the spin-$1/2$ field is
\be
D^{1/2}_{s\chi}(\phi,\theta,-\phi) = e^{-i s\phi} \, d^{1/2}_{s\chi}(\theta) \, e^{i\chi\phi} ,
\quad
d^{1/2}_{s\chi}(\theta) = \delta_{s\chi}\cos{\theta\over 2} - 2\, s\, \delta_{s,-\chi}\sin{\theta\over 2} .
\ee
With this notation, the bispinor (\ref{bispinor-PW-helicity}) takes yet another form:
\be
u_{{\bf k}\chi} = \sum_{s=\pm 1/2} e^{i(\chi-s)\phi} d^{1/2}_{s\chi}(\theta)U_{s}(E,\chi) ,\quad
U_{s}(E,\chi) = \left(\begin{array}{c}
\quad\sqrt{E_+}\;W_s \\
 2\chi \sqrt{E_-}\;W_s 
\end{array}\right) .
\ee
Unlike expressions (\ref{bispinor-PW-fixed}) and (\ref{bispinor-PW-helicity}),
the bispinor $U_s$ itself is now free from the $\phi$-dependence, 
which reappears only in the Wigner $D$-function.
This representation is convenient for calculating high-energy electron scattering processes,
as demonstrated in \cite{Serbo:2015}. Passing to the coordinate representation (\ref{Dirac-twisted}), we arrive at:
 \be
\Psi_{k_z\kappa \ell \chi}({\bf r}, t)=
{e^{i (k_z z- Et)} \over 2\pi}
\sum_{s=\pm 1/2} i^{(\chi-s)}
e^{i(\ell+\chi-s) \varphi}\, d^{1/2}_{s \chi}(\theta_0)\,
J_{\ell+\chi-s}(\kappa r ) \, U_{s}(E,\chi) ,
\label{helicity-basis-Dirac-Bessel}
\ee
which is the helicity-basis counterpart of Eqs.~(\ref{spinors-bliokh}).

\subsection{Vortex electrons in a laser field: Volkov--Bessel solutions}
\label{sec:Volkov}
\vspace{2mm}

The exact vortex solutions of the Dirac equation can also be found
for an electron moving in the field of an {\it electromagnetic (EM) wave}.
These solutions can be named {\it Volkov--Bessel beams} as they extend 
the well-known Volkov solutions \cite{Volkov1935,QEDbook} to the Bessel vortex electron.
These solutions were constructed and investigated first in \cite{Karlovets:2012}
and later in \cite{Hayrapetyan-et-al:2014,Bandyopadhyay:2015}, and 
they offer insights into modifications of the vortex electron properties in a strong laser field.
In this subsection, we denote the electron momentum by ${\bf p}$, reserving the letter ${\bf k}$ for the wave vector of the electromagnetic wave.

The Volkov--Bessel solutions are constructed in the same way as the usual Bessel states.
One uses the basis of plane-wave Volkov solutions with four-momentum $p^\mu$
to combine them as in (\ref{Dirac-twisted}) with the same Fourier amplitudes
$\tilde\psi_{\kappa \ell}({\bf p}_\perp)$.
The Dirac equation in the field of an EM wave reads \cite{QEDbook}:
\begin{equation}
[\gamma_\mu(\hat p^\mu - eA^\mu) - m_e]\Psi = 0,
\label{Dirac}
\end{equation}
where $\gamma_\mu$ are the Dirac matrices, $\hat p^\mu = (i\partial_t, -i{\nabla})$ is the electron canonical four-momentum operator, and $A^\mu$ is the electromagnetic four-potential.
The EM wave is described by the wave four-vector ${\mathcal k}^\mu = (\omega, {\mathbcal k})$, satisfying ${\mathcal k}^\mu {\mathcal k}_\mu =0$, 
and the four-potential $A^\mu$, in the traditionally-chosen Lorentz gauge, satisfies 
${\mathcal k}^\mu A_\mu =0$. 
The EM plane wave depends on the coordinates via the single phase variable 
$\xi \equiv {\mathcal k}^\mu r_\mu = \omega\, t - {\mathbcal k} \cdot {\bf r}$.
The Dirac equation (\ref{Dirac}) can be recast in the form of a second-order equation with a simpler spinorial structure:
\be
\left[ \hat{\mathcal p}_\mu \hat{\mathcal p}^\mu - m_e^2 -i e F_{\mu\nu}\sigma^{\mu\nu}\right]\Psi = 0.
\label{Dirac-2nd}
\ee
Here, $\hat{\mathcal p}^\mu = \hat p^\mu -e A^\mu$ is the kinetic electron four-momentum, $F_{\mu\nu} = \partial_\mu A_\nu - \partial_\nu A_\mu$ is the EM field tensor, and 
$\sigma^{\mu\nu} = (\gamma^\mu\gamma^\nu - \gamma^\nu\gamma^\mu)/2$.

\begin{figure}[!t]
\centering
\includegraphics[width=0.95\textwidth]{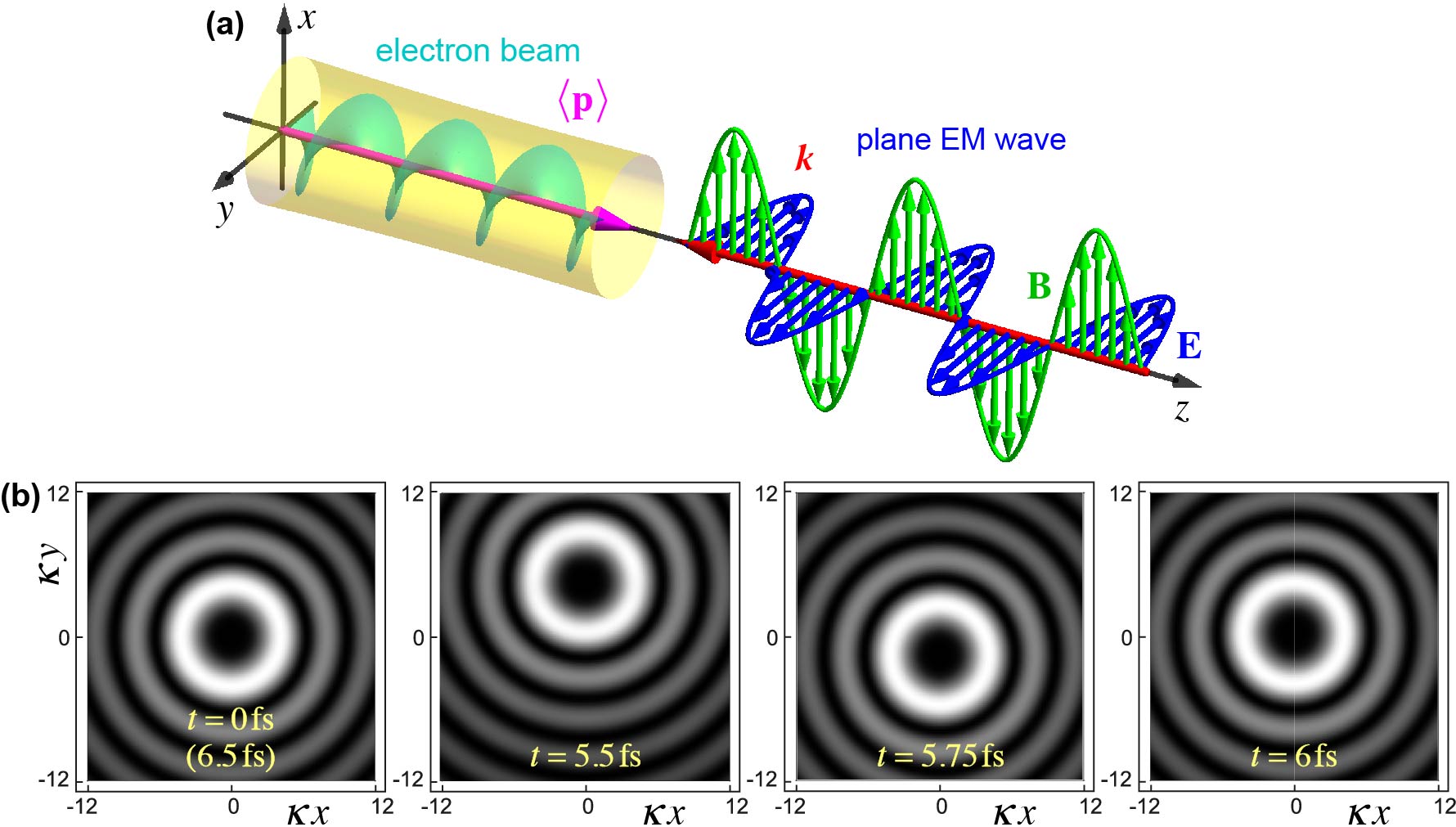}
\caption{(a) Schematics of the Volkov--Bessel problem: a Bessel-beam solution for the Dirac electron and the field of an electromagnetic (EM) plane wave. Here the electron vortex beam and the EM wave counterpropagate with the mean electron momentum $\langle {\bf p} \rangle$ and the EM wave vector ${\mathbcal k}$. The EM wave is linearly $y$-polarized. (b) Temporal variations of the transverse probability-density distribution in a paraxial Volkov--Bessel beam shown in the panel (a) \cite{Hayrapetyan-et-al:2014}. Here, the electron parameters are: azimuthal vortex index $\ell = 3$, energy $E=300$ keV, and opening angle $\theta_0 =0.02$ rad. Instead of a plane EM wave, here a five-cycle laser pulse was used with the central frequency $\omega_c = 10^{16}$ Hz and electric-field amplitude $E_0 = 10^8$ V/cm. The time range of 6.5 fs approximately corresponds to the duration of the pulse.}
\label{fig-Bessel-Volkov-temporal}
\end{figure}

The plane-wave Volkov solution for the electron with four-momentum $p^\mu = (E, {\bf p})$ satisfying $p^\mu p_\mu = m_e^2$ is \cite{Volkov1935,QEDbook}:
\be
\Psi_p(r)  \propto \left[1 + {e\over 2(p_\mu k^\mu)}(\gamma_\mu k^\mu)(\gamma_\mu A^\mu)\right] u_{p,s} \exp (i{\mathcal S}) ,
\label{Volkov}
\ee
where $u_{p,s}$ is the plane-wave Dirac bi-spinor (with any spin state ``$s$''), and ${\mathcal S}$ is the action:
\be
{\mathcal S} = - p_\mu r^\mu - {e\over (p_\mu {\mathcal k}^\mu)}\int d\xi\, (p_\mu A^\mu) + {e^2\over 2(p_\mu {\mathcal k}^\mu)}\int d\xi\, (A_\mu A^\mu) .
\label{classical-action-in-EM-wave}
\ee
In order to construct the Volkov--Bessel solution, one must specify the relative kinematics of the EM wave and the reference axis used to define the vortex electron.
The simplest convention is to take {\it counter-propagating} EM wave and 
electron vortex beam, Fig.~\ref{fig-Bessel-Volkov-temporal}(a). Assuming propagation of the vortex electron along the $z$-axis, the wave four-vector of the EM wave is ${\mathcal k}^\mu = \omega (1,\,0,\,0,\,-1)$.
One can consider either a circularly \cite{Karlovets:2012} or linearly \cite{Hayrapetyan-et-al:2014} polarized EM wave.
In either case, the expressions emerging from inserting Eq.~(\ref{Volkov}) into the Bessel-beam form (\ref{Dirac-twisted}) are simplified, and the integral over the azimuthal angle $\phi$ can be performed exactly.

An elegant way to evaluate this integral is to group the first two terms in (\ref{classical-action-in-EM-wave}). Then, the resulting integral becomes exactly as for the free Bessel electron, with the replacement of the transverse coordinates:
\be
{\bf r}_\perp \to {\bf R}_\perp =  {\bf r}_\perp  + {e \over (p_\mu {\mathcal k}^\mu)}\int d\xi\, {\bf A}_\perp .
\ee
One then obtains a sum of two Bessel functions [as in Eqs.~(\ref{spinors-bliokh}) or (\ref{helicity-basis-Dirac-Bessel}], depending on the chosen spin state), but with the substitutions $\kappa r \to \kappa  R_{\perp}$ in the Bessel fucntions and $\varphi \to \varphi_{\bf R}$ in the vortex phase factors (here $\varphi_{\bf R}$ is the azimuthal coordinate of ${\bf R}_{\perp}$). 
The last factor in Eq.~(\ref{classical-action-in-EM-wave}) represents, 
for a monochromatic plane wave, an integral over a constant, generating an extra phase 
$\propto (A_\mu A^\mu)\,\xi$. Just as for the usual Volkov solution, it amounts to the replacement 
of the energy and the longitudinal momentum by the {\it quasi}-energy and {\it quasi}-momentum.

One can now calculate the dynamical properties (e.g., the expectation values of the spin and OAM) of the Volkov--Bessel solutions \cite{Karlovets:2012}.
Since the effective radial coordinate ${R}_{\perp}$ explicitly depends on time, 
one can consider both instantaneous and effective time-averaged quantities (marked with the overbar here).
For the time-averaged quantities, the $z$-propagating Volkov plane-wave state with a well-defined helicity $\chi$ displays a reduced spin $\langle \bar{S}_z\rangle$:
\be
\langle \bar{S}_z\rangle = \chi\, {1- e^2 A_\perp^2 \omega /2E (p_\mu {\mathcal k}^\mu) 
\over 1+ e^2 A_\perp^2 \omega /2E (p_\mu {\mathcal k}^\mu)}.
\ee
%
This effective depolarization is a purely kinematic effect caused by the spin precession
in the external EM field, and its strength is governed by the dimensionless parameter
\be
\eta^2 = {e^2 A_\perp^2 \over m_e^2}.
\ee
In a strong laser field, $\eta \gsim 1$, $\langle \bar{S}_z\rangle$ is substantially reduced
and can even change sign with respect to the helicity $\chi$. 
For the Volkov--Bessel electron beams, the same effect is responsible for the shift of the time-averaged total angular momentum $\langle \bar{J}_z\rangle$, but the effect is always less than one, even for large $|\ell |$.
For ultrarelativistic electrons, the exact reversal point is shifted to higher $\eta$.
However in such strong fields, one needs to complement the picture with inelastic scattering processes.

An alternative approach to the azimuthal integrals for the Volkov--Bessel solutions
was advocated in \cite{Hayrapetyan-et-al:2014,Bandyopadhyay:2015}.
Considering an EM wave {\it linearly polarized along the $y$-axis}, $A^\mu = (0,0,A,0)$,
the exponential of the second term in Eq.~(\ref{classical-action-in-EM-wave}) can be expanded as
\be
e^{i f \sin \phi} = \sum_{n=-\infty}^{+\infty}\!J_n(f)\, e^{in\phi}\,,\quad
f = {e\kappa \over ({\mathcal k}_\mu p^\mu)}\int d\xi\, A(\xi)\,.
\ee
The Volkov--Bessel solution then takes the form of an infinite sum
of the Bessel modes with vortex number $(\ell+n)$ weighted with $i^n J_n(f)$.
Although this representation is not very convenient for an infinite EM plane wave,
it can be used to investigate the response of the Bessel electron to a strong few-cycle laser pulse \cite{Hayrapetyan-et-al:2014}.
In this case, only a few terms in the above summation are important.
The numerical analysis of \cite{Hayrapetyan-et-al:2014} demonstrated transverse $y$-oscillations of the Bessel-beam probability-density distribution as the laser pulse passes through the electron. Figure~\ref{fig-Bessel-Volkov-temporal}(b) illustrates such temporal variations of the transverse probability density distribution in a paraxial Volkov--Bessel beam.  

\subsection{Schemes for vortex beam collisions}
\vspace{2mm}

Nuclear and particle physics is another area where vortex electrons (and in general, vortex states of particles) can emerge as a novel tool for the experimental exploration
of fundamental interactions.
In virtually all situations involving collisions of particles, be it a fixed-target or a collider-like setting, are well described with plane waves.
The fact that real colliding particles are wavepackets is inessential; exceptions exist \cite{Kotkin:1992} but are extremely rare.
Vortex states bring in a new degree of freedom, the OAM, which can be exploited in collisions.
The instrumentation which would allow one to prepare, accelerate to high energies, and collide vortex electrons,
protons and other particles, does not exist yet, 
but first exploratory studies of accelerating twisted electrons to multi-MeV energies are underway in JLab \cite{Dutta:2014}. 
Anticipating that dedicated experimental efforts will eventually make such experiments possible,
it is timely to ask what opportunities this new instrument can offer for nuclear and high-energy physics.

This question leads us to the problem of the theoretical description of the scattering processes 
of high-energy vortex particles.
Below we will overview the general kinematic novelties which arise in collisions of vortex states,
and mention particular processes investigated so far.

\begin{figure}[!t]
\centering
\includegraphics[width=0.95\textwidth]{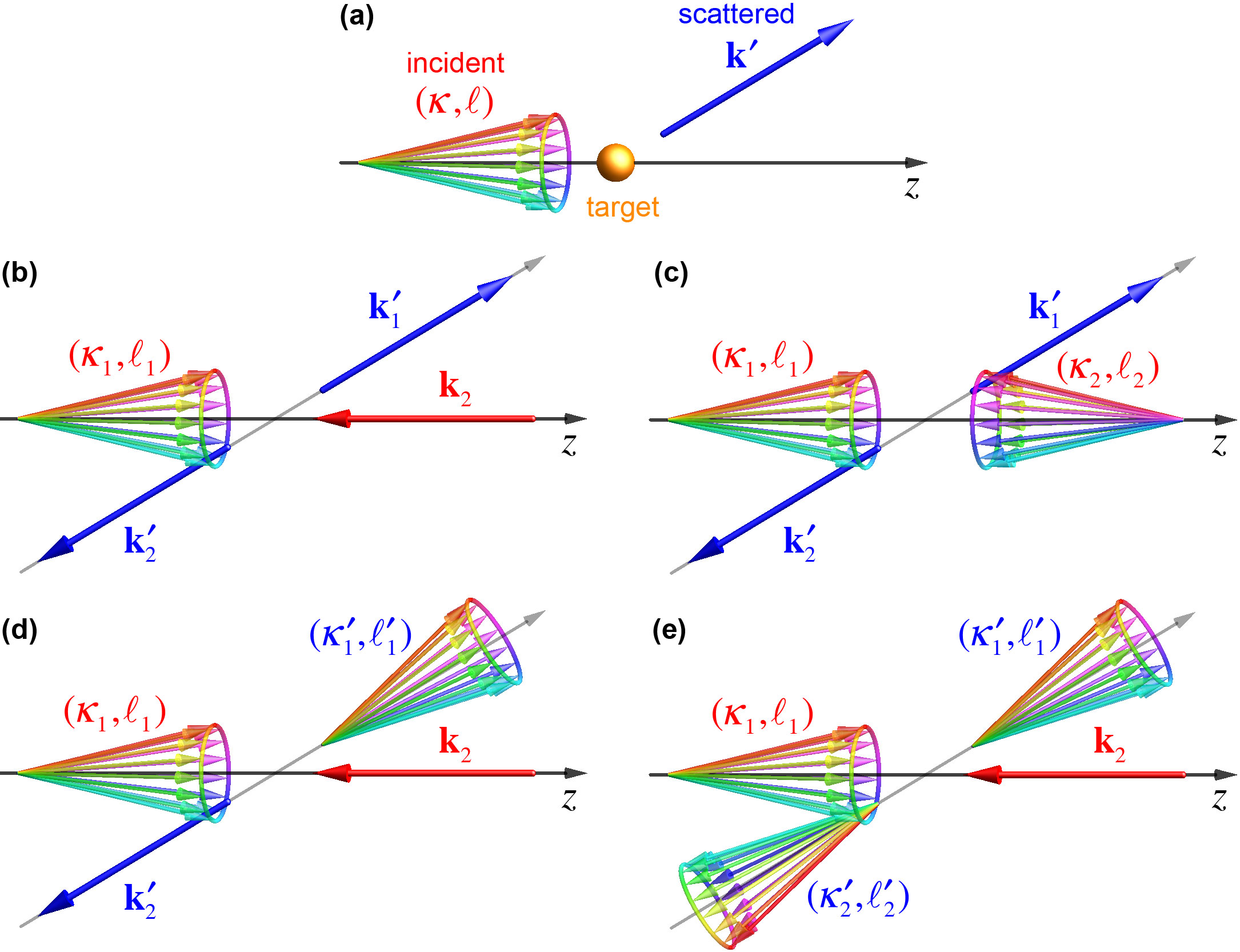}
\caption{Schematic diagrams of scattering processes involving vortex particles. (a) Fixed-target scattering of vortex (V) into a plane wave (PW). (b) Single-vortex scattering (collision) V+PW $\to$ PW+PW. (c) Vortex-vortex scattering into plane waves: V+V $\to$ PW+PW. (d,e) Vortex-into-vortex scattering V+PW $to$ V+PW and V+PW $to$ V+V.}
\label{fig-scattering-schemes}
\end{figure}

Several schemes for collisions of vortex states are possible, as shown in Fig.~\ref{fig-scattering-schemes}.
First, a {\it vortex state} (V) can scatter, either elastically or inelastically, on a fixed scattering center,
and the final state is usually assumed to be a {\it plane wave} (PW), Fig.~\ref{fig-scattering-schemes}(a).
The prototypical problem here is the (screened) Rutherford scattering 
of vortex electrons on atoms, either elastic \cite{VanBoxem_Rutherford,Serbo:2015} or inelastic \cite{Yuan:2013},
as well as radiative capture of vortex electrons by atoms \cite{Matula:2014,Zaytsev:2017}.
In these examples, a non-relativistic Schr\"odinger-equation treatment for incoming
spinless (scalar) vortex wave is sufficient to grasp the essential details, but at larger energies, the full relativistic treatment is needed.

Second, one can consider the scattering process in the collider-like kinematics,
when two incoming particles scatter into a certain final state. For such processes, the quantum-field-theoretic treatment
is more appropriate. 
Different vortex vs. plane waves settings have been considered: V+PW $\to$ PW+PW \cite{Ivanov:2011a,Seipt:2014}, V+V $\to$ PW+PW 
\cite{Ivanov:2011a,Ivanov:2011b,Ivanov:2012b,Ivanov:2012c,Ivanov:2016jzt,Ivanov:2016oue,Karlovets:2016jrd,Karlovets:2016dva},
V+PW $\to$ V+PW \cite{Jentschura:2011a,Jentschura:2011b,Ivanov:2011a,IvanovSerbo:2011},
and even V+PW$\to$ V+V \cite{Ivanov:2012a}, Figs.~\ref{fig-scattering-schemes}(b)--(e).
The fully-vortex scattering setting V+V$\to$V+V can be viewed as a particular case of the general formalism
of collision of arbitrarily shaped wavepackets \cite{Karlovets:2016dva}.
Each collision setting has its own challenges and novelties, both in the theoretical description and in the possible experimental realization.

\subsection{Fixed-target scattering}
\label{sec:fixed-target}
\vspace{2mm}

In the fixed-target scattering, V $\to$ PW [Fig.~\ref{fig-scattering-schemes}(a)], the scattering center can absorb any momentum transfer.
Since the initial state is a coherent superposition of many plane waves, 
individual plane wave scattering amplitudes with different momentum transfers
interfere in the total amplitude (for the general wavepacket scattering on atoms,
we refer to the recent pedagogical exposition \cite{Karlovets:2015nva}). 
For example, for the Bessel beam with parameters $\kappa$ and $\ell$,
we can write the V $\to$ PW scattering amplitude as
\be
f(\kappa,\ell; {\bf k}') =  \int {d^2 {\bf k}_\perp \over (2\pi)^2}\, \tilde\psi_{\kappa \ell}({\bf k}_\perp)\, f({\bf k}; {\bf k}')\,,
\label{fixed-target}
\ee
where $f({\bf k}; {\bf k}')$ denotes the usual plane wave scattering amplitude for the initial and final momenta 
${\bf k}$ and ${\bf k}'$. 
In the specific case of screened Rutherford scattering, the plane-wave scattering amplitude
is azimuthally symmetric, and one observes that $f(\kappa,\ell; {\bf k}') \propto \exp(i\ell\phi')$,
with $\phi'$ being the azimuthal angle of the final momentum ${\bf k}'$,
as a natural consequence of the OAM conservation \cite{VanBoxem_Rutherford}.
The polar angle distribution displays characteristic dependences on $\kappa$ 
and $\ell$ \cite{VanBoxem_Rutherford,Serbo:2015}.

In Eq.~(\ref{fixed-target}), the axis used to define the vortex state
passes exactly through the scattering center. 
The case when the reference axis is displaced from the scattering center by the impact parameter ${\bf b}_\perp$
can be treated either by including the extra factor $\exp(-i\,{\bf b }_\perp\! \cdot {\bf k}_\perp)$
in the integral (\ref{fixed-target}) \cite{Matula:2014,Serbo:2015} or by making use of the
Bessel addition theorem to rewrite this state in terms of aligned vortices.
The latter method was used in the elastic \cite{VanBoxem_Rutherford}
and inelastic \cite{Yuan:2013} vortex-electron-atom scattering, as well as  
in the analysis of atom excitation by off-axis vortex photons \cite{Afanasev:2013}.

In the case of a single scattering center, an experimental control of this transverse shift, 
at least within the transverse wavelength $1/\kappa$, will be a challenge to overcome.
Alternatively, one can consider an amorphous macroscopic fixed target, 
where a single vortex beam is scattered by many 
centers with uniformly distributed values of ${\bf b}_\perp$.
This setting is easy to realize experimentally, but averaging over ${\bf b}_\perp$
smears out the distributions and erases interesting signals.
For example, the properties of photons emitted in radiative capture of vortex electrons \cite{Matula:2014}
and of electrons elastic scattering \cite{Serbo:2015} by atoms
are insensitive to the electron vorticity $\ell$ and depend only on $\kappa$.
Fortunately, this uniform averaging does not apply to crystals and in particular to chiral crystals; 
in fact, elastic scattering of vortex electron beams can be used to investigate chirality in crystalline materials,
see Section~\ref{chiral}.

\subsection{Single-vortex scattering}
\vspace{2mm}

The single-vortex scattering, V+PW $\to$ PW+PW [Fig.~\ref{fig-scattering-schemes}(b)], is essentially identical to plane-wave scattering \cite{Ivanov:2011a,Seipt:2014}.
To describe it, we start with plane-wave two-particle scattering
with four-momenta $k^{\mu}_{1,2} = (E_{1,2},{\bf k}_{1,2})$ for the incoming particles and $k^{\mu\prime}_{1,2} = (E^{\prime}_{1,2},{\bf k}^{\prime}_{1,2})$ for the outgoing particles. 
The plane-wave $S$-matrix element can be written as
\be
S_{PW} = i (2\pi)^4\, \delta^{(4)}(k^{\mu}_1+k^{\mu}_2 - k^{\mu\prime}_1 - k^{\mu\prime}_2)\, {{\cal M}\!\left(k^{\mu}_1,k^{\mu}_2; k^{\mu\prime}_1,k^{\mu\prime}_2\right) \over\sqrt{16 E_1 E_2 E_1'E_2'}}\,.
\label{Spw}
\ee
The invariant amplitude ${\cal M}$ is calculated according to the standard Feynman rules. 
Transition to the vortex state is done by 
integrating (\ref{Spw}) over the plane-wave components of the initial vortex state \cite{Jentschura:2011a}. For example, for the pure Bessel state, similarly to Eqs.~(\ref{scalarlike-again}) and (\ref{a-kappa-ell}), we have:
\be
S_{V} = \int {d^2 {\bf k}_{1\perp} \over (2\pi)^2}\, \tilde\psi_{\kappa \ell}({\bf k}_{1\perp})\, S_{PW}\,.
\label{SV}
\ee
This simple expression exhibits an important effect. 
In contrast to the fixed-center scattering, now one integrates not only the scattering amplitude ${\cal M}$ but also the kinematical $\delta$-function in Eq.~(\ref{Spw}). 
The delta-function eliminates this integration and effectively removes the coherence of plane-wave components inside a vortex state.
As the result, the single-vortex cross-section is represented as the azimutally averaged plane wave
cross-section, $d\sigma_{V} = \int (d\phi_{k_1}/2\pi) d\sigma_{PW}({\bf k}_1)$, and
is {\it $\ell$-independent} \cite{Ivanov:2011a}.
The situation can become less trivial if the vortex state has an inhomogeneous polarization state with a polarization singularity \cite{Dennis2009},
or if a superposition of two $\ell$-states is used (see examples in Section~\ref{sec:magnetic-rotations}).
In the latter case, this scattering can be used, for example, 
to produce X-ray beams with accurately structured intensity distributions \cite{Seipt:2014}.

\subsection{Vortex-vortex scattering into plane waves}
\vspace{2mm}

The double-vortex scattering V+V $\to$ PW+PW [Fig.~\ref{fig-scattering-schemes}(c)] opens up novel physics opportunities,
as it allows one to measure quantities which are not observable in the usual plane-wave
collisions. 
Consider this process with two Bessel vortex states with parameters $\kappa_1,\, \ell_1$
and $\kappa_2,\, \ell_2$ defined with respect to the same axis.
Similarly to Eq.~(\ref{SV}), we now have:
\be
S_{VV} = \int {d^2 {\bf k}_{1\perp} \over (2\pi)^2} {d^2 {\bf k}_{2\perp} \over (2\pi)^2} 
\,\tilde\psi_{\kappa_1 \ell_1}({\bf k}_{1\perp})\, \tilde\psi_{\kappa_2, -\ell_2}({\bf k}_{2\perp})  \,S_{PW}\!\left(k^{\mu}_1,k^{\mu}_2; k^{\mu\prime}_1,k^{\mu\prime}_2\right)\,.
\label{SVV}
\ee
The standard procedure, based on Fermi's golden rule, to calculate the cross-section is 
to square the $S$-matrix element, regularize the squares of the delta-functions
with a finite volume and finite observation time,
normalize appropriately the initial and final states, 
divide the probability by the total observation time and the relative flux,
and finally integrate the result over final phase space.
For the vortex-vortex scattering, this calculation follows the same route but the expressions
differ significantly from the standard case \cite{Jentschura:2011a}.
The finite-volume normalization rules for the Bessel vortex states are different 
\cite{Jentschura:2011b,Ivanov:2011a,Karlovets:2012}.
Separation of the event rate into the flux and cross-section becomes ambiguous,
which is a generic feature of the wavepacket scattering formalism, 
see the classic description in \cite{Kotkin:1992} and the recent development in \cite{Karlovets:2016dva}.
One needs to define in a reasonable way the generalized flux and generalized cross-section.
In the works \cite{Jentschura:2011b,Ivanov:2011a,Karlovets:2012}, slightly different expressions were proposed,
but in the paraxial approximation, which is sufficient for all practical purposes,
and with a smoothly behaving invariant amplitude ${\cal M}$,
all these expressions coincide. 
Below, we will omit the word ``generalized'' when describing the cross-sections.

The most salient feature of the vortex-vortex collision is that the final state kinematics acquires a new degree of freedom with respect to the plane-wave collision.
In the all-plane-wave two-particle scattering, the total momentum is well-defined, 
${\bf K} = {\bf k}_1 + {\bf k}_2$ and is conserved during the process.
As a result, the final momenta ${\bf k}_1'$ and ${\bf k}_2'$ are maximally correlated:
if ${\bf k}_1'$ is fixed, ${\bf k}_2'$ has no freedom left,
as it must be equal to ${\bf k}_2' = {\bf K} - {\bf k}_1'$.
In the wavepacket collision, with a certain distributions over the initial momenta ${\bf k}_1$ and ${\bf k}_2$,
this correlation is relaxed. The final momenta ${\bf k}_1'$ and ${\bf k}_2'$ represent now 
independent, although partially correlated, degrees of freedom.
As a result, the cross-section is now differential in both ${\bf k}_1'$ and ${\bf k}_2'$,
or alternatively, differential in ${\bf k}_1'$ and ${\bf K}$. 
For generic normalized wavepackets, it can be represented as 
\be
d\sigma = d\sigma_0 \, R({\bf K})\, d^3{\bf K}\,,
\ee
where $d\sigma_0$ is the usual 
plane-wave cross-section taken together with its appropriate final phase space,
and $R({\bf K})$ is a certain function usually peaked at the sum of the average momenta of the two initial wavepackets \cite{Kotkin:1992,Ivanov:2012b}. In the plane-wave limit,
$R({\bf K}) \to \delta^{(3)}\!({\bf k}_1 + {\bf k}_2 - {\bf K})$, and $d\sigma_0$ is recovered.

\begin{figure}[!t]
   \centering
\includegraphics[width=0.8\textwidth]{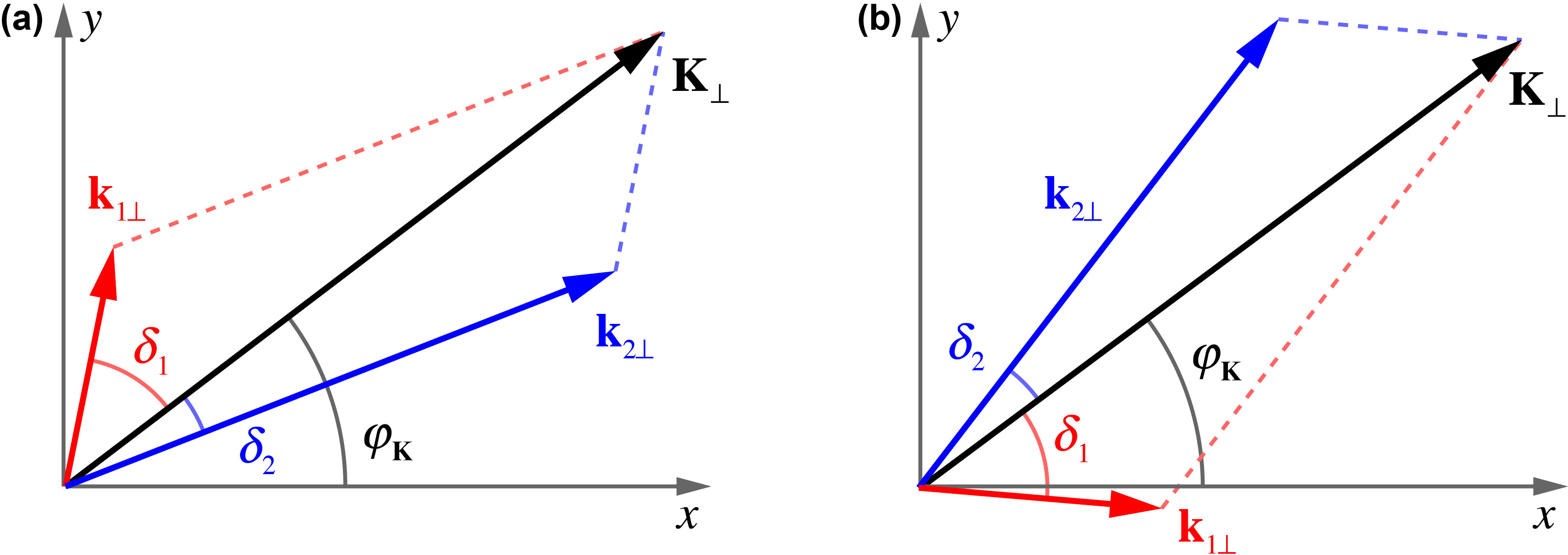}
\caption{\label{fig-two-configurations}Two kinematical configurations in the transverse plane contributing to the integral (\ref{Fperp})
for the final state with transverse momentum ${\bf K}_\perp$. 
The two invariant plane-wave amplitudes ${\cal M}_a$ and ${\cal M}_b$ differ from each other and interfere in (\ref{Fperp2}).}
\end{figure}

For pure Bessel beams, the non-trivial kinematical correlations concern 
only the transverse momenta: $d\sigma/d^2{\bf K}_\perp \propto |F_\perp|^2$, where 
\be
F_\perp = \int {d^2 {\bf k}_{1\perp} \over (2\pi)^2}{d^2 {\bf k}_{2\perp} \over (2\pi)^2}
\,\tilde\psi_{\kappa_1, \ell_1}({\bf k}_{1\perp})\, \tilde\psi_{\kappa_2,-\ell_2}({\bf k}_{2\perp})\,
\delta^{(2)}\!({\bf k}_{1\perp} + {\bf k}_{2\perp} - {\bf K}_{\perp})\, 
{\cal M}\!\left(k^{\mu}_1,k^{\mu}_2; k^{\mu\prime}_1,k^{\mu\prime}_2\right)\,.
\label{Fperp}
\ee
In this expression, the value of ${\bf K}$ is fixed by choice of the two final plane waves.
Since this integral contains four delta-functions and four integrations,
it can be evaluated exactly.
It is non-zero only when $\kappa_1$, $\kappa_2$, and $|{\bf K}_\perp|$
satisfy the triangle inequalities
$|\kappa_1 - \kappa_2| \le |{\bf K}_\perp| \le \kappa_1 + \kappa_2$,
and in this case it gets contributions from exactly two plane-wave configurations (a) and (b)
shown in Fig.~\ref{fig-two-configurations}. 

These two configurations are just reflections of each other with respect to the direction of ${\bf K}_\perp$.
Denoting the plane wave scattering amplitude evaluated at these two initial kinematics 
as ${\cal M}_a$ and ${\cal M}_b$, we obtain \cite{Ivanov:2012b}:
\be
F_\perp \propto {\cal M}_a \exp(i\ell_1\delta_1+i\ell_2\delta_2) + {\cal M}_b \exp(-i\ell_1\delta_1-i\ell_2\delta_2),
\label{Fperp2}
\ee
where $\delta_1$ and $\delta_2$ are the inner angles of the $(\kappa_1,\, \kappa_2,\, |{\bf K}_\perp|)$ triangle, 
see Fig.~\ref{fig-two-configurations}.
The net result is that the cross-section contains an additional term proportional to the interference
between two different plane wave amplitudes with equal final but different initial momenta:
\be
d\sigma \propto |{\cal M}_a|^2 + |{\cal M}_b|^2 + 
2\, \Re\left[{\cal M}_a{\cal M}_b^* \exp(2i\ell_1\delta_1+2i\ell_2\delta_2)\right].
\label{dsigma-int}
\ee
A more accurate analysis \cite{Ivanov:2012b} with normalized wavepackets of Bessel states
is necessary to regularize the end-point singularities,
and it shows that this interference term can be extracted via the azimuthal asymmetry of the cross-section.

The expression (\ref{dsigma-int}) was the starting point in demonstrating \cite{Ivanov:2012b,Ivanov:2012c,Ivanov:2016jzt}
that the scattering of two vortex states allows one to probe 
the overall {\it phase $\Phi_{\cal M}$ of the scattering amplitude} ${\cal M} = |{\cal M}|\exp(i\Phi_{\cal M})$.
In the usual plane-wave collision, the cross-section $d\sigma \propto |{\cal M}|^2$
is completely insensitive to the phase $\Phi_{\cal M}$ and its variation with kinematical parameters.
In vortex-vortex scattering, the two interfering plane-wave amplitudes ${\cal M}_a$ and ${\cal M}_b$ correspond to different initial and the same final momenta, which implies 
different momentum transfers. 
The phase $\Phi_{\cal M}$ can depend on this momentum transfer. For example, in the
elastic scattering of charged particles the amplitude acquires the well-known Coulomb phase,
which, for large energies and small scattering angles $\theta$, can be written as
\be
\Phi_{\cal M}(\theta) = \Phi_{\cal M 0} + 2\alpha\log(1/\theta),
\label{coulomb-phase}
\ee
where $\alpha$ is the fine-structure constant, and $\Phi_{\cal M 0}$ is an angle-independent quantity which, strictly speaking, requires infrared regularization and can be sensitive to the details of the process. 
Thus, the interference term in (\ref{dsigma-int}) is proportional to
$\cos[2\ell_1\delta_1+2\ell_2\delta_2 + \Phi_{\cal M}(\theta_a) - \Phi_{\cal M}(\theta_b)]$ and it is sensitive to the dependence $\Phi_{\cal M}(\theta)$.
This quantity can be extracted from the azimuthal asymmetry of the differential cross-section  \cite{Ivanov:2012b}.

\begin{figure}[!t]
\centering
\includegraphics[width=0.8\textwidth]{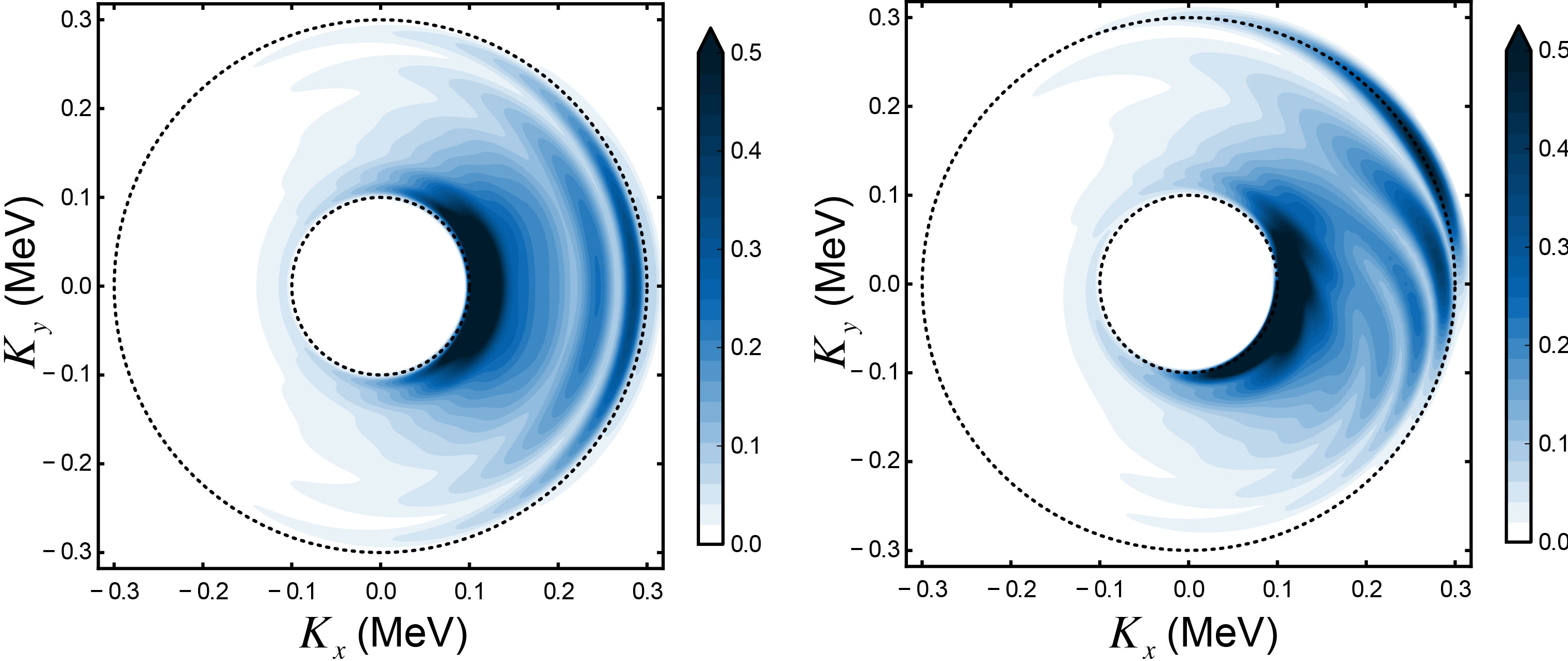}
\caption{Differential cross-section, in arbitrary units,
as a function of the total momentum ${\bf K} = {\bf k}_1 + {\bf k}_2$, for fixed ${\bf k}_1'$, for purely real Born-level elastic electron scattering amplitude (left), 
and for the momentum-transfer-dependent Coulomb phase (\ref{coulomb-phase}) (right) \cite{Ivanov:2016oue}. The parameters used here are: $E_1 = 2.1$ MeV, $k_1' = 500$ keV with ${\bf k}_1' = {k}_1' \bar{\bf x}$, $J_{z1} = 1/2$, $J_{z2} = 13/2$, and $\kappa_1$ and $\kappa_2$ are Gaussian-distributed around the values of 200 and 100 keV with the widths 10 and 5 keV, respectively. The results are averaged over different helicities of the incoming particles. In the right-hand panel, the fine-structure constant $\alpha$ is artificially set to 10 to enhance the visibility of the up-down asymmetry.}
\label{fig-rings}
\end{figure}

In Refs. \cite{Ivanov:2016jzt,Ivanov:2016oue}, this idea was investigated in detail with the example of the moderately relativistic elastic electron-electron scattering
using the Dirac electrons described by bispinors (\ref{bispinor-PW-helicity}) and their vortex combinations.
A typical ${\bf K}_\perp$-distribution of the cross-section is shown in Fig.~\ref{fig-rings}. 
The left and right plots correspond, respectively, to the purely real Born-level scattering amplitude and to the amplitude with the Coulomb phase $\Phi_{\cal M}(\theta)$ taken into account, with the value of $\alpha$ artificially set to 10 in (\ref{coulomb-phase}) for the purpose of illustration.
One can see the interference fringes arising from the interference between the two kinematical configurations of Fig.~\ref{fig-two-configurations}.
The left plot is symmetric with regard to the horizontal line (i.e., the direction of ${\bf k}'_{1\perp}$) because 
the cross-section contains no terms proportional to $\sin(\phi'_1-\phi_K)$ (with $\phi'_1$ and $\phi_K$ being the azimuthal angles of ${\bf k}^{\prime}_1$ and ${\bf K}$), 
while the right plot shows a distorted pattern.
This distortion can be quantified in terms of up-down asymmetry \cite{Ivanov:2016jzt,Ivanov:2016oue}.
It is this asymmetry that is proportional to the phase difference between the two contributions and can lead to a direct measurement of $\Phi_{\cal M}(\theta)$. 
For the realistic $\alpha \approx 1/137$, the asymmetry was found to be of the order of $10^{-4}$--$10^{-3}$,
which may be experimentally accessible with sufficient statistics.

In the works \cite{Karlovets:2016jrd,Karlovets:2016dva}, the above suggestion was considered as a particular case of a more general setting. In order to probe the overall phase of the scattering amplitude, one needs to collide states with manifestly broken azimuthal symmetry,
which necessarily goes beyond the plane-wave approximation.
Developing further the Wigner-function-based theoretical formalism for collision
of arbitrary wavepackets, Refs.~\cite{Karlovets:2016jrd,Karlovets:2016dva} showed that 
the cross-section contains a new phase-sensitive term proportional to the mean value 
of the following ``effective impact parameter'': 
\be
{\bf b}_{\rm eff} = {\bf b}_\perp - {\partial \Phi_1({\bf k}_{1\perp})\over \partial {\bf k}_{1\perp}} 
+ {\partial \Phi_2({\bf k}_{2\perp})\over \partial {\bf k}_{2\perp}}\,.
\ee
Here, ${\bf b}_\perp = {\bf r}_{1 \perp} - {\bf r}_{2 \perp}$ is the usual impact parameter,
i.e., the transverse separation of the centers of the two colliding wavepackets,
while $\Phi_i({\bf k}_{i\perp})$ describe their additional phases beyond the center-of-mass motion:
$\psi_i({\bf k}_i) \propto \exp[- i {\bf r}_i \cdot {\bf k}_i + i \Phi_i({\bf k}_{i\perp})]$.
This term can be extracted from the data via the asymmetry defined as the relative difference
between the cross-sections with ${\bf b}_{\rm eff}$ and $-{\bf b}_{\rm eff}$.
The average value of this operator can be non-zero either for an off-center collision
of wavepackets (${\bf b}_\perp \not = 0$) or for the head-on collision of wavepackets
with non-trivial phase fronts, such as vortex electron beams. 
In fact, with the definition of asymmetry adopted in \cite{Karlovets:2016jrd,Karlovets:2016dva}, 
the phase singularity needs to be shifted away from the collision axis in order to produce a non-zero asymmetry.
This development opens up several complementary ways to probe the phase 
of the scattering amplitude, which await experimental verification.

When experiments with vortex protons and other hadrons become possible,
the above method can be applied to hadronic processes. It will then offer additional information on hadronic interactions
which cannot be accessed in conventional experiments.
One example is the small-angle elastic $pp$ scattering with momentum transfer of the order of $0.1$--$1$ GeV.
At high energies, it is dominated by the exchange of the {\it Pomeron},
an emergent strongly-interacting object whose theoretical description is still debated \cite{PomeronModels}.
With vortex proton scattering, one can measure the dependence of the phase of the full amplitude 
in the momentum transfer. In this way, one gets a new observable 
against which the Pomeron models can be tested \cite{Ivanov:2012c,Karlovets:2016jrd,Karlovets:2016dva}.
So far, the Pomeron phase can be accessed only in the very narrow $t$-region via the strong-Coulomb interference,
and, in addition, it also relies on the good knowledge of the Coulomb phase.
With vortex protons, one should be able to probe this phase over the entire $t$ region,
including the diffraction dip region where a strong variation of the phase is expected in some models.
Another example is the intermediate-energy hadroproduction reactions such as $\gamma p \to K^+ \Lambda$,
which involve hadronic resonances in several competing partial waves \cite{photoproduction}.
Although the relative phases between these contributions can be accessed, 
disentangling them would become easier if the information on the overall phase were available.

\subsection{Vortex-into-vortex scattering}
\vspace{2mm}

The vortex-into-vortex scattering process V+PW $\to$ V+PW [Fig.~\ref{fig-scattering-schemes}(d)] brings new challenges.
The calculation of the strictly forward or backward scattering does not pose any difficulty \cite{Jentschura:2011a,Jentschura:2011b},
as one can use the same reference axis to describe the initial and final vortex states.
The orbital angular momentum is naturally transferred from the initial to the final vortex state: $\ell' = \ell$.
For off-forward scattering, the situation is more complicated.
While \cite{Jentschura:2011a} argued that one still has $\ell' \simeq \ell$,
the analysis of \cite{Ivanov:2011a} showed that the entire $\ell'$-region from $-\infty$ to $+\infty$ contributes
to the cross-section.
In \cite{IvanovSerbo:2011}, the origin of the discrepancy was traced back to the usage of non-normalizable
pure Bessel beams. If one uses normalizable vortex wavepackets and, in addition,
if one chooses its own reference axis for each vortex state,
then the controversy is resolved. This result stresses the usage of the {\it orbital helicity}
\cite{Bliokh2007} (i.e., the OAM component along the propagation axis) as the physically-relevant quantity rather than the OAM defined with respect to a fixed axis.

Additional difficulties arise if one views the process V+PW $\to$ V+PW or V+PW $\to$ V+V
[Fig.~\ref{fig-scattering-schemes}(d,e)] to produce high-energy vortex states \cite{Jentschura:2011a,Jentschura:2011b,Ivanov:2012a}.
Formally, the two outgoing waves are {\it momentum-entangled}, 
and there is no pre-existent way to label one particle as a vortex state and the other as a plane wave \cite{Ivanov:2012a}. Only after one particle is projected
on an approximate plane wave, and is measured by the detector with a momentum uncertainty less than $\kappa$, the other particle emerges in a vortex state. Whether this projection can be performed on an event-by-event basis
with the existing technology remains unclear.

Further on, the simultaneous energy and momentum
conservation in the off-forward V+PW $\to$ V+PW scattering implies that the outgoing
vortex state is {\it not monochromatic}. Indeed, different plane wave components of the final vortex state correspond to different energies not only of the vortex state itself, but also of the recoil plane wave. Thus, the coherence among the plane wave components required to form a vortex state is lost, or at best is hidden \cite{Ivanov:2011a,Seipt:2014}.

As far as specific high-energy processes are concerned, 
the only example considered in detail was the inverse Compton backscattering 
\cite{Jentschura:2011a,Jentschura:2011b,Seipt:2014}. 
Here, optical photons scatter almost backward off high-energy electrons and 
take a sizable portion of the electron energy. This process is well known and
is routinely used, for example, at the SPring-8 and HIgammaS facilities \cite{backscatteringuses} 
to produce GeV-range photons
for subsequent hadronic photoproduction experiments.
This process was calculated for vortex initial and final photons \cite{Jentschura:2011a,Jentschura:2011b}, while the electrons, both initial and final, were assumed to be plane waves.
In the strictly forward scattering, the final photon
is upconverted into the GeV energy range while retaining its OAM.
For slightly off-forward kinematics and with a due care mentioned above, 
the final orbital helicity is also close to the initial one \cite{IvanovSerbo:2011}.
Another application of this process was considered in \cite{Seipt:2014}.
Here, the plane wave photons scatter off energetic vortex electrons
and turn into a flux of energetic photons with structured intensity distribution in the transverse plane.
Preparing the initial vortex electrons in custom-tailored superpositions of different values of OAM, 
one can accurately shape the transverse distribution of the final X-ray pulse.

\section{Radiation by vortex electrons}
\label{sec:radiation}
\vspace{1mm}

Electrons can radiate. They emit electromagnetic (EM) radiation via {\it bremsstrahlung} 
when the electron trajectory is deflected by external fields
or through polarization radiation 
(an umbrella term including the Vavilov--Cherenkov radiation, diffraction radiation, transition radiation, etc.), 
when moving in or near a polarizable medium.
One can ask whether the radiation from vortex electrons differs in any aspect 
from the plane-wave case, and if so, what additional information it encodes.
For vortex electrons, two types of EM radiation have been investigated theoretically so far:
the {\it Vavilov--Cherenkov radiation} \cite{Kaminer-2015,ISZ-2016} 
and {\it transition radiation} \cite{IvanovKarlovets:2013a,IvanovKarlovets:2013b,KonkovPotylitsyn:2013}.
In both cases, the vortex nature of the electron leads to several distinct features of the radiation it emits. These features, if experimentally detected, should provide additional insight into the radiation process, and could serve as a complementary and convenient diagnostic
tool for measuring the parameters of vortex electrons.

\subsection{Vavilov--Cherenkov radiation}
\vspace{2mm}

The Vavilov--Cherenkov radiation \cite{Cherenkov:1934,Vavilov:1934,Bolotoskii2009}
from a vortex electron was investigated in \cite{Kaminer-2015,ISZ-2016}.
Both works treated the problem in the full quantum-electrodynamical approach and focused
on the spectral, $d\Gamma/d\omega$, and spectral-angular, $d^2\Gamma/d\omega d\Omega_\gamma$, distributions of the photon emission rate $\Gamma$ ($\omega$ is the photon frequency and $\Omega_\gamma$ is the solid angle spanned by the photon wave-vector directions), as well as on the polarization properties.

As a short reminder, within the standard quantum treatment of the Vavilov--Cherenkov radiation from a plane-wave electron, the emission process is described as a ``decay'' of the initial electron with the four-momentum momentum $p^{\mu}$ into the final electron $p^{\mu\prime}$ and the in-medium photon with momentum $\hbar {\mathcal k}^\mu$ \cite{Ginzburg-1940}. Note that in this section we restore the constants $\hbar$ and $c$. 
In what follows, $E$ stands for the initial relativistic energy of the electron, $\beta = v/c = pc/E$ is its dimensionless velocity, while $\hbar \omega$ is the energy of the emitted photon.
This radiation has the following spectral-angular distribution \cite{Ginzburg-1940}:
\be
{d\Gamma \over d\omega d\Omega_\gamma} =
 {\alpha \over 2\pi} \left[ \beta\sin^2\!\theta_{\rm Ch} + \frac{(\hbar \omega)^2}{2\beta E^2}\, ({\mathcal n}^2-1) \right] \delta\left(\cos\theta_{\bf {\mathbcal k}p}-\cos\theta_{\rm Ch}\right),
\label{VCh-angular}
\ee
where ${\mathcal n}(\omega)$ is the frequency-dependent refraction index of the medium, 
$\theta_{\bf {\mathbcal k}p}$ is the angle between the emitted photon and the initial electron, and $\theta_{\rm Ch}$ is the Cherenkov cone
opening angle given by
\be
\cos{\theta_{\rm Ch}}=\fr{1}{\beta {\mathcal n}}+\fr{\hbar \omega}{2E}\,\fr{n^2-1}{\beta {\mathcal n}}.
\label{VCh-angle}
\ee
Clearly, the emission angle satisfies $0 \le \theta_{\rm Ch} \le 90^\circ$, and the requirement that $\cos\theta_{\rm Ch} \le 1$
sets a natural cut-off to the spectrum, $\hbar\omega \le \hbar\omega_{\rm cutoff} = 2E(\beta {\mathcal n}-1)/({\mathcal n}^2-1)$. 
Under usual conditions, the cut-off photon energy is in the MeV range, 
far beyond the applicability range of the electrodynamics of the medium. However, 
by delicately adjusting the velocity of the electron, one could in principle bring this cut-off
into the optical/UV region, boosting the importance of the quantum effects in the Cherenkov radiation problem and exposing the spectral discontinuity right at this cut-off \cite{Kaminer-2015}.
Note that these spectral features are valid for a plane-wave electron and do not require it to be in any specially-designed wavepacket.
However, bringing them to the optical/UV region would strongly suppress the intensity of radiation and therefore make the direct observation of these cut-off effects extremely challenging \cite{ISZ-2016}.

Equation (\ref{VCh-angular}) is written for an unpolarized initial electron and after summation over final electron and photon polarizations.
Actually, the radiated Cherenkov photons are almost 100\% linearly polarized, with the polarization lying in the scattering plane. If the initial electron is polarized, then the Cherenkov light acquires a non-zero degree of circular polarization \cite{ISZ-2016}.

\begin{figure}[!t]
\centering
\includegraphics[width=0.55\textwidth]{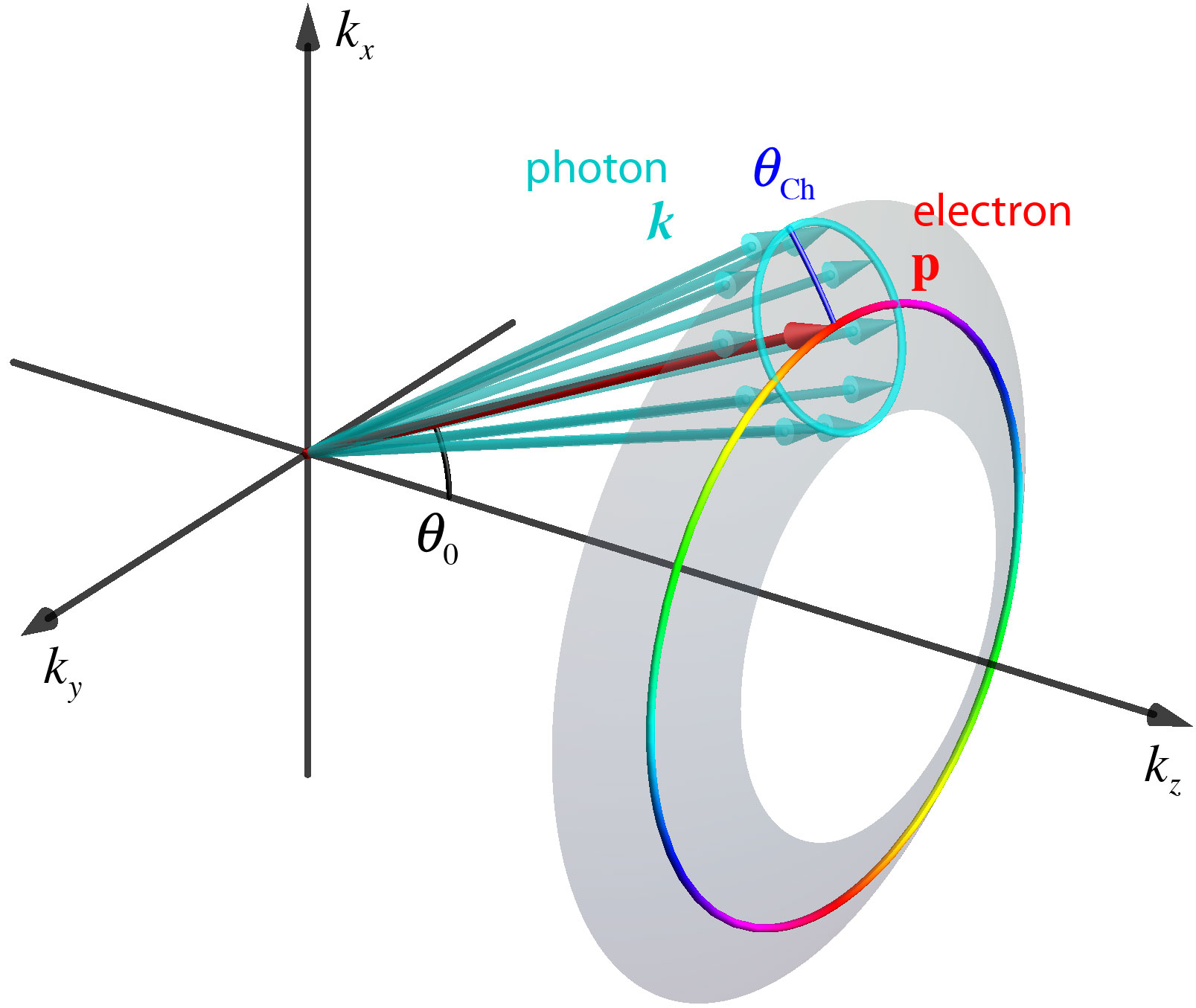}
\caption{Schematics of the Vavilov--Cherenkov radiation by a vortex electron \cite{Kaminer-2015,ISZ-2016}. 
Here, $\theta_0$ is the opening angle of the vortex electron, $\theta_{\rm Ch}$ is the opening angle of the Cherenkov cone for the plane-wave case, the gray ring represents the annular region of directions where the photons can be emitted.}
\label{fig-VCh-geometry} 
\end{figure}

If the initial electron is in a vortex state, the angular distribution of the Vavilov--Cherenkov radiation changes \cite{Kaminer-2015,ISZ-2016}.
A pure Bessel beam is a superposition of plane-wave electrons with a conical distribution of momenta ${\bf p}$, which is characterized by the polar angle $\theta_0$ ($\sin\theta_0 = \hbar \kappa/p$), Fig.~\ref{fig:Bessel}(a).
All these plane-wave components radiate, but if we do not detect the final electron 
and only study the photon angular distribution, this radiation adds up {\it incoherently}.
Indeed, in each plane-wave radiation subprocess the four-momentum conservation dictates $p^\mu = p^{\mu\prime} + \hbar {\mathcal k}^\mu$.
If one measures the intensity of radiation in a certain direction without detecting the final electron, one fixes ${{\mathcal k}^\mu}$ and integrates over all ${p^{\mu\prime}}$. But then different Fourier components of the initial vortex electron 
carry different values of $p^\mu$ and, due to the fixed ${\mathcal k}^\mu$, correspond to different final momenta $p^{\mu\prime}$.
Since the interference requires that all the final-state particles remain exactly in the same state, no interference is possible in this kinematics.
Thus, averaging the plane-wave intensity (\ref{VCh-angular})
over the appropriate initial electron momentum $p^\mu$, while keeping ${\mathcal k}^\mu$ fixed, we will obtain the spectral-angular distribution for the vortex electron. 

\begin{figure}[!t]
\centering
\includegraphics[width=0.8\textwidth]{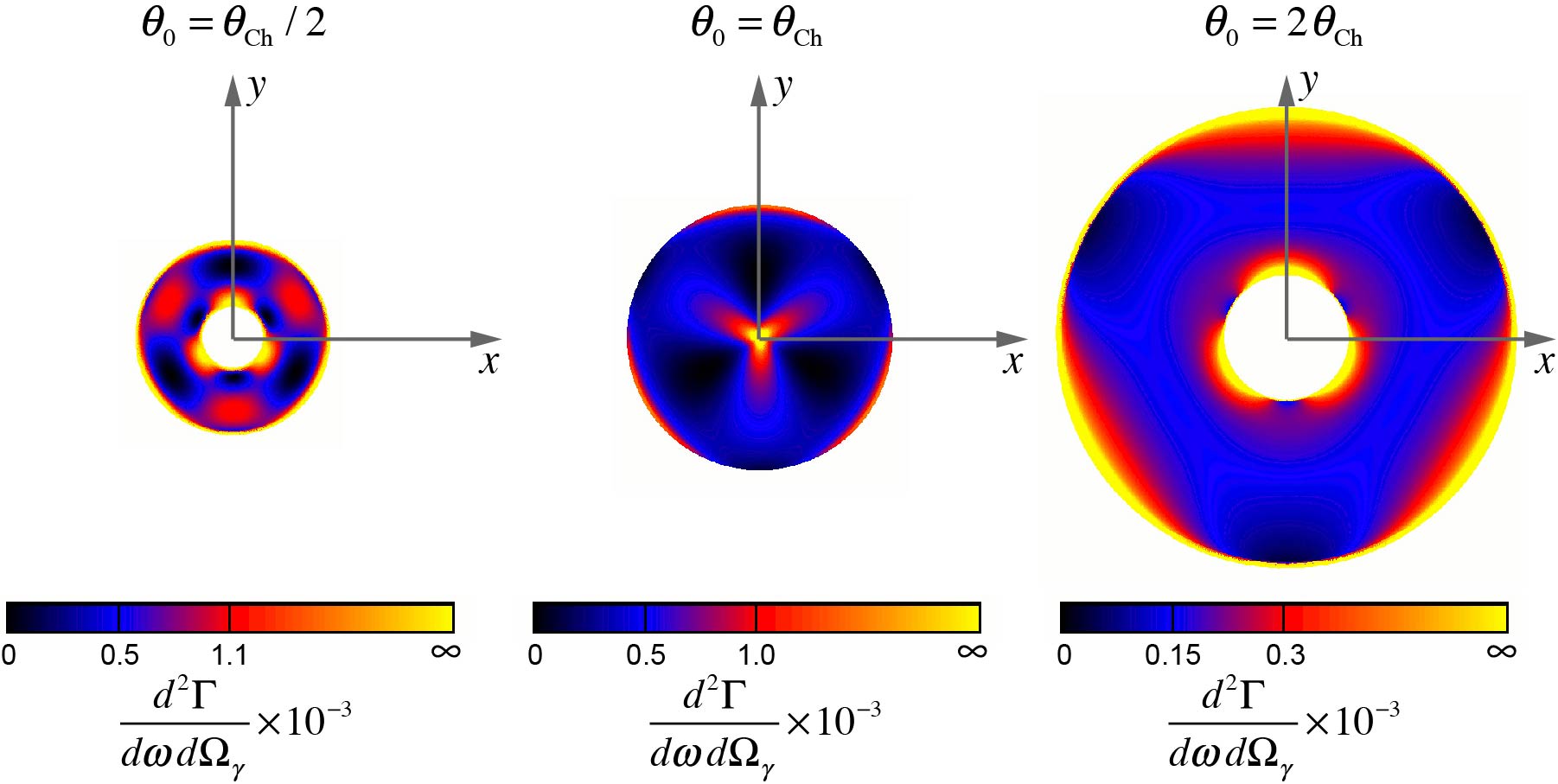}
\caption{The spectral-angular distribution of the photon emission rate $\Gamma$ as a function of the emitted-photon spherical angles $\theta_{\mathbcal k}$ and $\phi_{\mathbcal k}$ (mapped onto the $(x,y)$ plane of the far-field detector) for an electron in a superposition of two vortex states with $(J_z)_1-(J_z)_2 =3$ and equal amplitudes \cite{ISZ-2016}. The radii of the annular region of the radiation are determined by Eq.~(\ref{VCh-ring}).}
\label{fig-VCh-angular}
\end{figure}

In doing so, the most salient feature is the angular distribution.
It has a ring shape shown in Fig.~\ref{fig-VCh-geometry} and spans over the photon's polar angles $\theta_{\mathbcal k}$ with respect to the mean electron propagation direction (the $z$-axis):
\be
|\theta_{\rm Ch} - \theta_0| \le \theta_{\mathbcal k} \le \theta_{\rm Ch} + \theta_0.
\label{VCh-ring}
\ee
For a Bessel electron with a definite value of the $z$-component of the total angular momentum $J_z$, the angular distribution summed over the final helicities is azimuthally symmetric and grows towards the boundaries of the ring.
This leads to two remarkable phenomena which cannot be produced with plane-wave electrons.
First, if $\theta_{\rm Ch} + \theta_0 > \pi/2$, then some photons are emitted in the {\it backward} hemisphere with respect to the overall propagation direction of the vortex electrons \cite{Kaminer-2015}.
In order to observe this effect, one would need to achieve large opening angles of the vortex state or take a medium with a very large refractive index.
Second, if the two opening angles match, $\theta_{\rm Ch} = \theta_0$, 
the Cherenkov radiation strongly peaks in the {\it forward} direction, see the middle panel of Fig.~\ref{fig-VCh-angular}.
For perfect Bessel beams with a fixed value of the momentum polar angle $\theta_{\bf p} = \theta_0$, the intensity of the Vavilov-Cherenkov radiation will grow near the forward direction as $1/\theta_{\mathbcal k}$.
For realistic vortex electrons, $\theta_{\bf p}$ is not fixed but is distributed
over an angular region with width $\delta\theta_0$, Fig~\ref{fig:LG}(a),
this bright emission will smear over a spot of comparable angular size $\delta\theta_0$.
To observe this forward emission, one would need, first, vortex electrons with a sufficient transverse coherence length and exhibiting several radial intensity rings (originating from the Bessel distribution with $\theta_0$) to guarantee $\delta \theta_0/\theta_0 \ll 1$,
and, second, sufficient monochromaticity in order not to smear the very forward Cherenkov ring.
For the realistic value of $\theta_0 = 20$ mrad, which implies $1-\cos\theta_0 \simeq 2\cdot 10^{-4}$, one must adjust the electron velocity to the emission threshold velocity with the accuracy of $10^{-4}$.

After integration over all photon emission angles, the spectral distribution $d\Gamma/d\omega$ 
for the vortex electrons and, generically, for any monochromatic wavepacket, coincides with the plane-wave spectral density. This is a consequence of the incoherent summation of the radiation from different plane-wave components. Although the initial electron is a coherent superposition of different plane waves, once we integrate over the final-electron parameters, this coherence is lost.
In this respect, attributing a special role to the coherence as the origin of the spectral-angular features as done in \cite{Kaminer-2015} is unjustified. 

The polarization properties of the Cherenkov light also change when one switches
from the plane-wave to the vortex electron. The geometry of the problem shows that, for a pure Bessel state, the radiation intensity into any given direction inside the ring is an incoherent sum of two plane-wave intensities with different initial electron kinematics. As a result, the emitted light remains linearly polarized but its degree of polarization changes. In particular, the photons can become linearly polarized in the direction orthogonal to the scattering plane \cite{ISZ-2016}.

Finally, if instead of a single vortex state, the electron is taken in a superposition 
of two vortex states with total angular momenta $(J_z)_1$ and $(J_z)_2$, 
then the electron probability-density distribution becomes azimuthally-inhomogeneous, see Figs.~\ref{fig:bliokh8} and \ref{fig:bliokh9}. The multi-petal structure of the Vavilov--Cherenkov radiation from such structured electrons mimics the electron density profile in the focal plane, as shown in Fig.~\ref{fig-VCh-angular}. Thus, Cherenkov radiation emerges as a convenient macroscopic diagnostic tool for such electrons.

\subsection{Transition radiation}
\vspace{2mm}

The transition radiation occurs when an electron crosses the boundary separating two media with different permittivity or permeability \cite{GinzburgTsytovich:review,GinzburgTsytovich:book}. The simplest example
is the electron impinging from the vacuum onto a conductive plane.
In all cases studied experimentally, this radiation is associated with the particle's {\it electric charge}.
However, classical electrodynamics predicts that {\it magnetic moments} and higher-order multipoles can also radiate \cite{GinzburgTsytovich:review,GinzburgTsytovich:book}.
This contribution to radiation has never been detected for any kind of polarization radiation due to its weakness.

Vortex electrons with large OAM $\ell$ can make the observation of this contribution possible. Indeed, as we discussed in Sections~\ref{sect:EVexternal} and \ref{spinorbit}, paraxial free-space vortex electrons possess longitudinal magnetic moment \cite{Bliokh2007,Bliokh_Relativistic} ${\bf M} = (ec/2E) \left( \langle {\bf L} \rangle +2 \langle {\bf S} \rangle \right)$, Eqs.~(\ref{eq:bliokh20}), (\ref{eq:bliokh21}), and (\ref{eq:bliokh56}). Assuming longitudinal polarization, i.e., spin component $s=\pm 1/2$, the absolute value of the magnetic moment is $M = \gamma^{-1} |\ell + 2s| \mu_B$, where $\mu_B$ is the Bohr magneton and $\gamma$ is the Lorentz factor. 
Therefore, large values of $\ell$ would strongly enhance the magnetic-moment contribution to the transition radiation and, via interference with the electric-charge contribution, it can lead to an observable signal.

The essence of this idea can be explained in the following way
\cite{KonkovPotylitsyn:2012,KonkovPotylitsyn:2013,IvanovKarlovets:2013a,IvanovKarlovets:2013b}.
Within classical electrodynamics, an electron with velocity ${\bf v}$ can be viewed as a pointlike source equipped with the electric charge $e$ and magnetic moment ${\bf M}$. Accordingly, it is described with the electric and magnetic current densities ${\bf j}_e = e\, {\bf v}\,\delta({\bf r} - {\bf v} t)$ and
${\bf j}_{m} = \gamma^{-1} c\ \nabla \times [{\bf M}\, \delta({\bf r} - {\bf v}t)]$.
The curl leads to an extra factor $i\omega/c$ in the Fourier components of the radiation field.
As a result, the relative strength of the magnetic moment transition radiation always bears the following small factor \cite{GinzburgTsytovich:book}:
\be
\epsilon = \frac{M \omega}{\gamma\, e\, c}.
\label{xg}
\ee
The radiation energy contains this factor squared. For optical/UV photons and for moderately relativistic plane-wave electrons ($\ell =0$), $\epsilon = \hbar \omega/E \sim 10^{-5}$; for slower electrons, the radiation is much weaker. 
An additional difficulty arises from the fact that the quantum corrections
bear the same suppressing factor $\hbar \omega/E$. A calculation which keeps
the electron's magnetic moment contribution but neglects quantum effects is, strictly speaking, 
inconsistent, and a full quantum treatment is needed.
Thus, for non-vortex electrons carrying magnetic moment from the spin, the magnetic moment contribution is: (i) suppressed by many orders of magnitude with respect to the usual charge radiation, which makes it undetectable, and (ii) is not cleanly calculable within classical electrodynamics. 
However, for vortex electrons with large OAM, $|\ell | \gg 1$, 
the magnetic moment contribution to the polarization radiation is much larger and can become visible. 
At the same time, it justifies the quasi-classical approach to the radiation, as
the magnetic-moment contribution now dominates over quantum effects.

\begin{figure}[!t]
\centering
\includegraphics[width=0.95\textwidth]{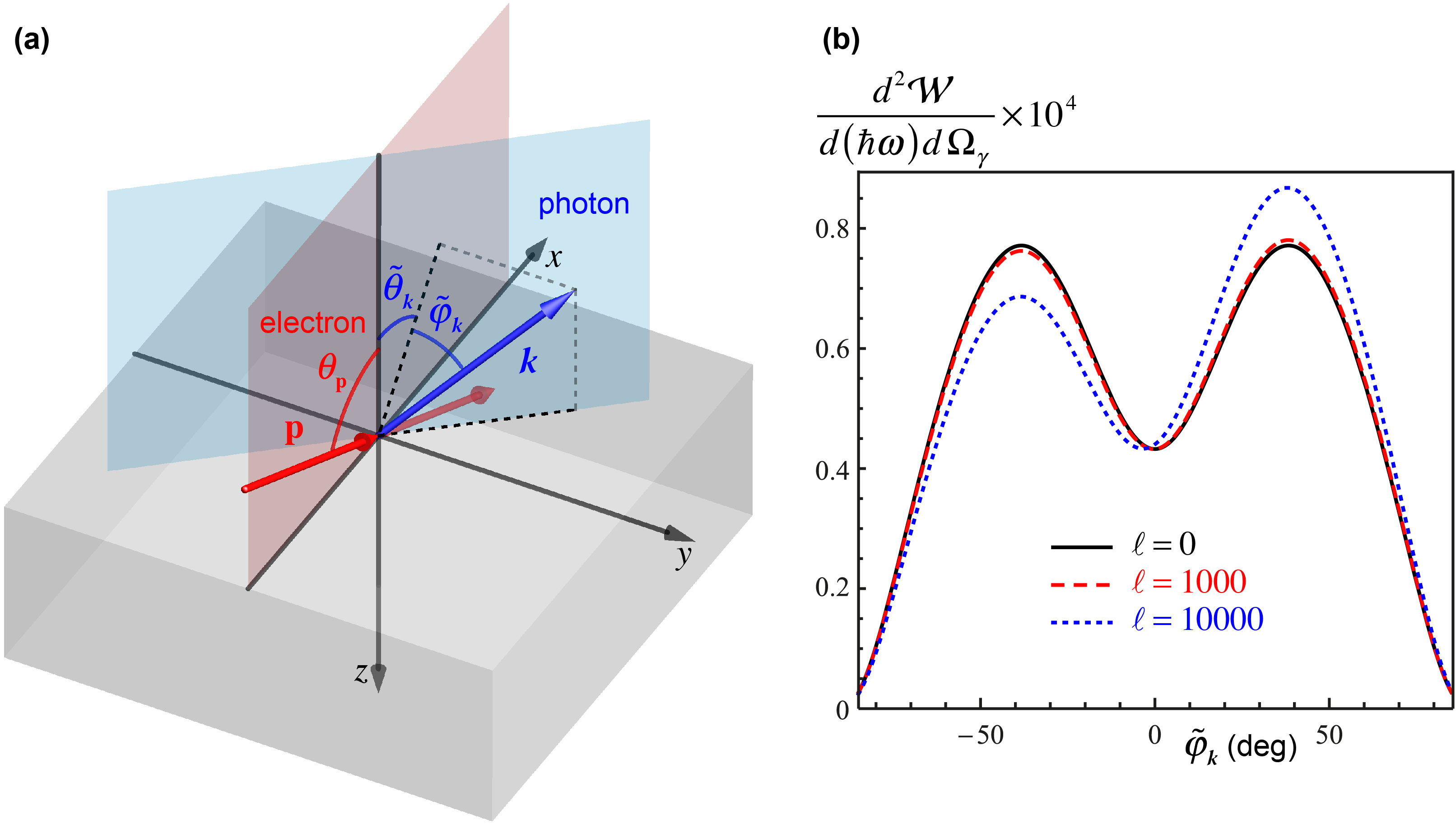}
\caption{(a) Schematics of the backward transition radiation produced by a vortex electron, carrying electric charge $e$ and magnetic moment ${\bf M} \parallel {\bf p}$, at oblique incidence on a metal surface $z=0$ \cite{IvanovKarlovets:2013a,IvanovKarlovets:2013b,KonkovPotylitsyn:2013}. The direction of the electron momentum ${\bf p}$ in the $(x,z)$ plane of incidence is determined by the polar angle $\theta_{\bf p}$, while the emitted out-of-plane photon is characterized by the wave vector ${\mathbcal k}$ with the angles $\tilde{\theta}_{\mathbcal k}$ and $\tilde{\varphi}_{\mathbcal k}$, as shown in the scheme. 
(b) The interference between the magnetic-moment and charge contributions leads to the left-right ($\tilde{\varphi}_{\mathbcal k} \to - \tilde{\varphi}_{\mathbcal k}$) asymmetry of the transition radiation intensity \cite{IvanovKarlovets:2013a}. The $\tilde{\varphi}_{\mathbcal k}$-dependence of the spectral-angular density of the transition radiation energy ${\mathcal W}$. The parameters are: $E = 300$ keV, $\hbar\omega = 5$ eV, $\theta_{\bf p} = 70^{\circ}$, and the observation angles are centered around $\tilde{\theta}_{\mathbcal k} = - 40^{\circ}$. The curves corresponds to $\ell = 0$ (solid black curve), $1000$ (dashed red curve) and $10000$ (blue dotted curve).}
\label{fig-TR}
\end{figure}

The total radiated energy can be split into electric-charge and magnetic-moment
contributions, as well as the interference term: $d{\mathcal W} = d{\mathcal W}_e + d{\mathcal W}_m + d{\mathcal W}_{em}$.
The quantity of interest is the interference term, which is linear in the small parameter $\epsilon$. There are two subtleties, which make the extraction of this interference intricate. First, the curl in the magnetic current leads to an extra $i$ factor in the EM field radiated by the magnetic moment, and as the result, the interference vanishes in the case of a transparent medium or an ideal conductor. Fortunately, for a realistic medium with complex permittivity, the interference term becomes non-zero.
Second, since ${\bf M}$ is a pseudovector, the interference term contains the triple scalar product $\bar{\mathbcal k} \cdot ({\bf M}\times \bar {\bf z})$,
where $\bar{\mathbcal k}$ is the direction of the emitted photon, and $\bar{\bf z}$ is the normal to the vacuum-metal interface, Fig.~\ref{fig-TR}(a).
It vanishes for the normal incidence, as well as for oblique incidence, after the full solid angle integration. It can be observed only at {\it oblique} incidence and only in the {\it differential} distribution. Refs.~\cite{IvanovKarlovets:2013a,IvanovKarlovets:2013b} proposed to extract the interference term via the left-right asymmetry:
\be
I_{LR} = {{\mathcal W}_L - {\mathcal W}_R \over {\mathcal W}_L + {\mathcal W}_R}\,, \quad 
{\mathcal W}_{L,R} = \int d\Omega_{\gamma L, R}\, {d^2{\mathcal W} \over d \hbar \omega d\Omega_\gamma}\,, 
\label{asymmetry}
\ee
where $d\Omega_{\gamma L}$ and $d\Omega_{\gamma R}$ 
indicate the two hemispheres lying to the left and to the right of the incidence plane, i.e., characterized by the angles $\tilde{\varphi}_{\mathbcal k} > 0$ and $\tilde{\varphi}_{\mathbcal k} < 0$, as shown in Fig.~\ref{fig-TR}(a).

Numerical calculations \cite{IvanovKarlovets:2013a,IvanovKarlovets:2013b} show that 300 keV vortex electrons with $\ell = 1000$ and impacting on an aluminium plate at large incidence angles produce a left-right asymmetry at the percent level,
and should be well observable for a 1 nA electron current.
For lower values of $\ell$, the asymmetry is proportionally small, but also seems to be within reach. The experimental detection of this asymmetry would provide the first ever direct demonstration that magnetic moments radiate.

In the works \cite{KonkovPotylitsyn:2012,KonkovPotylitsyn:2013}, an alternative
route to detecting the interference term $d{\mathcal W}_{em}$ was explored.
There, the degree of circular polarization in the emitted light was the quantity of interest.
Transition radiation of a point electric charge is {\it linearly} polarized \cite{GinzburgTsytovich:book,GinzburgTsytovich:review}. 
The interference term induces an {\it elliptical} polarization in the out-of-plane photons, quantified by another pseudoscalar quantity, the photon's helicity or the Stocks parameter ${\cal S}_3$ \cite{Azzam_book}. At generic angles, the value of ${\cal S}_3$ is very small, being suppressed by the same factor $\epsilon$. However, near particular directions, where the electric-charge radiation vanishes, the value of ${\cal S}_3$ can be very large. Thus, measuring ${\cal S}_3$ in the direction of the intensity minimum will also unveil the transition radiation emitted by the vortex electron's magnetic moment. The calculations reported in \cite{KonkovPotylitsyn:2012,KonkovPotylitsyn:2013}
show that for OAM $\ell = 100$, the value of ${\cal S}_3$ can be as large as $70\%$.
However this large elliptic polarization appears only for directions range $\sim 1$ arcminute near the minimum-intensity direction. Satisfying such severe angular cuts will definitely pose a challenge for a dedicated experiment.

\clearpage

\section{Concluding remarks}
\vspace{1mm}

\subsection{Summary of this work}
\vspace{2mm}

Ten years have passed since the prediction \cite{Bliokh2007} of free-electron vortex states carrying intrinsic orbital angular momentum (OAM) and their first generation \cite{Uchida2010,Verbeeck2010,McMorran2011} in transmission electron microscopes (TEMs) few years later. In this paper, we have reviewed the main theoretical and experimental achievements in investigations of vortex electrons during the first decade of this rapidly developing field.

First, we have provided a pedagogical introduction and a solid theoretical basis for the researchers starting their work in this emerging field. In particular, we have introduced the main concepts of phase singularities, angular momentum, and vortex wave beams/packets as applied to electron waves. We have also considered the main interaction phenomena involving vortex electrons, including: their nontrivial behaviour in external electromagnetic fields, spin-orbit interactions and other relativistic effects, a variety of elastic and inelastic scattering processes, radiation processes, etc.

Second, we have described the main features and peculiarities of TEM experiments with electron vortex beams. In particular, we provided a detailed analysis of various methods of their creation, a wealth of practical details, as well as of various ways of the OAM measurements in electron beams. Importantly, we have described numerous vortex-induced phenomena, which appear in the interactions of electron vortex beams with various kinds of samples in TEMs. This is the most promising direction for applications of vortex beams in electron microscopy, especially for the {\it atomic-resolution mapping of magnetic and chiral properties}.

Third, we discussed possible novel phenomena involving vortex electrons outside of the TEM context. The most exciting opportunities arise for higher energies, where interactions with strong laser fields, quantum particle collisions, and radiation phenomena can reveal new features depending on the OAM degrees of freedom. The generation of {\it vortex states of high-energy electrons or other quantum particles} is a future milestone to be achieved experimentally.

\subsection{Future prospects}
\vspace{2mm}

Vortex electrons can be considered as only one example in a much wider context of {\it structured states of quantum particles in free space}. Indeed, on the one hand, one can consider various structured modes, such as Hermite--Gaussian-like beams \cite{Schattschneider2012a,Guzzinati2016} and Airy beams \cite{VolochBloch2013, Karlovets:2014,Guzzinati2015,Karlovets:2016jrd}. On the other hand, similar vortex or non-vortex states can be explored for other quantum particles, including neutrons (already demonstrated experimentally) \cite{Clark_2015}, atoms \cite{Hayrapetyan2013,Lembessis2014} (demonstrated in confined BEC systems \cite{Andersen2006}), ions \cite{Idrobo2011}, and even macroscopic objects such as fullerene molecules \cite{Arndt1999,Brezger2002}. Prospects on why these new states of quantum matter waves would be useful and how this could be implemented experimentally are currently actively discussed.

In view of the substantial body of work that was already presented here, one could wonder if there remains much more to explore with respect to electron vortices in TEMs. It is our firm belief that these initial experiments only scratch the surface. For example, it became clear that a vortex detector that would measure and sort the electron OAM modes would be highly beneficial for electron energy loss spectroscopy (EELS) and magnetic chiral dichroism experiments, but the methods we presented are still a long way from such a versatile instrument. Also, in terms of signal-to-noise ratio and the source-size broadening effect, important steps need to be taken in order to bring atomic resolution mapping of magnetic states closer to reality. New technology breakthroughs in direct electron detectors and electron gun design are, however, slowly providing this progress. There is also a clear potential for using elastic scattering effects in obtaining magnetic and chirality information. Even though the theory seems well established, there are still many unexplored areas with potential for applications and further research.

In the domain of high energies, we are at the beginning of a long journey. The theoretical formalism for describing high-energy collisions with vortex electrons has been developed and applied to a few basic QED scattering processes. Calculations demonstrate that vortex electrons will give access to quantities which are difficult or impossible to measure in the usual collision settings. One can now apply this formalism to various inelastic processes such as {\it bremsstrahlung} by vortex electrons, production of hadrons by vortex photons colliding with protons, and eventually the deep inelastic scattering of ultrarelativistic vortex electrons on hadrons and nuclei, with the aim to access, in a radically different way, the nucleon dynamics inside nuclei and the spin and orbital angular momentum contributions to the proton's spin. One can also investigate what new venues in hadronic and nuclear physics will open up if protons, nuclei, and other particles can also be experimentally prepared in vortex states.

The experimental facilities in high-energy physics face big challenges to deal with vortex electrons: one needs to create and manipulate ultra-relativistic vortex electrons, transfer the vortex-electron know-how from electron microscopes to collider-like setting, and achieve an even stronger focusing. Therefore, dedicated experimental efforts are needed, and they will become worth investing when theorists prove that there will be a clear scientific payoff.

It should be noticed that bringing the {\it spin degrees of freedom} into play would considerably enrich physical phenomena involving vortex electron states. This is already well explored in optics \cite{Bliokh2015NP}, and spin-polarized electron sources are used in high-energy domain \cite{Hernandez-Garcia2008}. At the same time, electron microscopy only starts developing this direction \cite{Kuwahara2011,Yamamoto2011,Karimi2012}.

In the meanwhile, optical vortex beams and OAM states of photons, out of which vortex electrons were born, still form a very active research field. In general, higher energies require more expensive scientific instruments. Therefore, optics has an important advantage that optical technology is far more likely to appear in applications. In comparison, electron beam technologies are more likely to be limited to expensive scientific instruments, which typically take much longer to develop and affect the world around us.

Thus, electron vortex beams are still in an early stage of development, and many opportunities for future research are open. We hope that we provided a solid basis for the researchers venturing into this exciting direction, and that this review can help them to see the bigger picture and to avoid pitfalls that might occur along the way.

\section*{Acknowledgements}
\addcontentsline{toc}{section}{\protect\numberline{}Acknowledgements}

We acknowledge discussions with Mark R. Dennis and Andrei Afanasev.
This work was supported by the RIKEN Interdisciplinary Theoretical Science Research Group (iTHES) Project, the Multi-University Research Initiative (MURI) Center for Dynamic Magneto-Optics via the Air Force Office of Scientific Research (AFOSR) (Grant No. FA9550-14-1-0040), Grant-in-Aid for Scientific Research (A), Core Research for Evolutionary Science and Technology (CREST), the John Templeton Foundation, the Australian Research Council,
the Portuguese Fun\-da\-\c{c}\~{a}o para a Ci\^{e}ncia e a Tecnologia (FCT)
(contract IF/00989/2014/CP1214/CT0004 under the IF2014 Program), contracts UID/FIS/00777/2013 and CERN/FIS-NUC/0010/2015 (partially funded through POCTI, COMPETE, QREN, and the European Union),
Austrian Science Fund Grant No. I543-N20,  
the European Research Council under the 7th Framework Program (FP7) (ERC Starting Grant No. 278510 VORTEX), and FWO PhD Fellowship grants (Aspirant Fonds Wetenschappelijk Onderzoek-Vlaanderen).

\clearpage


\section*{Appendix A. Conventions and notations}
\addcontentsline{toc}{section}{\protect\numberline{}Appendix A. Conventions and notations}

\begin{table}[!h]
\centering
\begin{tabular}{l l}
\hline
\noalign{\vskip 2mm} 
    \textbf{Notation} & \textbf{Description} \\ 
\noalign{\vskip 2mm} 
\hline
\\
\multicolumn{2}{l}{\bf Abbreviations:}\\
\noalign{\vskip 2mm}
 
OAM & orbital angular momentum \\
SAM & spin angular momentum \\
SOI & spin-orbit interaction \\
LG  & Laguerre--Gaussian \\
TEM & transmission electron microscope \\
STEM & scanning transmission electron microscope \\
MIP & mean internal potential \\
EELS & electron energy loss spectroscopy \\
EMCD & energy loss magnetic chiral dichroism \\
FT & Fourier transform \\
EM & electromagnetic \\

& \\
\multicolumn{2}{l}{\bf Fundamental constants:} \\
\noalign{\vskip 2mm}

$\hbar$ & Planck's constant \\
$c$ & speed of light \\
$e = - |e|$ & electron's charge \\
$m_e$ & electron's mass \\
$\alpha$ & fine-structure constant \\

& \\
\multicolumn{2}{l}{\bf Units:} \\
\noalign{\vskip 2mm}

\multicolumn{2}{l}{Gaussian units are used throughout this review. In addition, the $\hbar = c = 1$ units are used in Section~\ref{sect:highenergy}.} \\

& \\
\multicolumn{2}{l}{\bf Conventions:} \\
\noalign{\vskip 2mm}

${\bf r}$, ${\bf p}$, ${\bf L}$, etc. & 3D vectors \\
${\bf r}_{\perp}$, ${\bf p}_{\perp}$, etc. & 2D vectors in the plane orthogonal to the main direction \\
$\hat{\bf p}$, $\hat{\bf L}$, etc. & quantum-mechanical operators of the corresponding quantities \\
$\langle {\bf r} \rangle$, $\langle {\bf p} \rangle$, $\langle {\bf L} \rangle$, etc. & expectation (mean) values of the corresponding operators/quantities \\
& (normalized per one electron) \\
$\bar{\bf x}$, $\bar{\bf y}$, $\bar{\bf z}$, $\bar{\bm \varphi}$, etc. & unit vectors of the corresponding coordinates \\
$\tilde{\psi}({\bf k})$, $\tilde{V}({\bf k})$, or ${\mathcal F} (\psi)$, ${\mathcal F} (V)$ & Fourier transforms of the corresponding functions $\psi({\bf r})$, $V({\bf r})$, etc. \\
$\tilde{\psi}({\bf k}) * \tilde{V}({\bf k})$ & convolution of functions \\

$r^\mu$, $k^\mu$, etc. & four-vectors in Minkowski spacetime \\
$(k_\mu r^\mu)$ & scalar product of four-vectors \\

& \\
\multicolumn{2}{l}{\bf Special functions:} \\
\noalign{\vskip 2mm}

$J_\ell$ & Bessel functions of the first kind \\
$L_{n}^{\ell}$ & Laguerre--Gaussian polinomials \\
$\Theta$ & Heaviside step function \\
$\delta$ ($\delta_{ab}$) & Dirac delta function (Kronecker delta) \\

\\
\hline
\end{tabular}
\caption{Abbreviations, conventions, and general notations used in this review.}
\label{table_I}
\end{table}

\begin{table}
\centering
\begin{tabular}{l l}
\hline
\noalign{\vskip 2mm} 
    \textbf{Notation} & \textbf{Description} \\ 
\noalign{\vskip 2mm} 
\hline
\\

${\bf r}$ & radius-vector \\
$(r, \varphi, z)$ & cylindrical coordinates (note that $r=r_\perp \neq |{\bf r}|$) \\
${\bf p}$ & momentum (canonical) \\
${\bf k}$ & wave vector \\
$(k_\perp, \phi, k_z)$ & cylindrical coordinates in the wave-vector space \\
$(\theta, \phi, k)$ & spherical coordinates in the wave-vector space \\
$\kappa$ & fixed radial component of the wave vector in Bessel beams \\
$\theta_0$ & fixed polar angle in Bessel beams \\
$\hat{H}$ & Hamiltonian \\
$\psi$ & scalar wave function \\
$\Psi$ & multi-component wave function \\
$\Phi$ & phase of the wave function \\
$E$ & energy (either kinetic or full-relativistic, depending on the problem) \\
$\rho$ & probability density or intensity ($\rho_e$: electric charge density) \\
${\bf j}$ & probability current (${\bf j}_e$: electric current) \\
${\bf L}$ & (canonical) orbital angular momentum \\
${\bf S}$ & spin angular momentum (${\bf s}$: non-relativistic spin angular momentum) \\
${\bf J}$ & total angular momentum \\
${\bf M}$ & magnetic moment \\
${\mathbcal p}$ & kinetic momentum in the presence of a vector-potential \\
${\mathbcal L}$ & kinetic orbital angular momentum in the presence of a vector-potential \\

${\bf v}$ & velocity \\
$\Omega$ & angular velocity (angular frequency) of electron's circular motion \\
& ($\Omega_L$: Larmor frequency, $\Omega_c$: cyclotron frequency) \\

$\ell$ & vortex topological charge (azimuthal quantum number) \\
$\ell_0$ & order of the fork-like dislocation in holograms generating vortex beams \\
$w$ & beam width ($w_0$: beam waist, $w_m$: transverse magnetic length) \\
$z_L$ & longitudinal magnetic (Larmor) length \\




${\bf A}$ & magnetic vector-potential \\
$V$ & scalar electric potential (voltage) \\
${\bf B}$ & magnetic field strength \\
${\bf E}$ & electric field strength \\
$\sigma$ & sign (direction) of the magnetic field \\
$\alpha_m$ & dimensionless magnetic flux or magnetic-monopole charge \\
$g$ & $g$-factor of electron's angular momentum in a magnetic field \\



${\bf q} = {\bf k} - {\bf k}'$ & mometum-transfer parameter in the ${\bf k} \to {\bf k}'$ scattering \\
$l,m$ & angular-momentum and magnetic quantum numbers for atomic orbitals \\

$\gamma$ & Lorentz factor \\
$\Lambda$ & dimensionless spin-orbit interaction parameter for the Dirac electron \\
$\chi$ & helicity of Dirac electrons \\
$\omega$ and ${\mathbcal k}$ & frequency and wave vector of electromagnetic waves (photons) in Sections~\ref{sec:Volkov} and \ref{sec:radiation} \\

    \\
    \hline
\end{tabular}
\caption{The main physical quantities and their notations used in this review.}
\label{table_II}
\end{table}


\clearpage

\section*{References}

\bibliographystyle{elsarticle-num}
\bibliography{vortex_PhysicsReports_v8}



\end{document}